\documentclass[twocolumn]{aastex63}
\usepackage{subfigure}
\newcommand\CO{$^{12}$CO }
\newcommand\COl{$^{13}$CO }
\newcommand\COll{C$^{18}$O }
\hypersetup{linkcolor=blue, citecolor=blue, filecolor=green, urlcolor=magenta}
\accepted{February 18, 2021}
\submitjournal{ApJS}
\shorttitle{Molecular clouds in the second quadrant}
\shortauthors{Ma et al.}
\begin{document}
\title{Molecular Clouds in the Second Quadrant of the Milky Way Mid-plane from l$=$104$.\!\!^{\circ}$75 to l=119$.\!\!^{\circ}$75 and b=$-$5$.\!\!^{\circ}$25 to b=5$.\!\!^{\circ}$25}
\correspondingauthor{Yuehui Ma, Hongchi Wang}
\email{mayh@pmo.ac.cn, hcwang@pmo.ac.cn}
\author[0000-0002-8051-5228]{Yuehui Ma}
\affil{Purple Mountain Observatory and Key Laboratory of Radio Astronomy, Chinese Academy of Sciences, 10 Yuanhua Road, Nanjing 210033, China}
\affil{University of Chinese Academy of Sciences, 19A Yuquan Road, Shijingshan District, Beijing 100049, China}

\author[0000-0003-0746-7968]{Hongchi Wang}
\affil{Purple Mountain Observatory and Key Laboratory of Radio Astronomy, Chinese Academy of Sciences, 10 Yuanhua Road, Nanjing 210033, China}
\affil{School of Astronomy and Space Science, University of Science and Technology of China, Hefei, Anhui 230026, China}

\author[0000-0003-2218-3437]{Chong Li}
\affil{Purple Mountain Observatory and Key Laboratory of Radio Astronomy, Chinese Academy of Sciences, 10 Yuanhua Road, Nanjing 210033, China}
\affil{University of Chinese Academy of Sciences, 19A Yuquan Road, Shijingshan District, Beijing 100049, China}

\author[0000-0001-8926-2215]{Lianghao Lin}
\affil{Purple Mountain Observatory and Key Laboratory of Radio Astronomy, Chinese Academy of Sciences, 10 Yuanhua Road, Nanjing 210033, China}
\affil{School of Astronomy and Space Science, University of Science and Technology of China, Hefei, Anhui 230026, China}

\author[0000-0002-3904-1622]{Yan Sun}
\affil{Purple Mountain Observatory and Key Laboratory of Radio Astronomy, Chinese Academy of Sciences, 10 Yuanhua Road, Nanjing 210033, China}

\author[0000-0001-7768-7320]{Ji Yang}
\affil{Purple Mountain Observatory and Key Laboratory of Radio Astronomy, Chinese Academy of Sciences, 10 Yuanhua Road, Nanjing 210033, China}

\begin{abstract}
We have studied the properties of molecular clouds in the second quadrant of the Milky Way Mid-plane from l$=$104$.\!\!^{\circ}$75 to l$=$119$.\!\!^{\circ}$75 and b$=-$5$.\!\!^{\circ}$25 to b$=$5$.\!\!^{\circ}$25 using the $^{12}$CO, $^{13}$CO, and C$^{18}$O $J=1-0$ emission line data from the Milky Way Imaging Scroll Painting project (MWISP). We have identified 857 and 300 clouds in the $^{12}$CO and $^{13}$CO spectral cubes, respectively, using the DENDROGRAM + SCIMES algorithms. The distances of the molecular clouds are estimated and the physical properties like masses, sizes, and surface densities of the clouds are tabulated. The molecular clouds in the Perseus arm are about 30$-$50 times more massive and 4$-$6 times larger than the clouds in the Local arm. This result, however, is likely biased by distance selection effects. The surface densities of the clouds are enhanced in the Perseus arm with an average value of $\sim$100 M$_{\odot}$ pc$^{-2}$. We selected the 40 most extended ($>$0.35 arcdeg$^2$) molecular clouds from the \CO catalog to build the H$_2$ column density probability distribution function (N-PDF). About 78\% of the N-PDFs of the selected molecular clouds are well fitted with log-normal functions with only small deviations at high-densities which correspond to star-forming regions with scales of $\sim$1-5 pc in the Local arm and $\sim$5-10 pc in the Perseus arm. About 18\% of the selected molecular clouds have power-law N-PDFs at high-densities. In these molecular clouds, the majority of the regions fitted with the power-law correspond to molecular clumps of sizes of $\sim$1 pc or filaments of widths of $\sim$1 pc. 
\end{abstract}


\section{Introduction} 
\label{sec1}

Molecular clouds are the birthplace of stars. They are the coldest and densest part of the turbulent and multi-phased interstellar medium (ISM). Since discovered by \cite{Wilson1970}, the CO molecule becomes the most widely used tracer of molecular clouds in the Milky Way and the galaxies. Because of the high abundance of the CO molecule in the ISM, the low-J transitions of \CO are usually optically thick, whereas those of \COl and \COll are relatively more optically thin and therefore can be effectively used to trace the regions of molecular clouds at higher densities. The knowledge of the distribution and physical properties of molecular clouds in the Milky Way mainly comes from large-scale surveys of the rotational transitional emission from CO and its two isotopologues \COl and \COll toward the Galactic plane \citep{Burton1975, Scoville1975, Gordon1976, Burton1978, Solomon1987, Dame2001, Heyer2001, Roman2010, Rice2016, Miville2017}. The masses of molecular clouds lie in the range from a few tens of solar masses to about 10$^6$ M$_{\odot}$, while the sizes vary from $\sim$10 to 150 pc. The typical velocity dispersion of molecular clouds revealed by the surveys is $\sim$1 km s$^{-1}$. The most massive molecular entities ($>$ 10$^4$ M$_{\odot}$) are called giant molecular clouds (GMCs), which have complex and hierarchical structures that can be further divided into clouds, clumps, and cores \citep{Blitz1999}. The cloud mass function (CMF) is found to follow a power-law distribution of exponents in the range from $\sim-$1.4 to $\sim-$2.2 \citep{Blitz1993, Kramer1998}. A power-law scaling relation exists between the cloud size and the velocity dispersion with an exponent of 0.38-0.5 \citep{Larson1981, Solomon1987}. There also exists a power-law correlation between the mass and size of molecular clouds, $M \propto R^{h_M}$, where $h_M$ depends on the inner density distribution of the molecular clouds \citep{Kauffmann2010a, Kauffmann2010b}. 

The hierarchical structure and the power-law behavior of the CMF and the $\sigma_v-r$ relation of molecular clouds are found to be the outcomes of large-scale turbulence in various numerical simulations ignoring self-gravitation \citep{Vazquez2001, Padoan2002, Federrath2009}. However, gravity is found to play an important role in the dynamics of the molecular clouds on small scales ($<$ 1 pc) \citep{Larson2003, Francesco2007}. Numerical simulations including gravity at the early evolutionary stage of molecular clouds also successfully recovered the mass and density distribution of molecular clouds \citep{Vazquez-Semadeni2007}. It is valuable to investigate the statistical properties and equilibrium states of molecular structures with surveys of molecular clouds that have high sensitivity, good spatial resolution, and wide sky coverage at the same time. This kind of surveys are capable of providing a sample of molecular clouds at different evolutionary stages in different environments and at different Galactocentric distances. It is also important that this kind of surveys allow us to investigate the physical properties and the scaling relations of molecular clouds in a high dynamic range of scales from the molecular cores to the GMCs.   

The ongoing Milky Way Imaging Scroll Painting (MWISP) project \citep{Su2019} provides us a good opportunity to study the statistical properties of molecular clouds. For this investigation, first we need to decompose the molecular line emission into individual clouds. For this pourpose we used the DENDROGRAM \footnote{\url{https://dendrograms.readthedocs.io}} \citep{Rosolowsky2008} and 
SCIMES\footnote{\url{https://scimes.readthedocs.io}} \citep{Colombo2015} algorithms. \cite{Colombo2015} have made detailed comparison of the performance between the DENDROGRAM+SCIMES algorithm and other algorithms such as CPROPS \citep{Rosolowsky2006} and ClumpFind \citep{Williams1994} frequently used for the automatic identification of extended structures in spectrometric data. They found that CPROPS and ClumpFind tend to overdivide the emission of molecular gas. Other traditional algorithms, like GAUSSCLUMPS \citep{Stutzki1990} and FELLWALKER \citep{Berry2015}, are more suitable for the identification of clumps or cores. 

The probability distribution function of H$_2$ column density (N-PDF) of molecular clouds is a useful statistical tool to investigate the underlying physics that influence the structure of the molecular clouds. A log-normal behavior of the N-PDF has been predicted in theoretical studies \citep{Vazquez-Semadeni1994, Padoan1997, Klessen2000} and is attributed to the supersonic turbulence in molecular clouds. Some following observations have confirmed the existence of the log-normal N-PDFs in relatively quiescent molecular clouds \citep{Goodman2009, Kainulainen2014}. Log-normal plus high-density power law tail or pure power-law distributions of the N-PDFs, have been observed in active star-forming regions with dust extinction or emission data \citep{Kainulainen2009, Schneider2013, Tremblin2014, Benedettini2015, Lombardi2015, Stutz2015}. However, \cite{Alves2017} propose that the completeness, i.e., the last closed contour, has a significant influence on the shape of N-PDFs and they conclude that, above the completeness limit, there is no observational evidence for log-normal N-PDFs in molecular clouds. The results of \cite{Tassis2010} also challenge the one-to-one correspondence between the shapes of the N-PDFs and the underlying physics in molecular clouds, as log-normal N-PDFs are observed in their modeled clouds without supersonic turbulence.

In this work, we use the DENDROGRAM+SCIMES algorithms to extract molecular clouds in the second quadrant of the Milky Way mid-plane from l$=$104$.\!\!^{\circ}$75 to l=119$.\!\!^{\circ}$75 and b=$-$5$.\!\!^{\circ}$25 to b=5$.\!\!^{\circ}$25 using the \CO and \COl datasets from the MWISP project and study their physical properties. The observations are introduced in Section \ref{sec2} and the results are presented in Section \ref{sec3}. We discuss the properties of the N-PDFs of a selected sub-sample of molecular clouds in Section \ref{sec4} and make a summary in Section \ref{sec5}.

\section{Observations}\label{sec2}
We have observed the \CO, \COl, and \COll $J=1-0$ line emission toward the Galactic plane with a sky coverage of 15\arcdeg$\times$10\arcdeg from l$=$104$.\!\!^{\circ}$75 to l=119$.\!\!^{\circ}$75 and b=$-$5$.\!\!^{\circ}$25 to b=5$.\!\!^{\circ}$25. The observation is part of the MWISP project, which is an unbiased simultaneous survey of the $J=1-0$ transitional emission of the three isotopologues of carbon monoxide toward the Galactic plane visible from the northern hemisphere. The detailed observational information and the data processing procedure of the MWISP project are introduced comprehensively in \cite{Su2019}. Here we only give a brief description of the observations and the dataset used in this work. The observations were taken  from 2012 March to 2018 October using the Purple Mountain Observatory (PMO) 13.7 m millimeter-wavelength telescope located at Delingha, China. The half-power beam width (HPBW) of the telescope is around 52$\arcsec$ and 50$\arcsec$ at 110 GHz and 115 GHz, respectively, and the pointing of the telescope has an accuracy of about 5$\arcsec$ during all the observational epochs. The telescope is equipped with a nine-beam Superconducting Spectroscopic Array Receiver (SSAR) \citep{Shan2012}. A two-sideband Superconductor-Insulator-Superconductor (SIS) mixer works as the front end of the receiver. The \CO, \COl, and \COll $J=1-0$ line emission data are obtained simultaneously with the \CO $J=1-0$ line emission being covered by the upper sideband while \COl $J=1-0$ and \COll $J=1-0$ lines by the lower sideband. The backend of the receiver is a Fast Fourier Transform Spectrometer (FFTS) with a total bandwidth of 1 GHz and 16,384 frequency channels, providing a spectral resolution of 61 kHz per channel which corresponds to a velocity resolution of 0.17 km s$^{-1}$ at 110 GHz. The antenna temperature is calibrated according to $\rm{T_{mb}=T^{*}_{A}/\eta_{mb}}$ during observation, where the $\eta_{mb}$ is the main beam efficiency and its value can be found in the annual status report \footnote{\url{http://english.dlh.pmo.cas.cn/fs/}} of the PMO-13.7 m millimeter telescope.

In the MWISP project, the survey area is divided into individual cells of size of 30$\arcmin\times$30$\arcmin$. For each cell, the observations were made in the position-switch on-the-fly (OTF) mode along the directions of Galactic longitude and Galactic latitude. The scanning rate is 50$\arcsec$ per second, and the dump time is 0.3 s. The data products of the MWISP survey are automatically pre-processed after observation. The pre-processing includes the rejection of prominent bad channels, subtraction of a first-order baseline from every spectrum, and the combination of the spectra obtained at different times for the same sky position. The final data are regrided into 30$\arcsec\times$30$\arcsec$ pixels in the directions of Galactic longitude and latitude. The required standard for the median RMS noise level in the MWISP project is below $\sim$0.5 K per channel at the \CO $J=1-0$ line wavelength and below $\sim$0.3 K at the \COl $J=1-0$ and \COll $J=1-0$ line wavelengths.

\section{Results}\label{sec3}
\subsection{Overall Distribution of Molecular Gas}\label{sec3.1}
\begin{figure*}[htb!]
	\centering
	\includegraphics[trim=0cm 0cm 0cm 1cm, width=0.6\linewidth, clip]{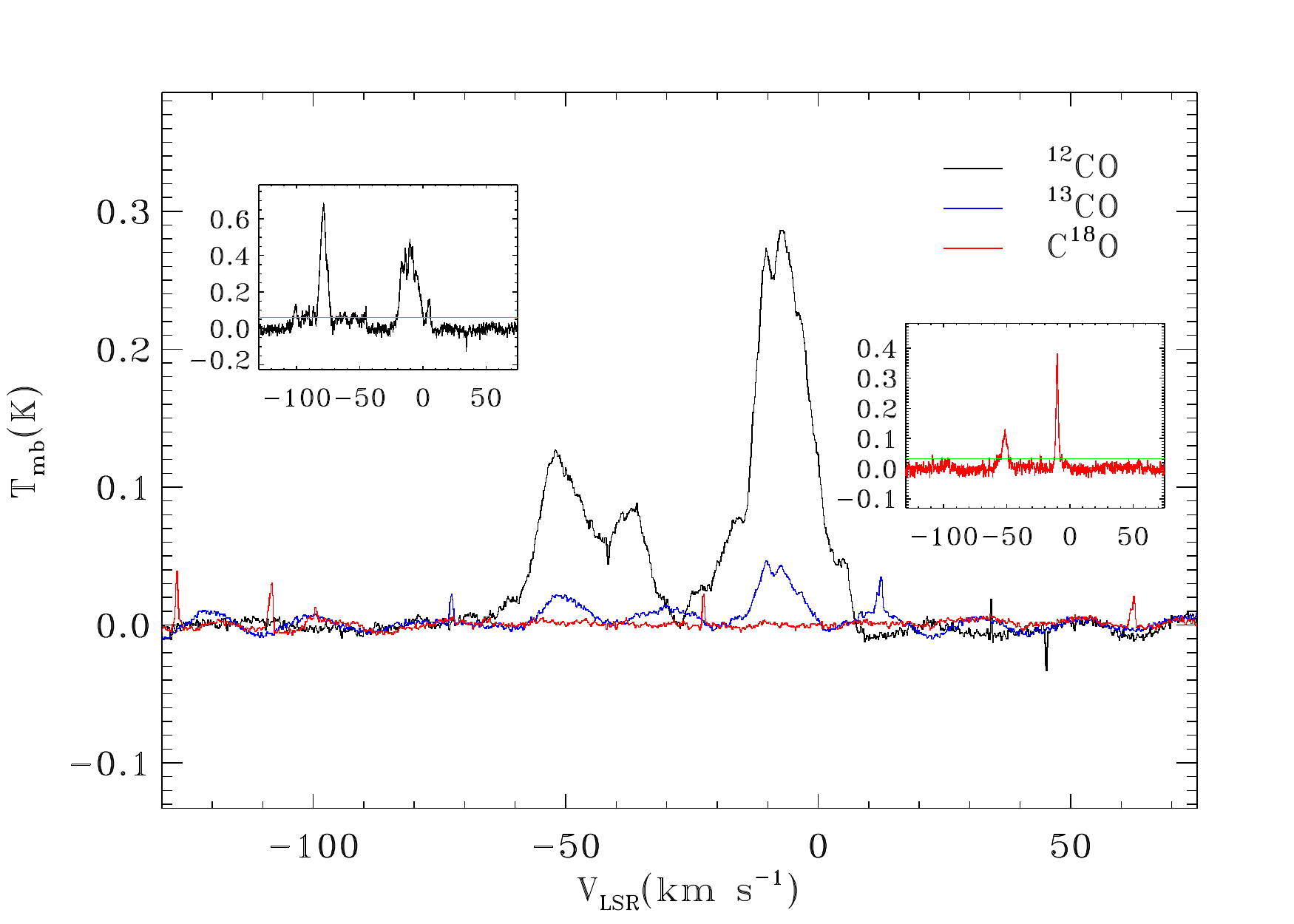}
	\caption{Average spectra of the \CO, \COl, and \COll $J = 1-0$ line emission in the surveyed region. The black, blue, and red lines correspond to the \CO, \COl, \COll spectra, respectively.  The spikes at $\sim34$ and $\sim$45 km s$^{-1}$ in the \CO spectrum and at $\sim-75$ and $\sim$10 km s$^{-1}$ in the \COl spectrum are due to bad channels, and so are the spikes at $\sim-127$, $\sim-110$, $\sim-23$, and $\sim60$ km s$^{-1}$ in the \COll spectrum. The left zoomed-in panel is the spectrum of the \CO spectra averaged over the spatial pixels that show emission in the velocity range from $-$115 to $-75$ km s$^{-1}$ in at least five contiguous channels with intensities above 2$\sigma$. The right zoomed-in panel is the average spectrum of \COll line emission of all positions that have at least three contiguous channels with intensities above 2$\sigma$. The green line represents the 3$\sigma$ noise level of the average spectrum.}
	\label{fig1}
\end{figure*}

The average spectra of the \CO, \COl, and the \COll $J=1-0$ line emission over the total surveyed area are presented in Figure \ref{fig1}. The molecular gas traced by \CO $J=1-0$ line has three main velocity components that are centered at $\sim-52$, $\sim-35$, and $\sim-$10 km s$^{-1}$, respectively, while there are only two velocity components, $-52$ and $-10$ km s$^{-1}$, in \COl $J=1-0$ line. There is also some \CO emission contained in the velocity range from $-$115 to $-75$ km s$^{-1}$, not visible in the total spectrum since it is confined in small regions. For clarity, we inserted in Figure \ref{fig1} the average spectrum of the \CO line at the positions where the \CO emission is detected in the velocity range from $-$115 to $-75$ km s$^{-1}$. The \CO, \COl, and \COll line emission is considered being detected at a position only when its spectrum show at least five, four, and three contiguous channels, respectively, with intensities above 2$\sigma_{\rm{RMS}}$. The \COll emission in the region is too weak to be identified in the average spectrum. However, it clearly emerges in the average spectrum of the positions where the \COll $J=1-0$ emission is above the detection criterion (see inset in Figure \ref{fig1}). The spikes at $\sim34$ and $\sim$45 km s$^{-1}$ in the \CO spectrum and at $\sim-75$ and $\sim$10 km s$^{-1}$ in the \COl spectrum are residual bad channels not properly removed from the automatic reduction pipeline, and so are the spikes at $\sim-127$, $\sim-110$, $\sim-23$, and $\sim60$ km s$^{-1}$ in the \COll spectrum. The bump in the velocity range from $-40$ to $-17$ km s$^{-1}$ in the \COl spectrum is caused by wavelike baselines in the spectra.

\begin{figure*}[htb!]
	\centering
	\includegraphics[trim=0.5cm 0.2cm 0.5cm 1cm, width=0.8\linewidth, clip]{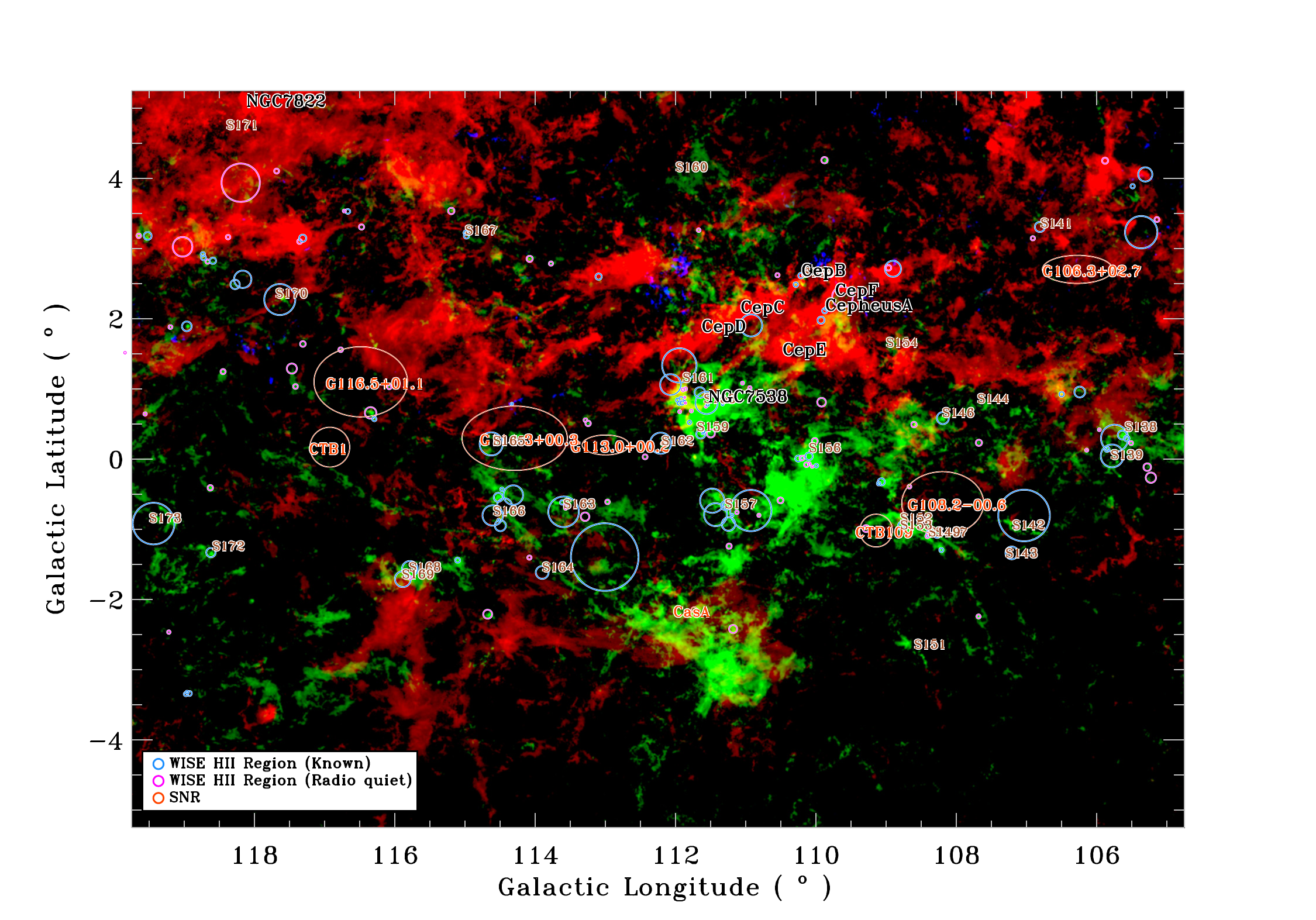}
	\caption{Color-coded intensity maps of the \CO emission in different velocity ranges. Red, green, and blue colors correspond to the \CO integrated intensities in the velocity ranges from $-$27 to 20, $-$75 to $-$27, and $-$115 to $-75$ km s$^{-1}$, respectively. The integrated intensity thresholds are 1.5 times $\sqrt{N}\sigma\delta v$, where N is the number of velocity channels in the integrated velocity range, $\sigma$ is the RMS noise per velocity channel, and $\delta v$ is the width of the velocity channel. The orange ellipses show the extents of the supernova remnants \citep{Green2014,Green2017} in the region. The \ion{H}{2} regions and the \ion{H}{2} region candidates in the WISE catalog \citep{Anderson2014} are indicated with blue and magenta circles, respectively. The corresponding names of the \ion{H}{2} regions in the Sharpless' catalog \citep{Sharpless1959} are shown with yellow letters, while the names of the supernova remnants are shown in red. Some active star-forming regions, like NGC 7822, are indicated with black letters.}
	\label{fig2}
\end{figure*}

\begin{figure*}[htb!]
	\centering
	\subfigure[]{
		\label{fig3a}
		\includegraphics[trim=4.5cm 2.5cm 4cm 4.5cm, width= 0.9\linewidth, clip]{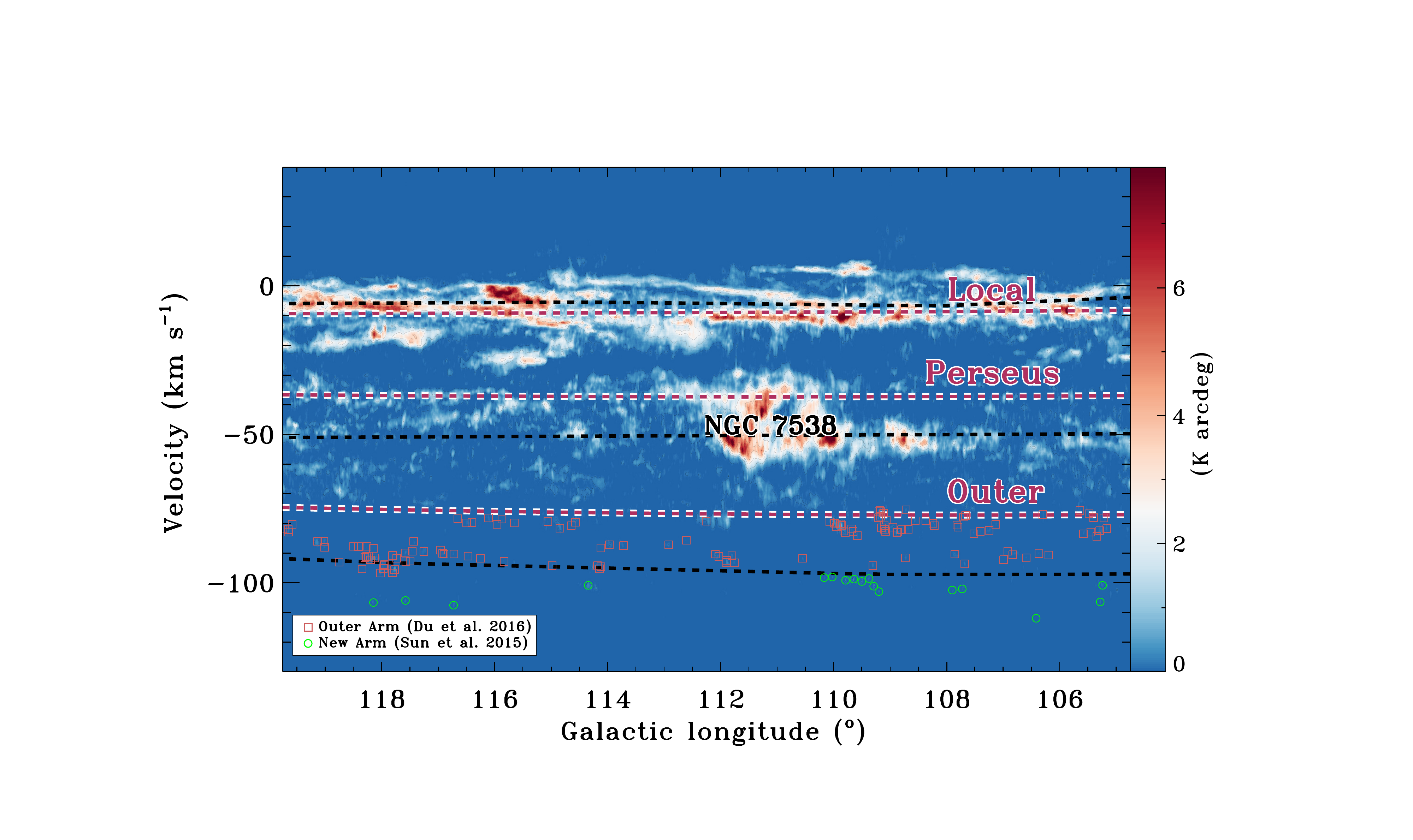}}
	\subfigure[]{
		\label{fig3b}
		\includegraphics[trim=2.5cm 3.5cm 2.5cm 5.5cm, width=0.6\linewidth, clip]{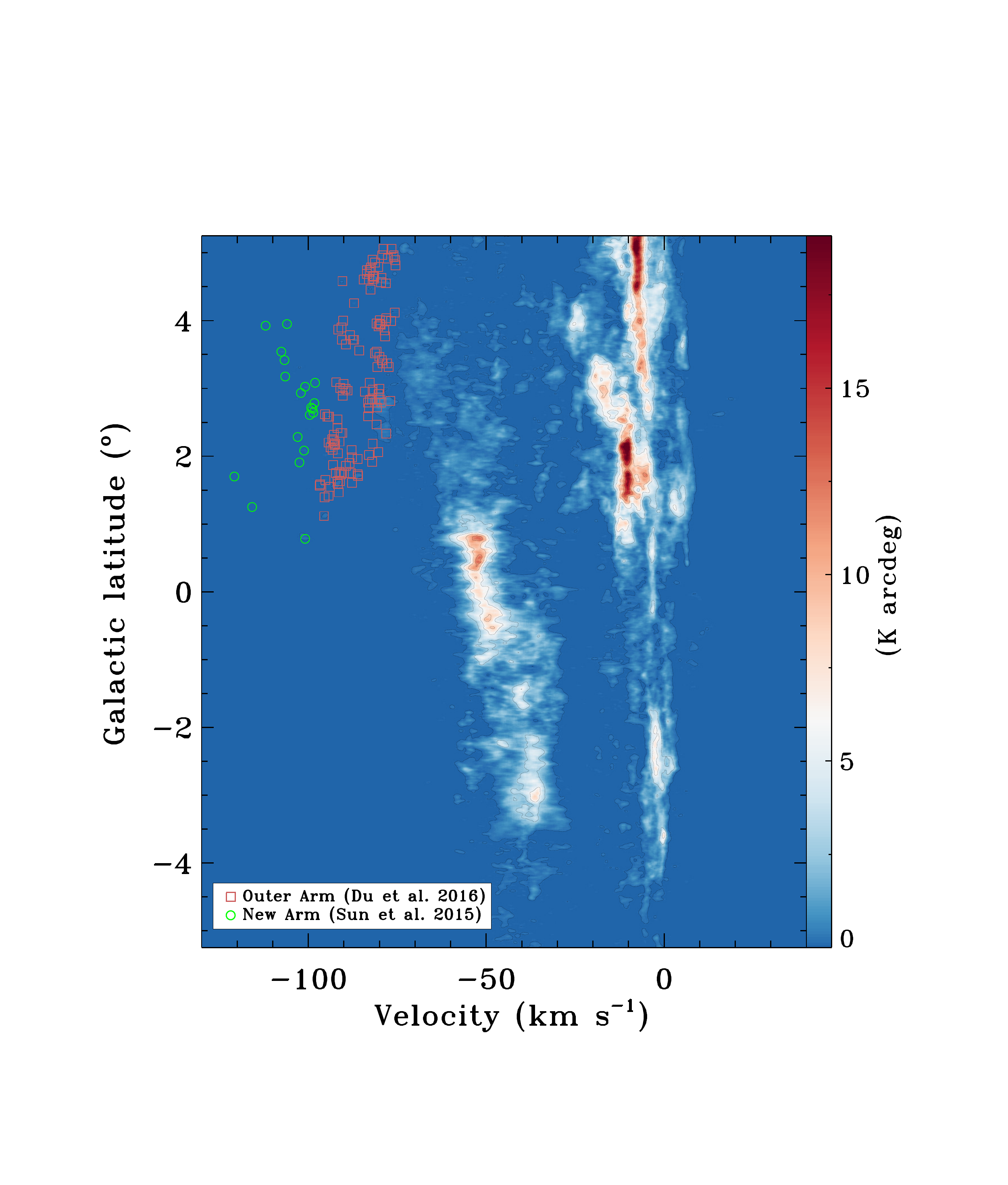}}
	\caption{(a) L$-$v diagram of \CO emission. The red dashed lines are the l$-$v curves of the Local, Perseus, and Outer spiral arms derived from model A5 from \cite{Reid2014}. The black dashed lines are the location of the spiral arms derived from CO and HI observations \citep{Weaver1970, Cohen1980}, which are used as the arm traces in \cite{Reid2016}. (b) B$-$v diagram of \CO emission. The contours in the two panels start at 1.5$\sigma$ and then increase to 0.7 times the emission peak in seven steps with the same intervals. The red squares and the green circles in the panels are the positions of the molecular clouds identified by \citep{Du2016} and \citep{Sun2015}, representing the Outer arm and the New (OSC) arm, respectively.}\label{fig3}
\end{figure*}

The spatial distribution of the molecular gas in different velocity ranges in the region is shown in Figure \ref{fig2}. The integrated intensity of \CO emission in velocity ranges from $-27$ to 20, $-75$ to $-27$, and $-115$ to $-75$ km s$^{-1}$ are indicated with red, green, and blue colors, respectively. In the observed portion of the outer Galaxy, more negative local standard of rest (LSR) velocity corresponds to farther distance. The nearby molecular gas shown in red color spreads over all the observed Galactic latitude range. The most distant gas, shown as the blue color, only distributes at positive Galactic latitudes which may be a result of the warp of the gas disk of our galaxy \citep{Westerhout1957, Wouterloot1990}. Several well known molecular clouds are located in this region, such as the nearby Cep GMC, shown in red in Figure \ref{fig2}, the GMC complex NGC 7538 and the Cas GMC, shown in green. The Cep GMC is a low-mass star-forming region, while the NGC 7538 complex is the birthplace of massive stars. The detailed analysis of the physical properties of Cas GMC was presented in our previous work \cite{Ma2019}. 

Figure \ref{fig3} shows the position velocity distribution of the \CO emission along the directions of the Galactic longitude and latitude. Within the span of Galactic longitude covered by this work, spiral arms of the Galaxy are identified as continuous curves in the l-v diagram. The red dashed lines in Figure \ref{fig3a} shows the projected l-v positions of the Local, Perseus, and the Outer arms that are derived according to the fitted log-periodic spirals of these arms in \cite{Reid2014}, which are based on the measured parallaxes of maser sources in high-mass star-forming regions and the Galactic rotation constants from their model A5. The black dashed lines are the l-v locations of the Local, Perseus, and Outer arms from the CfA CO observations \citep{Cohen1980}, which are presented in figures 8 and 9 in \cite{Reid2016} and are used as the spiral arm traces in the Bayesian distance estimator program in \cite{Reid2016}. Therefore, the previously known kinematic anomaly, which can be inferred from the large discrepancy between the kinematic distances and the luminosity/parallax distances of massive star-forming regions \citep{Xu2006}, of the Perseus spiral arm, is shown in Figure \ref{fig3a} as the separation between the red and black dashed ``Perseus" lines. There exists a velocity shear between the east and west parts of the Perseus arm, divided by the shell-like structure located at $l = 111\arcdeg, v = -45$ km s$^{-1}$, which corresponds to the NGC 7538 complex. The molecular gas to the east of the NGC 7538 complex is mainly concentrated at $-$37 km s $ ^{-1} $ while the gas to the west of the complex mainly at $-$52 km s $^{-1}$, which is the reason for the two peaks observed at these velocities in the \CO spectrum shown in Figure \ref{fig1}. The spiral arms in this region can be seen clearly in also the b-v diagram in Figure \ref{fig3b}. The velocity dispersion of the molecular gas in the Local arm is getting broader from south to north, which is caused by the fact that the majority of the gas in the Local arm is located at Galactic latitudes above b = 0\arcdeg. The velocity of the molecular gas in the Perseus arm is gradually blue-shifted along the south to north direction, as can be seen from Figure \ref{fig3b}. \cite{Sun2015} and \cite{Du2016} identified hundreds of molecular clouds in the ``New" arm, which is the extension of the Scutum-Centaurus arm in the outer Galaxy (OSC arm), and the Outer arm using the MWISP data. The locations of those clouds that fall in the region observed in this work are shown with green circles and red squares, respectively. Some of the identified clouds are indiscernible in our position-velocity diagrams, which is caused by the broad latitude range used in the integration. 

\subsection{Statistics of \CO, \COl, and \COll Emission in Different Spiral Arms}\label{sec3.2}
\begin{figure*}[htb!]
	\centering
	\includegraphics[trim=3cm 8cm 3cm 10cm, width=0.6\linewidth, clip]{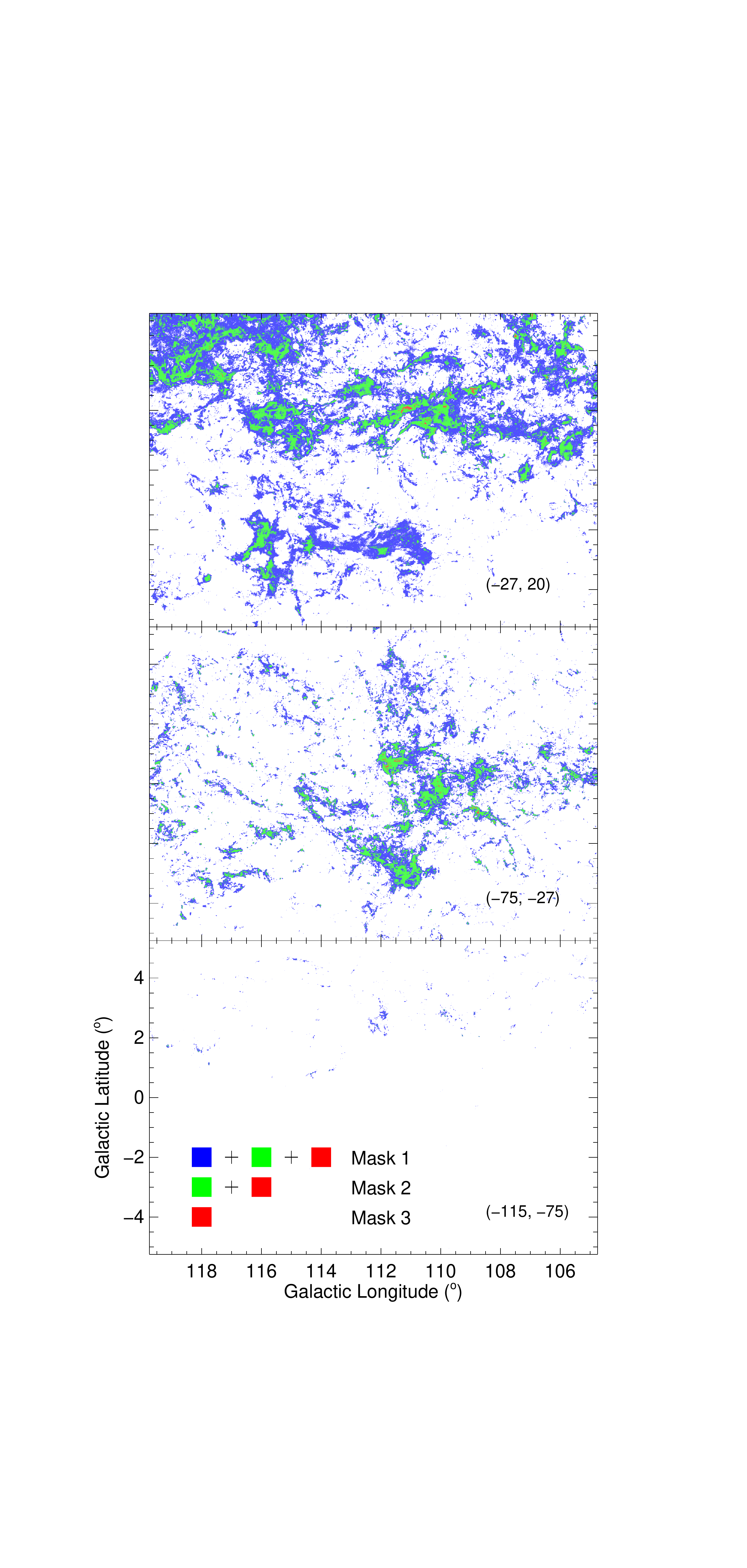}
	\caption{Spatial distributions of molecular gas in different Masks in the three velocity layers. Blue, green, and red colors correspond to the regions where \CO, \COl, and \COll emission is detected, respectively. Top, central and bottom panels correspond to the Local, Perseus, and Outer$+$OSC arms, respectively. The velocity ranges corresponding to each of the arms are given in the bottom right corner of each panel. The color compositions of the sky coverages of Masks 1-3 are shown in the lower-left corner of the bottom panel.} 
	\label{fig4}
\end{figure*}

\begin{figure*}[htb!]
	\centering
	\subfigure[]{
		\label{fig5a}
		\includegraphics[trim=0.8cm 0.5cm 1.2cm 1.3cm, width = 0.3\linewidth , clip]{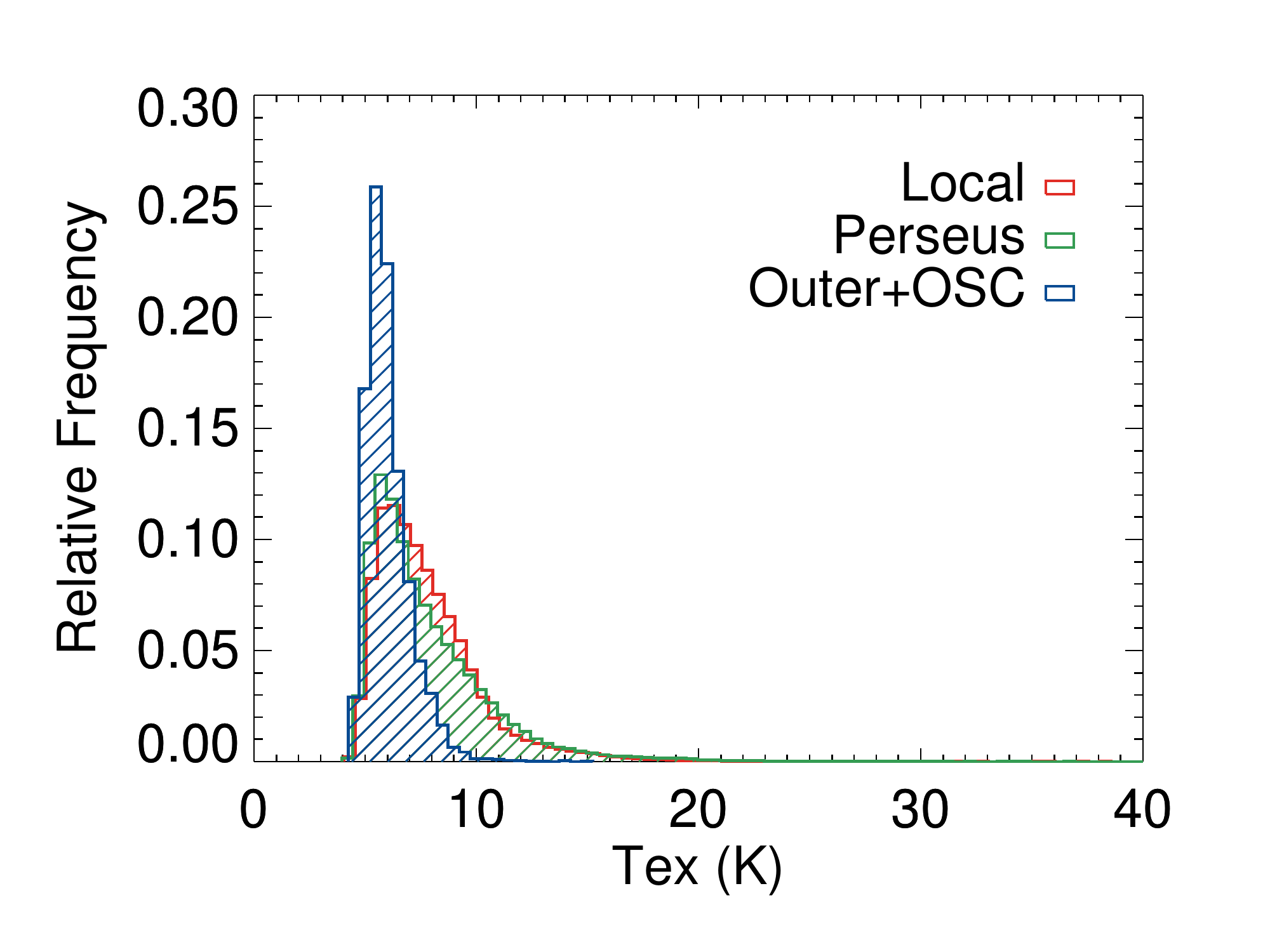}}
	\subfigure[]{
		\label{fig5b}
		\includegraphics[trim=0.8cm 0.5cm 1.2cm 1.3cm, width = 0.3\linewidth , clip]{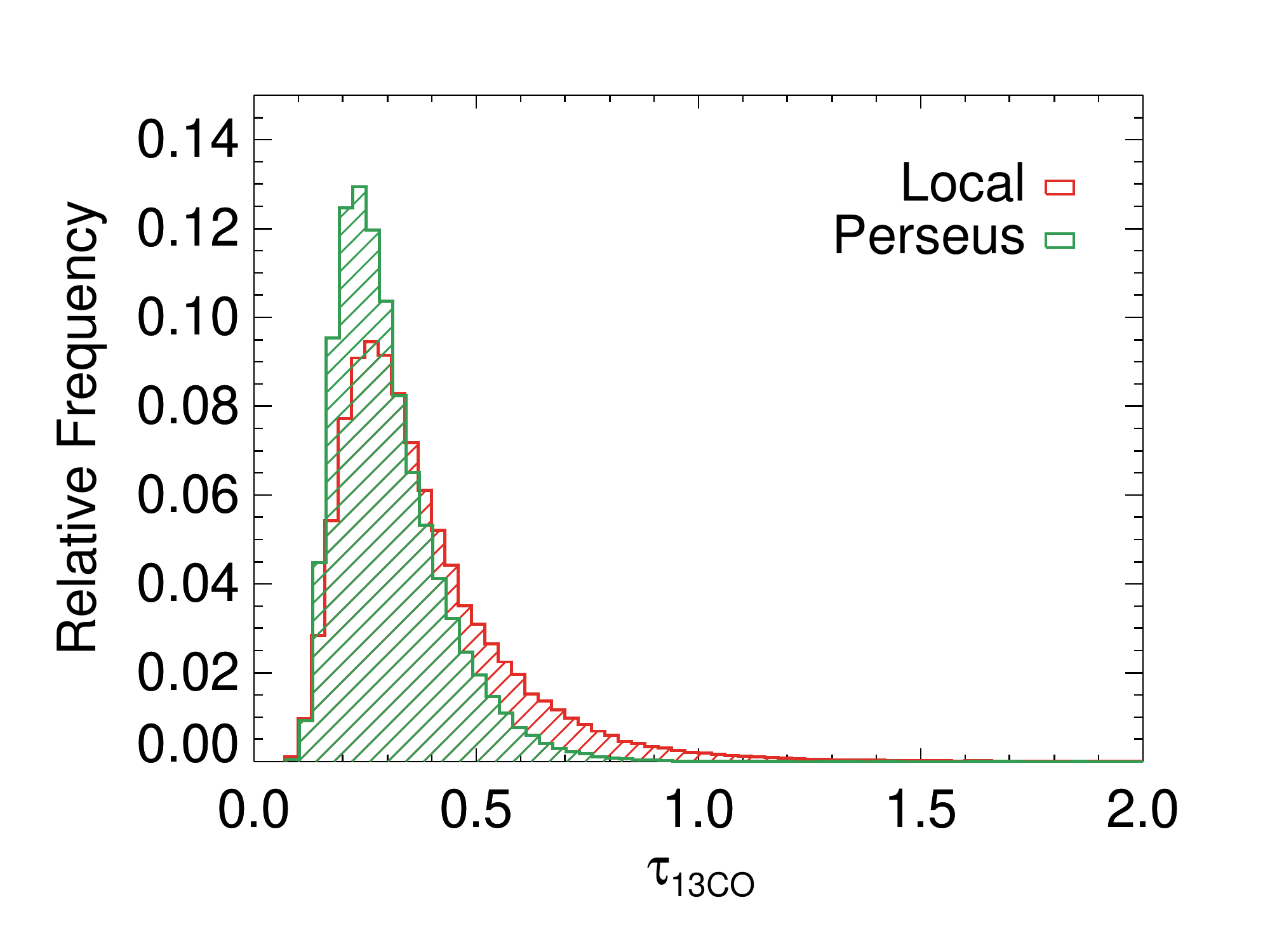}}
	\subfigure[]{
		\label{fig5c}
		\includegraphics[trim=0.8cm 0.5cm 1.2cm 1.3cm, width = 0.3\linewidth , clip]{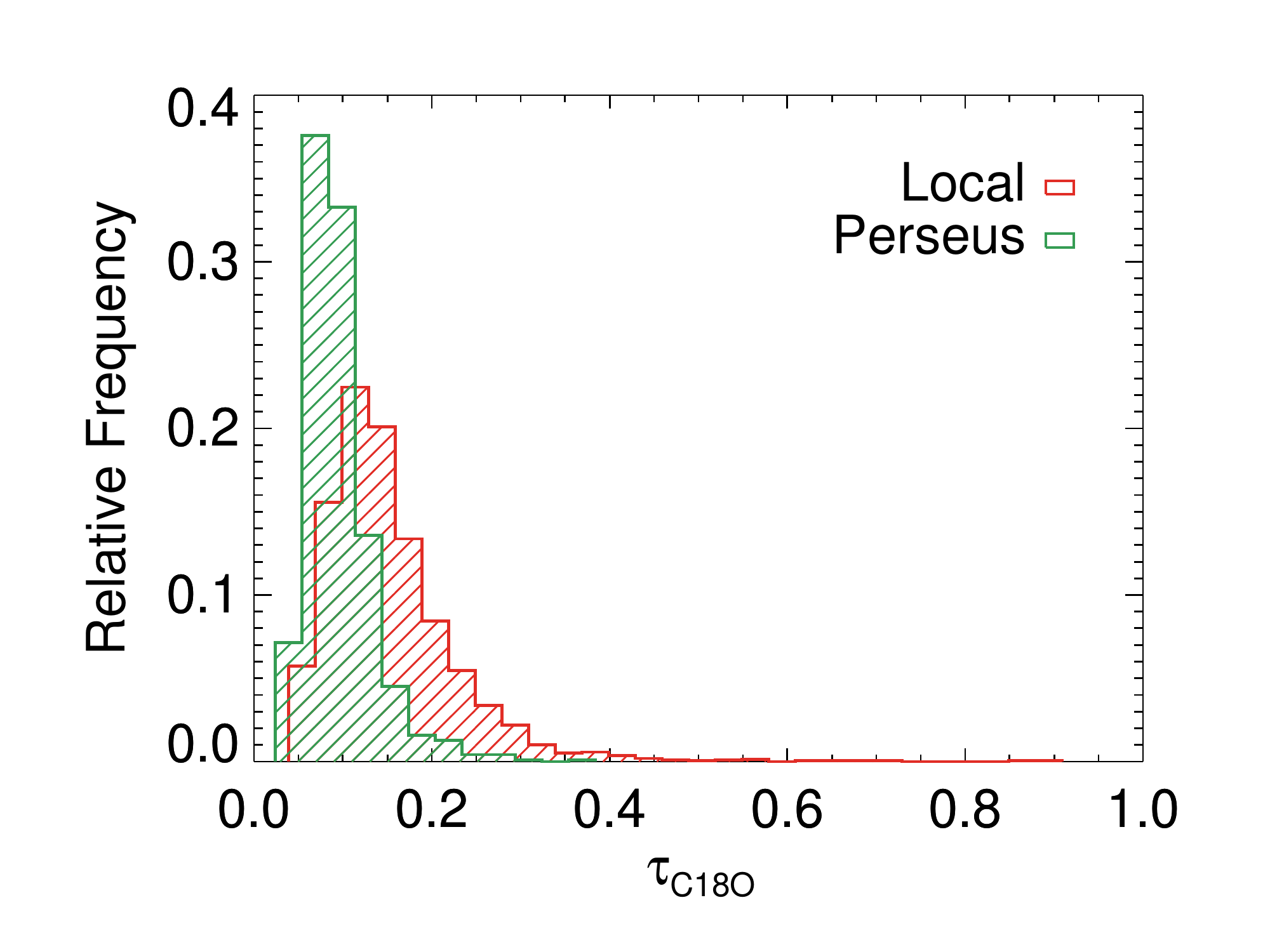}}
	\subfigure[]{
		\label{fig5d}
		\includegraphics[trim=0.8cm 0.5cm 1.2cm 1.3cm, width = 0.3\linewidth , clip]{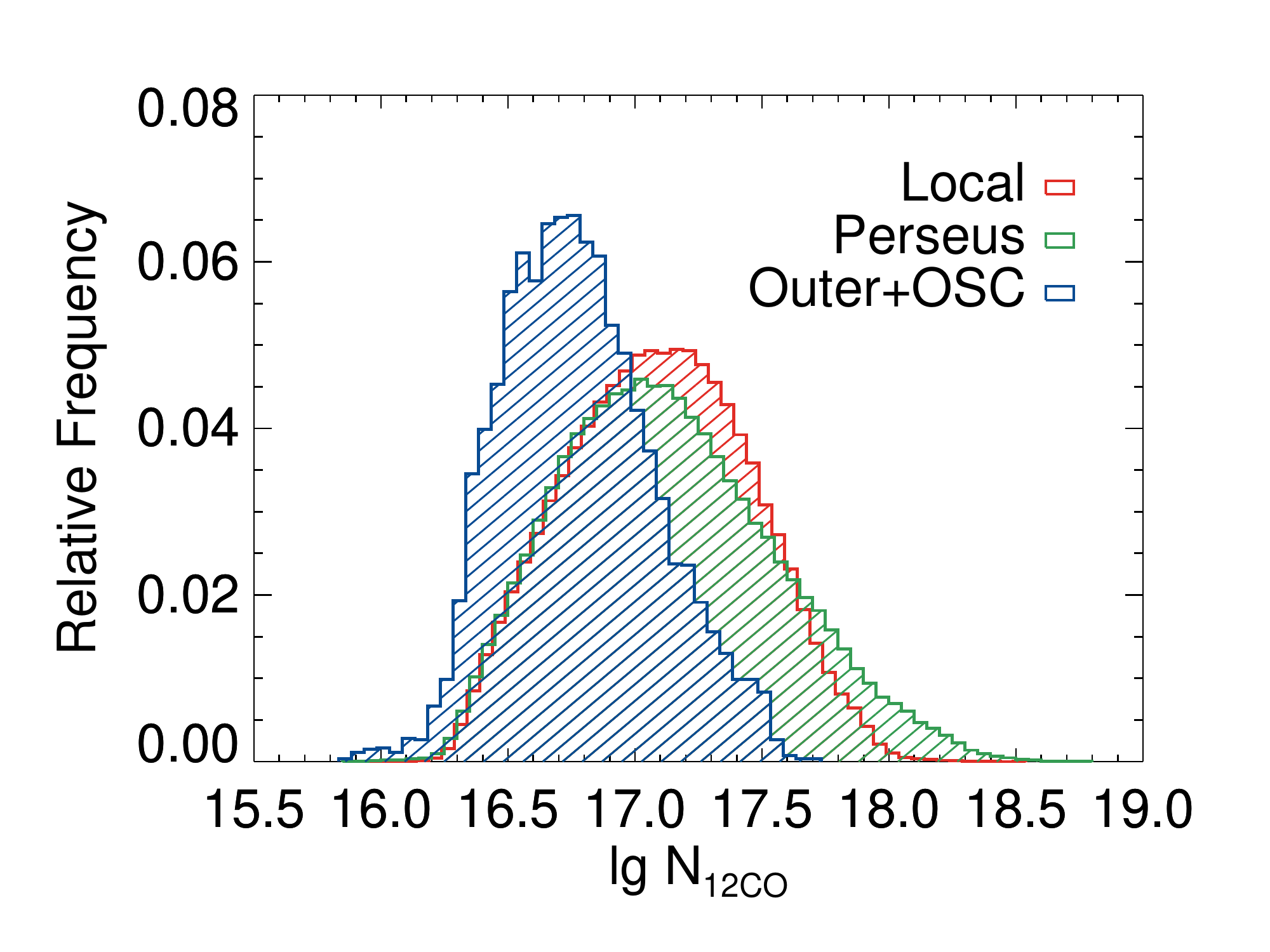}}
	\subfigure[]{
		\label{fig5e}
		\includegraphics[trim=0.8cm 0.5cm 1.2cm 1.3cm, width = 0.3\linewidth , clip]{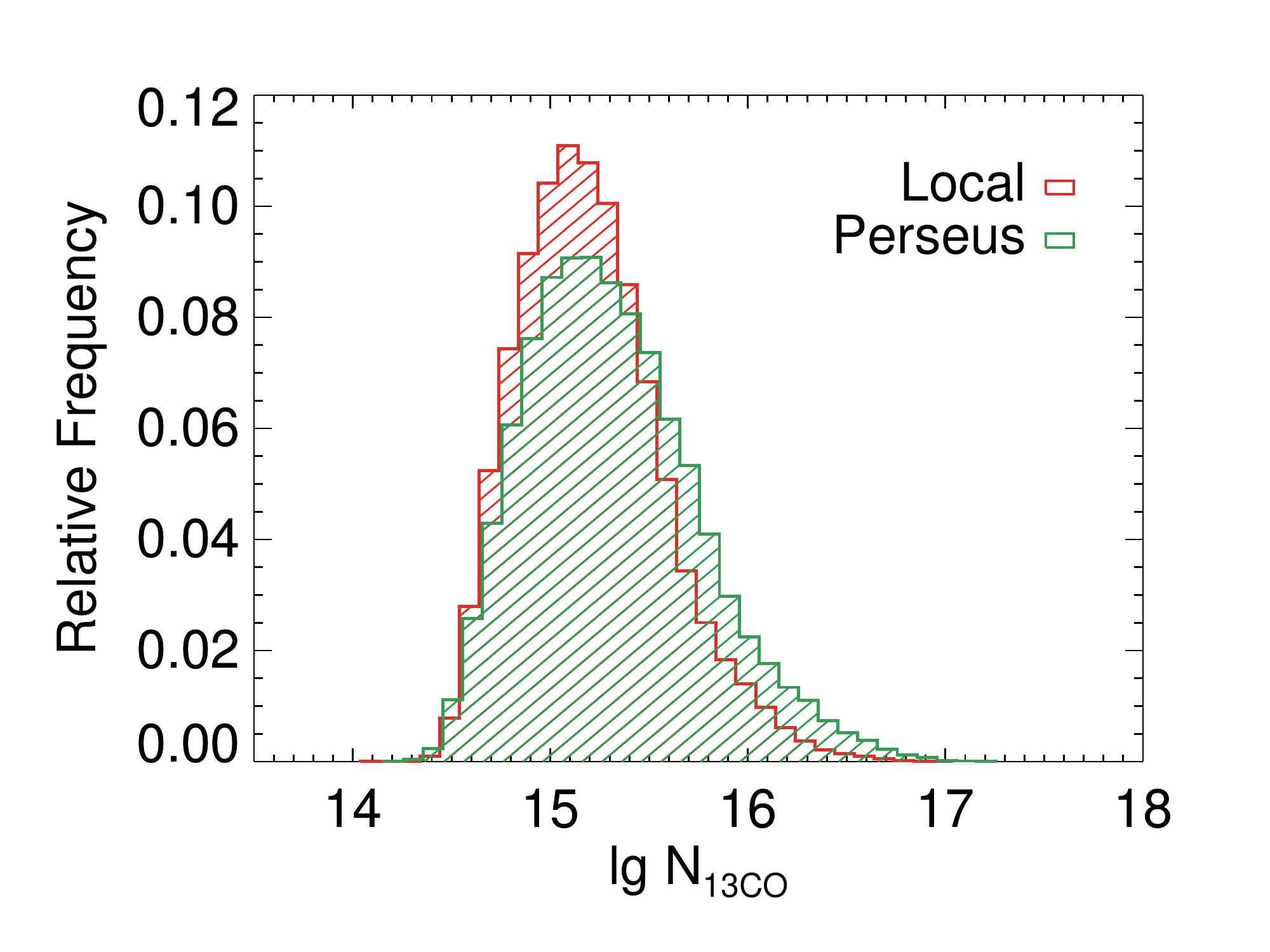}}
	\subfigure[]{
		\label{fig5f}
		\includegraphics[trim=0.8cm 0.5cm 1.3cm 1.3cm, width = 0.3\linewidth , clip]{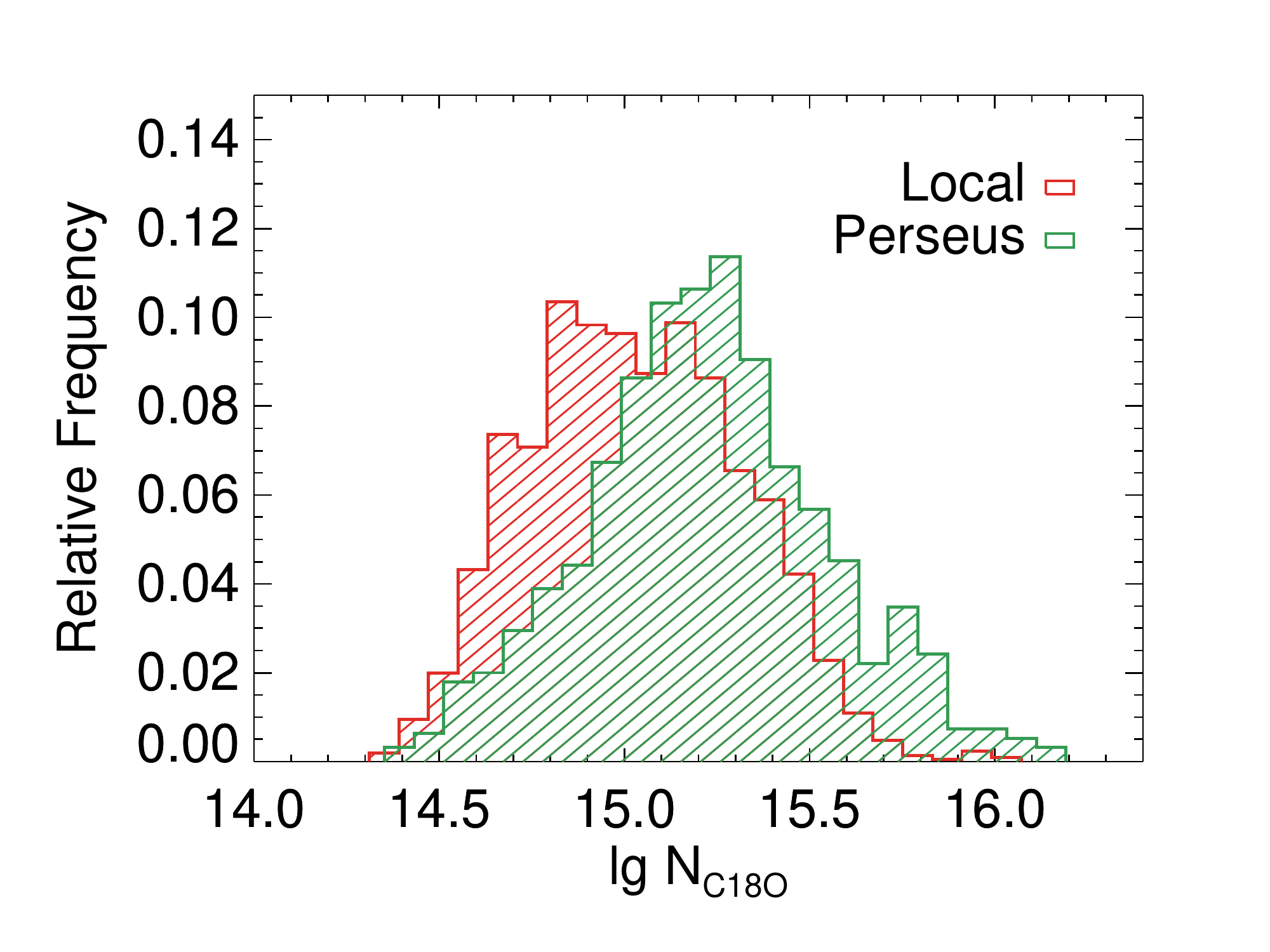}}
	\caption{Histograms of (a) excitation temperature in Mask 1 regions, (b) optical depth of \COl emission in Mask 2 regions, (c) optical depth of \COll emission in Mask 3 regions, (d) column density of \CO emission in Mask 1 regions, (e) column density of \COl emission in Mask 2 regions, and (f) column density of \COll emission in Mask 3 regions in different arms. The colors of the four arms are indicated in the upper-right corner of each panel.}\label{fig5}
\end{figure*}

According to the l-v distribution of the \CO emission and the spiral arm traces, the molecular gas in the observed region can be divided into three layers, the Local arm layer (from $-$27 to 20 km s$^{-1}$), the Perseus arm layer (from $-75$ to $-27$ km s$^{-1}$), and the Outer$+$OSC arm layer (from $-$115 to $-$75 km s$^{-1}$). Each layer can be further divided into three kinds of masks based on the detection of the \CO, \COl, and \COll emission. Mask 1 is defined to be the regions where \CO emission is detected. Mask 2 is the regions where both \CO and \COl emission are detected, while Mask 3 is the region where the emission from all three kinds of isotopologues is detected. According to this definition, the Mask 1 regions contain the Masks 2 and 3 regions, while the Mask 2 regions contain Mask 3 regions. The \CO, \COl, and \COll line emission is considered being detected at a position only when their spectra show at least five, four, and three contiguous channels, respectively, with intensities above 2$\sigma_{RMS}$. Figure \ref{fig4} presents the spatial distribution of Masks 1-3 in the three velocity layers. \CO and \COl emission is detected in all the spiral arms, while the \COll emission is only detected in very few pixels of the first two layers, corresponding to the densest part of active star-forming regions like the Cep GMC and NGC 7538 complexes. The total numbers and the percentages of the detection of \CO, \COl, and \COll emission are presented in Table \ref{tab1}. The \COl emission is detected among 22.2\% pixels with \CO detection in the Local arm, while this percentage is 24.1\% in the Perseus arm and only 3.8\% in the Outer$+$OSC arm. The \COll emission is detected among only about 0.31\% pixels with \CO detection in the Local and Perseus arms. We calculated the physical properties, such as the excitation temperature, optical depth of the \COl and \COll emission line, and the column density of the three isotopologues of the molecular gas for each pixel in Masks 1-3 in the Local and Persues arms. Since the pixel number with good detection within the masks 2 and 3 regions in the Outer+OSC arms are too small, their physical properties are not calculated. The column density of \CO molecule is directly converted from the H$_2$ column density using the abundance ratio [H$_2$/\CO] = 1.1$\times 10^4$ \citep{Frerking1982}, while the H$_2$ column density is obtained by multiplying the integrated intensity of the \CO emission by a conversion factor X$_{\rm CO} = 2.0\times10^{20}$ cm$^{-2}$ (K km$^{-1}$)$^{-1}$ \citep{Bolatto2013}. The excitation temperature of each pixel in each Mask and velocity layer is calculated using the peak brightness temperature of \CO in the corresponding velocity range of that Mask, using eq. 1 in \cite{Li2018}. The optical depth of \COl and \COll in Masks 2 and 3 are derived using the peak intensities of the \COl and \COll emission in the corresponding Masks and velocity ranges, using eqs. 2 and 3 of \cite{Li2018}, respectively. The column densities of \COl and \COll for the Local, Perseus, and Outer$+$OSC arms are calculated from the \COl and \COll integrated intensities in the corresponding velocity ranges using eqs. 5 and 6 in \cite{Li2018}. The statistics of the physical properties such as excitation temperature, optical depth, and column density of Masks 1-3 are summarized in Figure \ref{fig5} and Table \ref{tab1}. 

It can be seen from Table \ref{tab1} that the median excitation temperatures of Mask 1 in the four arms lie in the range from 6 to 8 K, while those of the  Mask 2 regions are relatively higher, lying in the range from 8 to 10 K, which are typical for molecular clouds where the gas is self-shielded from the interstellar radiation field. The range of the excitation temperatures in the Local and Perseus arms is quite large, from about 4 K to 47 K, indicating that large differences in the physical conditions and excitation of the gas are present within the molecular clouds at the spatial scales resolved by the MWISP survey. The densest regions, Mask 3, which correspond to the active star-forming regions like the Cep GMC, NGC 7822, L1188, and the NGC 7538 complex in the Local and Perseus arms, have much higher excitation temperatures, with the median values from 15 to 19 K. This is expected in regions where star formation is active since the new-born stars warm up their envelopes. Moreover, the radiation emitted by already formed stars, expecially if they are massive, can warm up their surrounding. The optical depth of \COl emission, $\tau_{\rm{13CO}}$, is less than 1 in most regions and has a median value in Mask 2 of about 0.3 for both the Local and Perseus arms. However, it reaches as high as 1.5 in some pixels, which may be small regions such as clumps or cores. The density in these regions can be much higher than that in the more diffuse part of the molecular clouds and consequently, the optical depth of the \COl lines increases so that the lines become optically thick. The \COll emission is optically thin both in the Local and the Perseus arms, but, on average, the pixels in the local arm have a \COll optical depth a factor of two higher than those in the Perseus arm. For the Local and Perseus arms, the distribution of N$_{\rm{12CO}}$ has a shape similar to that of N$_{\rm{13CO}}$. However, the median column densities of \CO molecules is reduced by a factor of $\sim$3 in the Outer$+$OSC arm compared with those in the Local and Perseus arms. This indicates that there is less material in the Outer$+$OSC arms of the Milky Way, the most external arms, than in the other two more internal arms. The column density of \COll molecule in the Perseus arm is greater than that in the Local arm by a factor of $\sim$1.5.

\begin{rotatetable*}
	\begin{deluxetable*}{lcrrlcccccc}
		\tablecaption{Statistical properties of molecular gas in the four arms of the Galaxy \label{tab1}}
		\tablecolumns{10}
		\tablewidth{0.6\linewidth}
		\tabletypesize{\tiny}
		\tablehead{
			\colhead{Arms} & \colhead{Mask} & \colhead{Pixel} & \colhead{Area} &  \colhead{$T_{ex}$} &
			 \colhead{$\tau_{^{13}CO}$} & \colhead{$\tau_{C^{18}O}$} &  \colhead{$lg(N_{^{12}CO)}$} & 
			\colhead{$lg(N_{^{13}CO})$} &  \colhead{$lg(N_{C^{18}O})$}   \\
			\colhead{Name} & \colhead{Name} & \colhead{Num (Per)} & \colhead{(deg$^{2}$)} &  
			\colhead{(K)} &  \colhead{} &  \colhead{} &
			 \colhead{$\rm{(cm^{-2})}$} &  \colhead{$\rm{(cm^{-2})}$} &  \colhead{$\rm{(cm^{-2})}$}\\
			 \colhead{(1)} & \colhead{(2)} & \colhead{(3)} & \colhead{(4)} &  \colhead{(5)} &
			 \colhead{(6)} & \colhead{(7)} &  \colhead{(8)} & 
			 \colhead{(9)} & \colhead{(10)}  
		}
		\startdata
		Local & Mask 1 &  658940 (29.01$\%$) & 45.760 & 4.05-44.71 (7.31) & ...& ... & 15.89-18.52 (17.10)& ...& ...\\
				& Mask 2 & 146517 (6.45$\%$) & 10.175 & 4.79-44.71 (9.83) & 0.07-7.18 (0.33)& ...&...& 14.04-17.03 (15.17)&...\\
				& Mask 3 & 2106 (0.09$\%$) & 0.146 & 5.65-43.57 (14.89) & ... & 0.04-0.88 (0.14)&...&...& 14.31-16.06 (15.02)\\
		\cline{1-10}
		Perseus & Mask 1 & 285143 (12.56$\%$) & 19.802 &  3.94-47.26 (7.08) &...&...& 15.85-18.76 (17.10)& ... & ... \\
				 & Mask 2 & 68766 (3.03$\%$) & 4.775  &  4.71-47.26 (10.58) & 0.07-2.19 (0.28) & ...&...&  14.16-17.25 (15.27)&...\\
				& Mask 3 & 950 (0.04$\%$) & 0.066 &  7.01-47.26 (18.72) &...&0.02-0.36 (0.09)& ...&...& 14.35-16.13 (15.22) \\
		\cline{1-10}
		Outer$+$OSC & Mask 1 &  5387 (0.23$\%$) & 0.374 &   4.23-14.74 (5.82)&...&...& 15.83-17.71 (16.76) &...&... \\
				& Mask 2 & 203 (0.00$\%$) & 0.014 & ... & ... & ... &...& ... & ...\\
				& Mask 3 & 1 (0.00$\%$)& 0. & ...&...&...&...&...&... \\
		\enddata
		\tablecomments{Col. 1: name of the arm. Col. 2: mask used, the definition of the masks is given in Sect. 3.2. Col. 3: number of pixels with good detection following the criteria explained in Sect. 3.1, in parenthesis the percentage of the good pixels with respect to the total observed pixels is given. Col. 4: total angular size of pixels with good detection. Col. 5: range and median value (in parenthesis) of the excitation temperature. Cols. 6-7: ranges and median values (in parenthesis) of the optical depths of \COl and \COll emission, respectively. Cols. 8-10: ranges and median values (in parenthesis) of the column density of \CO, \COl, and \COll, respectively.}
	\end{deluxetable*}
\end{rotatetable*}

\subsection{Catalogs of \CO and \COl Molecular Clouds}\label{sec3.3}
\subsubsection{Decomposition of the \CO and \COl Emission into Individual Clouds}\label{sec3.3.1}
With the \CO and \COl line emission data, we identify distinct \CO and \COl clouds in the surveyed region using the DENDROGRAM plus the SCIMES algorithms. 

In practice, the dendrogram algorithm is memory consuming when the dataset used is large. We smoothed the width of the velocity channels of \CO and \COl data into 0.5 km s$^{-1}$, resulting in new \CO and \COl datacubes with median noise levels of 0.31 and 0.17 K, respectively. Before the implementation of the DENDROGRAM algorithm, only those voxels within at least two consecutive velocity channels with intensities higher than 2$\sigma$ are selected and all other velocity channels are masked. The majority of the noise channels are removed under this criterion, but some contiguous bad channels were not removed. Nonetheless, if we set a higher noise threshold or broader velocity coverage, real signal could be removed. The DENDROGRAM algorithm is then implemented to the noise-masked and velocity-smoothed \CO and \COl datacubes. The minimum difference between two separate ``leaves" in a dendrogram tree is set to be 3$\sigma$, and the bottom threshold for detection is 2$\sigma$. The minimum voxel number of an individual ``leaf" is set to be 50, corresponding to 3.7 consecutive pixels along each of the l-b-v axes. The above settings corresponding to molecular cloud of size of 0.5$\times$0.5 and 1.6$\times$1.6 pc$^{2}$ for distances of 1.0 and 3.0 kpc, respectively, and velocity range of 1.8 km s $^{-1}$. We use ``volume" as the clustering criterion for the SCIMES algorithm. As a result, a mask cube that records the l-b-v positions and a catalog that contains the physical information of the output clusters are generated. The output ``clusters" are taken to be individual clouds in this work. Since some of the bad channels that produce spurious spikes in the spectra (see Figure. \ref{fig1}) still exist in the masked and smoothed data, they may cause false identification of molecular clouds. Therefore, after the cloud identification, we performed a manual check for each identified cloud to further ensure the identified molecular clouds are real structures and not caused by bad channels. We carried out molecular cloud identification on \CO and \COl data separately, and cross-matched the two resulting catalogs. If more than 85\% voxels of a \COl molecular cloud can be assigned to a single \CO cloud, then the \COl cloud is considered to match to the \CO cloud. As \COl usually traces the denser part of a molecular cloud, we find that one \CO cloud usually contains several \COl clouds.

\begin{figure*}[htb!]
	\centering
	\subfigure[]{
		\label{fig:Fig6a}
		\includegraphics[trim=0cm 0cm 2cm 1cm, width = 0.7\linewidth, clip]{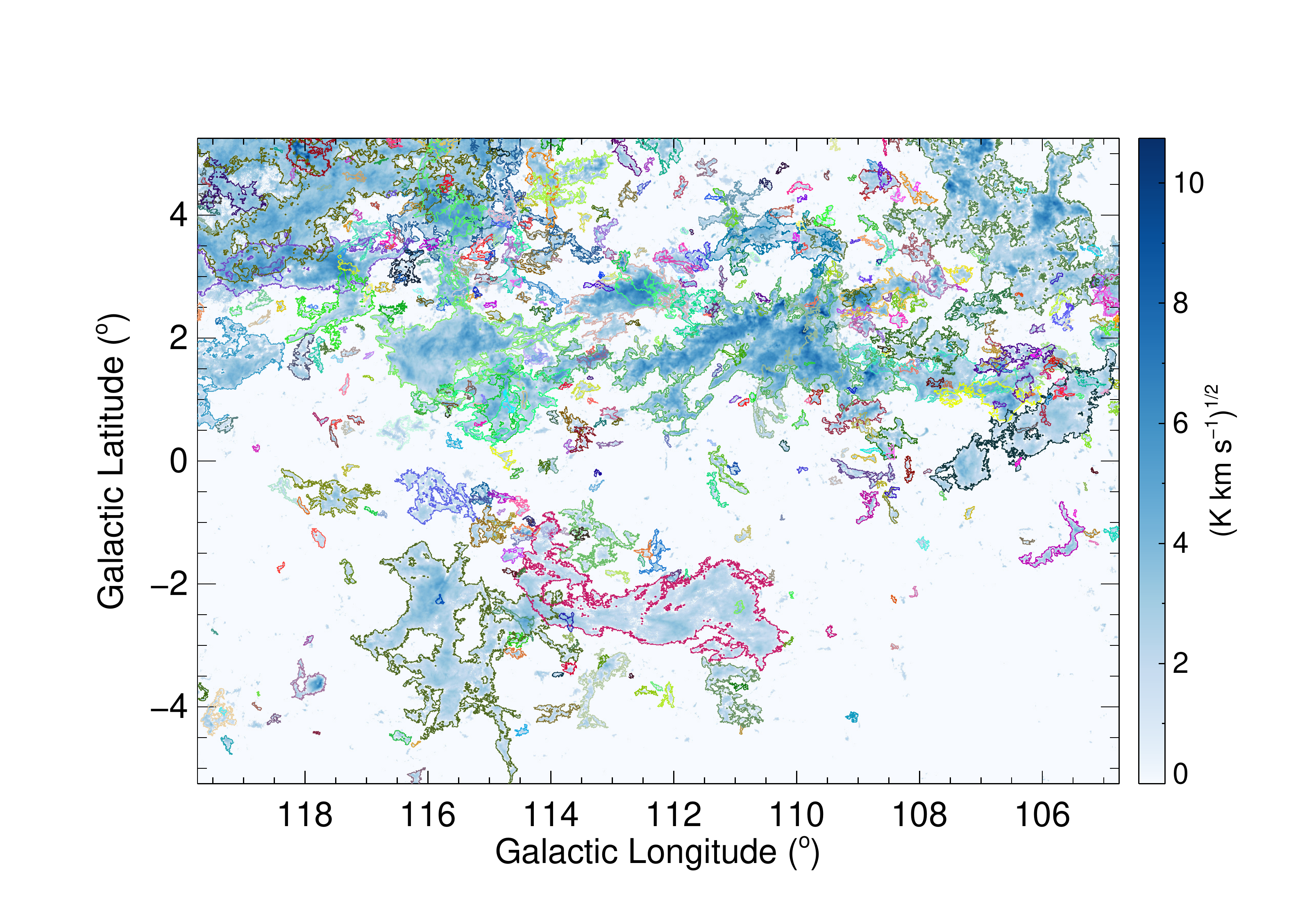}}
	\subfigure[]{
		\label{fig:Fig6b}
		\includegraphics[trim=1cm 0cm 4cm 2cm, width = 0.7\linewidth, clip]{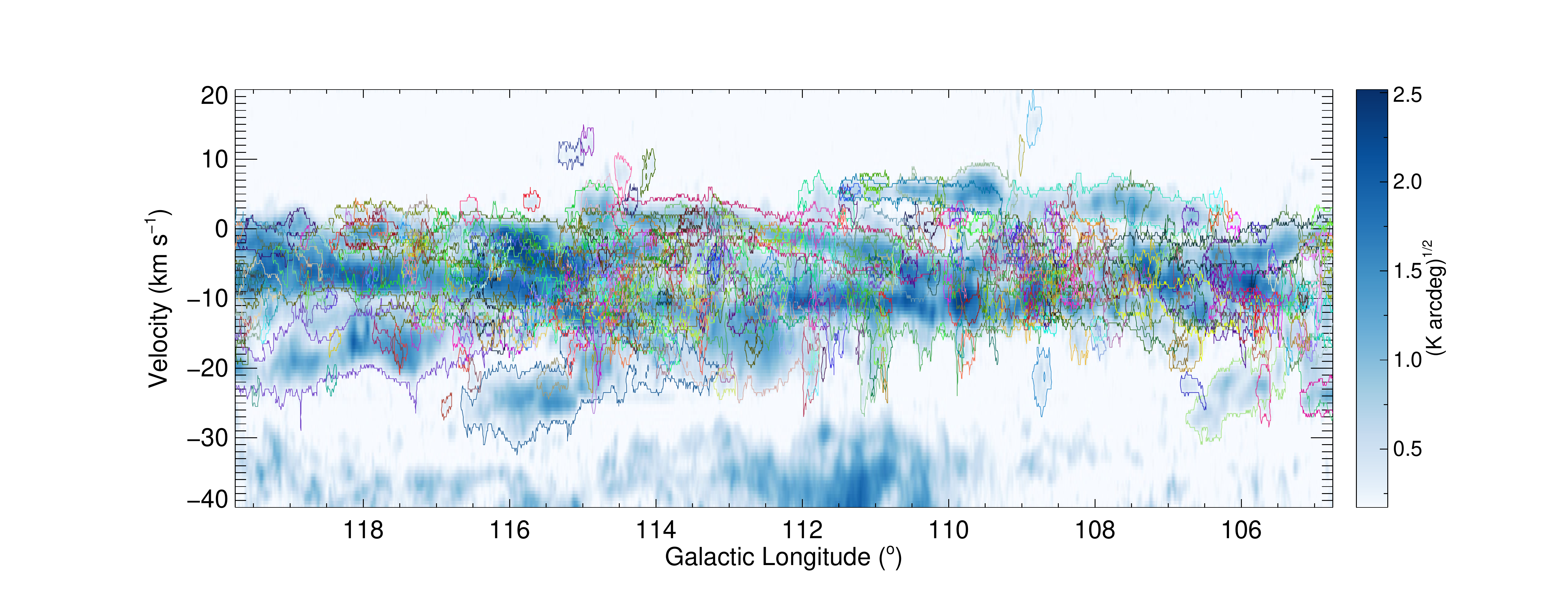}}
	\caption{Demonstration of the outlines, as the output of the SCIMES algorithm, of the identified \CO molecular clouds in the Local arm in (a) l-b space and (b) l-v space. Different colors in the two panels correspond to different molecular clouds. The background image in panel (a) is the integrated intensity of \CO in the velocity range from $-27$ to $20$ km s$^{-1}$, while that of the panel (b) is the l-v diagram integrated along the Galactic latitude from b=$-$5$.\!\!^{\circ}$25 to b=5$.\!\!^{\circ}$25. }\label{fig6}
\end{figure*}

\begin{figure*}[htb!]
	\centering
	\subfigure[]{
		\label{fig:Fig7a}
		\includegraphics[trim=0cm 0cm 2cm 1cm, width = 0.7\linewidth, clip]{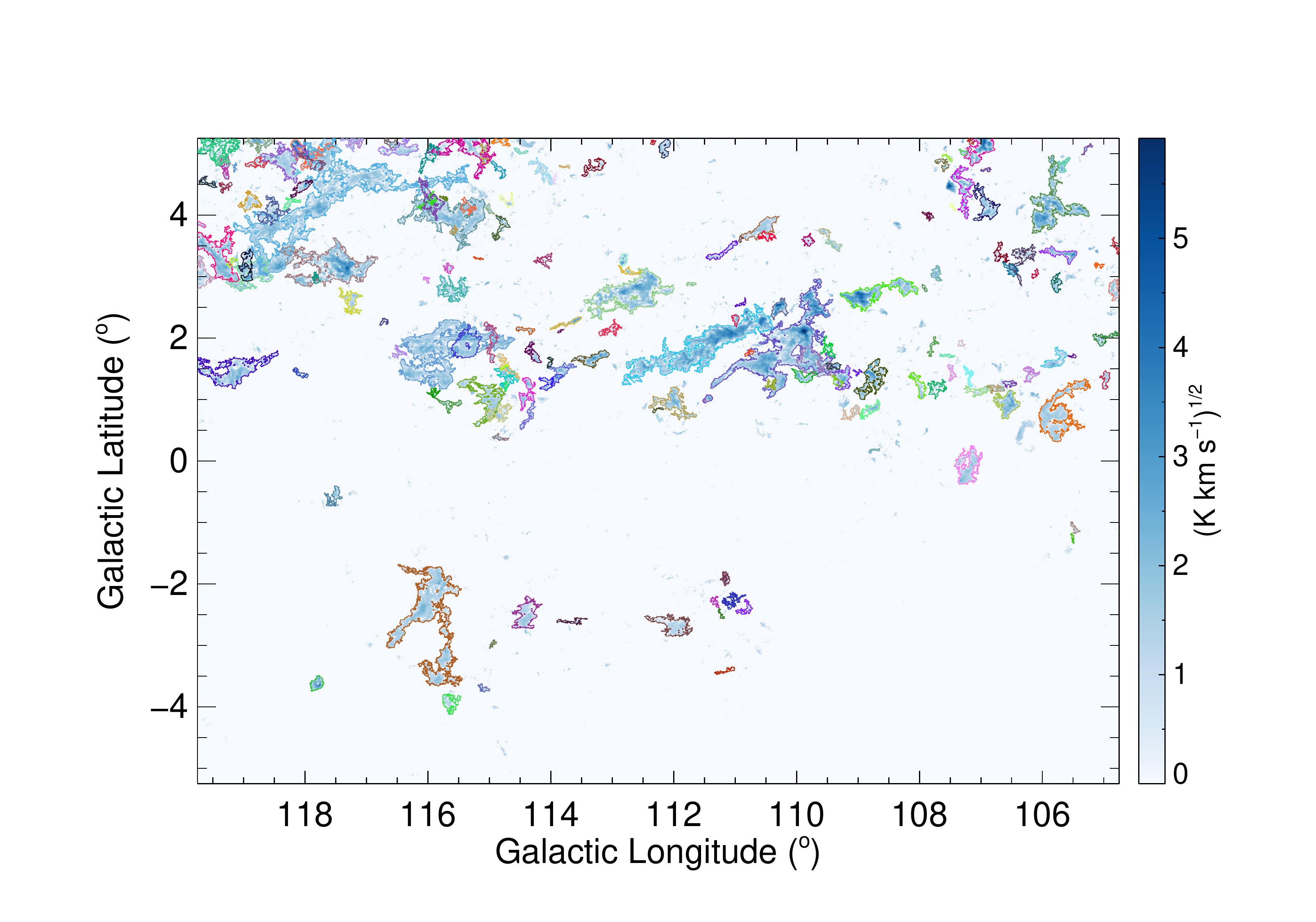}}
	\subfigure[]{
		\label{fig:Fig7b}
		\includegraphics[trim=1cm 0cm 4cm 2cm, width = 0.7\linewidth, clip]{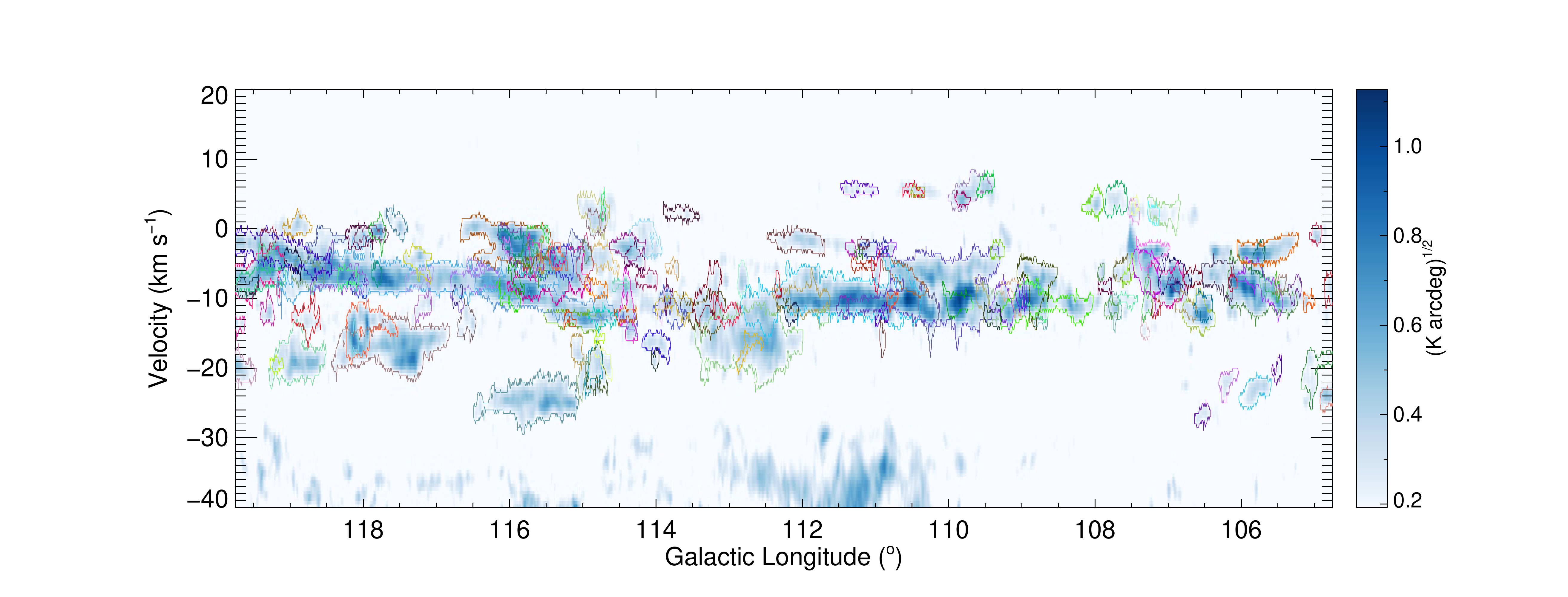}}
	\caption{Same as Figure \ref{fig6}, but for \COl clouds.}\label{fig7}
\end{figure*}

A total of 857 molecular clouds have been extracted from the \CO $J=1-0$ data, while 301 molecular clouds have been extracted from the \COl $J=1-0$ data. The molecular clouds that have centroid velocities in the velocity ranges from $-115$ to $-75$, $-75$ to $-27$, and $-27$ to $20$ km s$^{-1}$ are assigned to the Outer$+$OSC, Perseus, and Local arms, respectively. The total numbers of \CO molecular clouds in the Local, Perseus, and Outer$+$OSC arms are 440, 399, and 18, respectively, while the corresponding numbers of \COl molecular clouds are 176, 124, and 1. The \COl cloud in the Outer$+$OSC arm is found to be artificial, which is caused by bad velocity channels, therefore, is removed in the final catalog. According to the result of cross-matching, 279 \COl molecular clouds have counterparts of \CO clouds. The name of the matching \CO cloud for each \COl cloud is given in the final catalog (see Table \ref{tab3} in Section \ref{sec3.4}). Figures \ref{fig6} and \ref{fig7} present the demonstrations of the outlines of the \CO and \COl molecular clouds in the Local arm, respectively. The results of clouds identification in the Perseus and Outer$+$OSC arms are shown in the Appendix, Figures \ref{fig20}-\ref{fig22}.

\subsubsection{Distance estimation}\label{sec3.3.2}
Distance is critical for the calculation of the physical parameters of molecular clouds. For the \CO clouds with centroid velocities less than $-27$ km s$^{-1}$, which are beyond the Local Arm, we adopted the kinematic distances derived using a flat Galactic rotation curve, which is a good approximation for the outer Galaxy, with a solar galactocentric distance of $R_0$ = 8.34 kpc and a rotation velocity of $\Theta_0$ = 240 km s$^{-1}$ \citep{Reid2014}. Since we are looking toward the outer Galaxy, our distance estimate is not affected by the near or far ambiguity. For the nearby clouds, the peculiar motion of the gas and the internal velocity dispersion of molecular clouds may lead to large uncertainty in the estimation of the distances based on the kinematic method. Therefore for these clouds, we use an alternative method to estimate the distance. For molecular clouds within 2 kpc, we looked for the position of sharp steepening in the 3D reddening map within the projected area of each cloud to determine their distances. We used the three-dimensional map of dust reddening up to 5 kpc produced by \citet{Green2019} with the parallaxes from Gaia DR2 catalog and the stellar photometry from Pan-STARRS 1 and 2MASS.

\begin{figure*}[htb!]
	\centering
	\includegraphics[trim=0.2cm 0.2cm 0.5cm 0.5cm, width= \linewidth, clip]{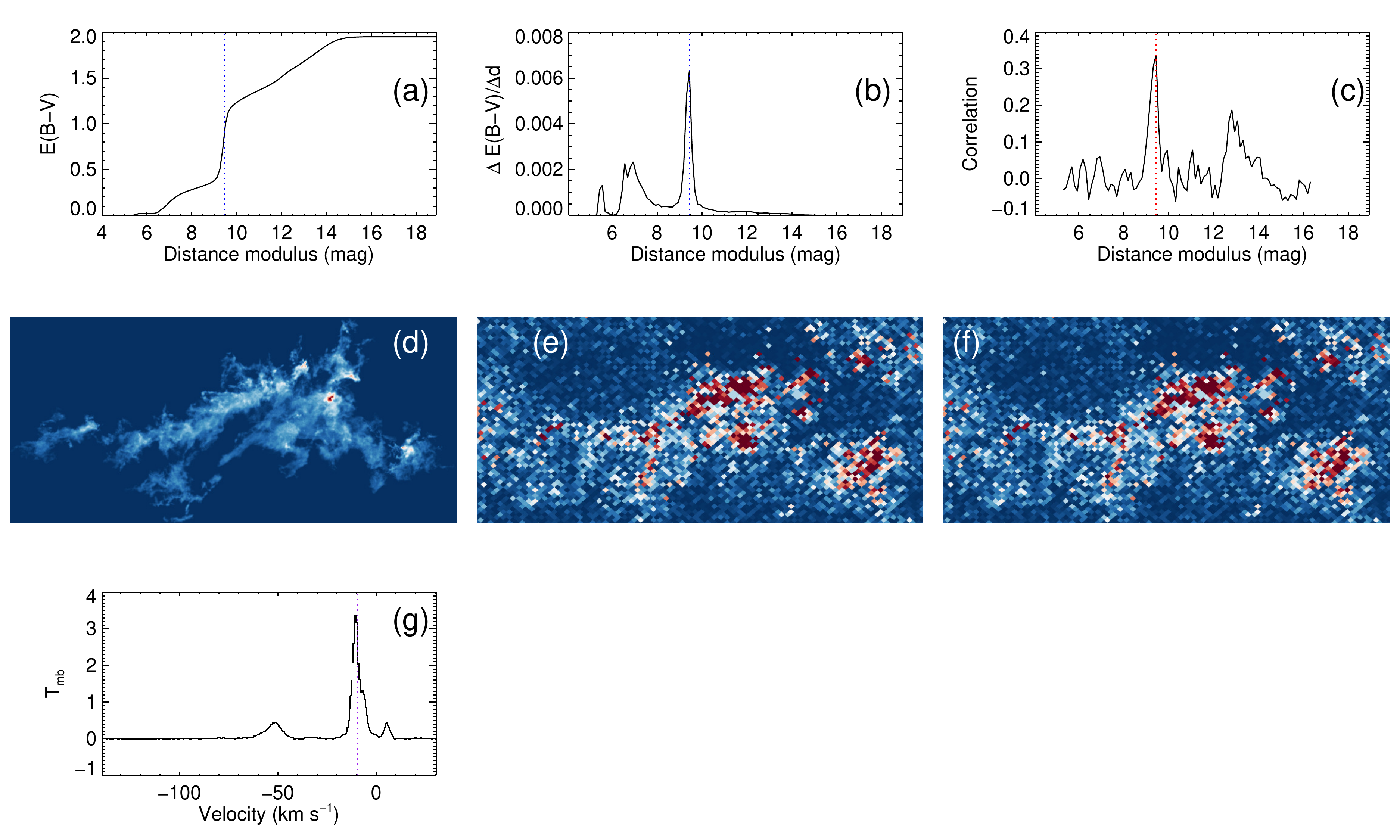}
	\caption{Example of distance estimate of molecular clouds using the 3D dust maps from \cite{Green2019}. (a) Average reddening versus distance modulus of the Cepheus GMC. The blue dashed line shows the distance of the Cepheus GMC. (b) Dust reddening density, $\Delta E(B-V)/\Delta d$, versus distance modulus. The blue dashed line marks the distance modulus that corresponds to the maximum of the $\Delta E(B-V)/\Delta d$ value. (c) Variation of the cross-correlation between the reddening density map and the \CO integrated intensity map of the cloud. The red dashed line shows the distance modulus that corresponds to the maximum of the cross-correlation. (d) \CO intensity map of Cepheus GMC integrated within the PPV mask resulting from the SCIMES algorithm. (e) Dust reddening density map at the distance modulus indicated by the blue dashed line in panel (b). (f) Dust reddening density map at the distance modulus as indicated by the red dashed line in panel (c). (g) Average spectrum of \CO emission within the projected boundary of Cepheus GMC.} 
	\label{fig8}
\end{figure*}

The public Python package ``dustmaps" is used to query the reddening map of \cite{Green2019}. We have queried a region of the ``median" dust reddening cube with the same pixel size and spatial extent as our CO datacube, which contains dust reddening along each line-of-sight from distance modulus from 4 mag to 18.725 mag in steps of 0.125 mag. For each \CO cloud, we derive the average dust reddening within the projected boundary of the cloud at different distance modulus, such as panel (a) in Figure \ref{fig8}. The distance of the molecular cloud is considered to correspond to the distance modulus where there is a sharp increase of the reddening curve, which also corresponds to a local maximum of the derivative of the reddening-distance modulus curve, defined as the dust reddening density. In practice, we use two methods to determine the ``step" in the average reddening-modulus curve of each cloud. The first method is to determine the distance modulus that corresponds to the minimum of the second derivative of the reddening versus the distance modulus, as indicated by the blue dashed line in panel (b) in Figure \ref{fig8}. The second method is to derive the modulus at which the cross-correlation between the integrated intensity map and the reddening density map reaches the maximum, as indicated by the red dashed line in panel (c) in Figure \ref{fig8}. For molecular clouds that have large angular sizes, such as the Cep GMC in Figure \ref{fig8}, the distances estimated by the two methods are consistent within 0.25 mag distance modulus. In this case, the averaged value from the two methods is adopted. For molecular clouds of small angular size, the distances obtained by the two methods are usually inconsistent. In this case, we choose the modulus found by the first method. However, in practice, for some clouds there are multiple ``steps" or no ``step" in the reddening-modulus curves. When there are multiple ``steps'' in the reddening-modulus curve, we check the average \CO spectrum of the cloud and match each step with different velocity components with the rule of more negative velocity corresponding to larger distance. For those clouds that no obvious ``step'' in the reddening-modulus curve is found, the cloud distance is not assigned. Figure \ref{fig8} is given as an example of the distance estimation. We have manually checked the figures for each of the 440 \CO clouds in the Local arm to estimate their distances. Finally, there are 267 out of the 440 nearby \CO molecular clouds with distance assigned by using the extinction method. 

\begin{figure*}[htb!]
	\centering
	\includegraphics[trim=0.2cm 0.2cm 0.5cm 0.5cm, width= 0.55\linewidth, clip]{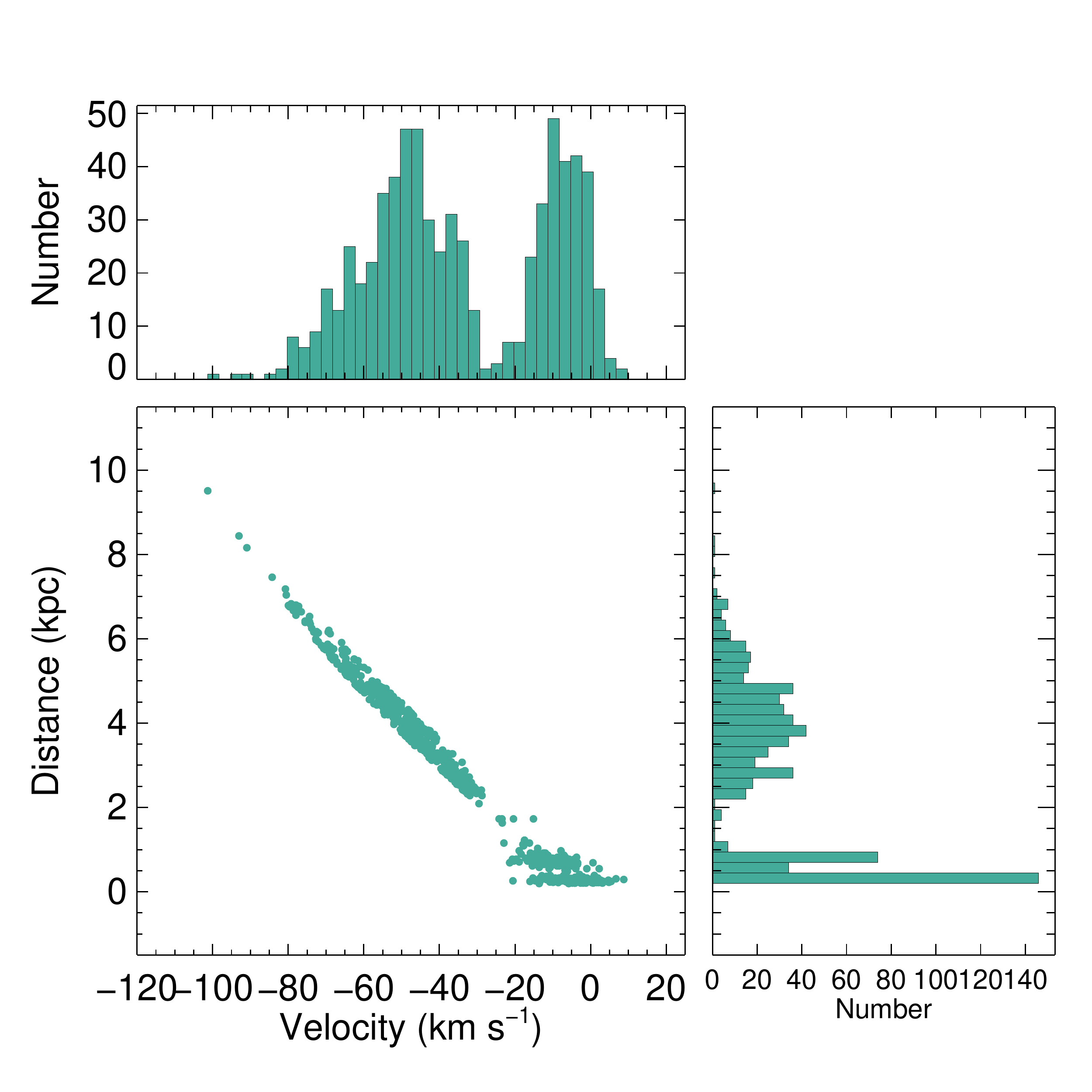}
	\caption{Relationship between the centroid velocities and the distances of the \CO clouds. The upper and right panels show the statistics of the centroid velocity and the distance, respectively. We adopt the extinction distances for the clouds with velocities $v>-27$ km s$^{-1}$ and the kinematic distances for the clouds with velocities $v<-27$ km s$^{-1}$.} 
	\label{fig9}
\end{figure*}

In summary, we estimated the distances of the molecular clouds of the Local arm with the 3D-extinction method, while for the clouds in the Perseus and Outer$+$OSC arms, we used the kinematic method. The distances of the \COl clouds are taken to be that of their counterpart \CO clouds. For the \COl clouds that do not have counterpart \CO clouds, distances are not assigned. The relationship of cloud distance with their centroid velocity and the corresponding histograms are given in Figure \ref{fig9}. As shown in Figure 9, the molecular clouds in the Local arm are mainly concentrated at two distances, $\sim$250 and $\sim$750 pc with small dispersions, while the distances of the molecular clouds in the Perseus arm distributed in a wider range from $\sim$2 to $\sim$6 kpc with the peak at $\sim$3.7 kpc. This different distribution of cloud distances in the two arms could be related to the different methods used to derive distance for the two arms. The molecular clouds in the Outer$+$OSC arms are located in the distance range from $\sim$6 to $\sim$9 kpc without any concentration. Very few molecular clouds are located in the interarm regions.  

\subsection{Physical Properties of the Identified Molecular Clouds} \label{sec3.4}
The statistical properties of molecular clouds in our Galaxy can provide insight into understanding their formation, evolution, and environments \citep{Dobbs2014}. In this section we derive the physical parameters of the identified \CO and \COl clouds. 
\subsubsection{Physical parameters}\label{sec3.4.1}
The centroid position and the velocity dispersion of the identified molecular cloud are given by the DENDROGRAM+SCIMES algorithms directly, which are defined as the brightness temperature-weighted first and second moments within the p-p-v mask of the cloud, respectively \citep{Colombo2015}. 

The effective radius of a cloud is derived as the geometric mean of the major and minor axes of the cloud given by the algorithm, and then deconvoluted with the HPBW of the PMO-13.7 m telescope,
\begin{equation} 
R_{eff} = \frac{1}{2}d(\theta_a\theta_b - \theta_{beam}^2)^{1/2}
\end{equation}
where $\theta_a$, $\theta_b$, are the fitted FWHM along the  major and minor axes of the cloud, and $\theta_{beam}$ is the HPBW of the PMO-13.7 m telescope, and d is the distance to the cloud. This method means that we approximate a molecular cloud to a Gaussian ellipsoid, and its projected area is an ellipse.

The total intensity and the exact area, in units of square arcseconds, of each cloud are tabulated in the output table of the DENDROGRAM algorithm, which can be used to calculate the averaged column density of a cloud as follows, 
\begin{equation}
N_0(H_2) = N_{tot}/num_{p}. 
\end{equation}
where $N_{tot}$ is the total column density of molecular hydrogen, and $num_p$ is the total pixel number of a cloud that can be obtained through dividing the exact area of a cloud by the pixel size of the data.	

For the molecular clouds extracted from the \CO $J=1-0$ data, the total H$_2$ column density was derived in the same way described in Section \ref{sec3.2} using the formula
\begin{equation}
N_{tot} = X_{\rm CO}I_{\rm CO}, 
\end{equation}
where $I_{\rm CO}$ is the \CO intensity integrated over the total area of the cloud and X$_{\rm CO} = 2.0\times10^{20}$ cm$^{-2}$ (K km$^{-1}$)$^{-1}$ \citep{Bolatto2013} is the conversion factor.
For the molecular clouds extracted from the \COl data, we used a different method to derive the total H$_2$ column density that takes into account the estimate of the gas excitation temperature, the \COl optical depth, and the variation of the $^{12}$C/$^{13}$C abundance ratio with the galactocentric distance. In particular, assuming the \COl molecules are under the Local Thermodynamic Equilibrium (LTE) condition, the total H$_2$ column density of the \COl clouds can be calculated according to
\begin{equation}
%\begin{split}
N_{tot} = A\times2.42\times10^{14}\frac{\tau(^{13}CO)}{1-e^{-\tau(^{13}CO)}}\\ \frac{1+0.88/{T_{ex}}}{1-e^{-5.29/T_{ex}}}I_{^{13}CO}, 
%\end{split}
\end{equation}
where $A$ is a constant related to the abundance ratio to convert the \COl column density to H$_2$ column density, $T_{ex}$ is the excitation temperature, $\tau(\rm{^{13}CO})$ is the optical depth at the peak intensity of the \COl emission within the boundary of a \COl cloud, and $I_{\rm{^{13}CO}}$ is the total \COl integrated intensity. The constant $A$ is the product of the abundance ratio [$^{12}$C/$^{13}$C] = 6.21$d_{GC}$ + 18.71 \citep{Milam2005} and H$_2$/\CO = $1.1\times10^4$ \citep{Frerking1982}, where $d_{GC}$ is the cloud distance from the Galactic center. The excitation temperature is calculated according to Eq. 1 in \cite{Li2018}, using the peak intensity of the \CO emission within the boundary of the \COl cloud, while the optical depth of the \COl emission is according to Eq. 3 in \cite{Li2018}.

The cloud mass, either for the \CO clouds or the \COl clouds, is eventually obtained through 
\begin{equation}
 M = N_{tot} d^2 \Omega \mu m_H, 
\end{equation}
where d is the cloud distance, $\Omega$ is the solid angle of each pixel, and $\mu$ = 2.8 is the atomic weight per molecular hydrogen. 

The surface densities and number densities of the clouds are derived as follows,
\begin{equation}
\Sigma = M/(\pi R_{eff}^2)
\end{equation}
\begin{equation}
n(H_2) = 3M/(4\pi R_{eff}^3\mu m_H).
\end{equation} 

The dynamical state of a molecular cloud is characterized by the virial parameter, which measures the ratio of the internal kinetic energy to the gravitational energy. In the literature, there are different forms of the virial parameter. In this work we follow the definition of \cite{Bertoldi1992}
\begin{equation}
\alpha_{\rm vir} = \frac{5\sigma_v^2R_{eff}}{GM},
\end{equation}
where G is the gravitational constant, $\sigma_v$ is the velocity dispersion. Theoretically, the critical value of $\alpha_{\rm vir}$ for a non-magnetized isothermal hydrostatic equilibrium sphere is 2 \citep{Kauffmann2013}. A cloud with a virial parameter above or below the critical value will eventually dissipate or collapse when not considering other physical mechanisms beside its self-gravity and internal pressure. We have derived the virial parameters for each $\rm{^{12}CO}$ and $\rm{^{13}CO}$ cloud. 

All above derived physical parameters of \CO and \COl clouds are tabulated in Tables \ref{tab2} and \ref{tab3}, respectively. In Table \ref{tab2} and Table \ref{tab3}, we have listed a total of 857 \CO molecular clouds and 300 \COl molecular clouds, respectively. However, the physical parameters, such as effective radius, mass, surface density, number density, and virial parameter, are only presented for the 684 \CO molecular clouds and 274 \COl molecular clouds that have assigned distances. The histograms of the physical parameters for the distance-assigned \CO and \COl clouds are presented in Figures \ref{fig11} and \ref{fig12}, respectively, and we also used the distance assigned clouds as the sample to study the scaling relations in Section \ref{sec3.5}. 

\begin{longrotatetable}
	\begin{deluxetable*}{lllrrrrrrrrrrrr}
		\tablecaption{Properties of $^{12}$CO clouds\label{tab2}}
		\tablewidth{\linewidth}
		\setlength\tabcolsep{8pt}
		\tabletypesize{\tiny}
		\tablehead{	\colhead{Name} & \colhead{l} &
			\colhead{b} & \colhead{$\theta_a$} &
			\colhead{$\theta_b$} & \colhead{PA} &
			\colhead{$v_{lsr}$} & \colhead{$\sigma_v$} &
			\colhead{$d$} & \colhead{$R_{eff}$} & \colhead{$N_0(H_2)$} &\colhead{$Mass$} &\colhead{$\Sigma$} &\colhead{n} &\colhead{$\alpha_{\rm vir}$}  \\
			\colhead{} & \colhead{$\arcdeg$} & \colhead{$\arcdeg$} & \colhead{$\arcsec$} &
			\colhead{$\arcsec$} & \colhead{$\arcdeg$} & \colhead{(km s$^{-1}$)} &
			\colhead{(km s$^{-1}$)} & \colhead{(kpc)} & \colhead{(pc)} & \colhead{(cm$^{-2}$)} & \colhead{(M$_{\sun}$)} & \colhead{(M$_{\sun}$ pc$^{-2}$)} &\colhead{(cm$^{-3}$)} & \colhead{}
			\\
			\colhead{(1)} & \colhead{(2)} & \colhead{(3)} & \colhead{(4)} & \colhead{(5)} & \colhead{(6)} & \colhead{(7)} & \colhead{(8)} & \colhead{(9)} & \colhead{(10)} & \colhead{(11)} & \colhead{(12)}& \colhead{(13)} & \colhead{(14)} & \colhead{(15)}
		}
		\startdata
		MWISP G104.767-03.132-041.32 & 104.767 &  -3.132 &      204 &       55 &    93 &   -41.32 &     1.08 &  3.73 &  2.21 &   6.5e+20 &       520.0 &  33.8 &      166 &     5.7\\
		MWISP G104.780+02.258-002.08 & 104.780 &   2.258 &      200 &       79 &    95 &    -2.08 &     0.49 & ... & ... &   3.3e+20 & ... & ... & ... & ...\\
		MWISP G104.790+01.372-048.22 & 104.790 &   1.372 &      109 &       40 &   162 &   -48.22 &     0.62 &  4.32 &  1.55 &   5.5e+20 &       244.2 &  32.4 &      228 &     2.8\\
		MWISP G104.815+01.257-007.89 & 104.815 &   1.257 &      206 &      101 &   127 &    -7.89 &     0.74 & ... & ... &   4.8e+20 & ... & ... & ... & ...\\
		MWISP G104.827+00.885-001.44 & 104.827 &   0.885 &      182 &      149 &    75 &    -1.44 &     1.16 &  0.23 &  0.22 &   5.1e+20 &         4.8 &  32.7 &     1655 &    70.1\\
		MWISP G104.833+00.155-065.87 & 104.833 &   0.155 &      354 &      152 &    55 &   -65.87 &     1.33 &  5.91 &  7.82 &   7.7e+20 &      8169.8 &  42.5 &       59 &     2.0\\
		MWISP G104.845-01.155-013.37 & 104.845 &  -1.155 &      224 &      195 &  -161 &   -13.37 &     1.27 &  0.24 &  0.29 &   7.4e+20 &         8.5 &  31.9 &     1199 &    64.0\\
		MWISP G104.860+02.192+001.67 & 104.860 &   2.192 &      274 &      147 &  -165 &     1.67 &     0.53 & ... & ... &   5.5e+20 & ... & ... & ... & ...\\
		MWISP G104.863+01.003-050.03 & 104.863 &   1.003 &      111 &       90 &   124 &   -50.03 &     1.05 &  4.48 &  2.51 &   5.6e+20 &       722.7 &  36.5 &      158 &     4.4\\
		MWISP G104.894+03.260-011.29 & 104.894 &   3.260 &      616 &      367 &    62 &   -11.29 &     2.47 & ... & ... &   1.2e+21 & ... & ... & ... & ...\\
		MWISP G104.913+02.753-024.20 & 104.913 &   2.753 &      607 &      412 &   106 &   -24.20 &     0.82 &  1.73 &  4.93 &   9.8e+20 &      4327.6 &  56.6 &      125 &     0.9\\
		MWISP G104.922+02.575-002.77 & 104.922 &   2.575 &      450 &      246 &   122 &    -2.77 &     2.14 & ... & ... &   4.8e+20 & ... & ... & ... & ...\\
		MWISP G104.950+01.082-058.92 & 104.950 &   1.082 &      146 &       45 &   126 &   -58.92 &     1.31 &  5.26 &  2.37 &   3.6e+20 &       254.7 &  14.4 &       66 &    18.4\\
		MWISP G104.978+01.130-007.80 & 104.978 &   1.130 &      136 &       53 &  -152 &    -7.80 &     0.51 & ... & ... &   3.8e+20 & ... & ... & ... & ...\\
		MWISP G104.983-02.678-042.50 & 104.983 &  -2.678 &      301 &       79 &   123 &   -42.50 &     1.20 &  3.82 &  3.34 &   8.0e+20 &      1117.6 &  31.8 &      103 &     5.0\\
		MWISP G104.985+02.011-020.40 & 104.985 &   2.011 &      409 &      272 &  -169 &   -20.40 &     2.23 &  1.73 &  3.29 &   1.6e+21 &      5038.7 & 148.2 &      492 &     3.8\\
		MWISP G104.998+00.735-054.11 & 104.998 &   0.735 &      221 &       55 &   175 &   -54.11 &     0.77 &  4.82 &  2.99 &   4.3e+20 &       587.6 &  20.9 &       76 &     3.5\\
		MWISP G105.057+02.266-002.05 & 105.057 &   2.266 &      287 &      101 &   159 &    -2.05 &     0.43 &  0.35 &  0.34 &   3.3e+20 &         7.6 &  20.7 &      659 &     9.5\\
		MWISP G105.090+02.907-014.19 & 105.090 &   2.907 &      238 &      123 &  -140 &   -14.19 &     1.77 & ... & ... &   6.2e+20 & ... & ... & ... & ...\\
		MWISP G105.094+01.123-047.81 & 105.094 &   1.123 &      496 &      281 &  -176 &   -47.81 &     1.90 &  4.27 &  9.11 &   1.4e+21 &     21978.9 &  84.3 &      101 &     1.7\\
		MWISP G105.094-04.029-046.86 & 105.094 &  -4.029 &       82 &       62 &    68 &   -46.86 &     1.90 &  4.18 &  1.64 &   7.0e+20 &       521.1 &  61.4 &      408 &    13.1\\
		MWISP G105.098+00.853-047.35 & 105.098 &   0.853 &      132 &       99 &   133 &   -47.35 &     1.05 &  4.23 &  2.72 &   6.5e+20 &       781.2 &  33.6 &      135 &     4.5\\
		MWISP G105.142+00.322-069.25 & 105.142 &   0.322 &      192 &      119 &  -140 &   -69.25 &     1.16 &  6.20 &  5.32 &   9.9e+20 &      5337.9 &  59.9 &      122 &     1.6\\
		MWISP G105.151+03.434-009.98 & 105.151 &   3.434 &      212 &       84 &   152 &    -9.98 &     0.53 &  0.24 &  0.19 &   1.3e+21 &        13.9 & 129.2 &     7631 &     4.3\\
		MWISP G105.151-01.327-016.15 & 105.151 &  -1.327 &      168 &       30 &   158 &   -16.15 &     1.04 & ... & ... &   4.0e+20 & ... & ... & ... & ...\\
		MWISP G105.153-00.170-010.70 & 105.153 &  -0.170 &      145 &       48 &   145 &   -10.70 &     0.70 & ... & ... &   3.9e+20 & ... & ... & ... & ...\\
		MWISP G105.155+02.907-009.07 & 105.155 &   2.907 &      171 &      105 &    70 &    -9.07 &     0.62 & ... & ... &   4.6e+20 & ... & ... & ... & ...\\
		MWISP G105.212+00.698-061.54 & 105.212 &   0.698 &      120 &       51 &   178 &   -61.54 &     0.92 &  5.48 &  2.37 &   4.0e+20 &       445.7 &  25.2 &      116 &     5.2\\
		MWISP G105.212-01.181-014.72 & 105.212 &  -1.181 &      222 &       85 &  -174 &   -14.72 &     0.90 &  0.65 &  0.50 &   6.7e+20 &        30.4 &  38.0 &      822 &    15.5\\
		MWISP G105.230+00.787-052.97 & 105.230 &   0.787 &      263 &       93 &  -173 &   -52.97 &     1.14 &  4.71 &  4.19 &   7.7e+20 &      3044.2 &  55.1 &      143 &     2.1\\
		MWISP G105.231+00.479-051.08 & 105.231 &   0.479 &      124 &       71 &    68 &   -51.08 &     1.50 &  4.54 &  2.38 &   4.7e+20 &       585.4 &  32.8 &      150 &    10.6\\
		MWISP G105.247-00.284-034.00 & 105.247 &  -0.284 &       98 &       67 &   176 &   -34.00 &     0.66 &  3.07 &  1.38 &   7.6e+20 &       313.4 &  52.1 &      411 &     2.3\\
		MWISP G105.262+01.329-009.19 & 105.262 &   1.329 &      392 &      187 &   152 &    -9.19 &     1.21 & ... & ... &   7.8e+20 & ... & ... & ... & ...\\
		MWISP G105.275+00.360-011.72 & 105.275 &   0.360 &      219 &      114 &   120 &   -11.72 &     0.67 & ... & ... &   5.9e+20 & ... & ... & ... & ...\\
		MWISP G105.276-03.925-042.92 & 105.276 &  -3.925 &       80 &       58 &   117 &   -42.92 &     1.16 &  3.83 &  1.43 &   4.8e+20 &       216.8 &  33.9 &      259 &    10.2\\
		\enddata
		\tablecomments{The source name is defined under the MWISP standard. According to the MWISP standard for nomenclature, molecular clouds are named after their centroid positions and velocities. Specifically, the names start with ``MWISP" and then contain the spatial coordinates of the molecular clouds accurate to three decimal places and the centroid velocities accurate to two decimal places. The accuracy is set according to the pointing accuracy and the velocity resolution of the PMO-13.7m telescope. Columns 2$-$6 give the centroid positions, the intensity-weighted major and minor axes, and the position angles of the clouds. The centroid velocity, the velocity dispersion, and the distance of the clouds are presented in columns 7$-$9. Columns 10$-$15 list the effective radius, average column density, mass, surface density, number density, and the virial parameters of the clouds, respectively. \\This table is available in its entirety in machine$-$readable form in the online material.}
	\end{deluxetable*}
\end{longrotatetable}

\begin{longrotatetable}
	\begin{deluxetable*}{lllrrrrrrrrrrrrrrl}
		\tablecaption{Properties of $^{13}$CO clouds\label{tab3}}
		\tablewidth{\linewidth}
		\setlength\tabcolsep{3pt}
		\tabletypesize{\tiny}
		\tablehead{	\colhead{Name} & \colhead{l} &
			\colhead{b} & \colhead{$\theta_a$} &
			\colhead{$\theta_b$} & \colhead{PA} &
			\colhead{$v_{lsr}$} & \colhead{$\sigma_v$} &
			\colhead{$d$} & \colhead{$R_{eff}$} & \colhead{$T_{ex}$} & $\tau_{^{13}CO}$ & \colhead{$N_0(H_2)$} &\colhead{$Mass$} &\colhead{$\Sigma$} &\colhead{n} &\colhead{$\alpha_{\rm vir}$} & \colhead{Name of matching \CO cloud}
			\\
			\colhead{} & \colhead{$\arcdeg$} & \colhead{$\arcdeg$} & \colhead{$\arcsec$} &
			\colhead{$\arcsec$} & \colhead{$\arcdeg$} & \colhead{(km s$^{-1}$)} &
			\colhead{(km s$^{-1}$)} & \colhead{(kpc)} & \colhead{(pc)} & \colhead{(K)}& \colhead{} & \colhead{cm$^{-2}$} & \colhead{(M$_{\sun}$)} & \colhead{(M$_{\sun}$ pc$^{-2}$)} &\colhead{cm$^{-3}$} & \colhead{}& \colhead{}
			\\
			\colhead{(1)} & \colhead{(2)} & \colhead{(3)} & \colhead{(4)} & \colhead{(5)} & \colhead{(6)} & \colhead{(7)} & \colhead{(8)} & \colhead{(9)} & \colhead{(10)} & \colhead{(11)} & \colhead{(12)}& \colhead{(13)} & \colhead{(14)} & \colhead{(15)} & \colhead{(16)} & \colhead{(17)} & \colhead{(18)}
		}
		\startdata
		MWISP G104.794+03.498-008.57 & 104.794 &   3.498 &      199 &       85 &   114 &    -8.57 &     1.04 & ... & ... & ... & ... & ... & ... & ... & ... & ... & ...\\
		MWISP G104.806+00.169-052.16 & 104.806 &   0.169 &      153 &      104 &   154 &   -52.16 &     0.74 &  4.63 &  3.30 &  12.6 &   0.4 &   1.3e+21 &      3767.9 &   110.1 &      364 &     0.5 & MWISP G105.389+00.275-052.19\\
		MWISP G104.829+02.795-024.17 & 104.829 &   2.795 &      283 &      145 &    91 &   -24.17 &     0.57 &  1.73 &  1.99 &  10.1 &   0.7 &   2.5e+21 &       991.6 &    79.5 &      435 &     0.8 & MWISP G104.913+02.753-024.20\\
		MWISP G104.967+01.333-000.91 & 104.967 &   1.333 &      261 &      109 &    82 &    -0.91 &     0.55 &  0.41 &  0.39 &  10.2 &   0.6 &   1.3e+21 &        29.3 &    60.3 &     1675 &     4.7 & MWISP G106.122+00.579-002.93\\
		MWISP G105.022+01.990-020.39 & 105.022 &   1.990 &      347 &      156 &  -164 &   -20.39 &     1.90 &  1.73 &  2.29 &   9.5 &   0.4 &   1.8e+21 &       723.6 &    43.9 &      209 &    13.2 & MWISP G104.985+02.011-020.40\\
		MWISP G105.031+00.105-049.45 & 105.031 &   0.105 &      177 &       48 &   168 &   -49.44 &     0.55 &  4.63 &  2.40 &   9.6 &   0.2 &   2.0e+20 &       564.0 &    31.2 &      142 &     1.5 & MWISP G105.389+00.275-052.19\\
		MWISP G105.088+03.180-010.45 & 105.088 &   3.180 &      161 &       84 &   173 &   -10.45 &     0.74 & ... & ... & ... & ... & ... & ... & ... & ... & ... & ...\\
		MWISP G105.102+01.089-046.99 & 105.102 &   1.089 &      202 &      178 &    50 &   -46.99 &     1.19 &  4.27 &  4.60 &  11.7 &   0.6 &   3.2e+21 &      7846.6 &   118.1 &      280 &     1.0 & MWISP G105.094+01.123-047.81\\
		MWISP G105.237+00.781-052.64 & 105.237 &   0.781 &      183 &       60 &  -172 &   -52.64 &     0.84 &  4.71 &  2.78 &   7.8 &   0.3 &   2.4e+20 &       699.0 &    28.7 &      112 &     3.2 & MWISP G105.230+00.787-052.97\\
		MWISP G105.269+05.089-008.95 & 105.269 &   5.089 &      200 &       65 &  -137 &    -8.95 &     0.95 &  0.87 &  0.56 &  14.8 &   0.3 &   6.1e+20 &        60.7 &    62.2 &     1219 &     9.6 & MWISP G106.529+04.081-007.04\\
		MWISP G105.471-01.101-007.97 & 105.471 &  -1.101 &      206 &       67 &   123 &    -7.97 &     0.57 &  0.26 &  0.17 &  12.2 &   0.3 &   4.1e+20 &         3.8 &    40.1 &     2520 &    17.5 & MWISP G105.691-01.341-008.22\\
		MWISP G105.504+01.705-020.42 & 105.504 &   1.705 &      126 &       54 &  -178 &   -20.42 &     0.83 &  1.73 &  0.79 &   8.5 &   0.3 &   1.3e+20 &        51.1 &    26.0 &      359 &    12.4 & MWISP G106.064+01.491-023.45\\
		MWISP G105.505-01.254-008.59 & 105.505 &  -1.254 &      174 &       37 &   100 &    -8.59 &     0.90 &  0.26 &  0.12 &  10.8 &   0.2 &   1.2e+20 &         1.1 &    25.6 &     2393 &    99.4 & MWISP G105.691-01.341-008.22\\
		MWISP G105.630+03.365-008.76 & 105.630 &   3.365 &      524 &      148 &   168 &    -8.76 &     1.67 &  0.87 &  1.38 &  20.5 &   0.6 &   1.1e+22 &      1076.0 &   180.4 &     1429 &     4.2 & MWISP G106.529+04.081-007.04\\
		MWISP G105.632+00.341-052.38 & 105.632 &   0.341 &      649 &      210 &   145 &   -52.38 &     1.38 &  4.63 &  9.76 &  25.7 &   0.4 &   1.9e+22 &     54287.3 &   181.3 &      202 &     0.4 & MWISP G105.389+00.275-052.19\\
		MWISP G105.665+04.784-008.20 & 105.665 &   4.784 &      221 &      106 &    66 &    -8.20 &     0.48 &  0.87 &  0.75 &  17.7 &   0.5 &   1.8e+21 &       185.0 &   104.5 &     1520 &     1.1 & MWISP G106.529+04.081-007.04\\
		MWISP G105.751+00.778-003.26 & 105.751 &   0.778 &      986 &      515 &    61 &    -3.26 &     0.70 &  0.41 &  1.67 &  10.8 &   0.8 &   2.0e+22 &       451.3 &    51.7 &      338 &     2.1 & MWISP G106.122+00.579-002.93\\
		MWISP G105.774+02.876-004.31 & 105.774 &   2.876 &      265 &      132 &    87 &    -4.31 &     0.91 &  0.87 &  0.92 &   9.7 &   0.6 &   1.4e+21 &       144.9 &    54.5 &      647 &     6.1 & MWISP G106.529+04.081-007.04\\
		MWISP G105.792+01.622-023.18 & 105.792 &   1.622 &      390 &      174 &  -141 &   -23.18 &     0.93 &  1.73 &  2.57 &  11.2 &   0.7 &   4.3e+21 &      1719.0 &    83.0 &      353 &     1.5 & MWISP G106.064+01.491-023.45\\
		MWISP G105.793-00.577-037.50 & 105.793 &  -0.577 &      132 &       49 &   102 &   -37.50 &     1.01 &  3.27 &  1.46 &  11.9 &   0.4 &   5.7e+20 &       807.9 &   120.3 &      898 &     2.1 & MWISP G105.798-00.603-036.76\\
		MWISP G105.821-00.098-054.00 & 105.821 &  -0.098 &      182 &       32 &  -178 &   -53.99 &     0.40 &  4.63 &  1.94 &  11.7 &   0.3 &   2.1e+20 &       598.0 &    50.5 &      284 &     0.6 & MWISP G105.389+00.275-052.19\\
		MWISP G105.837+00.260-051.09 & 105.837 &   0.260 &      182 &       69 &    82 &   -51.09 &     0.83 &  4.63 &  2.92 &  17.4 &   0.3 &   9.3e+20 &      2655.8 &    99.0 &      370 &     0.9 & MWISP G105.389+00.275-052.19\\
		MWISP G105.855+04.123-008.87 & 105.855 &   4.123 &      897 &      702 &    67 &    -8.87 &     1.47 &  0.87 &  3.92 &  31.4 &   0.4 &   9.0e+22 &      8987.6 &   185.8 &      517 &     1.1 & MWISP G106.529+04.081-007.04\\
		MWISP G106.116+03.034-007.74 & 106.116 &   3.034 &      158 &       71 &    54 &    -7.74 &     0.50 &  0.87 &  0.52 &   9.4 &   0.3 &   2.0e+20 &        20.2 &    24.3 &      514 &     7.4 & MWISP G106.529+04.081-007.04\\
		MWISP G106.196+01.399-022.05 & 106.196 &   1.399 &      219 &      199 &    72 &   -22.06 &     1.01 &  1.73 &  2.05 &   9.4 &   0.5 &   1.0e+21 &       397.9 &    30.2 &      160 &     6.1 & MWISP G106.064+01.491-023.45\\
		MWISP G106.267+00.046-056.07 & 106.267 &   0.046 &      310 &      162 &    46 &   -56.07 &     1.03 &  4.87 &  6.20 &  14.8 &   0.3 &   1.7e+21 &      5260.1 &    43.5 &       76 &     1.5 & MWISP G106.342-00.084-055.73\\
		MWISP G106.325+01.239-028.09 & 106.325 &   1.239 &      126 &       53 &    59 &   -28.09 &     0.65 &  1.73 &  0.78 &   8.8 &   0.3 &   1.1e+20 &        44.8 &    23.2 &      323 &     8.5 & MWISP G106.064+01.491-023.45\\
		MWISP G106.339+03.276-006.58 & 106.339 &   3.276 &      582 &      309 &    56 &    -6.58 &     0.96 &  0.87 &  2.10 &  12.6 &   0.3 &   3.4e+21 &       344.2 &    24.9 &      129 &     6.6 & MWISP G106.529+04.081-007.04\\
		MWISP G106.460+00.896-053.92 & 106.460 &   0.896 &      217 &      103 &   135 &   -53.92 &     0.72 &  4.70 &  3.98 &  12.7 &   0.5 &   1.7e+21 &      5034.3 &   101.0 &      277 &     0.5 & MWISP G106.475+00.931-053.87\\
		MWISP G106.481+03.112-011.21 & 106.481 &   3.112 &      201 &      110 &   147 &   -11.21 &     0.67 &  0.87 &  0.73 &  17.2 &   0.8 &   5.4e+21 &       545.0 &   325.6 &     4873 &     0.7 & MWISP G106.529+04.081-007.04\\
		MWISP G106.523+00.507-056.03 & 106.523 &   0.507 &      157 &       72 &   170 &   -56.03 &     0.55 &  4.87 &  2.92 &   8.9 &   0.4 &   3.8e+20 &      1207.1 &    45.1 &      168 &     0.8 & MWISP G106.440+00.388-055.76\\
		MWISP G106.534+01.262-026.82 & 106.534 &   1.262 &      192 &       88 &   175 &   -26.82 &     0.69 &  1.73 &  1.27 &   9.2 &   0.5 &   7.6e+20 &       302.1 &    59.8 &      515 &     2.3 & MWISP G106.064+01.491-023.45\\
		MWISP G106.554+00.932-012.42 & 106.554 &   0.932 &      350 &      275 &    81 &   -12.42 &     0.78 &  0.65 &  1.15 &  15.0 &   0.7 &   1.4e+22 &       781.1 &   188.1 &     1787 &     1.0 & MWISP G106.663+01.012-011.66\\
		MWISP G106.562+01.048-061.18 & 106.562 &   1.048 &      310 &      196 &   157 &   -61.18 &     0.87 &  5.34 &  7.50 &  19.4 &   0.3 &   4.2e+21 &     16173.9 &    91.5 &      133 &     0.4 & MWISP G106.492+01.037-061.09\\
		MWISP G106.693+03.364-007.26 & 106.693 &   3.364 &      304 &      204 &   106 &    -7.26 &     0.53 &  0.87 &  1.23 &   8.6 &   0.5 &   8.1e+20 &        81.4 &    17.2 &      152 &     4.8 & MWISP G106.529+04.081-007.04\\
		\enddata
		\tablecomments{Same as Table \ref{tab2} but for \COl clouds. Columns 2$-$6 give the centroid positions, the intensity-weighted major and minor axes, and the position angles of the clouds. The centroid velocity, the velocity dispersion, and the distance of the clouds are presented in columns 7$-$9. Columns 10$-$17 list the derived parameters of the clouds, i.e., effective radius, excitation temperature, the optical depth of \COl emission, average column density, mass, surface density, number density, and the virial parameters of the clouds, respectively. The last column gives the name of the \CO cloud that matches the \COl cloud in column 1. \\This table is available in its entirety in machine$-$readable form in the online material.}
	\end{deluxetable*}
\end{longrotatetable}

\subsubsection{Size and Mass of the Clouds}\label{sec3.4.2}
The histograms of the effective radius of the \CO and \COl clouds are given in Figures \ref{fig11b} and \ref{fig12b}, respectively. More than half of the \CO and \COl clouds have sub-pc sizes. In the conventional classification scheme of molecular gas structures, these molecular entities should be classified as molecular clumps, but, for simplicity, we still refer to them as ``clouds" in this work. The median radius of the \CO molecular clouds in the Local and Perseus arms are $\sim$0.5 and $\sim$3.0 pc, respectively, while the values are $\sim$0.7 and $\sim$2.8 pc for the \COl molecular clouds. The distributions of the mass of the \CO and \COl clouds are presented in Figures \ref{fig11c} and \ref{fig12c}. The median mass of \CO molecular clouds in the Local, Perseus, and Outer$+$OSC arms is $\sim$20 M$_{\odot}$, $1\times10^3$ M$_{\odot}$, and $2\times10^3$ M$_{\odot}$, respectively. The median mass of the \COl clouds in the Local and Perseus arms is 70 and $2\times10^3$ M$_{\odot}$, respectively, which is moderately larger than that of the \CO clouds. Generally, the molecular clouds in the Perseus and Outer$+$OSC arms are much larger and more massive than the molecular clouds in the Local arm. The possible reasons for the systematic difference between the median radii and masses in the Local and Perseus arms are as follows.

\begin{figure*}[htb!]
	\centering
	\subfigure[]{
		\label{fig10a}
		\includegraphics[trim=0cm 1cm 0cm 1cm, width = 0.4\linewidth , clip]{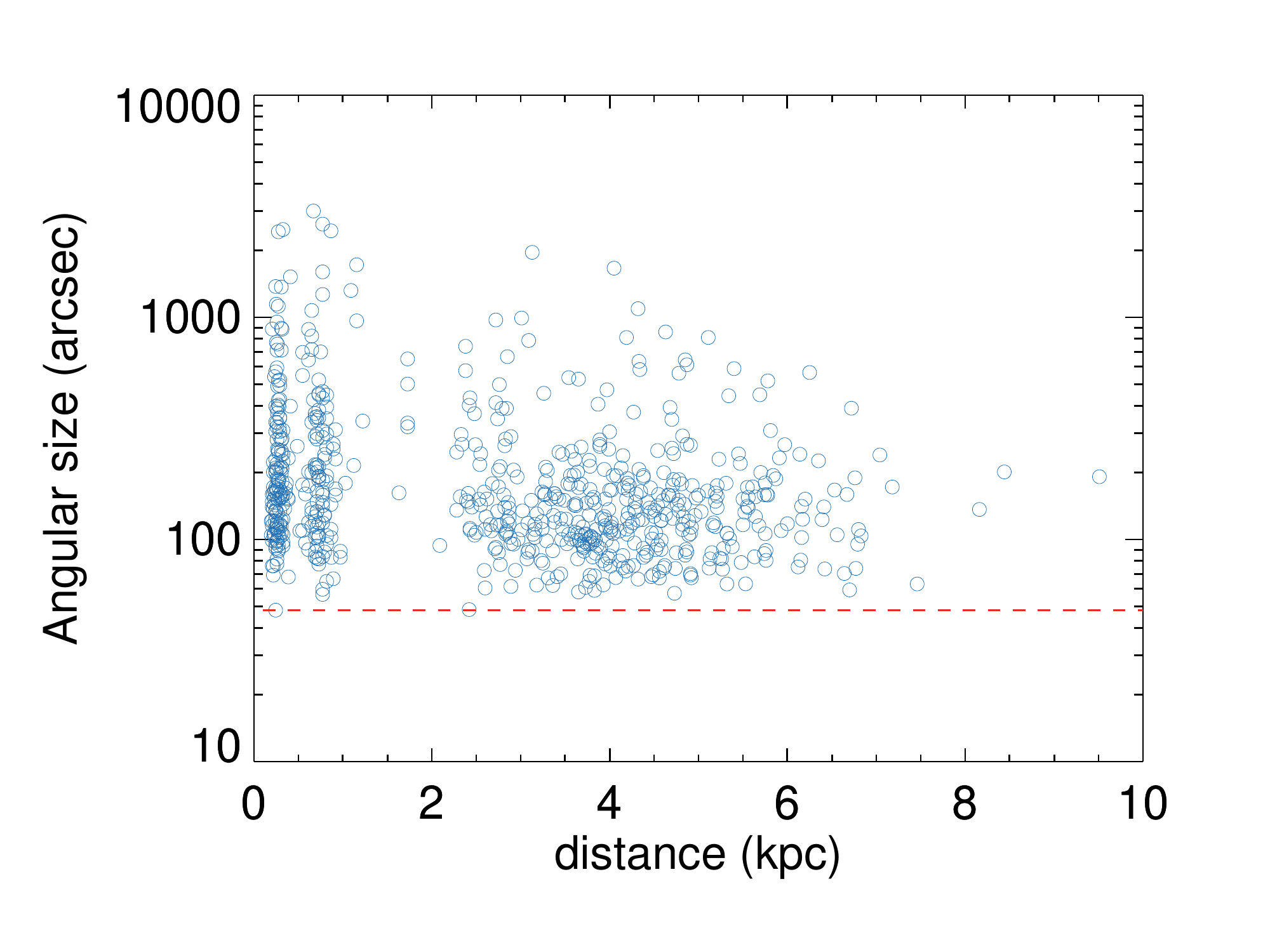}}
	\subfigure[]{
		\label{fig10b}
		\includegraphics[trim=0cm 1cm 0cm 1cm, width = 0.4\linewidth , clip]{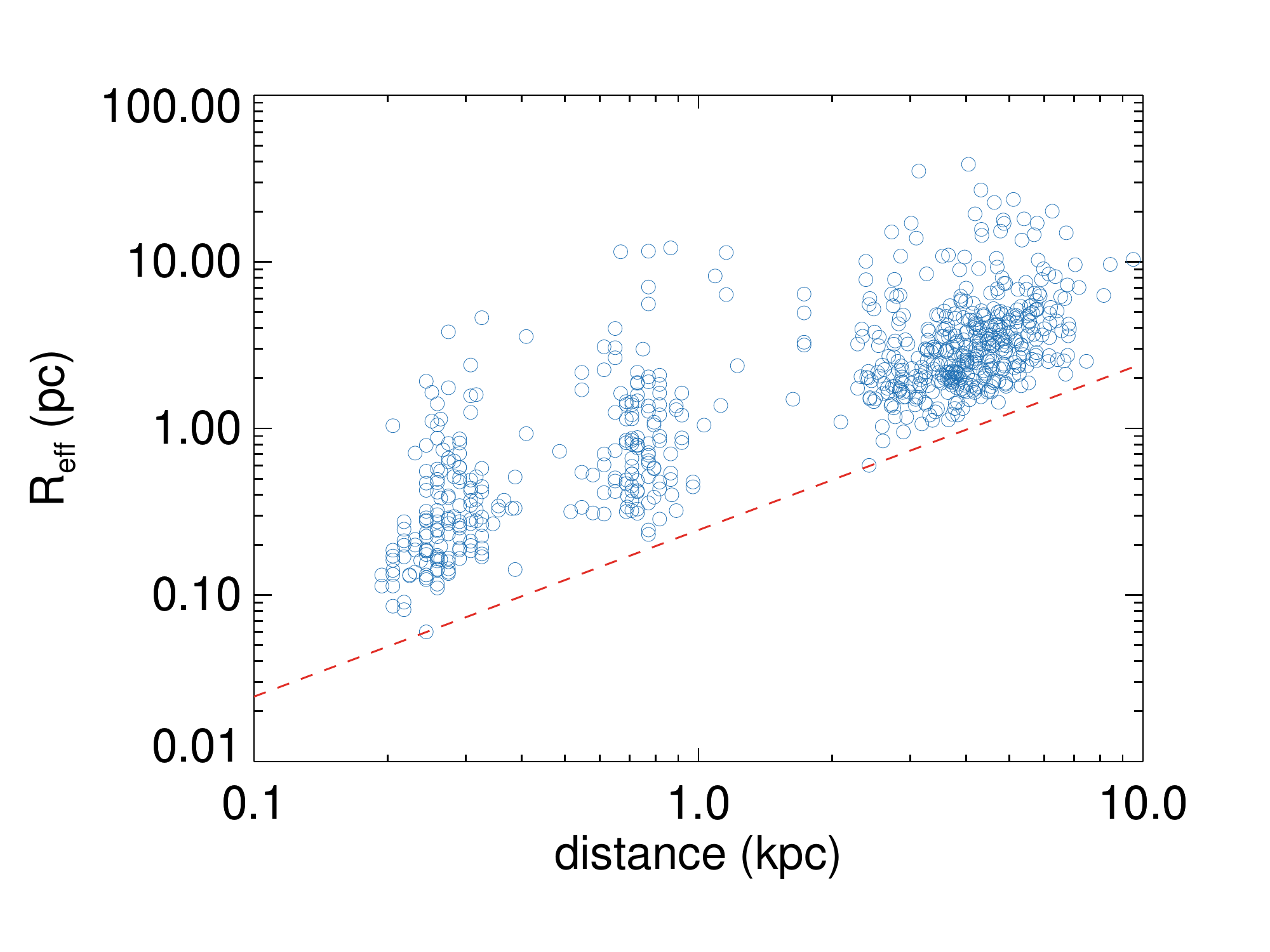}}
	\subfigure[]{
		\label{fig10c}
		\includegraphics[trim=0cm 1cm 0cm 1cm, width = 0.4\linewidth , clip]{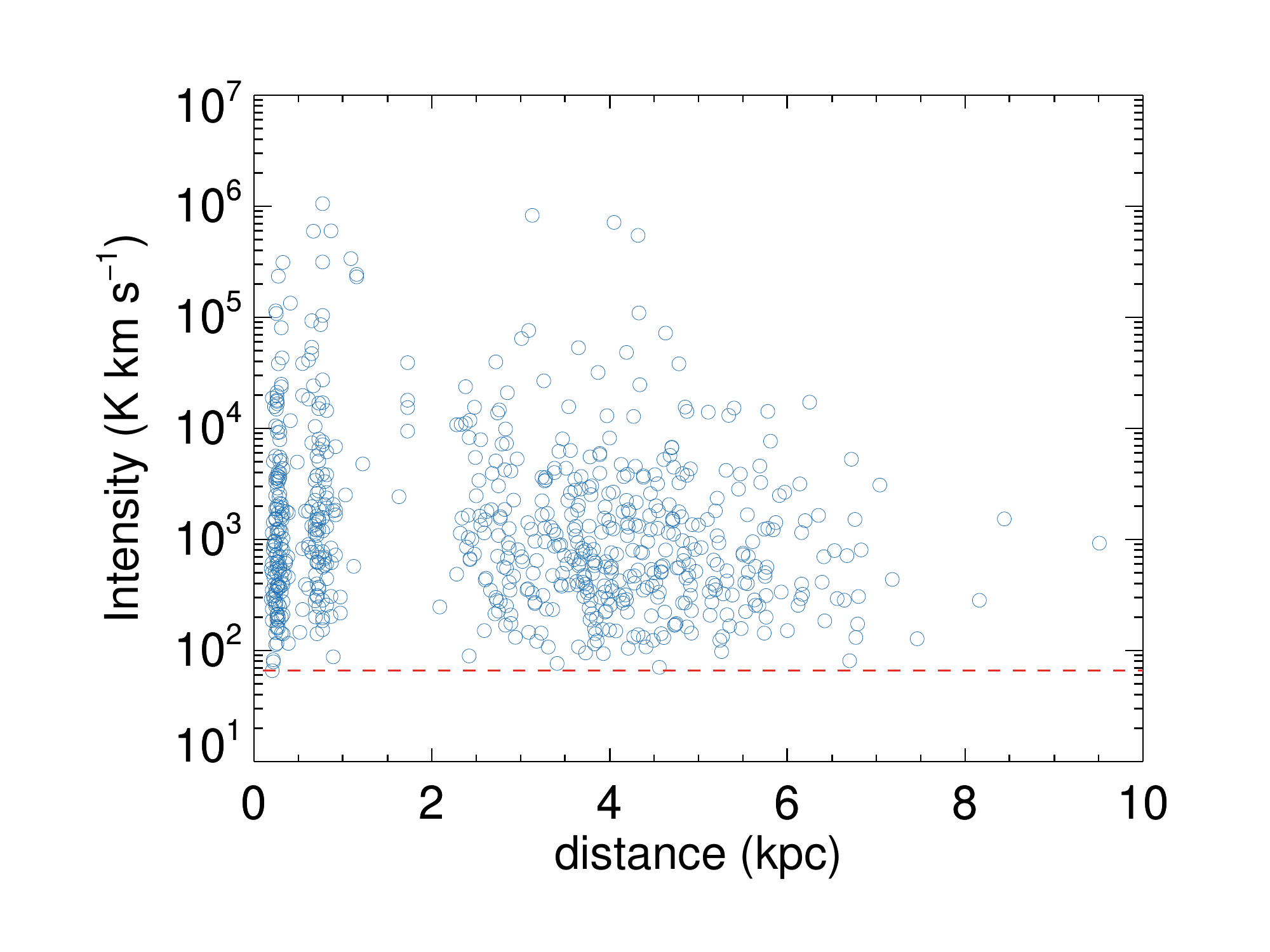}}
	\subfigure[]{
		\label{fig10d}
		\includegraphics[trim=0cm 1cm 0cm 1cm, width = 0.4\linewidth , clip]{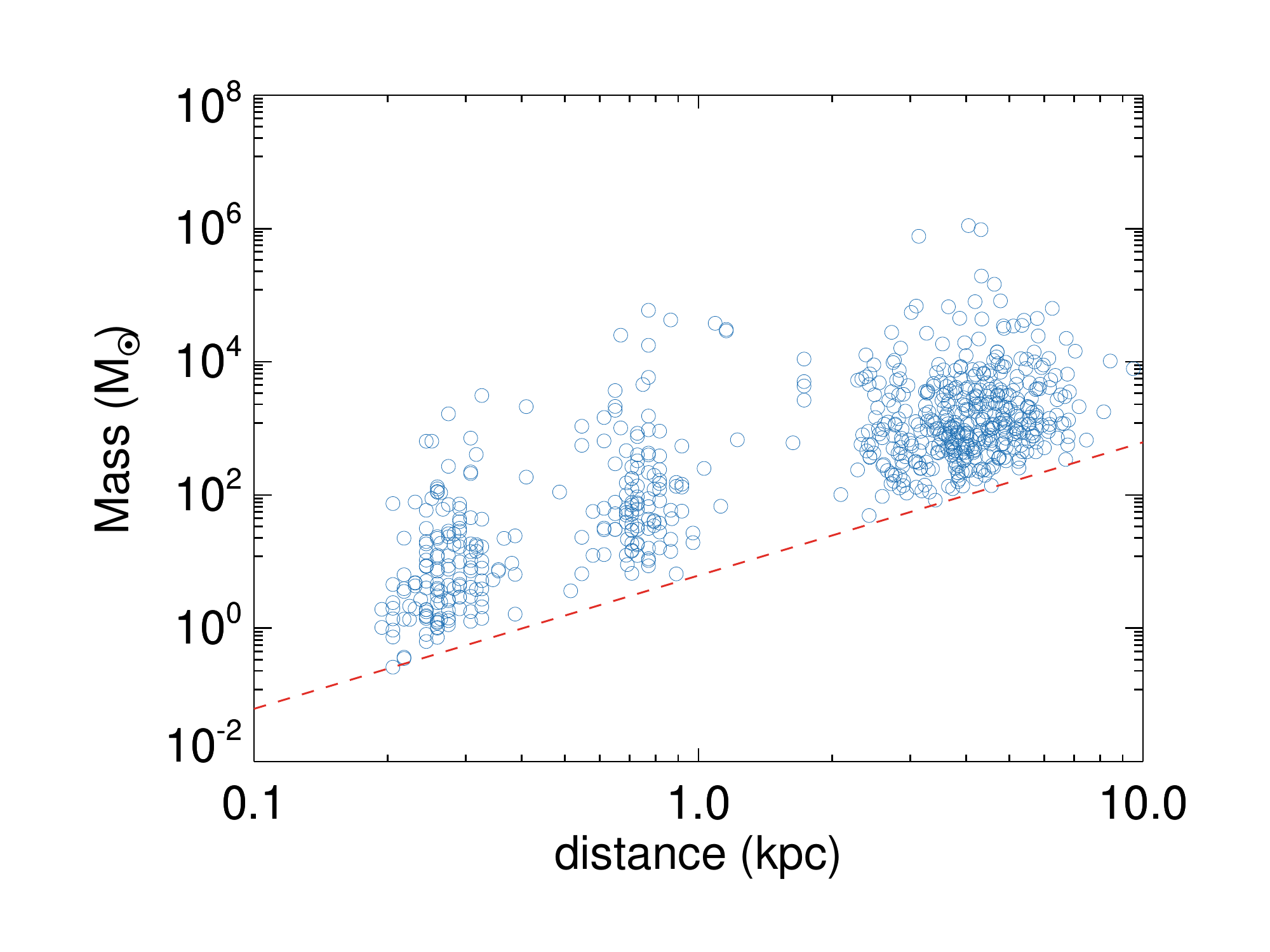}}
	\caption{Variation versus distance of (a) the angular size, which is defined as the geometric mean of the intensity weighted second-moment of the spatial scale along the major and minor axes, (b) the effective radius, (c) the total integrated intensity, and (d) the mass of the \CO clouds. The red dashed lines in panels (a) and (c) represent the minimum angular size and minimum total integrated-intensity that detected from nearby to a distance of 7.5 kpc. The red dashed lines in panels (b) and (d) are the minimum effective radius and minimum mass corresponding to the minimum angular size and total integrated-intensity at different distances.}\label{fig10}
\end{figure*}

\begin{figure*}[htb!]
	\centering
	\subfigure[]{
		\label{fig11b}
		\includegraphics[trim=0cm 0cm 0cm 0cm, width = 0.32\linewidth, clip]{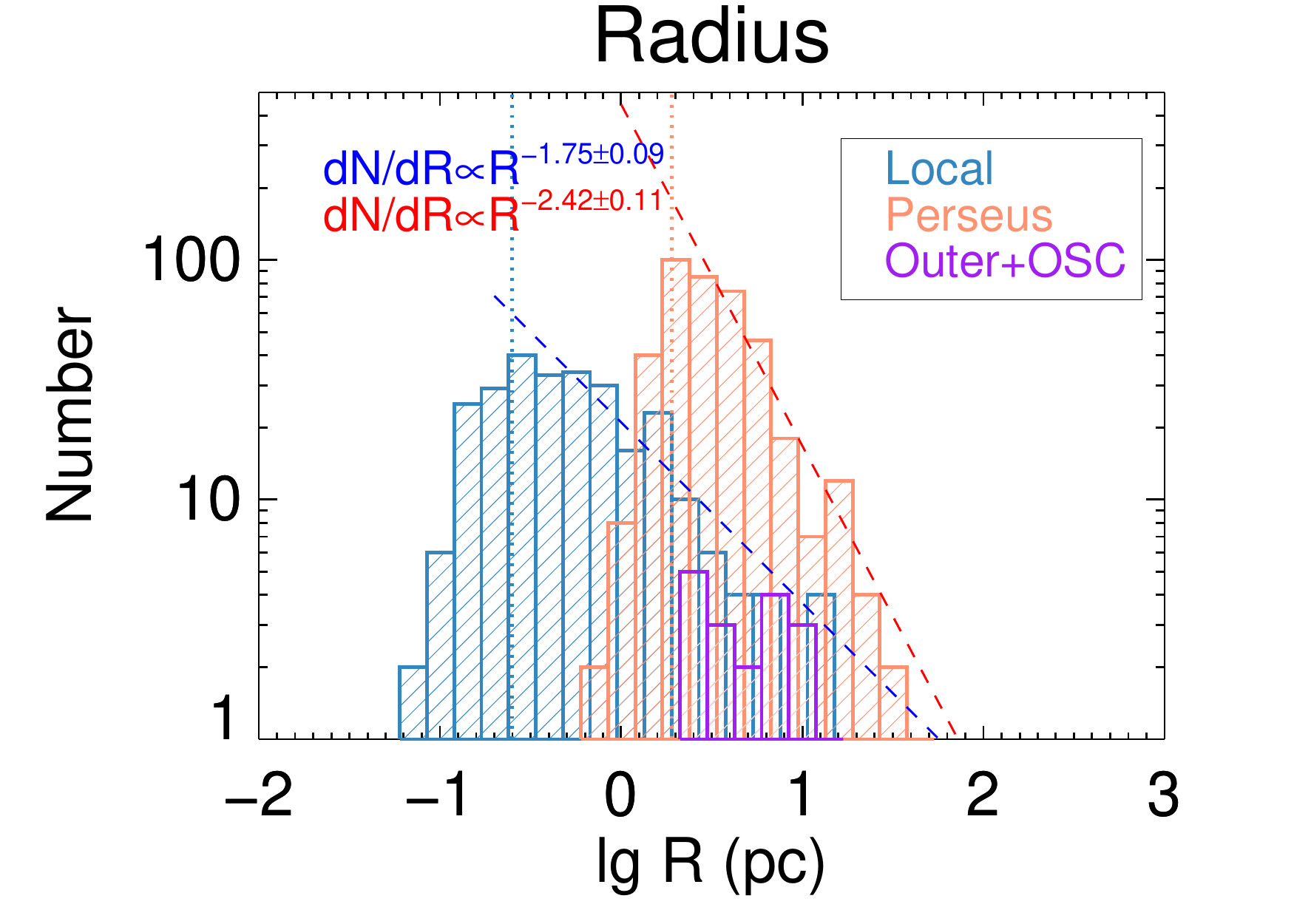}}
	\subfigure[]{
		\label{fig11c}
		\includegraphics[trim=0cm 0cm 0cm 0cm, width = 0.32\linewidth, clip]{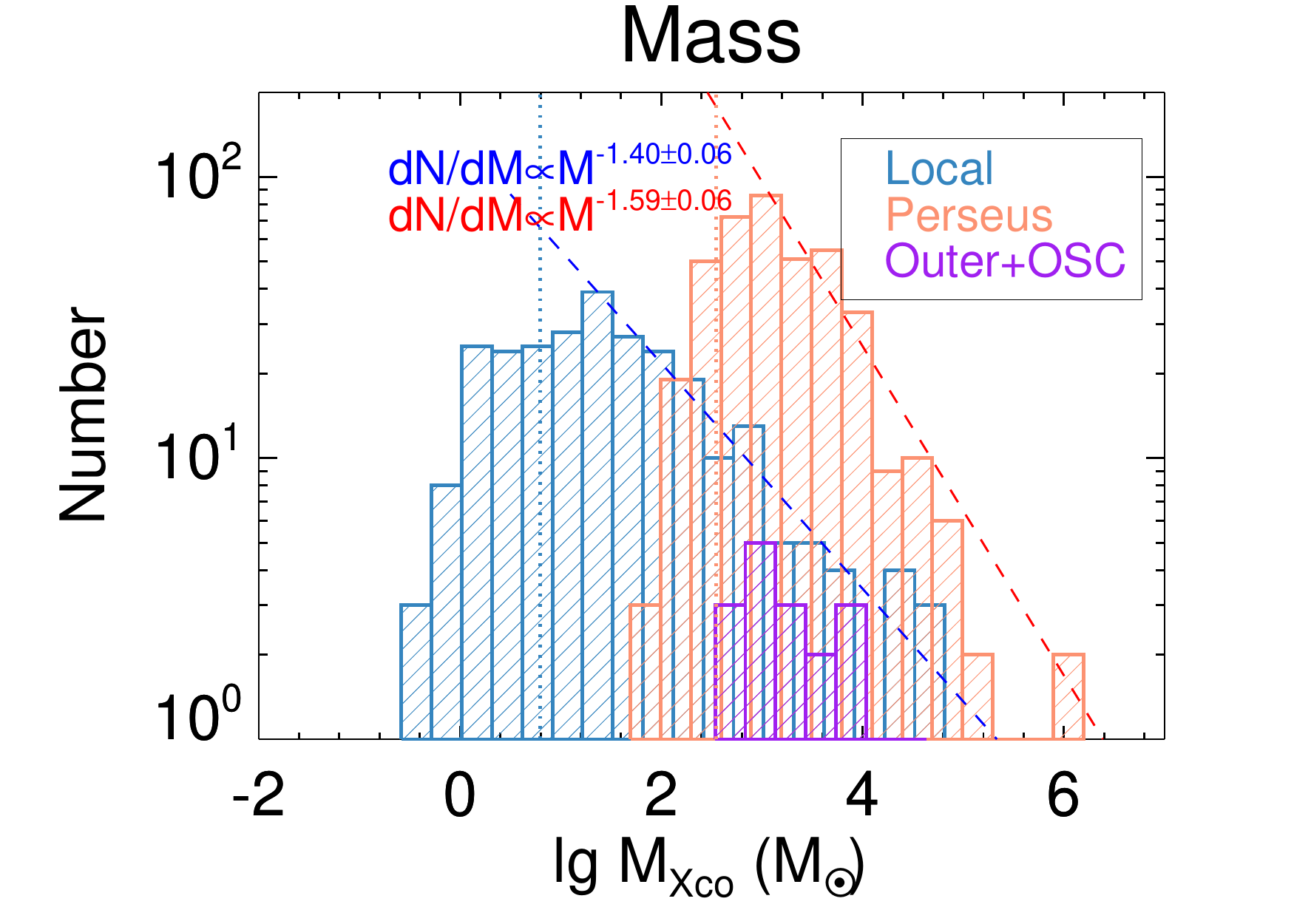}}
	\subfigure[]{
		\label{fig11d}
		\includegraphics[trim=0cm 0cm 0cm 0cm, width = 0.32\linewidth, clip]{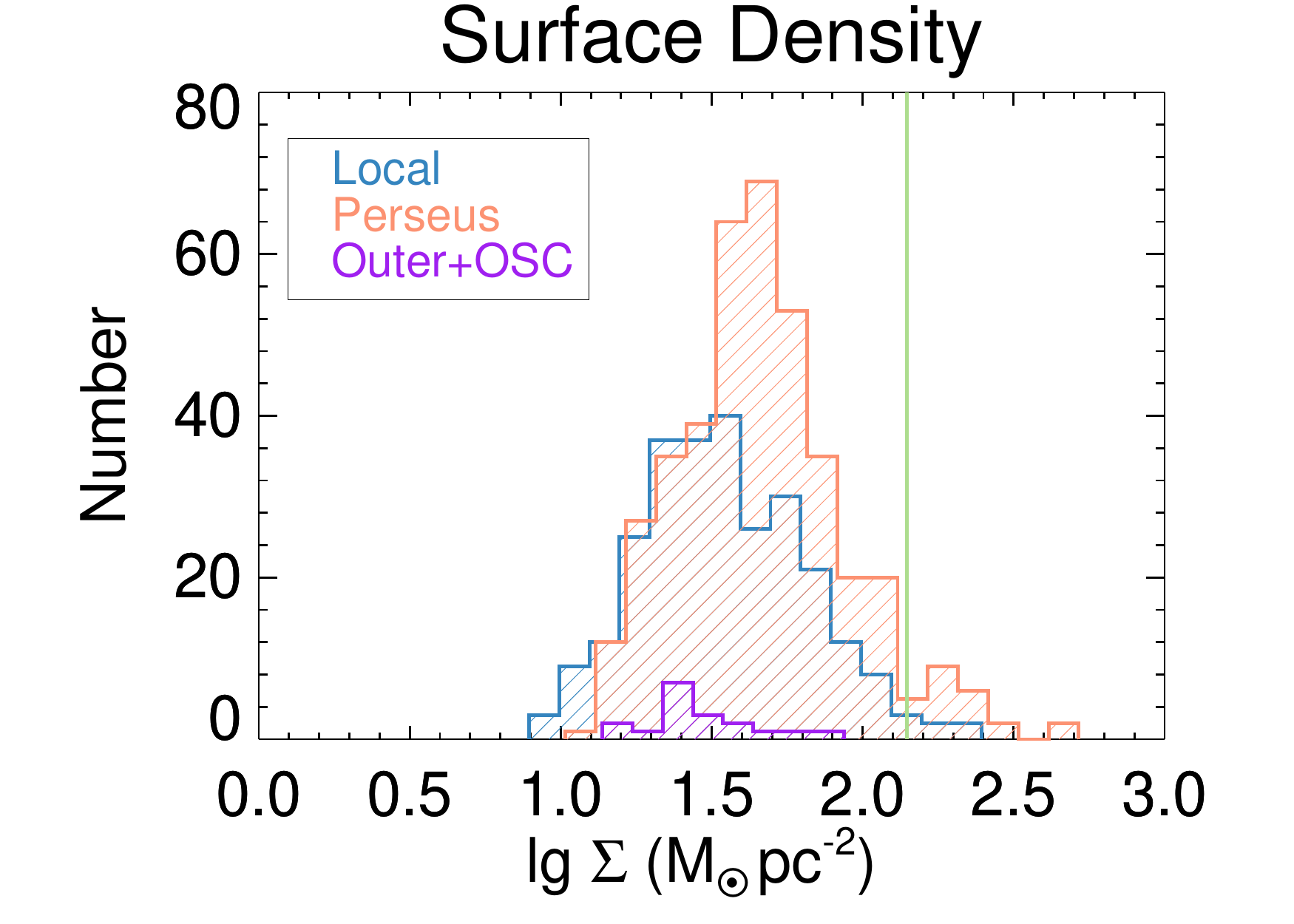}}
	\subfigure[]{
		\label{fig11e}
		\includegraphics[trim=0cm 0cm 0cm 0cm, width = 0.32\linewidth, clip]{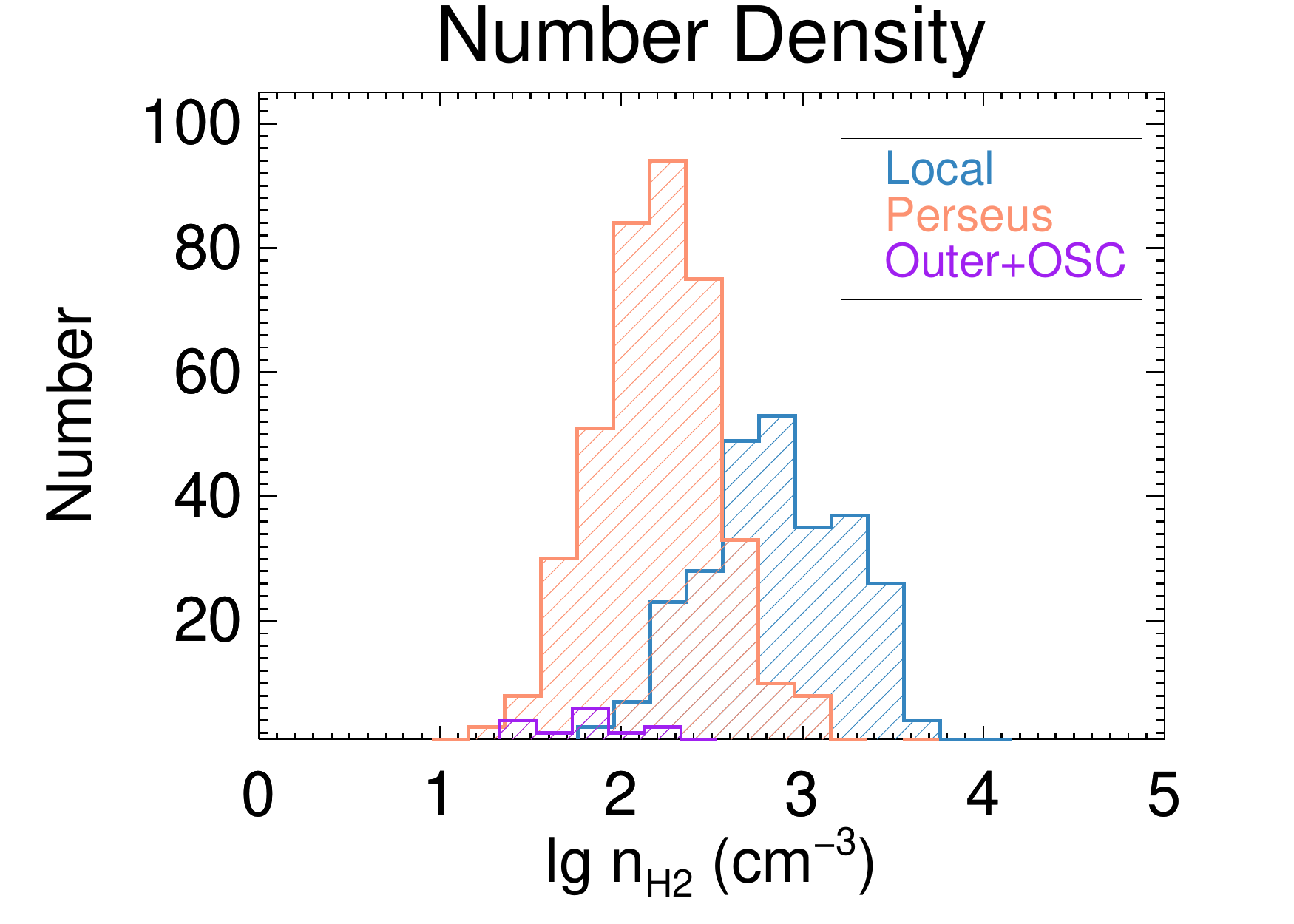}}
	\subfigure[]{
		\label{fig11a}
		\includegraphics[trim=0cm 0cm 0cm 0cm, width = 0.32\linewidth, clip]{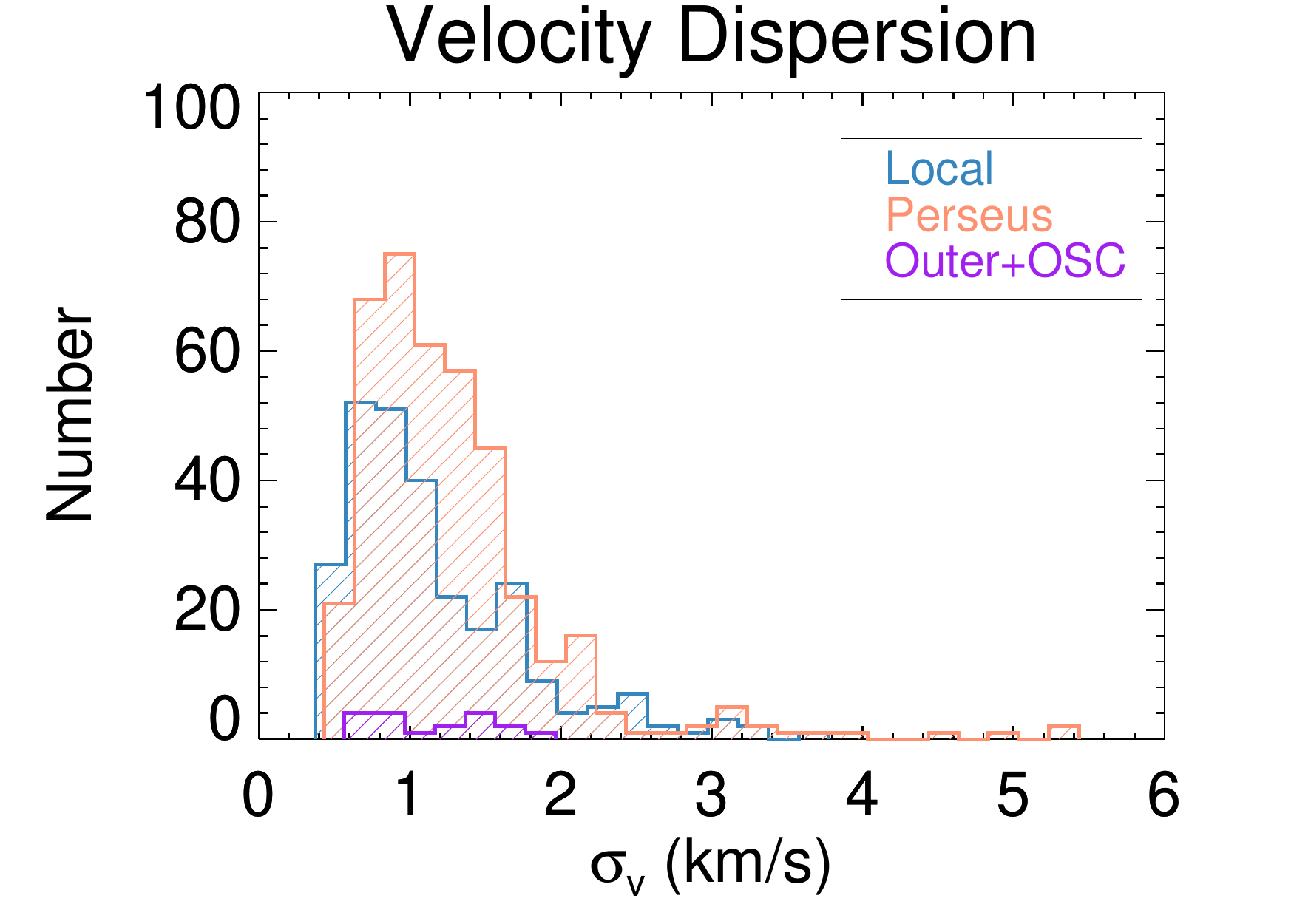}}
	\caption{Distributions of (a) effective radii, (b) masses from the X$_{\rm CO}$ method, (c) mass surface densities, (d) number densities, and (e) velocity dispersions, of the identified \CO clouds. The red and the blue dashed lines in panels (a) and (b) give the power-law fitting of the distribution of the radius and the mass of the clouds respectively, with the blue dashed lines for the clouds in the Local arm and the red dashed lines for the clouds in the Perseus arm. The blue and red dotted vertical lines in panel (a) and (b) indicate the minimum values of the radius and mass at the distances of the clouds in the Local and the Perseus arms, respectively. The green vertical line in panel (c) marks the surface density corresponding to the star formation threshold of column density, i.e., $6.3\times10^{21}$ cm$^{-2}$ \citep{Johnstone2004, Lada2010, Kainulainen2014}.}\label{fig11}
\end{figure*}

\begin{figure*}[htb!]
	\centering
	\subfigure[]{
		\label{fig12b}
		\includegraphics[trim=0cm 0cm 0cm 0cm, width = 0.32\linewidth , clip]{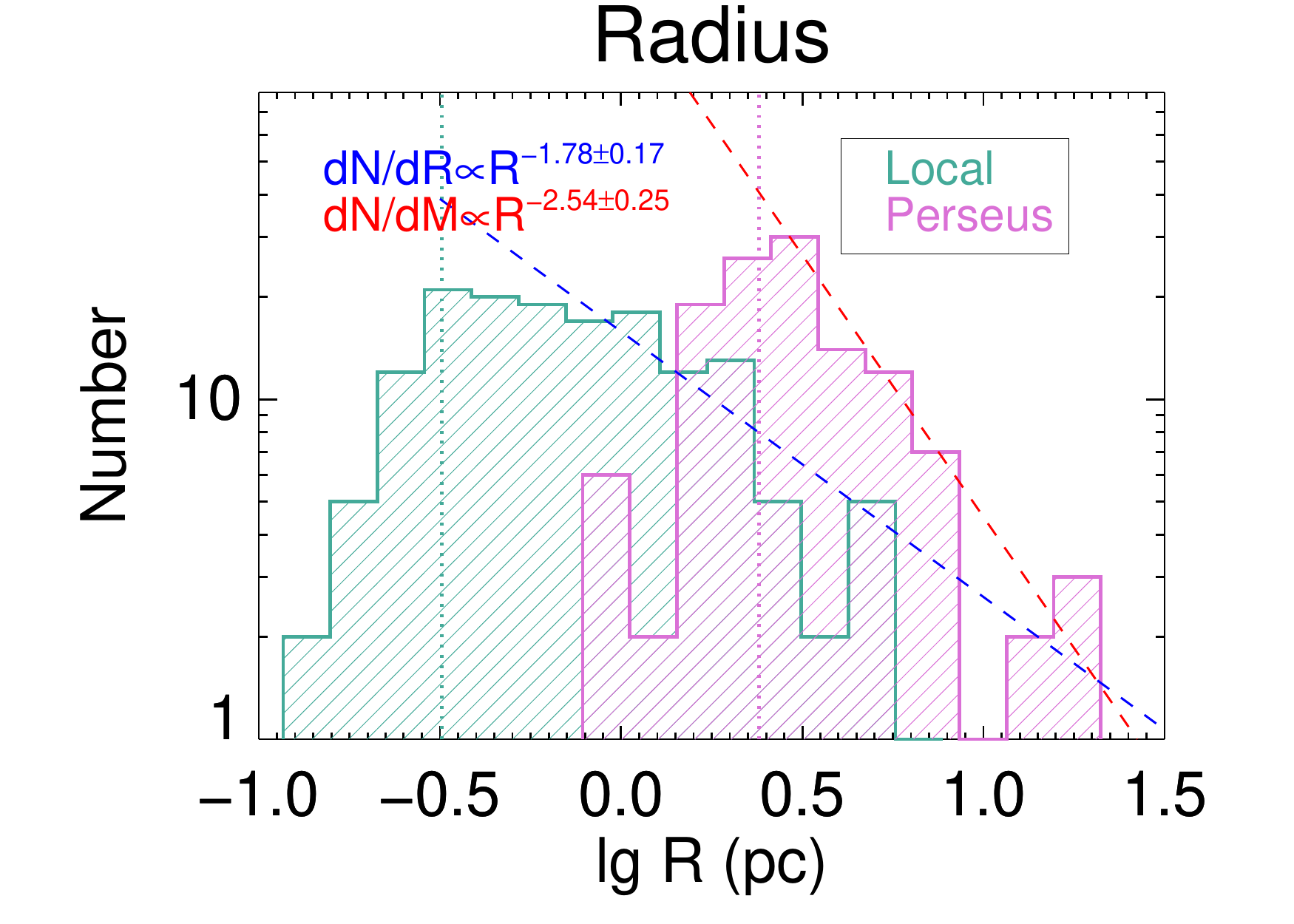}}
	\subfigure[]{
		\label{fig12c}
		\includegraphics[trim=0cm 0cm 0cm 0cm, width = 0.32\linewidth , clip]{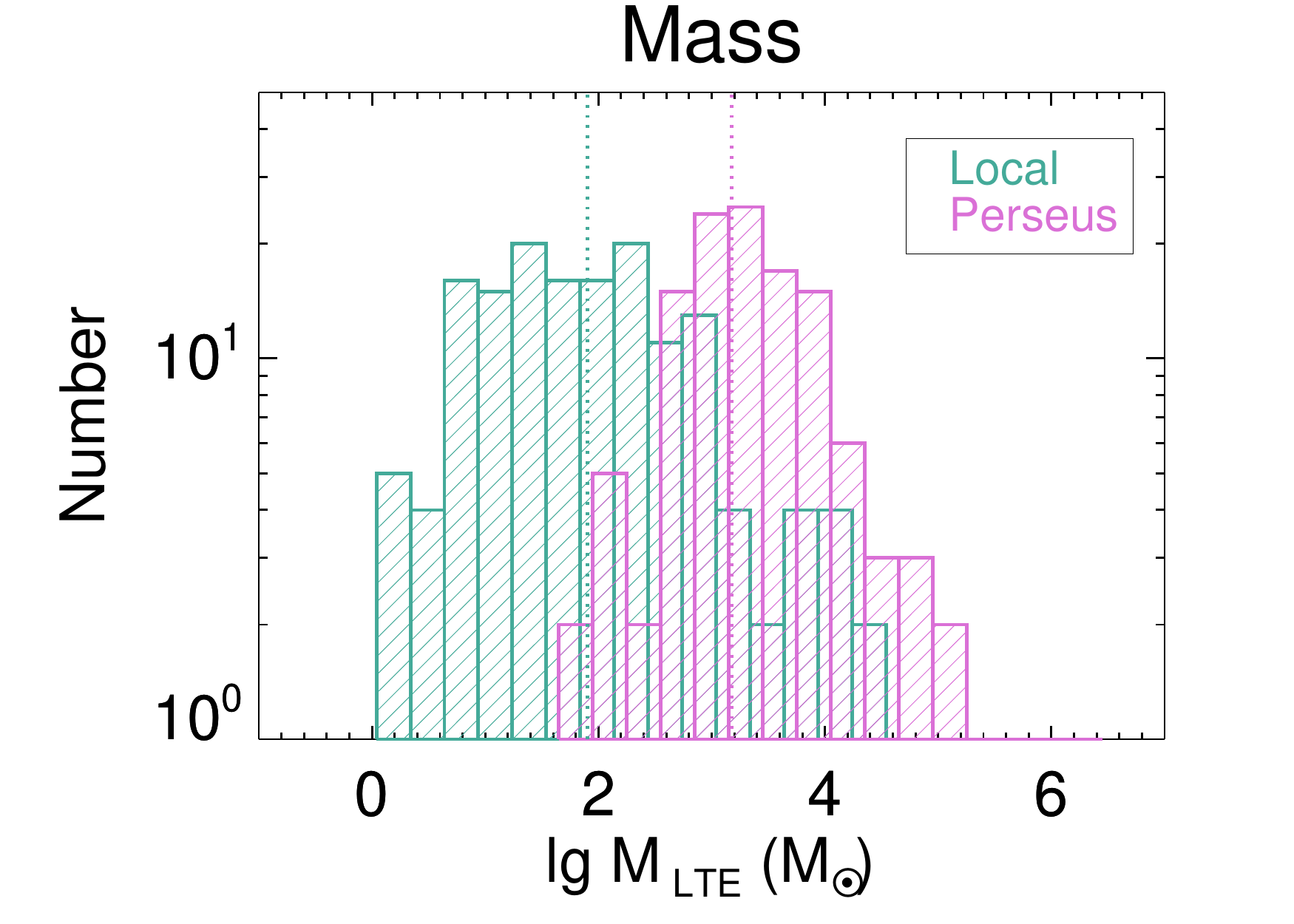}}
	\subfigure[]{
		\label{fig12d}
		\includegraphics[trim=0cm 0cm 0cm 0cm, width = 0.32\linewidth , clip]{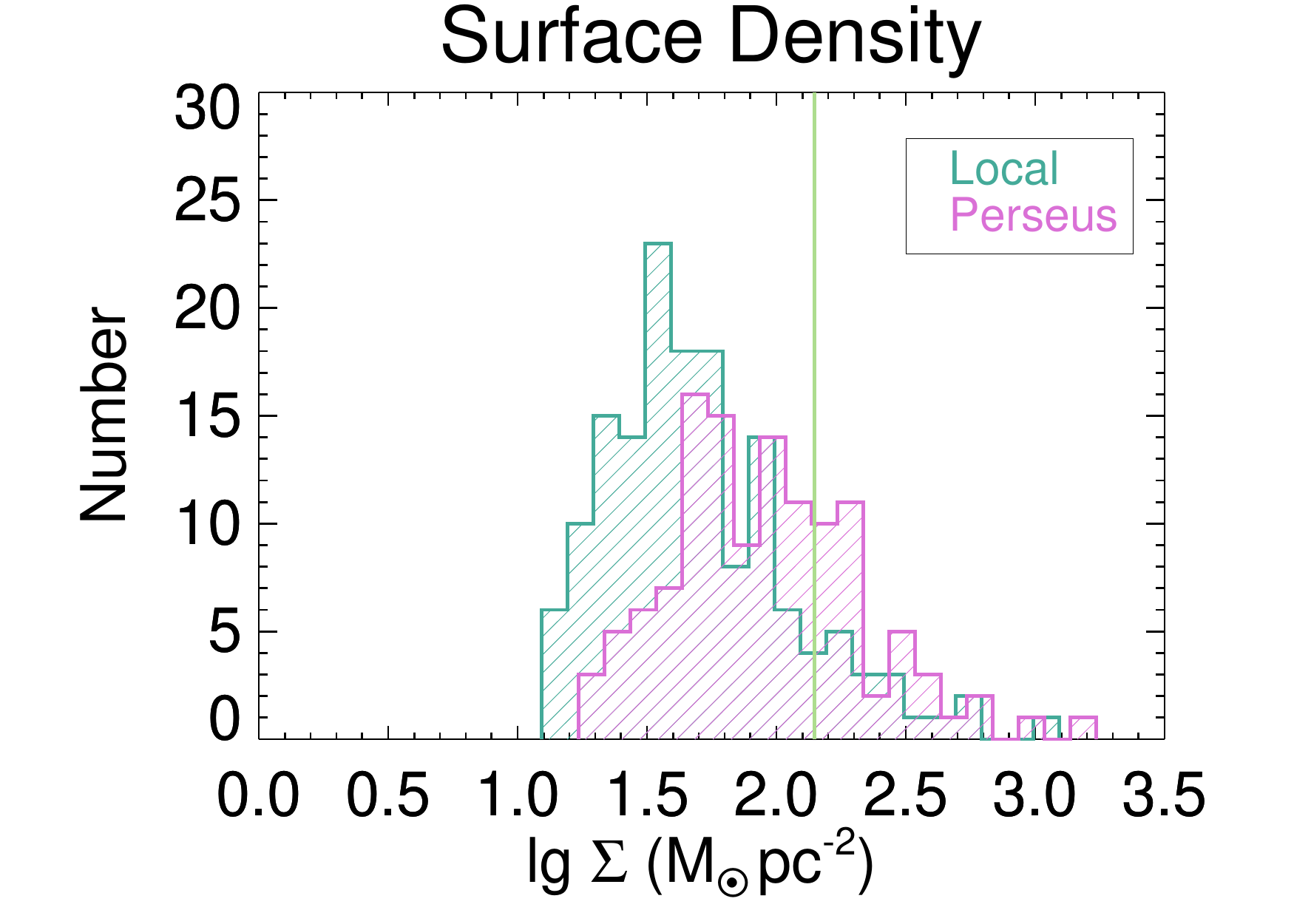}}
	\subfigure[]{
		\label{fig12e}
		\includegraphics[trim=0cm 0cm 0cm 0cm, width = 0.32\linewidth , clip]{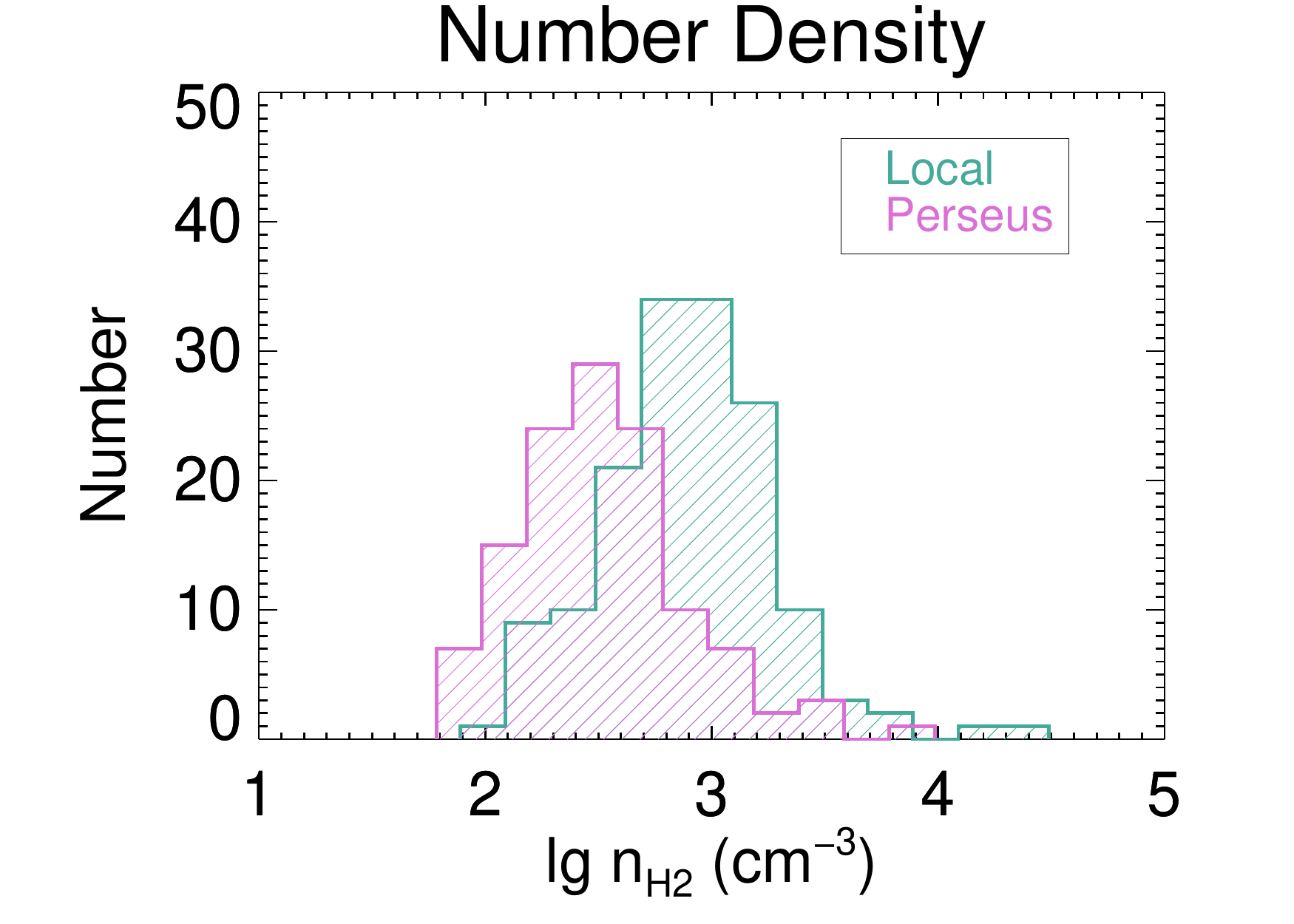}}
	\subfigure[]{
		\label{fig12a}
		\includegraphics[trim=0cm 0cm 0cm 0cm, width = 0.32\linewidth , clip]{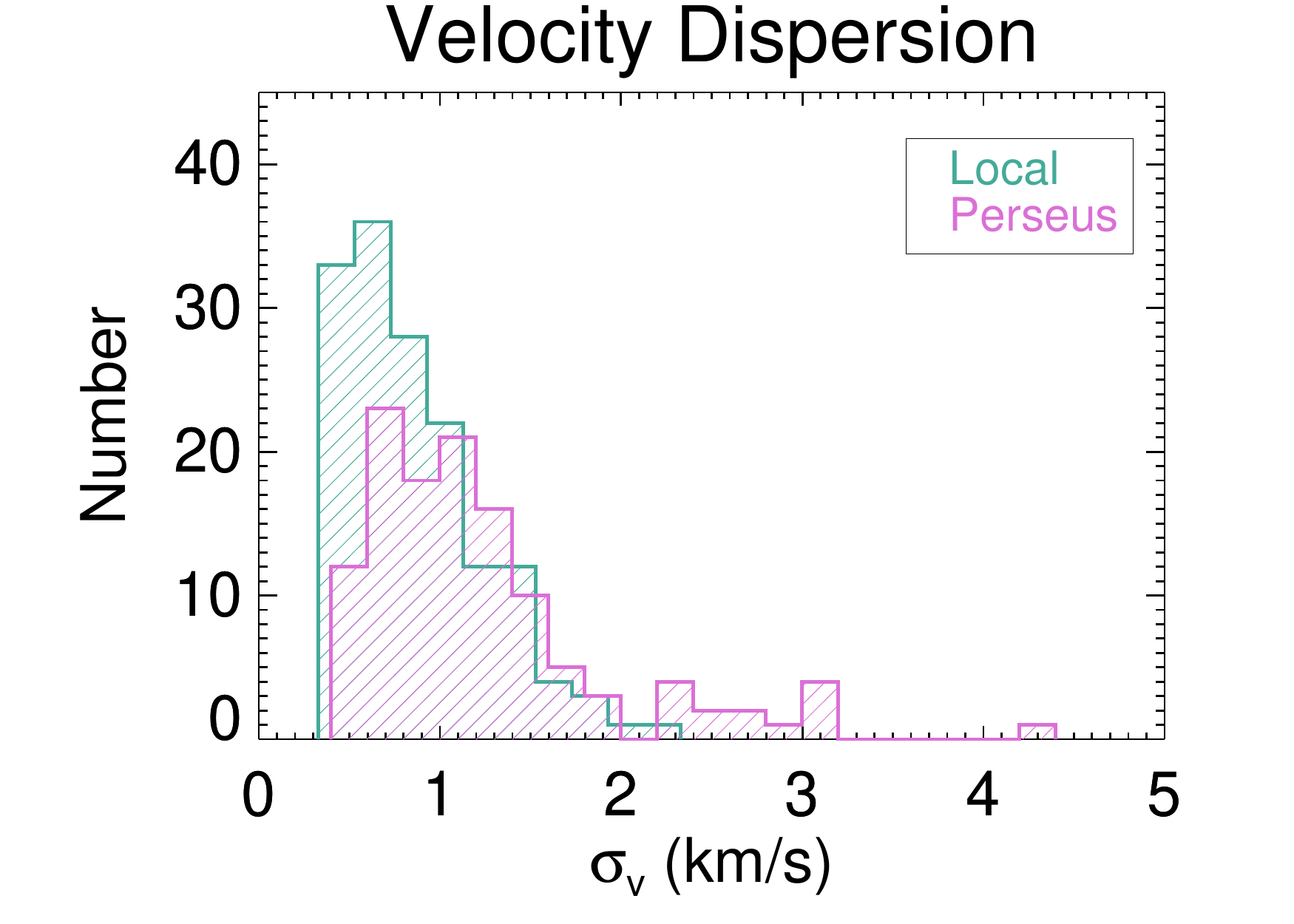}}
	\caption{Same as Figure \ref{fig11}, but for \COl clouds. The mass of the \COl clouds is calculated using the LTE method.}\label{fig12}
\end{figure*}

First, the distance selection effect, i.e., only the molecular clouds that are larger and more massive than the sensitivity limit can be detected, regardless of their distance. The output molecular clouds of the SCIMES algorithm have similar minimum angular sizes, $\sim$50 \arcsec, and similar minimum total integrated-intensities, $\sim$66 K km s$^{-1}$, at different distances, as shown in Figures \ref{fig10a} and \ref{fig10c}. The minimum angular size and integrated-intensity correspond to the minimum effective radius and mass that increase with distance, therefore causes the non-detection of the smaller and less massive clouds at the higher distances. Figures \ref{fig10b} and \ref{fig10d} shows the variation of the effective radius and the mass of the \CO clouds as a function of the cloud distance. The minimum effective radius at 1 and 7.5 kpc, taken as the reference upper limit distances of the Local and Perseus arms, is $\sim$0.25 and $\sim$1.9 pc, respectively, as shown in Figure \ref{fig10b}, while the minimum mass at these distances is $\sim$6 and $\sim$349 M$_{\sun}$, respectively. The finite spatial resolution of our observations introduces an additional bias due to the fact that the minimum projected angular distance on the sky between two clouds that allows us to identify them as separate structures increases with their distance. For this reason, clouds close to each other are seen as one larger cloud at higher distances. Second, different methods are used in the distance estimation for the clouds in the two arms. We find that the distance estimated with the dust extinction map is usually smaller than the kinematic distance. In addition, the distances of the molecular clouds in the Perseus arm are also overestimated because of the kinematic abnormality of the Perseus arm. Third, same settings and criteria are used with the SCIMES algorithm to identify clouds in different spiral arms. It can be seen from Figures \ref{fig10a} and \ref{fig10c} that the clouds identified by the SCIMES algorithm have similar angular sizes and total integrated intensities in the Local and Perseus arms, therefore, the molecular clouds are inevitably less massive in the Local arm than in the Perseus arm. 

In the concept that molecular clouds are overdense sub-structures of the turbulent, multi-phase fractal ISM \citep{Scalo1988}, the distributions both of the size and the mass of the clouds possess power-law forms, $dN/dR\propto R^{\alpha_{R}}$ $dN/dM\propto M^{\alpha_{M}}$, where $dN$ is the number of clouds in the interval of $dR$ or $dM$. Instead of fitting $dN$ in equally separated radius bins, we fitted the linear relationship between $\lg dN$ and $\lg R$ (and $\lg M$) taking account of statistic errors, where $dN$, in this case, is the number in each logarithmic bin. The resulting exponent in the relation $dN/d(lg R)\propto R^{\beta}$ is related to $\alpha_{R}$ through $\beta = 1 + \alpha_R$, and the same for the mass. The minimum effective radii (or mass) of the \CO clouds at the reference position of the Local and Pesrus arms are near the peaks of their $\lg R$ (or $\lg M$) distributions. Therefore, we consider them as a reasonable estimate of the completeness limit of the radius (or mass) distribution, and fitted the distribution from the center of the bin with the peak values of $dN$ to the center of the bin of the largest $\lg R$ (or $\lg M$). 

As shown in Figure \ref{fig11b}, the measured exponent, $\alpha_{R}$, in the Local arm is $-1.75$, while in the Perseus arm is $-2.42$. The exponents for the radius distribution of the \COl clouds are slightly smaller than the above values, which are $-1.78$ and $-2.54$ for the Local and Perseus arms, respectively. The best-fit of the power-law distribution $dN/dM \propto M^\alpha_M$ is $\alpha_M = -1.40$ for the Local arm and $\alpha_M = -1.59$ for the Perseus arm. The radius distributions derived in this work are shallower than those of the OGS and GRS survey, $\alpha_{R}=-$3.2 in the OGS survey \citep{Heyer2001} and $\alpha_{R}=-$3.9 in the GRS survey \citep{Roman2010}, and also shallower than the result $\alpha_{R}\sim-3.3$ from \cite{Elmegreen1996} by compiling size measurements of clumps and clouds. The mass distribution is also shallower than those obtained by \cite{Rice2016} from molecular clouds in the outer Galaxy, $\alpha_M$ = $-$2.2. A possible explanation for the difference between our results and the above previous results is that different methods of cloud identification are used in these studies. For example, the CLUMPFIND algorithm used by \cite{Roman2010} tends to separate molecular clouds into several small-scale structures \citep{Li2020}. On the contrary, in this work, we aim at identifying large structures that contain at least 50 voxels in p-p-v space. Besides, the SCIMES algorithm uses a clustering process to merge the connecting ``leaves" into a cloud, tending to connect small structures into a large one. 

\subsubsection{Surface Density and Number Density}\label{sec3.4.4}
Figures \ref{fig11d} and \ref{fig11e} present the distributions of $\Sigma$ and $n_{H_2}$ of the \CO molecular clouds, and the corresponding distributions of the \COl molecular clouds are given in Figure \ref{fig12d} and \ref{fig12e}. The median mass surface densities of the \CO clouds in the Local, Perseus, and Outer$+$OSC arms are $\sim$33, $\sim$45, and $\sim$26 $\rm{M_{\sun}\ pc^{-2}}$, respectively, and the corresponding values of the \COl clouds in the Local and the Perseus arms are $\sim$44 and $\sim$89 $\rm{M_{\sun}\ pc^{-2}}$. In the Local arm, about 63\% of the total mass of the \CO clouds is contained in the clouds with surface densities above the threshold for star formation, $\rm{\sim 140\ M_{\sun}\ pc^{-2}}$ (corresponds to N$\rm_{H2}{\sim6.3\times10^{21}}$ cm$^{-2}$) \citep{Johnstone2004, Lada2010, Kainulainen2014}, and the percentage is 88\% in the Perseus arm. 

The maximum surface densities of the \CO and \COl molecular clouds are 431 and 1703 $\rm{M_{\sun}\ pc^{-2}}$, which correspond to H$_2$ column densities of 1.9$\times\ 10^{22}$ and 7.7$\times\ 10^{22}$ cm$^{-2}$, respectively. The maximum densities are located in the NGC 7538 GMC. \cite{Urquhart2013} suggested a lower limit of mass surface density of 0.05 g cm$^{-2}$ for massive star formation, which corresponds to H$_2$ column density $\sim$1.1$\rm{\times10^{22}\ cm^{-2}}$. The mass surface density of the NGC 7538 GMC is much higher than this limit, which is consistent with the observed concentration of CH$_3$OH masers within it \citep{Moscadelli2014}, testifying the presence of high-mass star formation activity in this region. We also checked the positions of other five \CO molecular clouds with mass surface densities above 0.05 g cm$^{-2}$ and found that these molecular clouds are all located in the Perseus arm and that they are associated with \ion{H}{2} regions or \ion{H}{2} region candidates \citep{Anderson2014}.

The distributions of the number density of the \CO and \COl molecular clouds in different spiral arms are shown in Figures \ref{fig11e} and \ref{fig12e}, respectively. The median values of $n_{H_2}$ of \CO molecular clouds in the Local, Perseus, and Outer$+$OSC arms are 648, 157, and 61 cm$^{-3}$, respectively, while the corresponding values of \COl molecular clouds in the Local and Perseus arms are 786 and 326 cm$^{-3}$. The measured number densities of molecular clouds in the Local arm is systematically higher than that in the Perseus and the Outer$+$OSC arms with both \CO and \COl as the tracer, which may indicate a bias and/or a systematic effect in the estimation of this parameter. This consideration is also supported by the fact that the median number density of the distant (d$>$2 kpc) molecular clouds is much below the critical density of the \CO and \COl $J = 1-0$ transition, $\sim$700 cm$^{-3}$. One possible reason is that the number density is a distance-dependent parameter. Therefore the same caveats listed in the previus subsection for mass and distance are valid also for the number density. In particular, it scales with $R_{eff}$ as $n\sim R_{eff}^{p}$, where $p$ is in the range from $\sim-$0.6 to $\sim-$0.8 in this work (see the scaling relations in Section \ref{sec3.5}). The dependence means that when the distance, so is the $R_{eff}$, is overestimated by a factor of two, which is the case for the massive star-forming region W3(OH) in the Perseus arm \citep{Xu2006}, the number density could be underestimated for $\sim$34\% to $\sim$43\%. However, even if this effect is corrected, the number density in the Perseus arm is still much less than the \CO critical density. Another possible factor to cause underestimation is that the volume filling factor of the \CO and \COl emission for the molecular clouds in the Perseus arm is low. In other words, there are fine internal structures in the molecular clouds that can not be resolved by the beam of the PMO-13.7m telescope at the distance of the Perseus arm. Metaphorically speaking, there could be ``holes" in the molecular clouds that cause dilution of the molecular line emission in the beam of the telescope.

\subsubsection{Velocity Dispersion}\label{sec3.4.5}
The derived total velocity dispersion within the boundary of a given cloud includes the contributions from the internal thermal motion and non-thermal turbulence. As shown in Figure \ref{fig11a}, the median velocity dispersions of \CO molecular clouds in the two spiral arms are similar, $\sim1.1\ \rm{km\ s^{-1}}$. However the proportion of the molecular clouds with velocity dispersions greater than 1.5 km s$^{-1}$ is larger in the Perseus arm than that in the Local arm. Most of the molecular clouds in the Perseus arm with $\sigma_v > 4\ \rm{km\ s^{-1}}$ are associated with or located in the vicinity of \ion{H}{2} regions like S 163, S 157, and NGC 7538, indicating complex dynamics in these regions. The median values of $\sigma_v$ of the \COl molecular clouds in the Local and Perseus arms are 0.8 and 1.1 km s$^{-1}$, respectively, similar to that of the \CO molecular clouds. The distribution and median values of $\sigma_v$ from this work is consistent with that of the molecular clouds in the solar circle \citep{Roman2010}, and the results from the OGS survey \citep{Heyer2001}. \cite{Miville2017} used a multi-Gaussian decomposition method to re-analysis the dataset from \cite{Dame2001}, and their results are nearly $\sim$2.1 times as large as our results. This difference may be caused by the coarse velocity resolution (1.3 km s$^{-1}$) of the CfA survey.

\subsubsection{Surface Density Variation Across the Galactocentric Distance}\label{sec3.4.6}
\begin{figure*}[htb!]
	\centering
	\includegraphics[trim=0cm 0.2cm 0.5cm 0.5cm, width= 0.6\linewidth, clip]{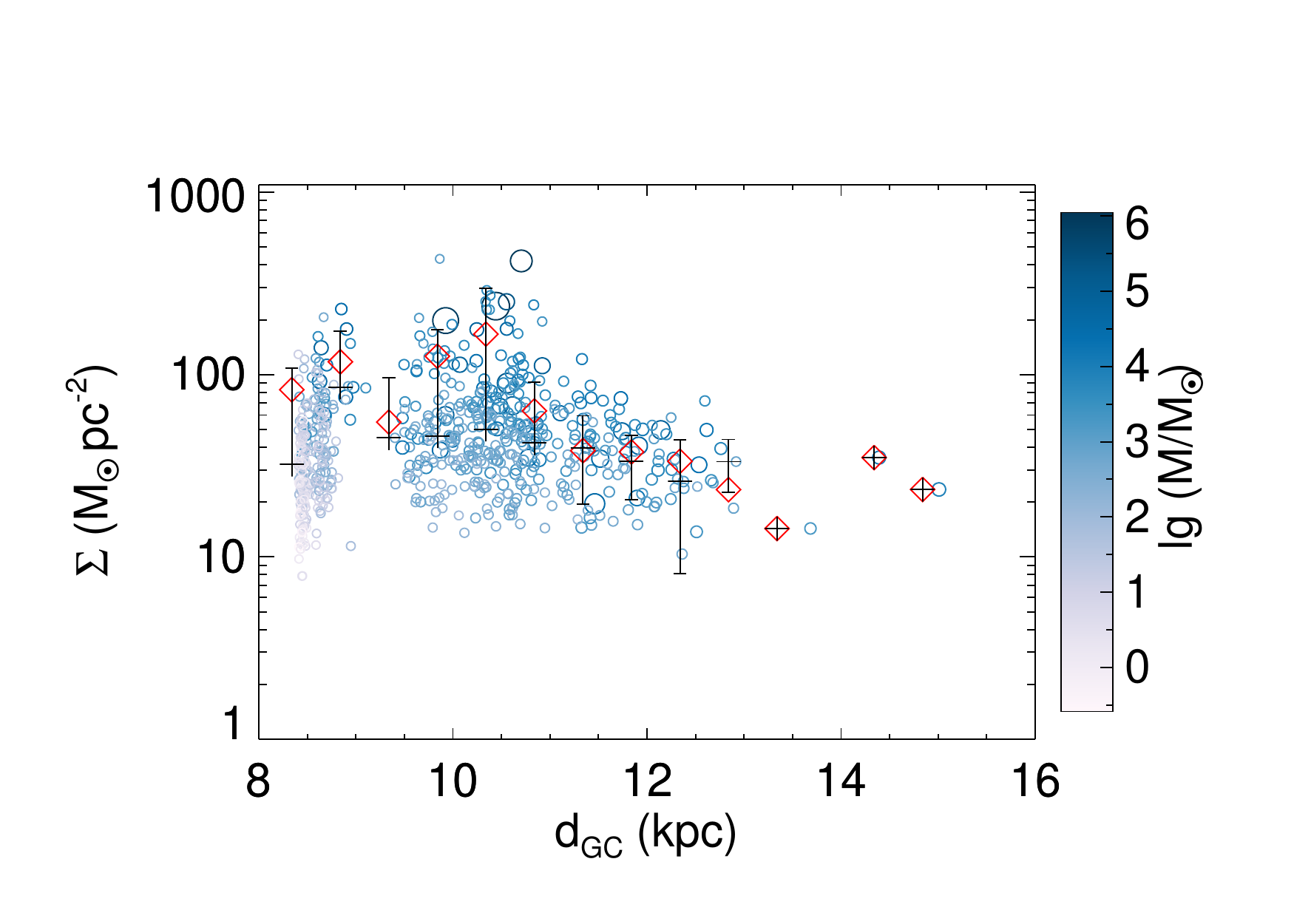}
	\caption{Variation of the surface densities of molecular clouds as a function of the galactocentric distances. The radii of the circles correspond to the sizes of the clouds and the color shades correspond to the masses of the clouds. The red diamonds are the average surface densities calculated through dividing the total mass with the total area of all the clouds in each interval of galactocentric distance of 0.5 kpc. The black pluses show the medium surface density in each interval. The scatter of the medium surface density in each interval is indicated in the figure, which is defined as the square root of the mass-weighted mean squared deviation from the medium surface density in each interval. The lower-ends of the error bars of the first six median values are negative and are not displayed.} 
	\label{fig13}
\end{figure*}
\cite{Heyer2015} have made from the literature a compilation of the cloud surface density $\Sigma$ across the Galactocentric distance and found that the surface densities of the molecular clouds in the inner Galaxy are significantly higher than those in the outer Galaxy. Observations have shown that $\Sigma$ in the outer Galaxy decreases exponentially as a function of the Galactocentric distance \citep{Wouterloot1990}. This kind of radial distribution of $\Sigma$ toward the outer Galaxy is confirmed by the re-decomposition of the CO data from the CfA survey \citep{Miville2017}. The surface densities of the \CO clouds toward the observed region are shown as a function of the galactocentric distance in Figure \ref{fig13}. The median surface density of the clouds in each 0.5 kpc bin first increases from $\sim$30 to $\sim$50 $\rm{M_{\sun}\ pc^{-2}}$ from d$_{GC}=$8.5 to 9 kpc, remains at around $\sim$45 $\rm{M_{\sun}\ pc^{-2}}$ from 9 to 11 kpc, and then slightly decreases to $\sim$30 $\rm{M_{\sun}\ pc^{-2}}$ at 12.5 kpc. The scatter of $\Sigma$ in each d$_{GC}$ bin is significant. The median $\Sigma$ obtained in this work is comparable to the compiled results in \cite{Heyer2015} and substantially lower than the surface density in the inner galaxy, $\sim$170 $\rm{M_{\sun}\ pc^{-2}}$ \citep{Roman2010, Heyer2015}. The average surface density in galactocentric distance interval of 0.5 kpc is calculated as the total mass of the clouds divided by the total area of the clouds. We can see that the average surface density is higher than the median $\Sigma$ in the range of d$_{GC}<11.5$ kpc, especially at the distance of the Perseus arm ($d_{GC} \sim 9.5-10.5\ kpc$). This difference is reasonable as we can see in Figure \ref{fig13} that the large and massive molecular clouds at these distances also have higher surface densities than the smaller and less massive ones.

\subsection{Scaling Relations and Equilibrium States of the molecular clouds}\label{sec3.5}
In this section, we present the scaling relations between the measured physical parameters of molecular clouds. 
\begin{figure*}[htb!]
	\centering
	\subfigure[]{
		\label{fig14a}
		\includegraphics[trim=0cm 0cm 0cm 0cm, width = 0.32\linewidth , clip]{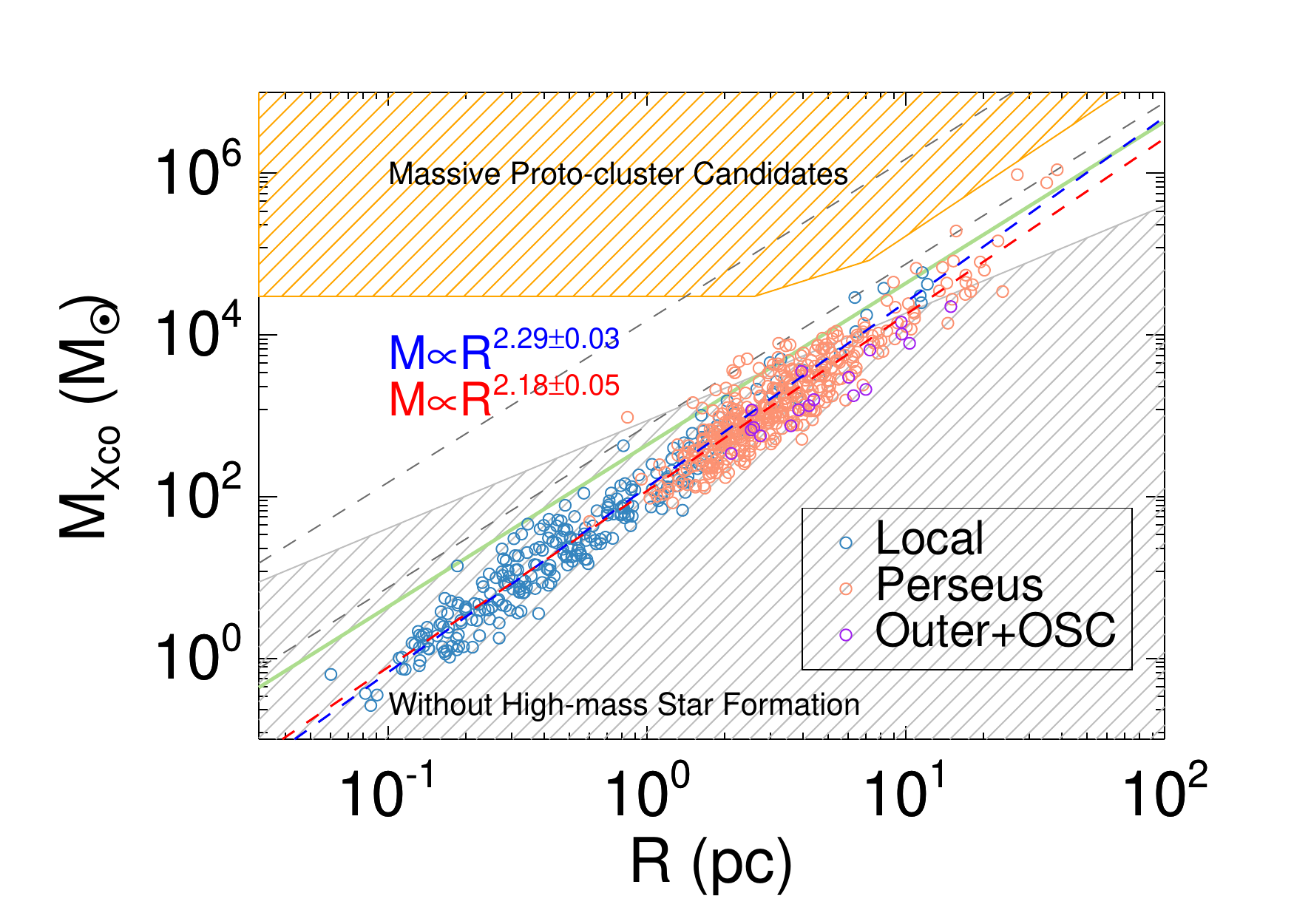}}
	\subfigure[]{
		\label{fig14b}
		\includegraphics[trim=0cm 0cm 0cm 0cm, width = 0.32\linewidth , clip]{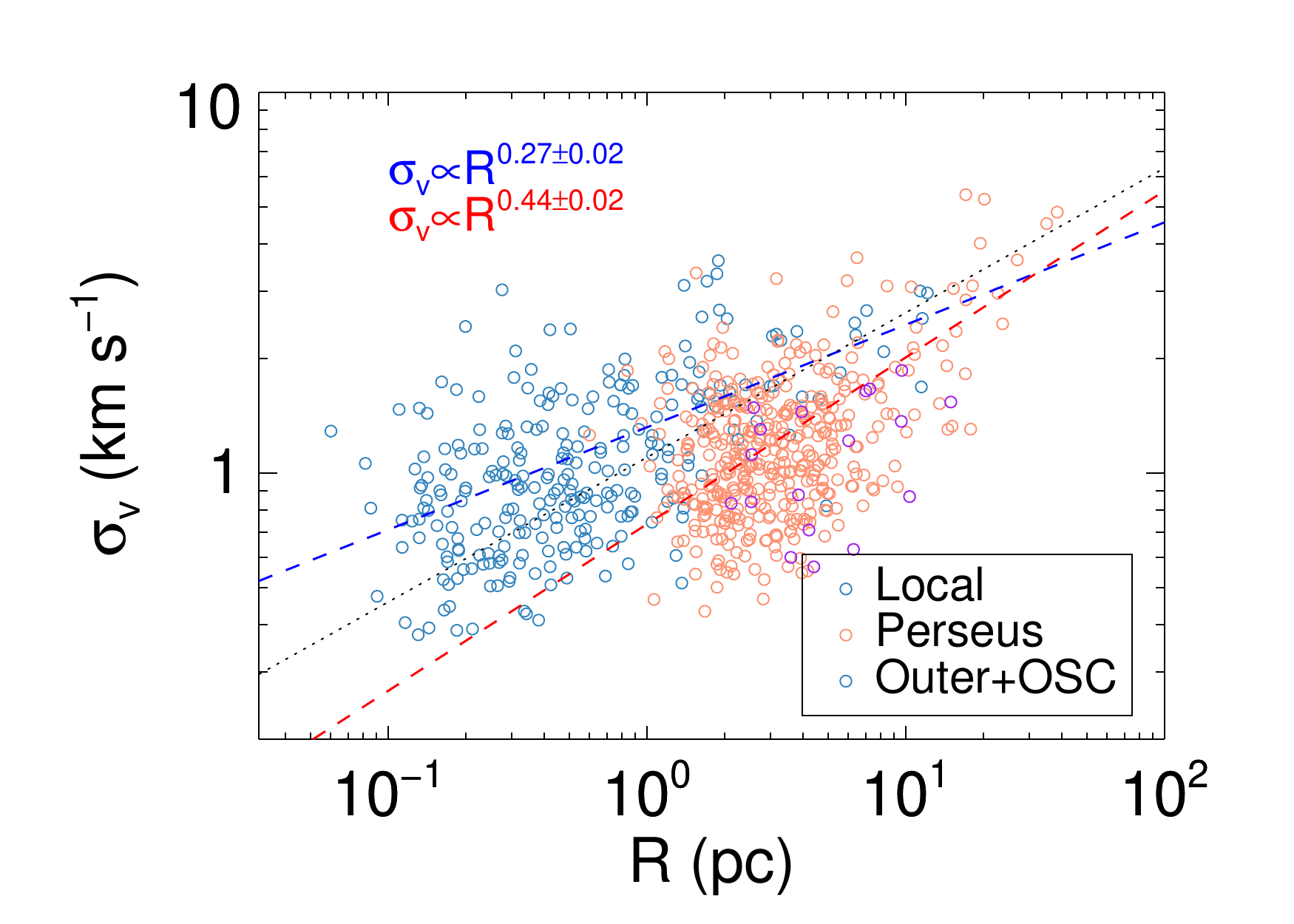}}
	\subfigure[]{
		\label{fig14c}
		\includegraphics[trim=0cm 0cm 0cm 0cm, width = 0.32\linewidth , clip]{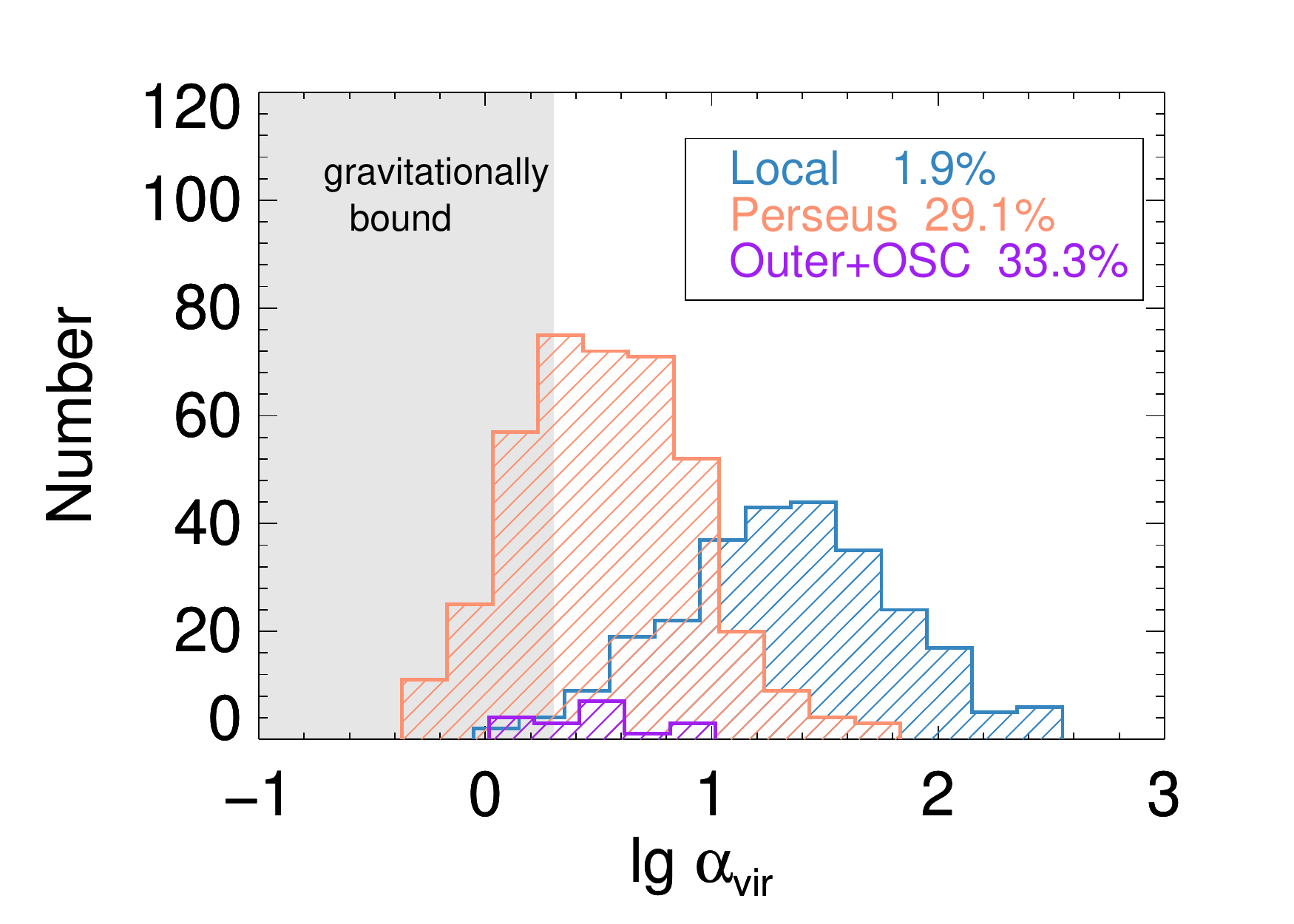}}
	\subfigure[]{
		\label{fig14d}
		\includegraphics[trim=0cm 0cm 0cm 0cm, width = 0.32\linewidth , clip]{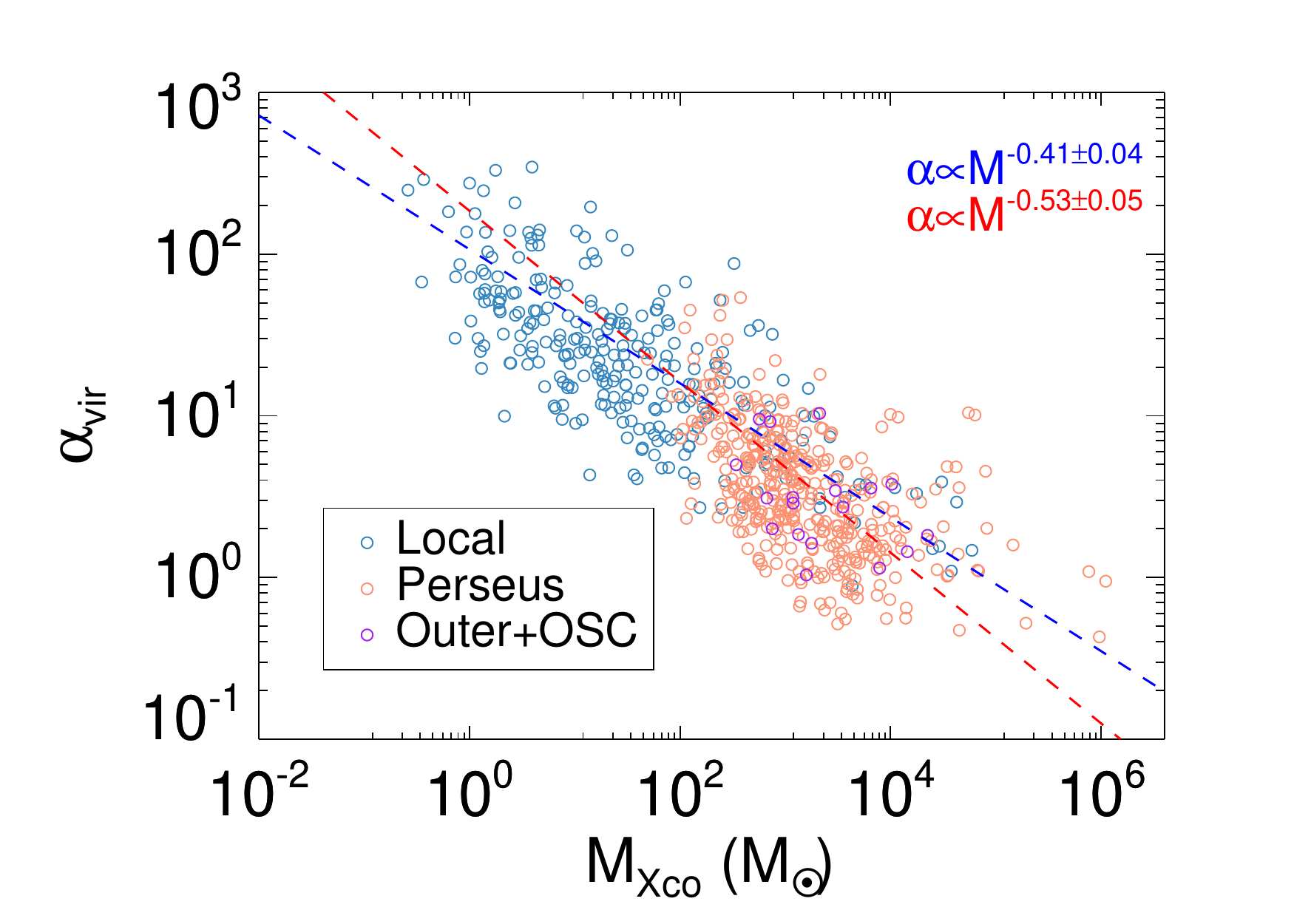}}
	\subfigure[]{
		\label{fig14e}
		\includegraphics[trim=0cm 0cm 0cm 0cm, width = 0.32\linewidth , clip]{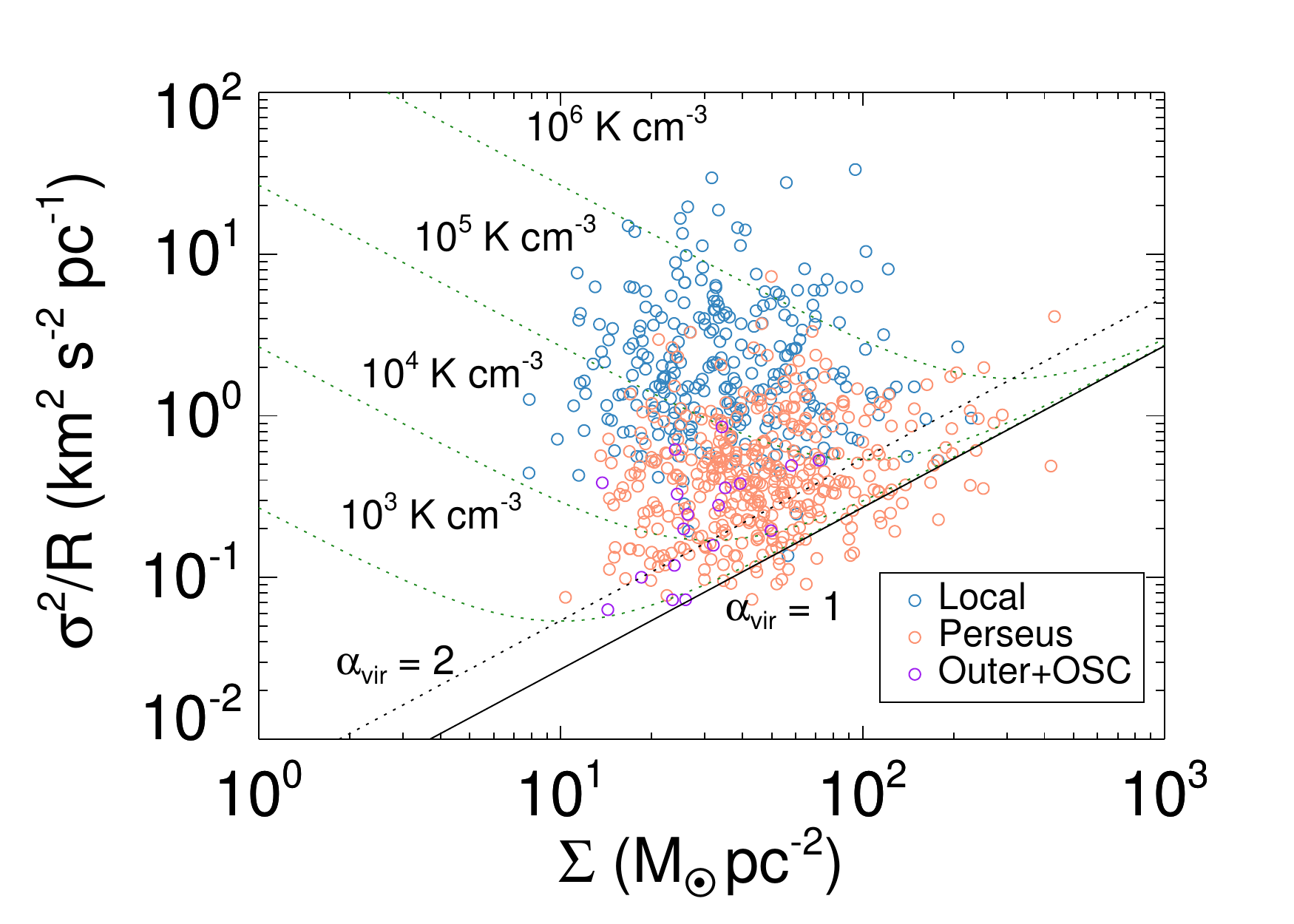}}
	\caption{(a) M$-$R relation of the \CO clouds, (b) line-width$-$size relation, (c) distribution of the virial parameters, (d) the relationship between the virial parameters and the masses, and (e) the relationship between $\sigma_v^2/R$ and surface density of the \CO clouds. The orange shaded area in panel (a) is the region defined by the M-R relations, found by \cite{Bressert2012}, for the molecular clouds capable of forming massive proto clusters. The grey shaded area in panel (a) is the region where the clouds cannot form high-mass stars according to the relation $M(R) \leqslant 870M_\sun(r/pc)^{1.33}$ of \cite{Kauffmann2010b}. The grey dashed lines in this panel indicate the empirical upper and lower bounds of the cloud surface density of 1 g cm$^{-2}$ and 0.05 g cm$^{-2}$ for massive star formation \citep{Urquhart2013}. The grey shaded area in panel (c) indicates $\alpha_{\rm vir} < 2$. The fraction of the clouds with $\alpha_{\rm vir}$ less than 2 are shown in the top$-$right corner in panel (c). The fitted M$-$R, $\sigma_v-$R, and M$-\alpha$ relations of the clouds in the Local and the Perseus arms are shown in blue and in red in panels (a), (b), and (d), respectively.}\label{fig14}
\end{figure*}

\begin{figure*}[htb!]
	\centering
	\subfigure[]{
		\label{fig15a}
		\includegraphics[trim=0cm 0cm 0cm 0cm, width = 0.32\linewidth , clip]{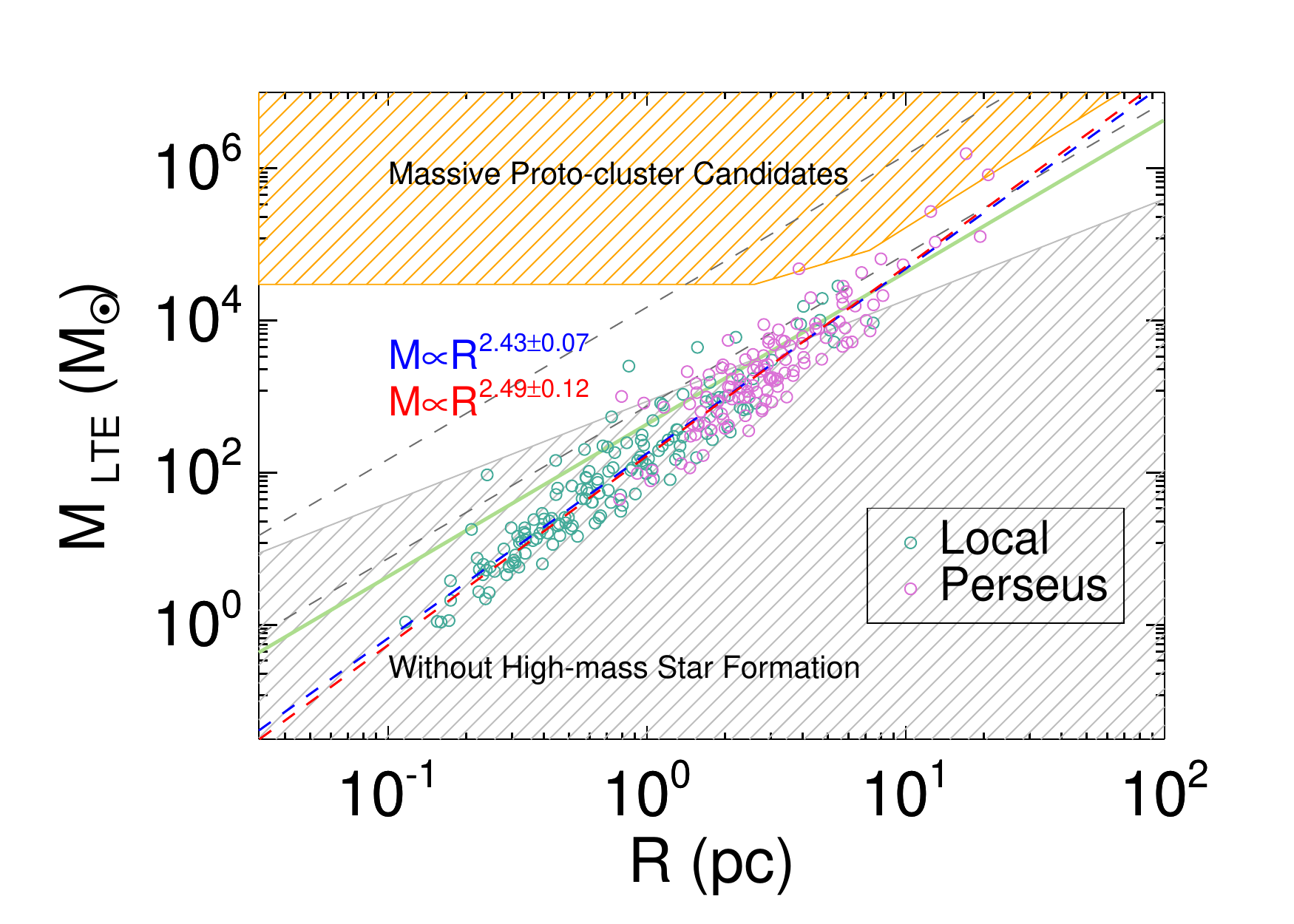}}
	\subfigure[]{
		\label{fig15b}
		\includegraphics[trim=0cm 0cm 0cm 0cm, width = 0.32\linewidth , clip]{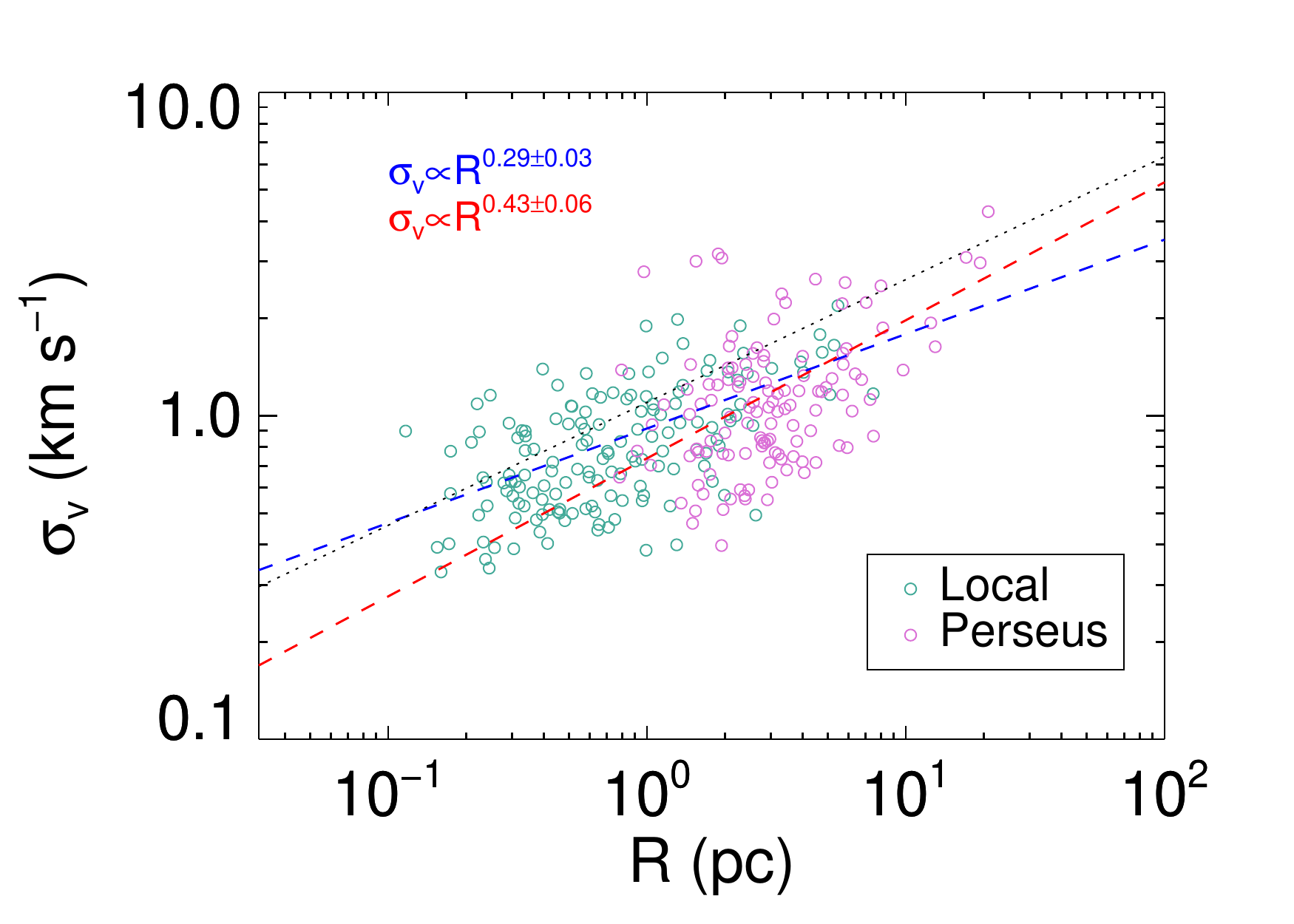}}
	\subfigure[]{
		\label{fig15c}
		\includegraphics[trim=0cm 0cm 0cm 0cm, width = 0.32\linewidth , clip]{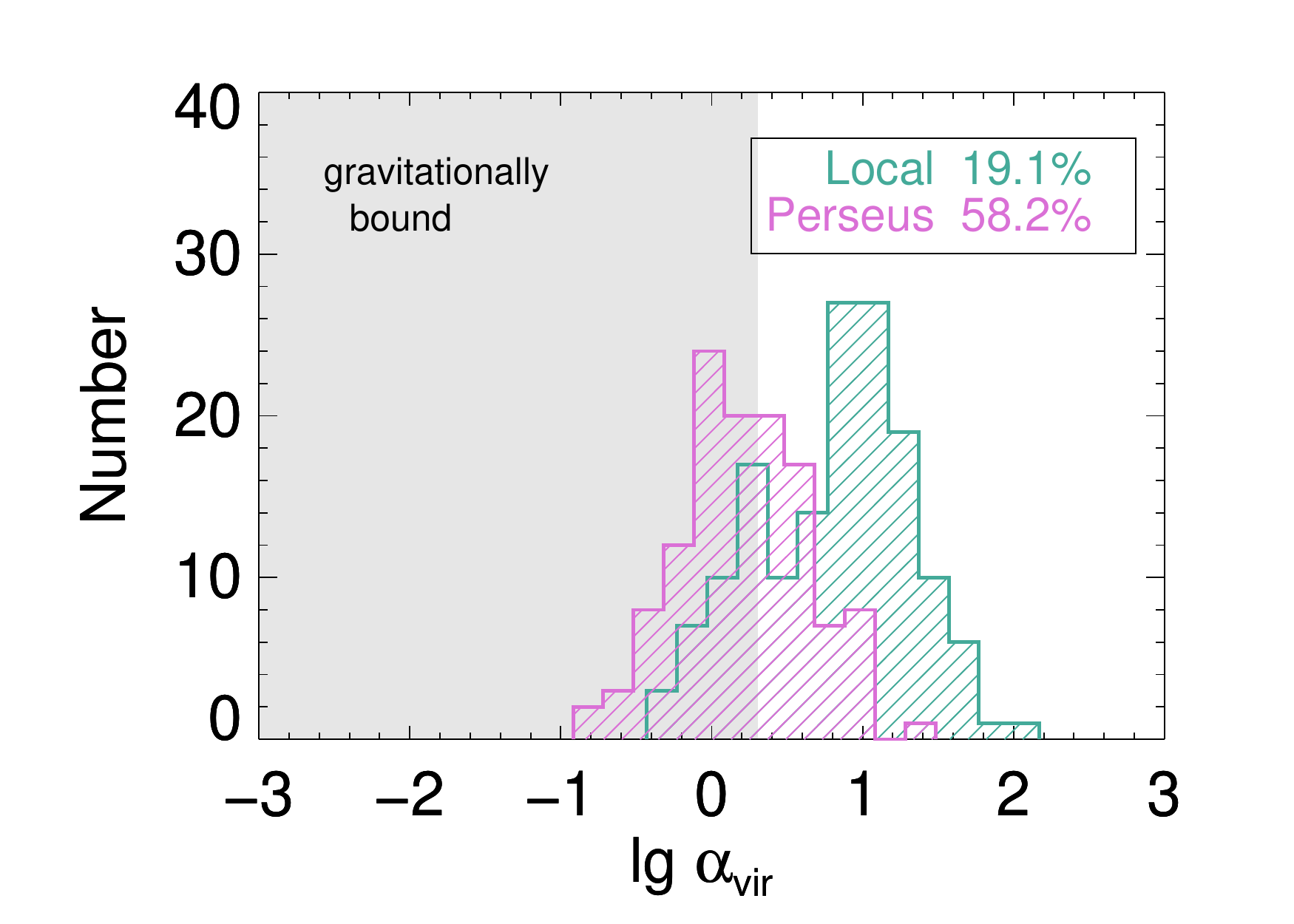}}
	\subfigure[]{
		\label{fig15d}
		\includegraphics[trim=0cm 0cm 0cm 0cm, width = 0.32\linewidth , clip]{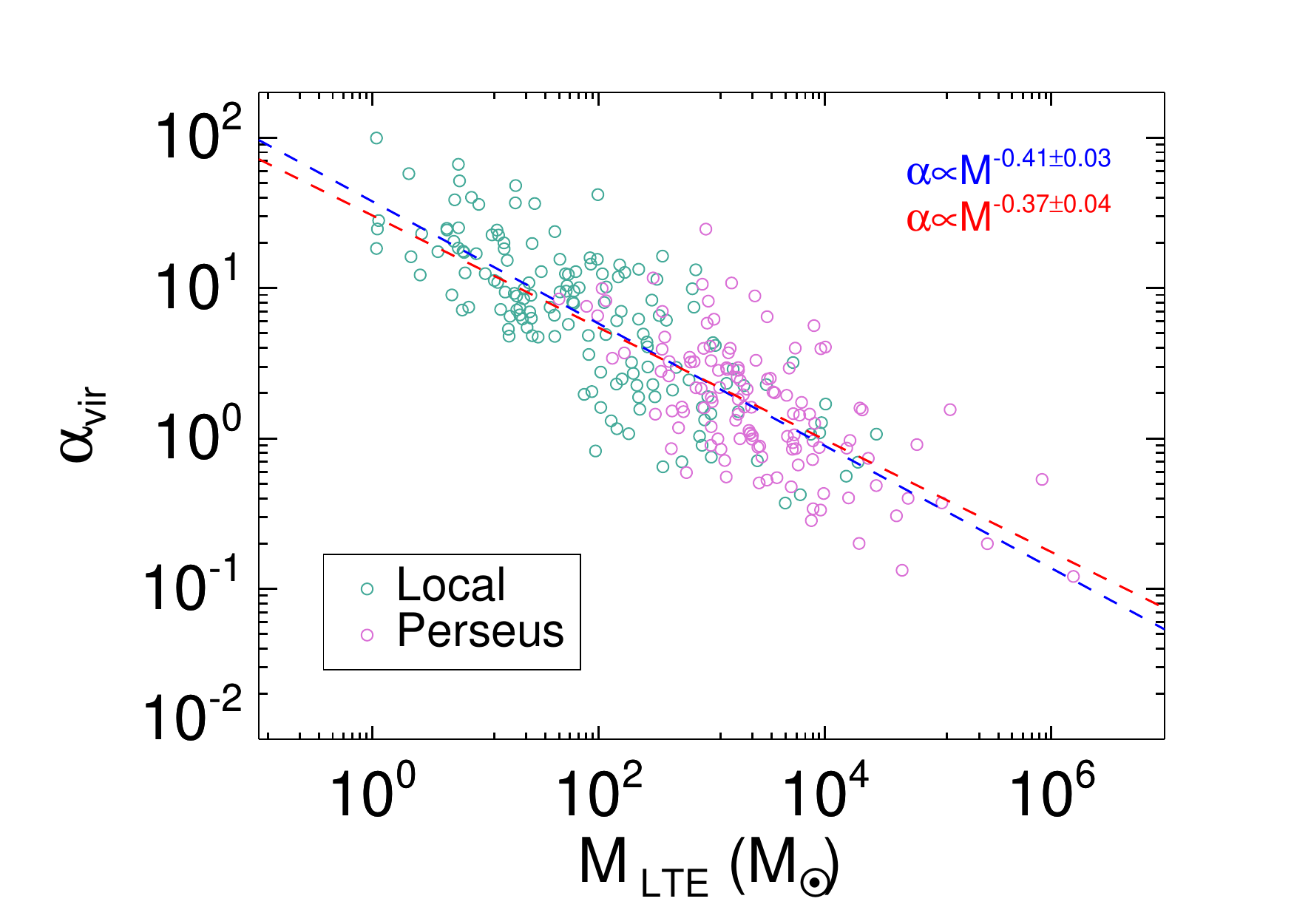}}
	\subfigure[]{
		\label{fig15e}
		\includegraphics[trim=0cm 0cm 0cm 0cm, width = 0.32\linewidth , clip]{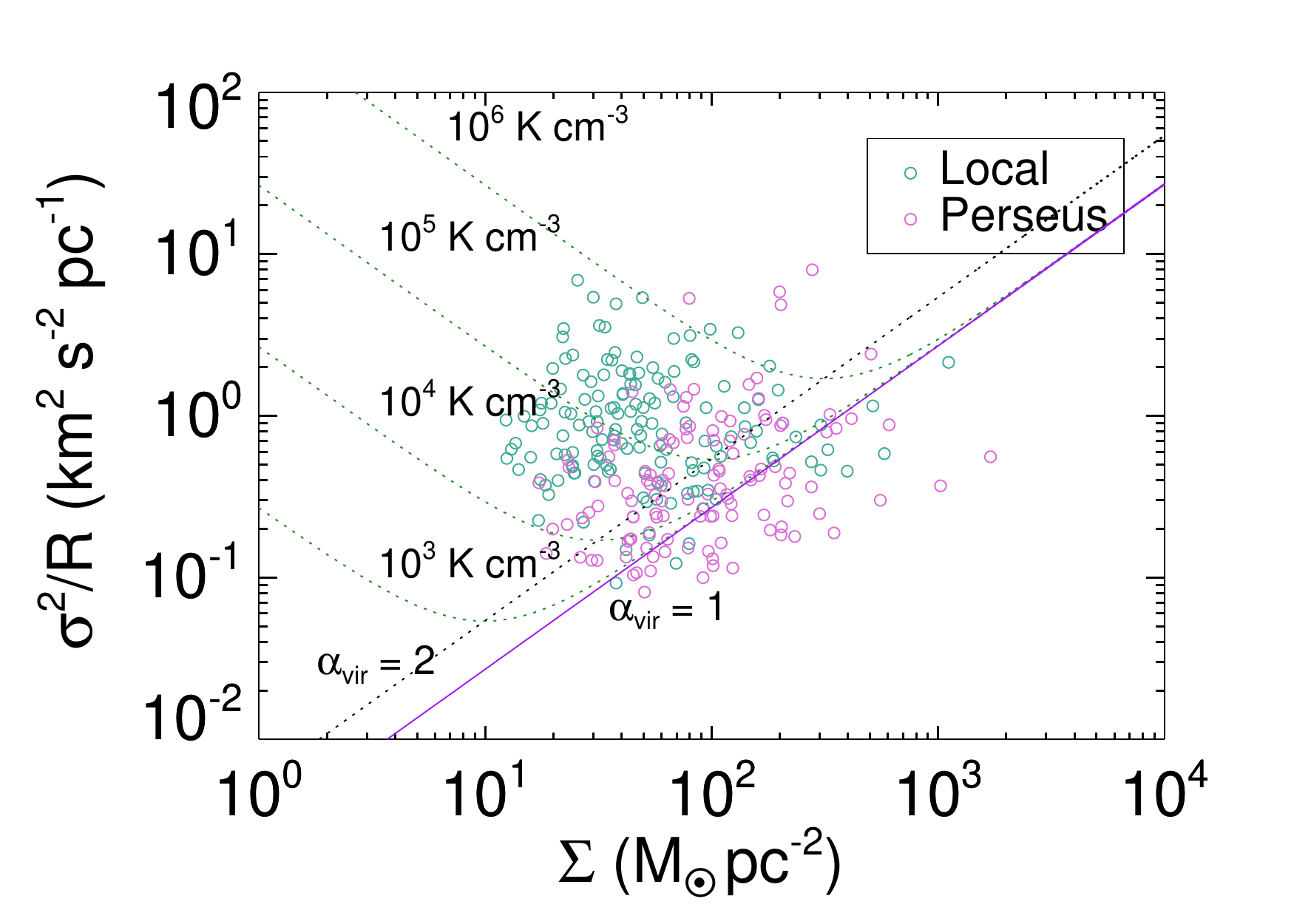}}
	\caption{Same as Figure \ref{fig14}, but for \COl clouds.}\label{fig15}
\end{figure*}

\subsubsection{Mass-radius relation} \label{sec3.5.1}
Larson's third law \citep{Larson1981} states that molecular clouds have a constant surface density which means that molecular clouds have a scaling relationship between their masses and radii of the form of $M\propto R^2$. However, the power-law exponent in the M-R relation is found to be a probe of the internal density distribution of molecular fragments on the sub-pc to pc scales \citep{Kauffmann2010a, Kauffmann2010b}. Observations of molecular clouds have shown that the stellar cluster forming clumps are more massive than those clumps devoid of stellar clusters although they have the same size \citep{Rathborne2006, Portegies2010, Walsh2011, Bressert2012}. \cite{Kauffmann2010b} obtained the empirical relationship $M(R)\geqslant 870M_\sun(R/pc)^{1.33}$ for molecular clouds capable to form massive stars. 

The M-R relations of the \CO clouds in the Local and Perseus arms are presented in Figure \ref{fig14a}. They can be well fitted with power-law functions of exponents $\sim$2.2 in a broad range of radius from $\sim0.06$ to $\sim40$ pc. None of the \CO molecular clouds are distributed in the region of the M-R parameter space capable of hosting massive proto clusters following the prescription of \cite{Bressert2012}. However, five molecular clouds have surface mass densities above the lower limit of massive star formation given by \cite{Urquhart2013} and they are all associated with \ion{H}{2} regions. The M-R relations for the \COl clouds are presented in Figure \ref{fig15a} and the fitted power-law exponents of the \COl clouds in the Local and Perseus arms are $\sim2.4$, suggesting non-uniform inner densities. Most of the \COl molecular clouds located above both the grey shaded area and the 0.05 g cm$^{-2}$ limit for possible massive star formation \citep{Urquhart2013} in Figure \ref{fig15a} correspond to active star-forming regions in the Local arm, such as Cep OB3 and L1188 GMCs, and the molecular clouds in the Perseus arm that are associated with \ion{H}{2} regions. Notably, the four molecular clouds that fall in the region for massive proto-cluster candidates are all associated with \ion{H}{2} regions, namely NGC 7538, S157, and S152. These GMCs are candidate regions for massive cluster formation.       

\subsubsection{Velocity Dispersion-Size relation} \label{sec3.5.2}
The resemblance of the exponent in the scaling relation $\sigma_v\propto R^{0.38}$, found by \cite{Larson1981}, to that in the velocity structure-function of incompressible turbulence \citep{Kolmogorov1941} reminds us of the important role of turbulence in molecular clouds. Observationally, \cite{Solomon1987} found that the scaling relationship between $\sigma_v$ and $R$ is better confined with a steeper power-law with an exponent of 0.5. This exponent has been further confirmed by \cite{Heyer2009} using the GRS data with improved spatial and spectral resolution. The interpretation of this relationship is attributed to the simple virial equilibrium state of the molecular cloud. 

The $\sigma_v-R$ relationship of the \CO and \COl clouds are presented in Figures \ref{fig14b} and \ref{fig15b}, respectively. The power-law exponents of the \CO and \COl clouds in the Local arm are 0.27 and 0.29, respectively. The exponents of the clouds in the Perseus arm are larger, 0.44 and 0.43 for \CO and \COl clouds, respectively. The Pearson correlation coefficients of the fitted $\sigma_v-R$ relations for the two arms lie in the range from 0.42 to 0.55. The Larson's relation with the exponent of 0.38 is also drawn in Figures \ref{fig14b} and \ref{fig15b} with grey dotted lines for comparison. The difference between the scaling exponents of the molecular clouds in the Local and Perseus arms possibly indicates different importance of turbulence in molecular clouds at different Galactocentric distances. \cite{Traficante2018} and \cite{Benedettini2020} suggested the presence of two different regimes in the $\sigma_v-R$ relationship, with a quite flat power law for the smaller clouds with radius below $\sim$2 pc and a steeper power law for the larger clouds with radius above $\sim$2 pc. The exponents obtained in this work are consistent with their suggestions in that the Local arm (smaller) clouds exhibit a smaller exponent (0.27) than the Perseus arm (larger) clouds (0.44). 

\subsubsection{Equilibrium State of the Molecular Clouds}\label{sec3.5.3}
Figures \ref{fig14c} and \ref{fig15c} display the distribution of the virial parameters of the \CO and \COl clouds, respectively, and also show the portion of the gravitationally bound ($\alpha_{\rm vir}\le 2$) molecular clouds in each arm. A subcritical ($\alpha_{\rm vir} > 2$) cloud will dissipate unless an external pressure from the environment can help to confine the cloud, while a supercritical ($\alpha_{\rm vir}\ll 2$) cloud will collapse within a few free fall time scales unless some other physical processes like the magnetic field are included in supporting the cloud against gravity. The virial parameters of the \CO clouds in the Local arm span a broad range from $\sim$1 to $\sim$1000, while those of the \CO clouds in the Perseus arm lie in the range from $\sim$0.3 to $\sim$60. However, the virial parameters of the \COl clouds are much smaller, from $\sim$0.4 to $\sim$99 with an average of $\sim$7.5 for the clouds in the Local arm and $\sim$0.1 to $\sim$25 with an average of $\sim$1.6 for the clouds in the Perseus arm. The virial parameters of the \CO clouds in the Outer$+$OSC arm distributed within the range of $\alpha_{\rm vir}$ from $\sim$1 to $\sim$10. The majority, $\sim$98.1\%, of the \CO clouds in the Local arm are gravitationally unbound, whereas in the Local and Perseus arms, 19.1\% and 58.2\% of the \COl clouds are gravitationally bound, respectively. The systematic difference of the virial parameter between the clouds in the Local and the Perseus arms may indicate different dynamical states of molecular clouds in the two arms. However, the percentage of the gravitationally bound clouds is distance-dependent, since the detection of small and less massive molecular clouds is not complete in the Perseus arm. Besides, based on the scaling relation we have obtained in Figures \ref{fig14} and \ref{fig15}, the virial parameters of the molecular clouds in the Perseus arm are proportional to $R_{eff}^{-0.3\sim-0.6}$. If the distances of the molecular clouds in the Perseus arm are overestimated by a factor of two, the virial parameters would be underestimated by a factor of $\sim 19\%$ to $\sim34\%$. This means the true fraction of the gravitationally bound molecular clouds in the Perseus arm could be lower than the values shown in Figures \ref{fig14c} and \ref{fig15c}. Nevertheless, even considering the possible underestimation of $\alpha_{\rm vir}$, the percentage of the gravitationally bound clouds in the Perseus arm is still higher than that in the Local arm. The overall difference of the masses of the molecular clouds in the two spiral arms is another potential cause for the difference in virial parameters. As discussed in the next paragraph, there is an anti-correlation between the masses of the clouds and their virial parameter. The molecular clouds in the Perseus arm usually have larger masses, therefore, tending to have smaller virial parameters.
  
\cite{Bertoldi1992} have presented a theoretical argument that when self-gravity is unimportant, the virial parameter is expected to be correlated with $M^{-2/3}$ for the pressure-confined clumps. \cite{Kauffmann2013} have reevaluated virial parameters for dense cores and clouds using data compiled from the literature. The power-law behavior $\alpha_{\rm vir}\propto M^{h_{\alpha}}$ is confirmed for all the different samples of dense cores and clouds, and the exponents are tabulated in their table 2. We also find the $\alpha_{\rm vir}\propto M^{h_{\alpha}}$ behaviour both for the \CO and \COl clouds in our survey (Figures \ref{fig14d} and \ref{fig15d}). The power-law exponents of the $\alpha_{\rm vir}-M$ relation of \CO clouds in the Local and the Perseus arms are $-0.41$ and $-0.53$, respectively, within the mass range from $\sim$0.1 M$_{\odot}$ to $\sim 10^4$ M$_{\odot}$. The exponents in $\alpha_{\rm vir}-M$ relation for \COl clouds in the Local and the Perseus arms are  $-0.41$ and $-0.37$, respectively. The results obtained in this work resemble those derived by \cite{Kauffmann2013} for the GRS survey \citep{Heyer2009, Roman2010}.

Subcritical molecular clouds need other confining sources to keep the equilibrium state. The virial theorem for a uniform, isothermal, non-magnetized, spherical cloud immersed in the interstellar environment with pressure $P_e$ can be written as \citep{Spitzer1978, Field2011}, 
\begin{equation}
\frac{\sigma^2}{R} = \frac{1}{3} (\frac{3\pi G\Sigma}{5} + \frac{4 P_e}{\Sigma})
\label{Eq:2}
\end{equation}
The solution for the pressure-confined virial equilibrium state (PVE) is the V-shaped lines in Figures \ref{fig14e} and \ref{fig15e} for different external pressures. The relation between $\sigma^2/R$ and $\Sigma$ under the simple virial equilibrium condition (SVE) (when the internal pressure balances with the self-gravity) with $\alpha_{\rm vir} = 1$ and $\alpha_{\rm vir} = 2$ are also shown in Figures \ref{fig14e} and \ref{fig15e}. In Figure \ref{fig14e}, the subcritical \CO molecular clouds in the Local and the Perseus arms show poor correlation of $\sigma^2/R$ with $\Sigma$ and they are clustered in the regime of external pressures from $P_e/k\sim\ 10^4$ K cm$^{-3}$ to $P_e/k\sim10^6$ K cm$^{-3}$, with a trend that the \CO clouds in the Local arm have higher $P_e$ than the clouds in the Perseus arm. Some of the Perseus arm clouds are located below the $\alpha_{\rm vir}\sim 2$ SVE line, indicating the important role of self-gravity in these clouds. The \COl molecular clouds in the Local arm also show concentration in the $\sigma^2/R$ versus $\Sigma$ diagram (Figure \ref{fig15e}). The external pressure needed to confine the \COl molecular clouds in the Local arm is comparable to that of the \CO clouds.

\section{Discussion} \label{sec4}

\subsection{Probability Distribution Functions of H$_2$ Column Density of Large molecular clouds}\label{sec4.1}

In this section, we investigate the N-PDFs of molecular clouds in the surveyed region. Since the abundance of \CO is much higher than \COl in molecular clouds and the \CO emission is more extended than that of \COl, a \CO cloud usually corresponds to several \COl clouds that can be considered as the denser parts of the same and more extend structure traced by the \CO emission. Therefore, the boundary extracted from the \CO emission is more representative of the edge of a molecular cloud in the N-PDF analysis. To have sufficient pixels to derive a robust N-PDF for a cloud, we selected molecular clouds of relatively large projected area. According to the exact projected-area in the \CO catalog, forty clouds with area greater than 0.35 arcdeg$^2$, i.e., 5000 pixels, are selected, among which 29 clouds belong to the Local arm and 11 clouds to the Perseus arm. Since the \COl emission has a lower optical depth than the \CO emission, we used the \COl emission within a \CO cloud boundary to estimate the H$_2$ column density. The H$_2$ column density is calculated as described in Section \ref{sec3.4.1} at pixels where the peak brightness of the \COl spectrum is at least above 4$\sigma_{RMS}$. An N-PDF is simply the histogram of the logarithm of the normalized column density, $s=\ln(N_{H_2}/<N_{H_2}>)$. If the column density is log-normally distributed, the PDF of $s$ follows the formula
\begin{equation}
p(s) = \frac{1}{\sqrt{2\pi}\sigma}\exp[{-\frac{(s-\mu)^2}{2\sigma^2}}],
\end{equation} 
where $\mu$ and $\sigma$ are the mean and dispersion of the normal distribution of s. If the column density is power-law-distributed, which means $p(s) \propto N_{H_2}^\alpha$, the PDF of $s$ follows
\begin{equation}
\ln p(s) = \alpha s + c, 
\end{equation} 
where $\alpha$ is the power-law index and c is a constant related to the probability of the starting location of the fitting. In the log-log space, the N-PDFs that have log-normal shapes show as parabolas, while those that have power-law shapes show as straight lines.

\begin{figure*}[htb!]
	\centering
	\includegraphics[trim=0cm 0cm 0cm 0cm, width=\linewidth , clip]{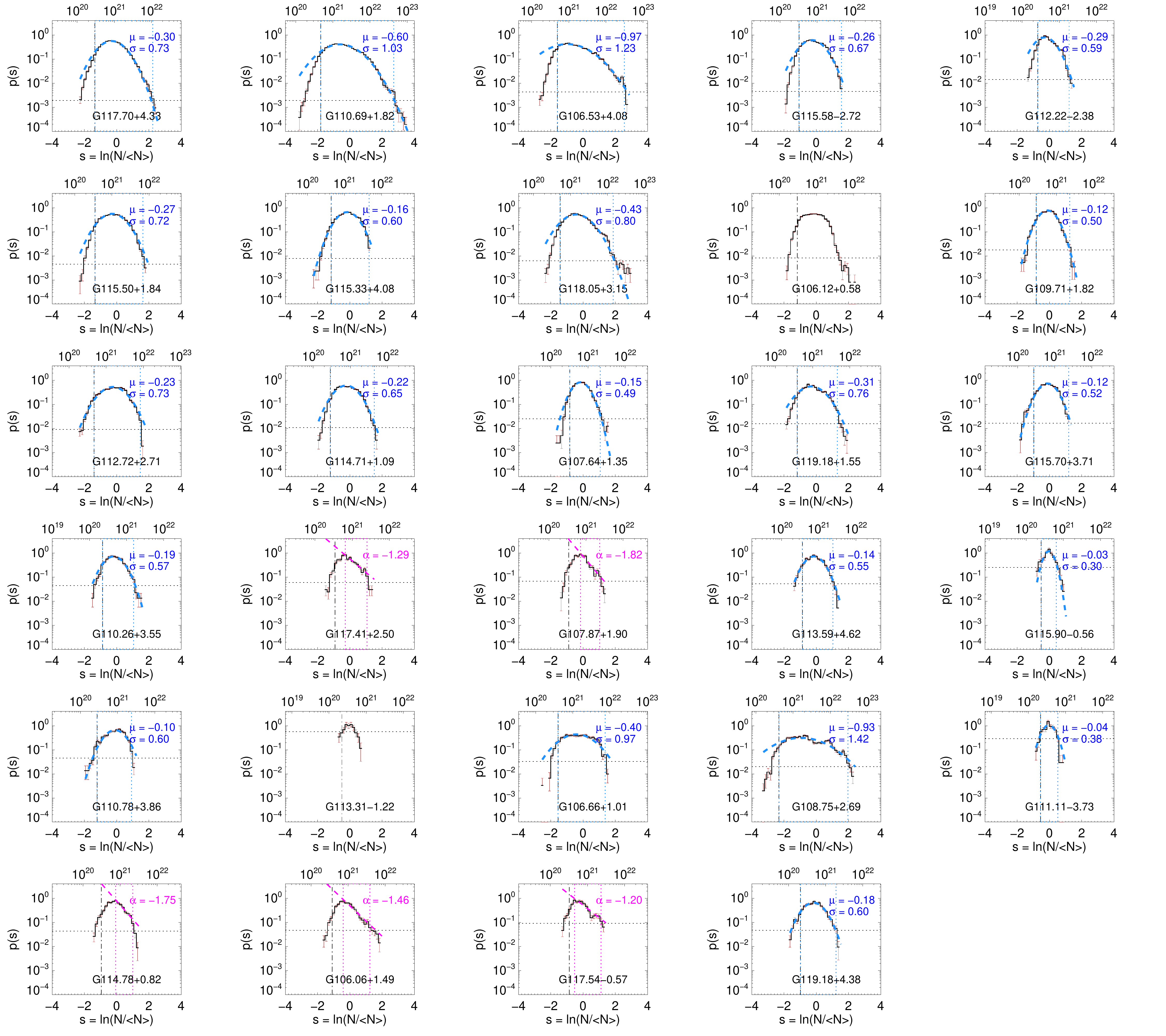}
	\caption{N-PDFs of the twenty-nine selected molecular clouds in the Local arm. The column densities of the clouds are normalized by their mean, and the upper axes give the corresponding values of N$_{H_2}$. The vertical black dash-dotted lines in each panel indicate the reference detection completeness limit of the column density of each cloud. The horizontal dashed lines mark the $p(s)$ at which the count in a bin is ten. The blue dashed curves are the Gaussian fittings of the N-PDFs, while the magenta curves are the power-law fittings. The corresponding fitting ranges are marked with purple (Gaussian fittings) or blue (power-law fittings) dotted lines in the panels of the fitted N-PDFs. The fitted parameters $\mu$ and $\sigma$ of the Gaussian function are indicated in blue, and the fitted exponents of the power-law function $p(s) \propto N_{H_2}^\alpha$ are indicated in magenta. The red bars mark the statistical errors in each bin of s.}
	\label{fig16}
\end{figure*}

\begin{figure*}[htb!]
	\centering
	\includegraphics[trim=0cm 0cm 0cm 0cm, width=\linewidth , clip]{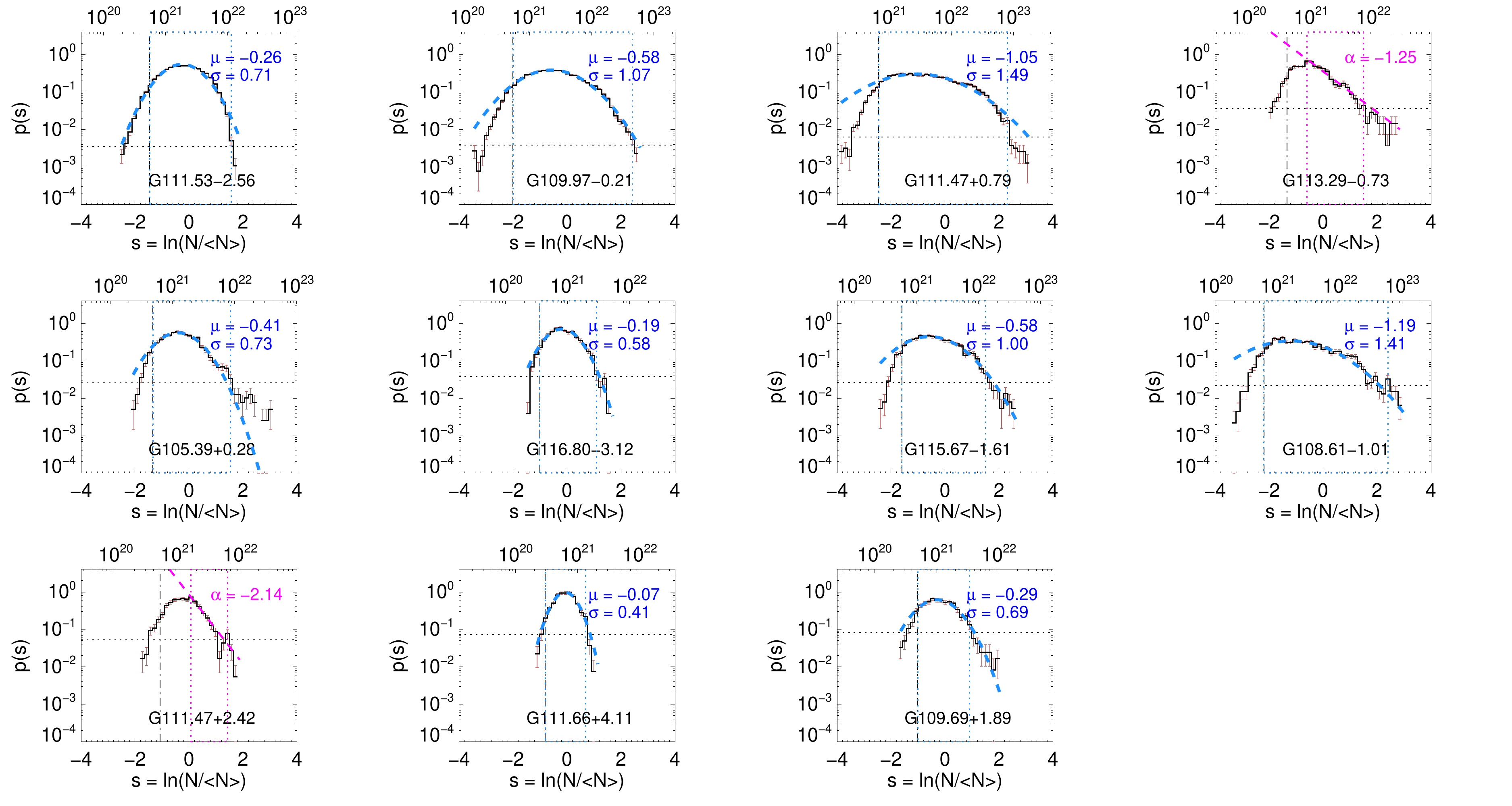}
	\caption{Same as Figure \ref{fig16}, but for the eleven selected clouds in the Perseus arm.}
	\label{fig17}
\end{figure*}

The derived N-PDFs for the selected clouds in the Local and the Perseus arms are presented in Figures \ref{fig16} and \ref{fig17}, respectively, in the order of the angular sizes of the \CO boundaries. Because the names of the molecular clouds under the MWISP standard, like those in Table \ref{tab2}, are too long to conveniently legend in Figures \ref{fig16} and \ref{fig17}, we use in this section the shortened names, which start with a letter ``G" and followed by the spatial coordinates accurate to two decimal places. Except for Section \ref{sec4.1}, the identified molecular clouds are all named according to the MWISP standard. The N-PDF of each cloud is fitted with a log-normal function or a power-law, depending on the shape. For the N-PDFs that both a log-normal function and a power-law function can be fit, we use the reduced chi-squared of the fittings as the criterion to choose the better form of the fitting, with the fitting form of smaller reduced chi-squared being chosen. We fixed the upper limit of the fitting in the higher bin with at least ten counts. The lower limit for the power-law fitting is fixed at the peak of the distribution. For the log-normal fitting, ideally, the detection completeness limit should be used as the lower limit in the fitting. However, an accurate value of the detection completeness limit for H$_2$ column density is hard to determine. In this work, we calculated the median uncertainty of H$_2$ column density within the \COl cloud and take three times the median uncertainty, which we refer to as reference detection completeness limit, as the lower limit in the log-normal fitting. By manual examination, we found that the reference detection completeness limit defines a closed contour that is nearly identical to the emission cutoff of the \COl cloud (see, e.g., Figure \ref{fig18}). Therefore, we believe that the reference completeness limit is a reasonable value for the lower limit in the log-normal fitting. The mean column densities, statistical, and fitted parameters of the N-PDFs of the selected molecular clouds are tabulated in Tables \ref{tab4} and \ref{tab5}. 

\begin{deluxetable*}{cccccccccc}[htb!]
	\setlength{\tabcolsep}{8pt}
	\tablecaption{Properties of the N-PDFs of the Clouds in the Local Arm \label{tab4}}
	\tablecolumns{10}
	\tablehead{
		\colhead{Name} & \colhead{d} & \colhead{M$_{LTE}$} & \colhead{$<N_{H_2}>$} & \colhead{$\mu_{data}$} & \colhead{$\sigma_{data}$} & \colhead{$\mu_{fit}$} & \colhead{$\sigma_{fit}$} & \colhead{$\alpha$} & Shape \\
		& \colhead{kpc} & \colhead{(M$_{\odot}$)} & \colhead{(cm$^{-2}$)} &  &  & & & & \colhead{(LN/LN$^*$/PL)} \\
		\colhead{(1)} & \colhead{(2)} & \colhead{(3)} & \colhead{(4)} & \colhead{(5)} & \colhead{(6)} & \colhead{(7)} & \colhead{(8)} & \colhead{(9)} & \colhead{(10)}
     }	
	\startdata
	G117.70+4.33&   0.67&   8.48$\times10^{3}$&   1.15$\times10^{21}$&  $-$0.24&   0.68&  $-$0.30&   0.73&...&LN\\
	G110.69+1.82&   0.77&   2.79$\times10^{4}$&   2.86$\times10^{21}$&  $-$0.46&   0.91&  $-$0.60&   1.03&...&LN$^*$\\
	G106.53+4.08&   0.87&   1.31$\times10^{4}$&   2.41$\times10^{21}$&  $-$0.48&   0.92&  $-$0.97&   1.23&...&LN$^*$\\
	G115.58$-$2.72&   0.33&   6.83$\times10^{2}$&   9.46$\times10^{20}$&  $-$0.19&   0.61&  $-$0.26&   0.67&...&LN\\
	G112.22$-$2.38&   0.27&   8.99$\times10^{1}$&   5.43$\times10^{20}$&  $-$0.12&   0.48&  $-$0.29&   0.59&...&LN\\
	G115.50+1.84&   0.77&   5.54$\times10^{3}$&   1.34$\times10^{21}$&  $-$0.23&   0.67&  $-$0.27&   0.72&...&LN\\
	G115.33+4.08&   1.16&   7.34$\times10^{3}$&   1.39$\times10^{21}$&  $-$0.17&   0.59&  $-$0.16&   0.60&...&LN\\
	G118.05+3.15&   1.09&   1.30$\times10^{4}$&   2.18$\times10^{21}$&  $-$0.31&   0.73&  $-$0.43&   0.80&...&LN$^*$\\
	G106.12+0.58&	0.41&	6.36$\times10^{2}$&	9.80$\times10^{20}$&  $-$0.18&	0.61&   ...&	...& ...&...\\
	G109.71+1.82&   0.24&   6.87$\times10^{1}$&   6.48$\times10^{20}$&  $-$0.12&   0.50&  $-$0.12&   0.50&...&LN\\
	G112.72+2.71&   1.16&   8.62$\times10^{3}$&   1.91$\times10^{21}$&  $-$0.24&   0.71&  $-$0.23&   0.73&...&LN\\
	G114.71+1.09&   0.77&   1.63$\times10^{3}$&   9.66$\times10^{20}$&  $-$0.19&   0.61&  $-$0.22&   0.65&...&LN\\
	G107.64+1.35&   0.25&   5.71$\times10^{1}$&   7.35$\times10^{20}$&  $-$0.11&   0.46&  $-$0.15&   0.49&...&LN\\
	G119.18+1.55&   0.65&   1.03$\times10^{3}$&   1.25$\times10^{21}$&  $-$0.22&   0.65&  $-$0.31&   0.76&...&LN\\
	G115.70+3.71&   0.31&   1.27$\times10^{2}$&   7.20$\times10^{20}$&  $-$0.13&   0.52&  $-$0.12&   0.52&...&LN\\
	G110.26+3.55&   0.27&   2.82$\times10^{1}$&   5.40$\times10^{20}$&  $-$0.14&   0.52&  $-$0.19&   0.57&...&LN\\
	G117.41+2.50&   0.32&   4.43$\times10^{1}$&   8.62$\times10^{20}$&  $-$0.16&   0.55&...&...&  $-$1.29&PL\\
	G107.87+1.90&   0.61&   1.44$\times10^{2}$&   8.30$\times10^{20}$&  $-$0.13&   0.49&...&...&  $-$1.82&PL\\
	G113.59+4.62&   0.65&   2.13$\times10^{2}$&   8.54$\times10^{20}$&  $-$0.13&   0.51&  $-$0.14&   0.55&...&LN\\
	G115.90$-$0.56&   0.31&   4.89$\times10^{0}$&   4.21$\times10^{20}$&  $-$0.06&   0.33&  $-$0.03&   0.30&...&LN\\
	G110.78+3.86&   0.19&   2.32$\times10^{1}$&   9.42$\times10^{20}$&  $-$0.15&   0.57&  $-$0.10&   0.60&...&LN\\
	G113.31$-$1.22&	0.26&	1.34$\times10^{0}$&	3.64$\times10^{20}$&  $-$0.05&	0.32&	...&	...& ...&...\\
	G106.66+1.01&   0.65&   7.47$\times10^{2}$&   1.89$\times10^{21}$&  $-$0.31&   0.80&  $-$0.40&   0.97&...&LN$^*$\\
	G108.75+2.69&   0.75&   3.72$\times10^{3}$&   4.18$\times10^{21}$&  $-$0.63&   1.10&  $-$0.93&   1.42&...&LN$^*$\\
	G111.11$-$3.73&   0.20&   1.65$\times10^{0}$&   3.68$\times10^{20}$&  $-$0.05&   0.33&  $-$0.04&   0.38&...&LN\\
	G114.78+0.82&   0.55&   1.71$\times10^{2}$&   8.07$\times10^{20}$&  $-$0.13&   0.51&...&...&  $-$1.75&PL\\
	G106.06+1.49&   1.73&   2.22$\times10^{3}$&   1.11$\times10^{21}$&  $-$0.20&   0.59&...&...&  $-$1.46&PL\\
	G117.54$-$0.57&   0.26&   1.72$\times10^{1}$&   7.61$\times10^{20}$&  $-$0.16&   0.53&...&...&  $-$1.20&PL\\
	G119.18+4.38&   0.31&   3.47$\times10^{1}$&   5.57$\times10^{20}$&  $-$0.16&   0.56&  $-$0.18&   0.60&...&LN\\
	\enddata
	\tablecomments{Columns 1$-$4 give the name, distance, mass, and the mean $H_2$ column density. The mass and the column density are derived using the \COl emission. The statistical mean and dispersion of $s = \ln N_{H_2}/<N_{H_2}>$ of the clouds are presented in columns 5-6, while columns 7-8 give the corresponding fitted parameters. Column 9 presents the exponents, $\alpha$, of the fitted power-law distributions. Column ten is the description of the shape of the fitted N-PDFs, where LN means log-normal, LN$^{*}$ means log-normal distribution with slight excesses at the high-density end, and PL means power-law.}
\end{deluxetable*}

\begin{deluxetable*}{cccccccccc}[htb!]
	\setlength{\tabcolsep}{8pt}
	\tablecaption{Properties of the N-PDFs of the Clouds in the Perseus Arm \label{tab5}}
	\tablecolumns{10}
	\tablehead{
		\colhead{Name} & \colhead{d} & \colhead{M$_{LTE}$} & \colhead{$<N_{H_2}>$} & \colhead{$\mu_{data}$} & \colhead{$\sigma_{data}$} & \colhead{$\mu_{fit}$} & \colhead{$\sigma_{fit}$} & \colhead{$\alpha$} & Shape \\
		& \colhead{kpc} & \colhead{(M$_{\odot}$)} & \colhead{(cm$^{-2}$)} &  &  & & & & \colhead{(LN/LN$^*$/PL)}\\
	    \colhead{(1)} & \colhead{(2)} & \colhead{(3)} & \colhead{(4)} & \colhead{(5)} & \colhead{(6)} & \colhead{(7)} & \colhead{(8)} & \colhead{(9)} & \colhead{(10)}
         }
	\startdata
	G111.53$-$2.56&   3.13&   2.06$\times10^{5}$&   2.38$\times10^{21}$&  $-$0.24&   0.71&  $-$0.26&   0.71&...&LN\\
	G109.97$-$0.21&   4.05&   5.35$\times10^{5}$&   4.02$\times10^{21}$&  $-$0.49&   0.98&  $-$0.58&   1.07&...&LN\\
	G111.47+0.79&   4.32&   7.38$\times10^{5}$&   7.96$\times10^{21}$&  $-$0.74&   1.20&  $-$1.05&   1.49&...&LN$^*$\\
	G113.29$-$0.73&   3.01&   1.21$\times10^{4}$&   1.55$\times10^{21}$&  $-$0.37&   0.76&...&...&  $-$1.25&PL\\
	G105.39+0.28&   4.63&   4.87$\times10^{4}$&   1.85$\times10^{21}$&  $-$0.30&   0.71&  $-$0.41&   0.73&...&LN$^*$\\
	G116.80$-$3.12&   2.72&   5.92$\times10^{3}$&   9.79$\times10^{20}$&  $-$0.15&   0.54&  $-$0.19&   0.58&...&LN$^*$\\
	G115.67$-$1.61&   3.09&   3.28$\times10^{4}$&   2.90$\times10^{21}$&  $-$0.39&   0.84&  $-$0.58&   1.00&...&LN\\
	G108.61$-$1.01&   4.33&   1.44$\times10^{5}$&   5.32$\times10^{21}$&  $-$0.72&   1.12&  $-$1.19&   1.41&...&LN$^*$\\
	G111.47+2.42&   4.19&   1.52$\times10^{4}$&   1.51$\times10^{21}$&  $-$0.18&   0.59&...&...&  $-$2.14&PL\\
	G111.66+4.11&   2.38&   1.60$\times10^{3}$&   6.72$\times10^{20}$&  $-$0.07&   0.38&  $-$0.07&   0.41&...&LN\\
	G109.69+1.89&   4.78&   1.19$\times10^{4}$&   1.36$\times10^{21}$&  $-$0.20&   0.61&  $-$0.29&   0.69&...&LN\\
	\enddata
	\tablecomments{Same as Table \ref{tab4} but for the clouds in the Perseus arm.}
\end{deluxetable*}

Thirty-one (77.5\%) of the forty selected clouds have log-normal N-PDFs above the reference completeness limit of column density without significant excesses at the high density end, while seven of the selected clouds have power-law N-PDFs and the N-PDFs of two clouds can not be fitted with either log-normal or power-law functions. Some molecular clouds have log-normal N-PDFs with minor excesses at the high column density ends. Most of the molecular clouds in our results have log-normal N-PDFs, which is different from the results obtained by \cite{Alves2017}. We note that the reference detection completeness limit, which we used in the fitting of log-normal N-PDF, is very close to the last closed N(H$_2$) isocontour of the cloud in \cite{Alves2017}. Although the detailed shape of the N-PDF at the low column density end may be affected by the selected different closed isocontours, the existence of the turn-over in the N-PDF, which is the important signature of log-normal N-PDF, can not be altered by the small difference between the reference detection completeness limit in this work and the last closed isocontour in \cite{Alves2017}. The molecular clouds can be divided into three categories according to the shapes of their N-PDFs, i.e., pure log-normal (LN), log-normal with minor excess (LN*), and power-law (PL). 

\cite{Kainulainen2009} derived the N-PDFs for 23 molecular clouds in the solar neighborhood, d$\sim$ 250-700 pc, using the near-infrared dust extinction as the tracer of H$_2$ column density. Their results suggest that molecular clouds that are active in star formation have log-normal forms of N-PDFs at low column density ends, but show significant excesses above log-normal distributions or power-law distributions at high column density ends. Similar distributions have also been obtained in Herschel observations, such as the results of \cite{Schneider2013}, \cite{Schneider2015}, and \cite{Pokhrel2016}. The $\sigma$ parameters obtained by \cite{Kainulainen2009} from the fitting of the log-normal components of the N-PDFs lie between $\sim$0.3$-$0.5. The fitted dispersions of $s$, $\sigma_{fit}$ (in order to distinguish from $\sigma_{data}$ in the following text), in this work, however, lie between $\sim0.3$ and $\sim1.5$ with a median value of $\sim$0.7, which are approximately 1.7 times the widths of the N-PDFs of \cite{Kainulainen2009}. This is in part due to the different fitting ranges used in the two studies. The log-normal fitting in \cite{Kainulainen2009} is limited to s $=-$0.5$-$1. Therefore, the high-density part of the N-PDF will significantly exceed the log-normal distribution and the resulting $\sigma$ is then narrow. The fitting ranges are much broader in this work, which may result in broader log-normal N-PDFs. Nevertheless, the dispersion of $s$ $(\sigma_{data})$ calculated directly from the data for the N-PDFs of active star-forming regions in \cite{Kainulainen2009} are all greater than 1. The $\sigma_{data}$ parameters in this work are similar to those of \cite{Kainulainen2009}.

\subsection{Relation Between the Shapes of the N-PDFs and the Star Forming Activities in the molecular clouds}
We compare the spatial distribution of the H$_2$ column densities of the selected clouds with the infrared, WISE 3/4 (12/22 $\mu m$) band images, to examine the relation between the N-PDF forms and the star formation activities in these molecular clouds. The WISE 3 and 4 bands are good indicator of star formation since they contain the polycyclic aromatic hydrocarbon (PAH) emission and the emission from the warm dust heated by star formation activities.

\begin{figure*}[htb!]
	\centering
	\begin{minipage}[t]{0.6\linewidth}
		\centering
		\includegraphics[trim=0cm 6cm 4cm 4cm, width = \linewidth, clip]{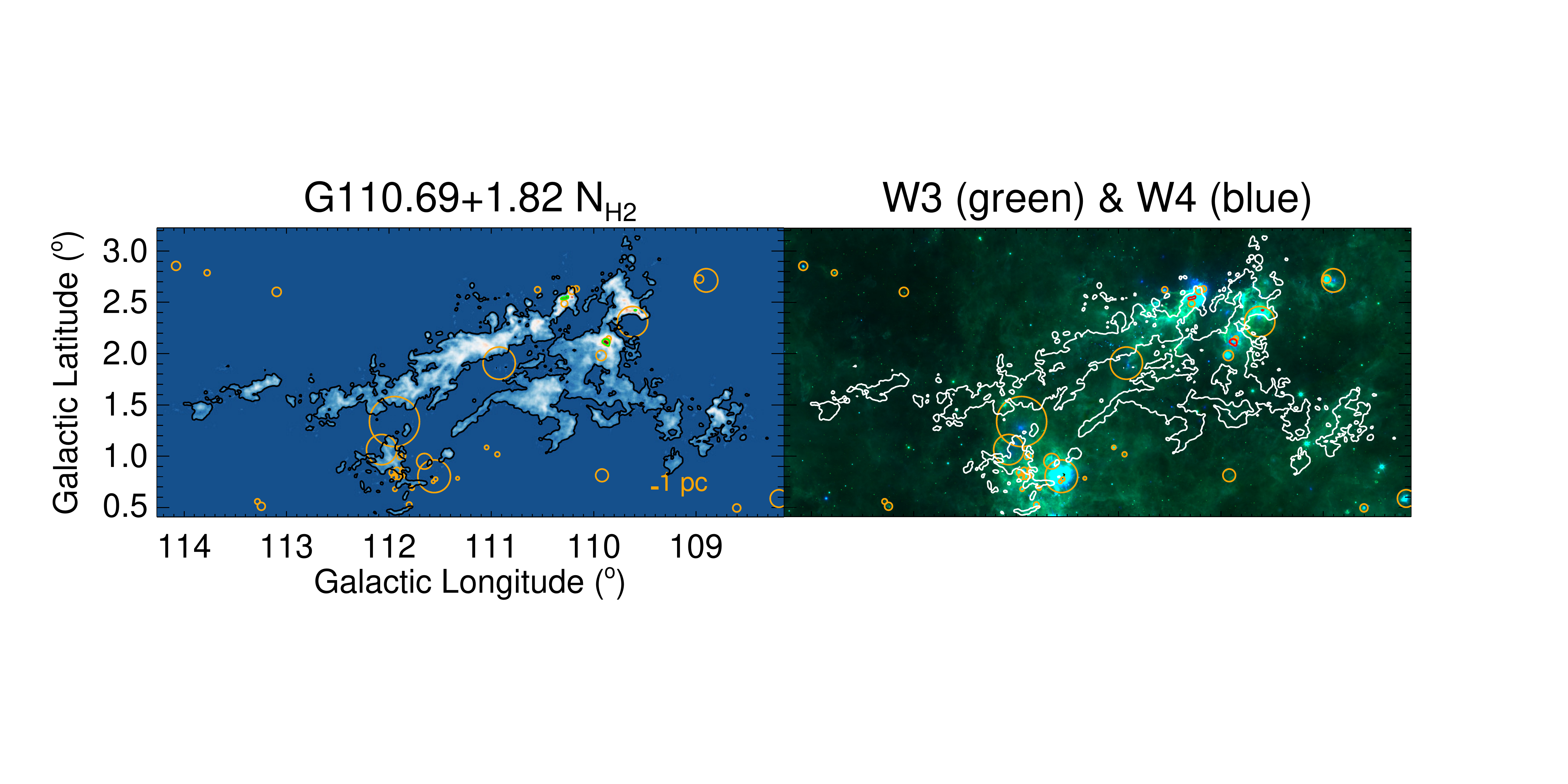}
		\label{fig18a}
		\\(a) 
	\end{minipage}
	\begin{minipage}[t]{0.6\textwidth}
		\centering
		\includegraphics[trim=0cm 6cm 4cm 4cm, width = \linewidth, clip]{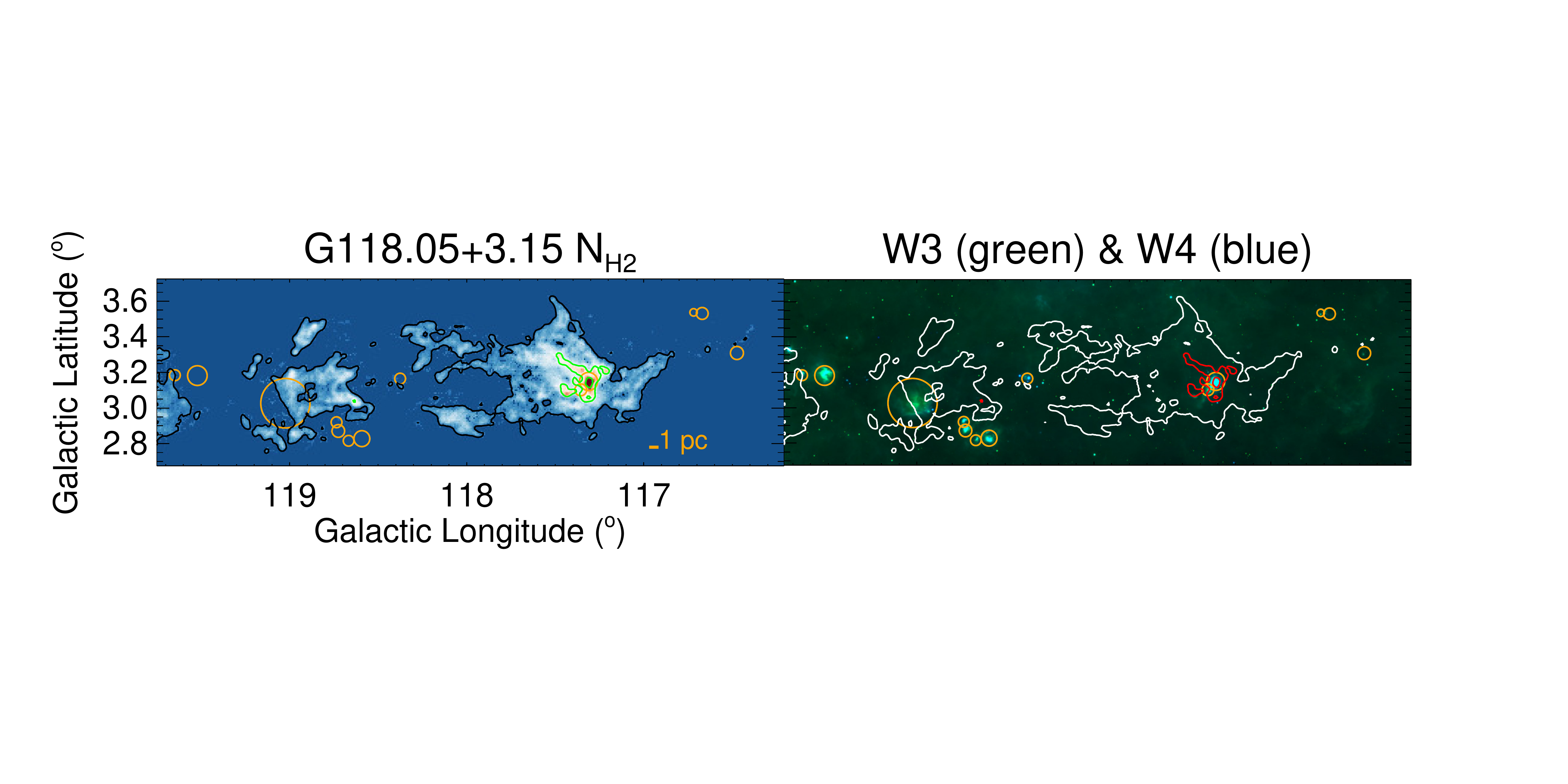}
		\label{fig18b}
		\\(b)
	\end{minipage}
	\begin{minipage}[t]{0.6\textwidth}
		\centering
		\includegraphics[trim=0cm 4cm 2cm 2cm, width = \linewidth, clip]{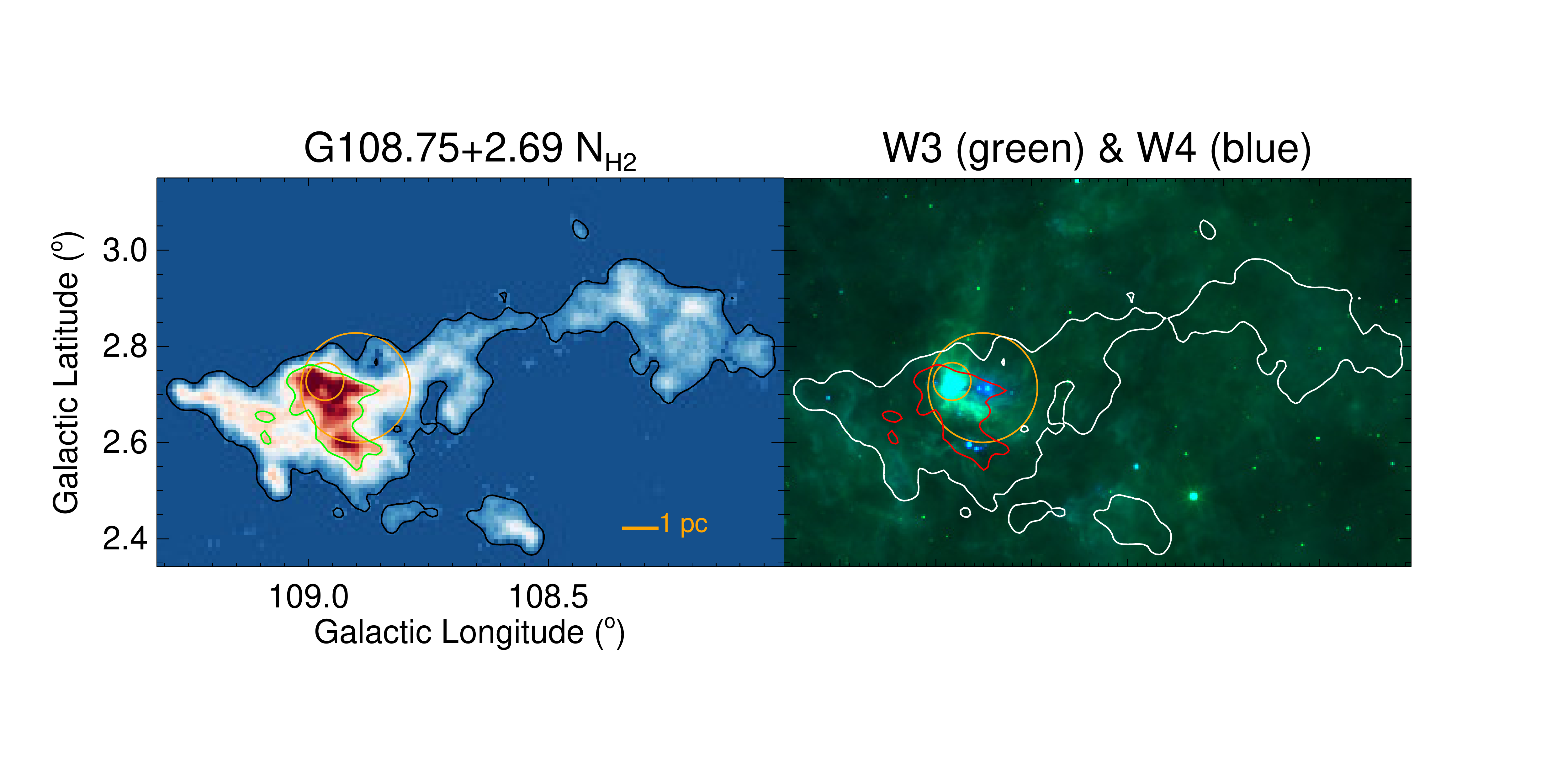}
		\label{fig18c}
		\\(c)
	\end{minipage}
	\begin{minipage}[t]{0.6\textwidth}
		\centering
		\includegraphics[trim=0cm 4cm 3cm 3cm, width = \linewidth, clip]{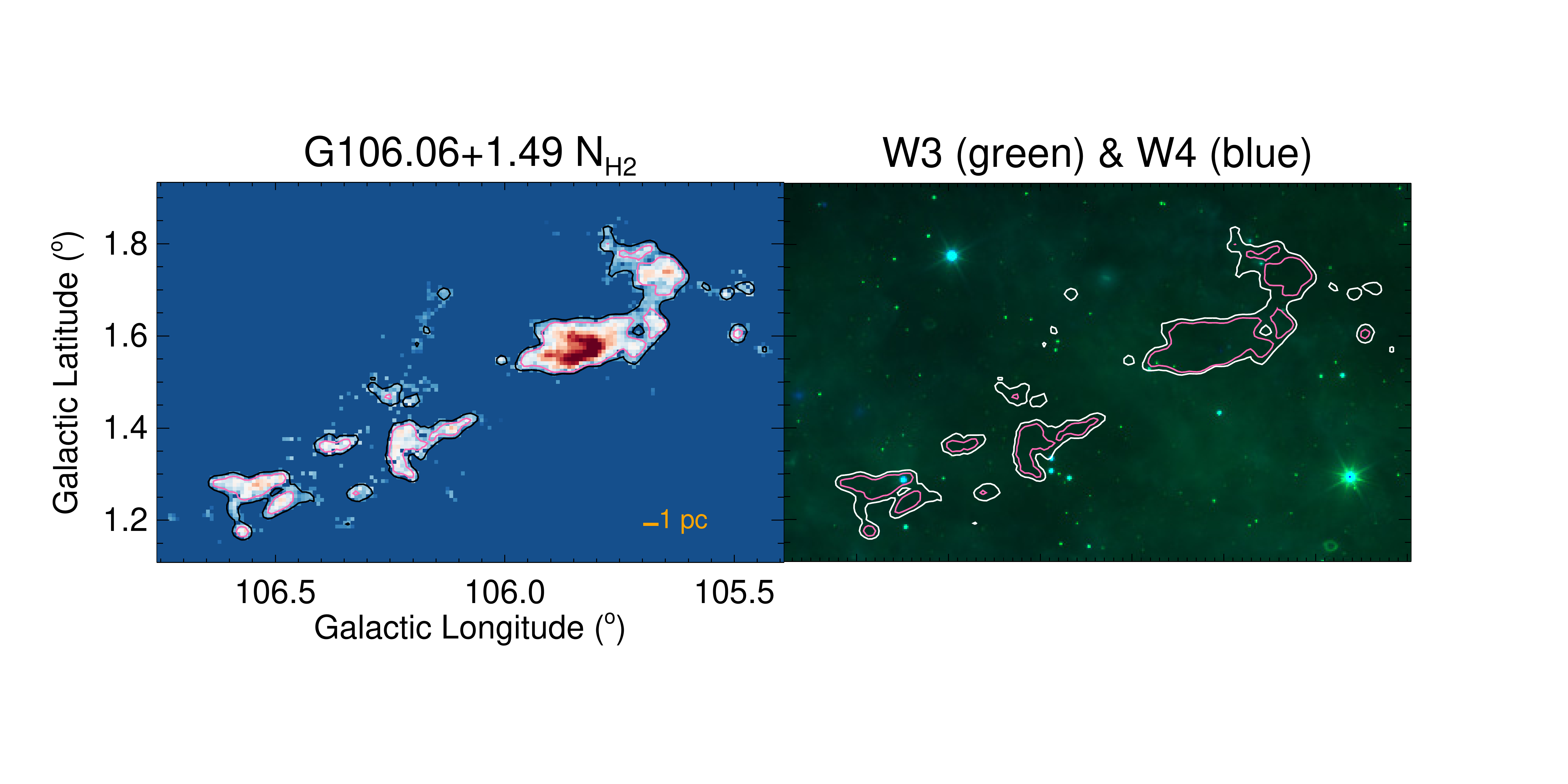}
		\label{fig18d}
		\\(d)
	\end{minipage}
	\caption{Column density distribution (left) and the WISE 3/4 band images (right) for the selected molecular clouds in the Local arm. Panels (a)$-$(c) are three examples of the clouds that have log-normal N-PDFs with slight excesses at the high-density end. The black contours in the left panels and the white contours in the right panels show the reference detection completeness limit of the H$_2$ column density of the clouds. The green contours in the left panels and the red contours in the right panels in (a)-(c) correspond to the column densities where the N-PDFs start to show excesses above log-normal distributions. Panel (d) is an example of the molecular clouds with power-law N-PDFs. The magenta contours in panel (d) show the column density peak occurrence. The orange circles in these figures mark the positions of the \ion{H}{2} regions or \ion{H}{2} region candidates in the WISE catalog.}
	\label{fig18}
\end{figure*}

\begin{figure*}[htb!]
	\centering
	\begin{minipage}[t]{0.6\linewidth}
		\centering
		\includegraphics[trim=0cm 0cm 4cm 1cm, width = \linewidth, clip]{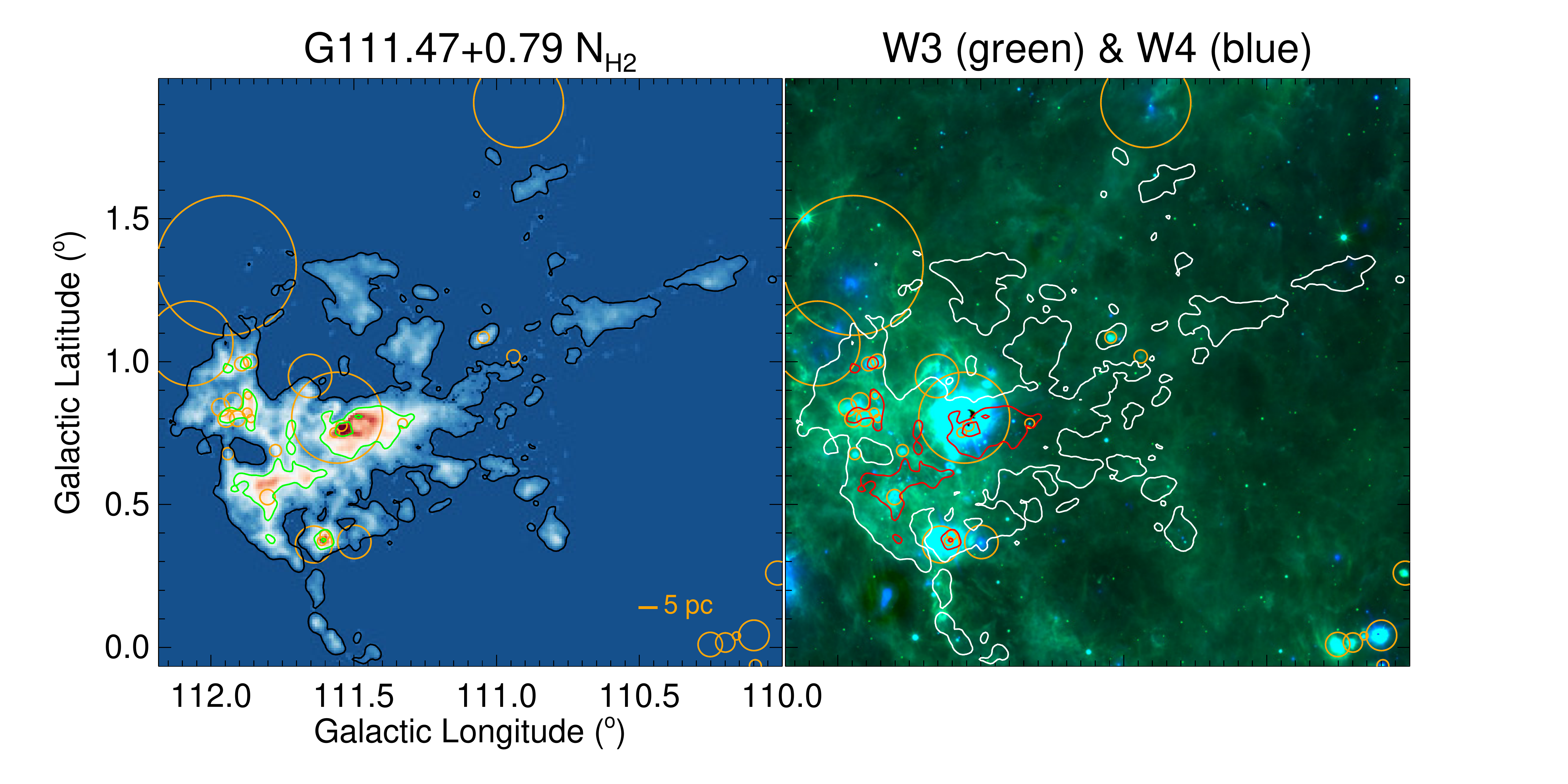}
		\label{fig19a}
		\\(a)
	\end{minipage}
	\begin{minipage}[t]{0.6\textwidth}
		\centering
		\includegraphics[trim=0cm 2cm 2cm 3cm, width = \linewidth, clip]{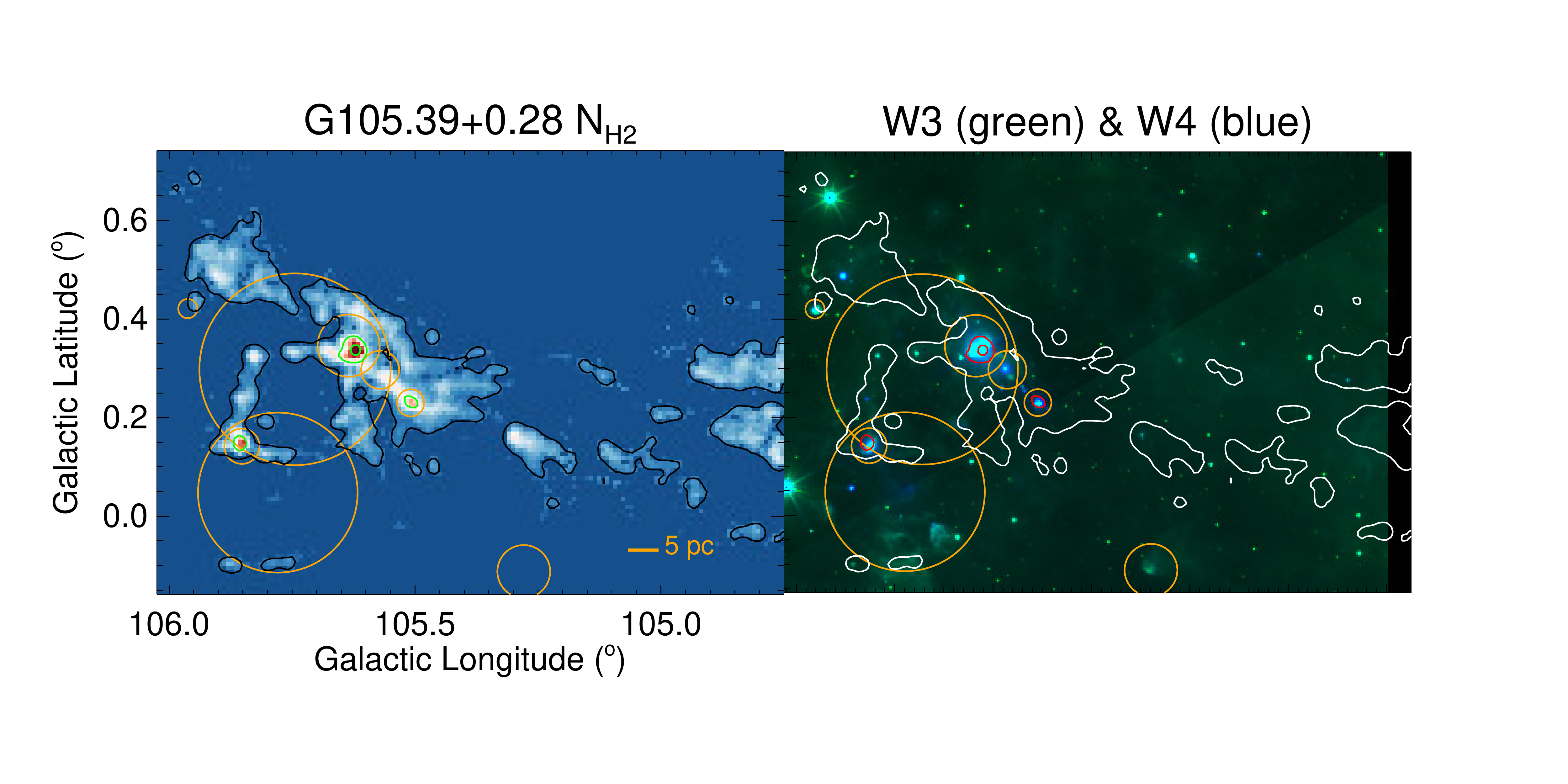}
		\label{fig19b}
		\\(b)
	\end{minipage}
	\begin{minipage}[t]{0.6\textwidth}
		\centering
		\includegraphics[trim=0cm 4cm 4cm 4cm, width = \linewidth, clip]{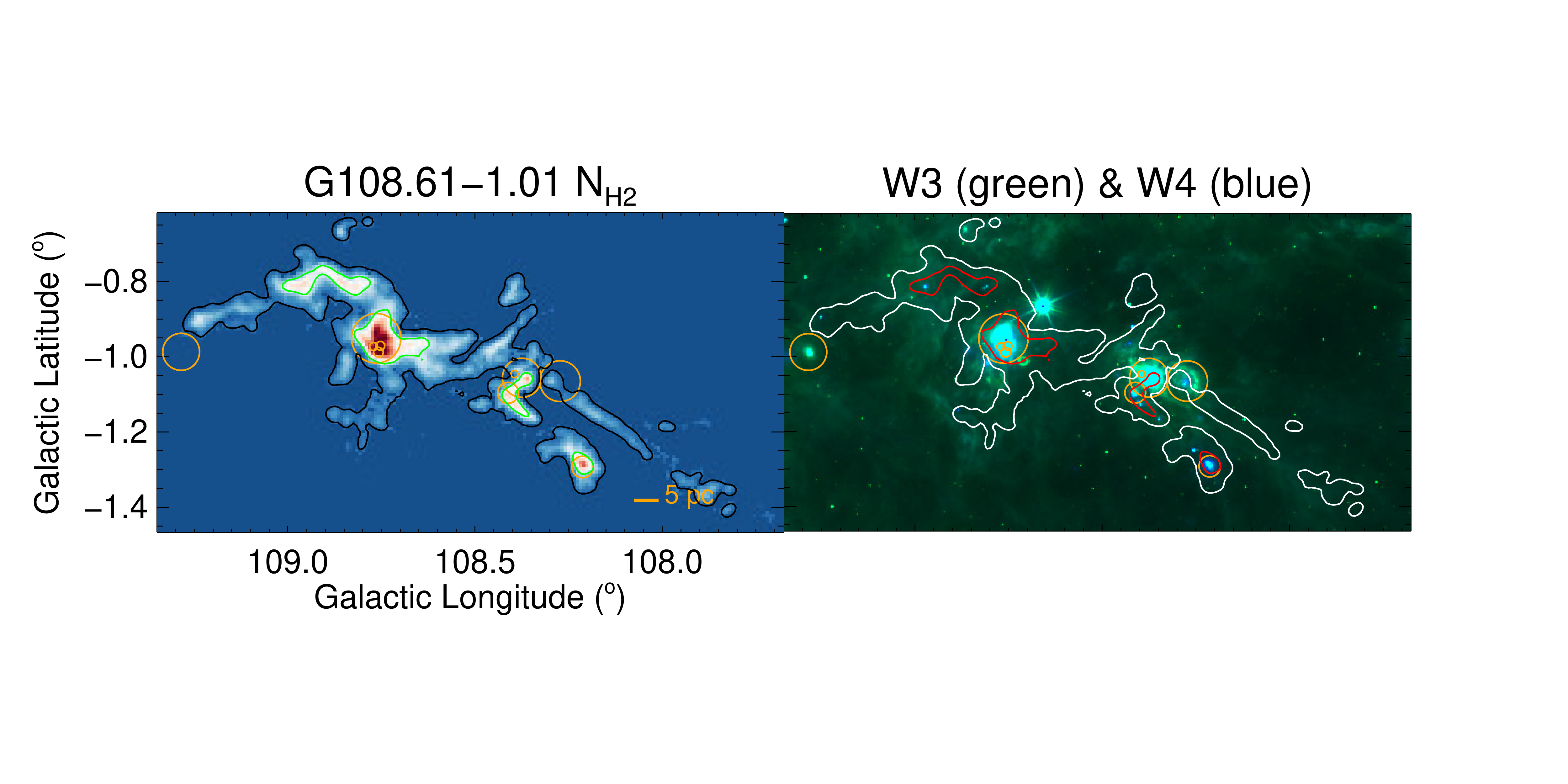}
		\label{fig19c}
		\\(c)
	\end{minipage}
	\begin{minipage}[t]{0.6\textwidth}
		\centering
		\includegraphics[trim=0cm 4cm 4cm 4cm, width = \linewidth, clip]{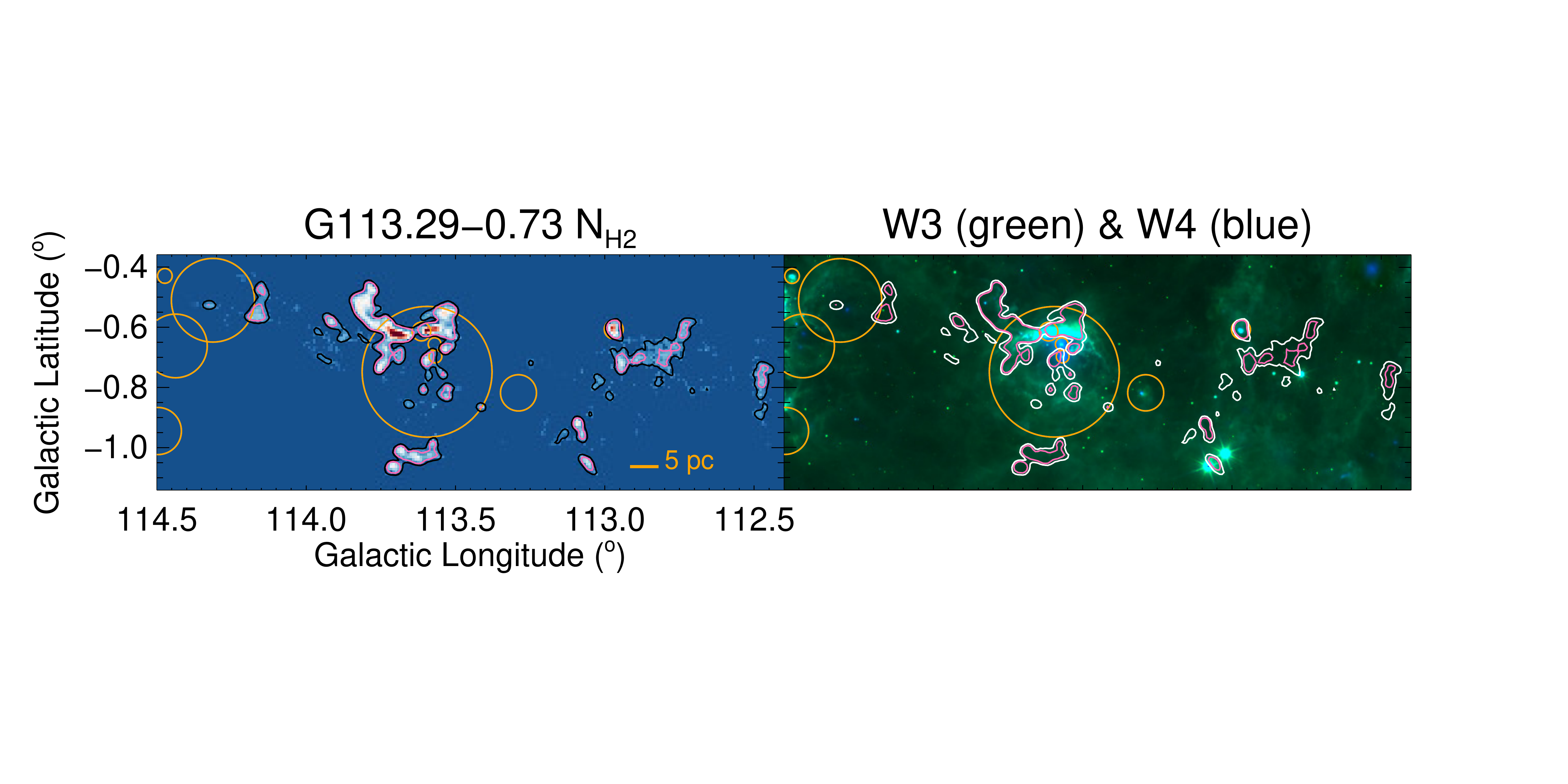}
		\label{fig19d}
		\\(d)
	\end{minipage}
	\caption{Same as Figure \ref{fig18} but for the clouds in the Perseus arm. The first three panels are three examples of LN$^{*}$ clouds, and the last panel is an example of PL clouds.}
	\label{fig19}
\end{figure*}
\clearpage

Figures \ref{fig18} presents the H$_2$ column density maps and the WISE 3/4 images of three LN$^{*}$ molecular clouds and one PL clouds in the Local arm and Figure \ref{fig19} presents the example clouds in the Perseus arm. The H$_2$ column density maps and WISE 3/4 images of other selected molecular clouds are given in Figures \ref{fig23}-\ref{fig31} in the Appendix. The green contours in the left panel and the red contours in the right panel of Figure \ref{fig18}(a) correspond to N$\sim3\times$10$^{22}$ cm$^{-2}$, at which the cloud G110.69+1.82 (Cep GMC) has slight excesses above its log-normal N-PDF. The contours coincide with three intensity peaks (Cep A, Cep B, and Cep F) in the column density map, of sizes of $\leqslant$1 pc. The N-PDF of the LN$^{*}$ cloud G118.05+3.15 show excess above log-normal distribution at two column densities, N$_{H_2}\sim7\times10^{21}$ cm$^{-2}$ (within $<$5 pc) and N$_{H_2}\sim2\times10^{22}$ cm$^{-2}$ (within $<$1 pc), shown as the two green contours in Figure \ref{fig18}(b) and this region is associated with a WISE \ion{H}{2} region. 

The N-PDF of LN$^{*}$ molecular cloud G108.75+2.69 exhibits slight excesses above the log-normal distribution at the column density N$_{H_2}\sim1\times10^{22}$ cm$^{-2}$, shown as the green and red contours in Figure \ref{fig18}(c). These contours coincide with a WISE \ion{H}{2} region ($\sim$ 2pc). In the Local arm, the column densities that correspondes to the excesses above the log-normal N-PDFs are mainly concentrated in clumps of sizes of $\sim 1-5$ pc. Similar to the case of the Local arm, the slight high-density excesses of the N-PDF of the LN$^*$ molecular clouds in the Perseus arm are mainly concentrated in regions of sizes of $\sim 5-10$ pc that are associated with \ion{H}{2} regions (Figure \ref{fig19}). 

The images for the cloud G106.06+1.49 which has the PL form of N-PDF are shown in Figure \ref{fig18}(d). We can see that this cloud does not have signature of star formation activity. The other five molecular clouds with PL N-PDFs in our survey also do not exhibit star formation activity (see Figures \ref{fig26c}, \ref{fig26d}, \ref{fig28d}, \ref{fig29a}, and \ref{fig31b} in the Appendix). The power-law parts of the N-PDFs of the PL molecular clouds are concentrated in clumps or filaments of size or width of $\sim0.2-1$ pc. For example, the PL cloud G114.78+0.82 is composed of several filamentary structures of size $\sim$1 pc. The PL portion of the N-PDF of cloud G113.29-0.73 corresponds to several filamentary structures around the \ion{H}{2} region S 163. The PL cloud G111.47+2.42 consists of a few filamentary structures with lengths of $\sim$5 pc and widths of $\sim$1 pc, and its WISE 3/4 images do not show active star formation (see Figure \ref{fig31b}).

\section{Summary}\label{sec5}
We have conducted a comprehensive study of the properties of molecular clouds in a 15\arcdeg$\times$10.5$\arcdeg$ region in the second quadrant of the Milky Way mid-plane using the \CO, \COl, and \COll data from the MWISP survey. The distribution and basic statistics of the physical properties of the molecular gas are presented. Using the DENDROGRAM based SCIMES algorithm, we used the \CO and \COl line emission to identify molecular clouds and studied the statistical properties of these clouds. The scaling relations between the physical parameters are investigated, and comparisons of the scaling relations between different spiral arms are discussed. Forty clouds are selected as a sub-sample to study the properties of the N-PDFs using the \COl emission line as the tracer of H$_2$ column density. The main results are presented as follows.
\begin{enumerate}
\item Under the influence of the distance selection effect, we have identified molecular clouds in the Local arm above the size limit of $\sim$0.25 pc and mass limit of $\sim$6 M$_{\sun}$, and large and massive molecular clouds in the Perseus arm above the size limit of $\sim$1.9 pc and mass limit of $\sim$349 M$_{\sun}$. With this bias, the median mass of the identified \CO and \COl molecular clouds in the Perseus arm is $\sim$50 and $\sim$30 times that of the Local arm, respectively, and the molecular clouds in the Perseus arm are $\sim$6 times larger than those of the Local arm, as measured by \CO emission, while $\sim$4 times larger, as measured by \COl emission. The surface density of molecular clouds is significantly enhanced in the Perseus arm, up to $\sim$100 M$_{\odot}$ pc$^{-2}$. 

\item The exponent of the $\sigma_v-R$ relation is $\sim$0.29 for the molecular clouds in the Local arm, while it is $\sim$0.44 for those in the Perseus arm, showing that turbulence of molecular clouds may vary with Galactocentric distance. 

\item The percentage of gravitationally bound molecular clouds in the Perseus arm (58.2\%) is much higher than that in the Local arm (19.1\%) in our results, partly due to the distance selection effect. An external pressure P$_e$/k$\sim$10$^4-10^6$ cm$^{-3}$ is needed for the molecular clouds in the Local arm to stay in equilibrium.

\item The N-PDFs derived with the \COl emission are dominated by log-normal distributions with few or only minor excesses above the log-normal distribution. The excesses at high-density correspond to star-forming regions of scales $\sim1-5$ pc for the Local arm and of scales $\sim5-10$ pc for the Perseus arm. The majority of the clouds that have power-law N-PDFs correspond to molecular clumps of sizes of $\sim$1 pc or filaments of widths of $\sim$1 pc.
\end{enumerate}

\acknowledgments
We thank the PMO-13.7 m telescope staffs for their supports during the observation and the staffs of the MWISP scientific group for their valuable suggestions. We thank the anonymous referee for his/her constructive suggestions that help to improve this manuscript. Y. M. thanks Fujun Du for his helpful discussion. This work is supported by the National Key R$\&$D Program of China (NO. 2017YFA0402701) and Key Research Program of Frontier Sciences of CAS under grant QYZDJ-SSW-SLH047. We acknowledge the support by NSFC grant 11973091. Y.M. acknowledges financial supports by the Natural Science Foundation of Jiangsu Province of China (Grant No. BK20181513) and by the Natural Science Foundation of China (Grant No. 11973090). Y.S. acknowledges supports by NSFC grant 11773077. This work makes use of the SIMBAD database, operated at CDS, Strasbourg, France. This research made use of SCIMES, a Python package to find relevant structures into dendrograms of molecular gas emission using the spectral clustering approach.
\clearpage

\begin{figure*}[htb!]
	\centering
	\subfigure[]{
		\label{fig20a}
		\includegraphics[trim=2cm 1cm 1cm 1cm, width = 0.8\linewidth, clip]{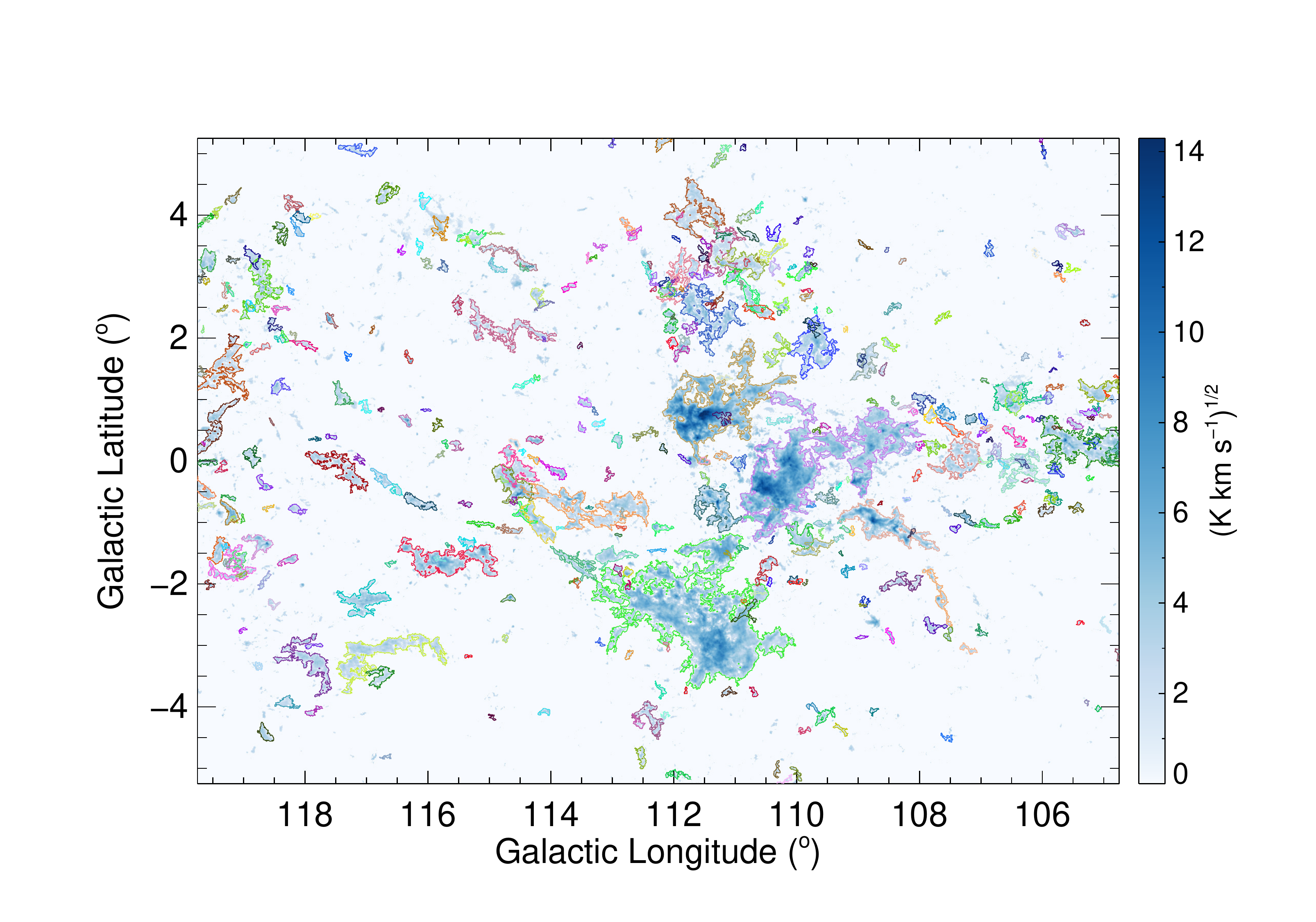}}
	\subfigure[]{
		\label{fig20b}
		\includegraphics[trim=3cm 1cm 3cm 2cm, width = 0.8\linewidth, clip]{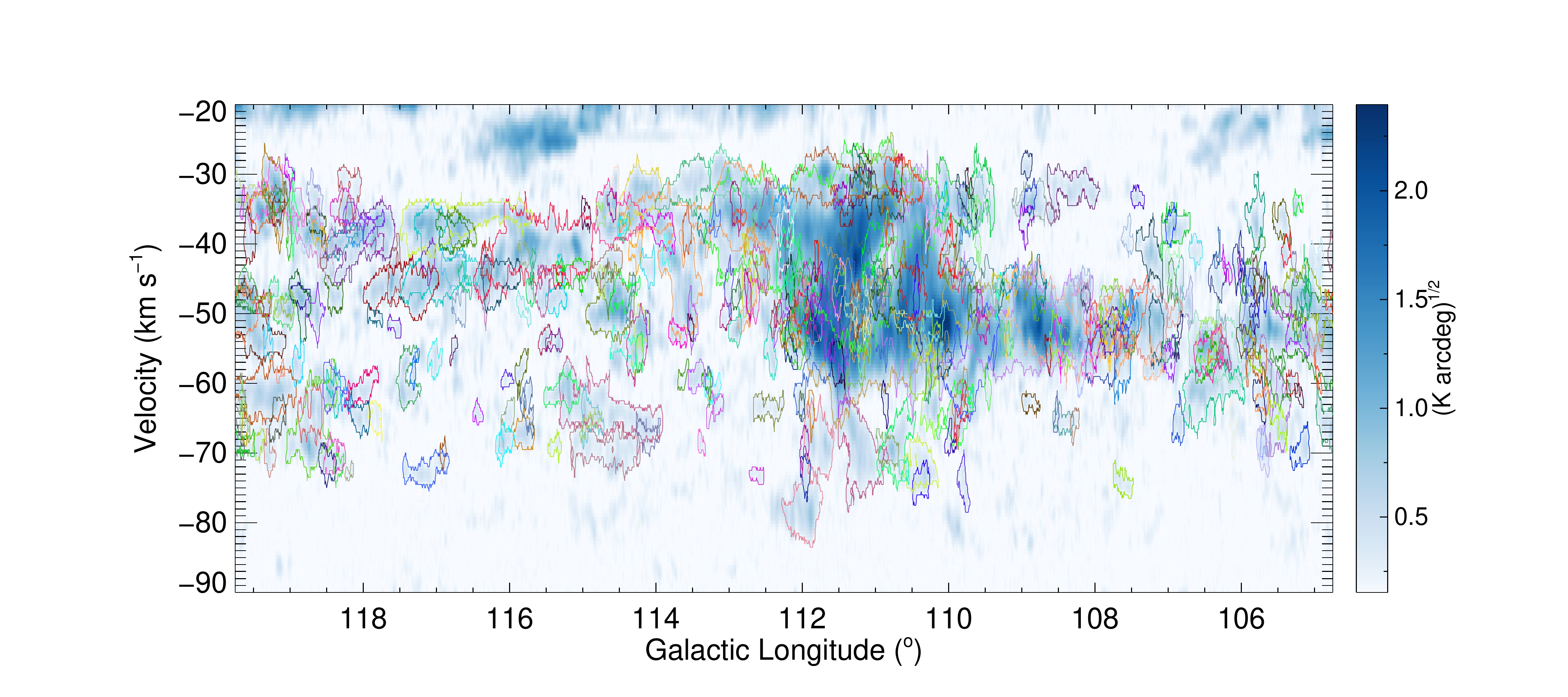}}
	\caption{Demonstration of the outlines of the identified \CO molecular clouds in the Perseus arm in the (a) l-b space and the (b) l-v space. Different colors correspond to different molecular clouds.}
	\label{fig20}
\end{figure*}

\begin{figure*}[htb!]
	\centering
	\subfigure[]{
		\label{fig21a}
		\includegraphics[trim=2cm 1cm 1cm 1cm, width = 0.8\linewidth , clip]{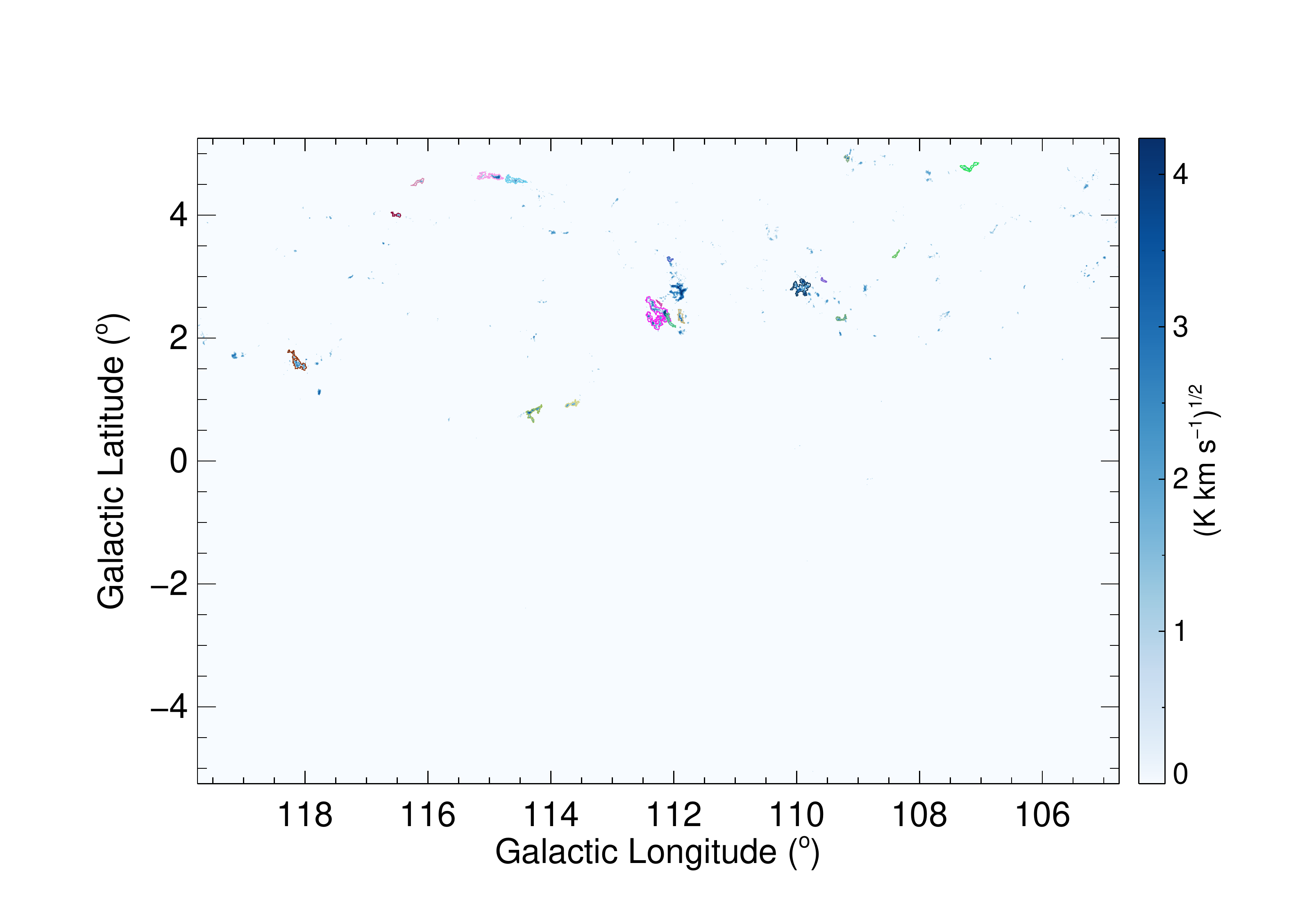}}
	\subfigure[]{
		\label{fig21b}
		\includegraphics[trim=3cm 1cm 2cm 2cm, width = 0.8\linewidth , clip]{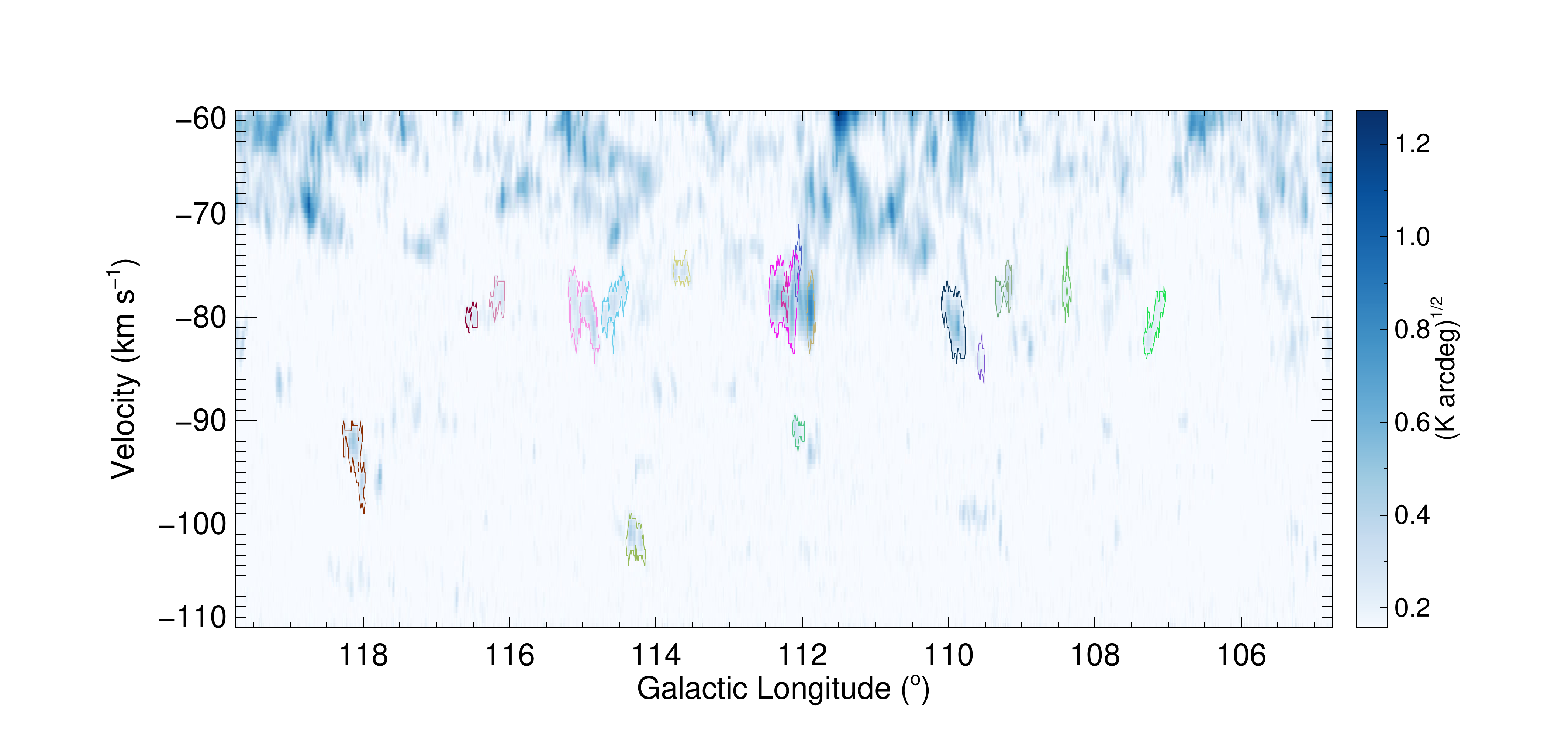}}
	\caption{Same as Figure \ref{fig20}, but for the \CO clouds in the Outer$+$OSC arm.}
	\label{fig21}
\end{figure*}

\begin{figure*}[htb!]
	\centering
	\subfigure[]{
		\label{fig22a}
		\includegraphics[trim=2cm 1cm 1cm 1cm, width = 0.8\linewidth , clip]{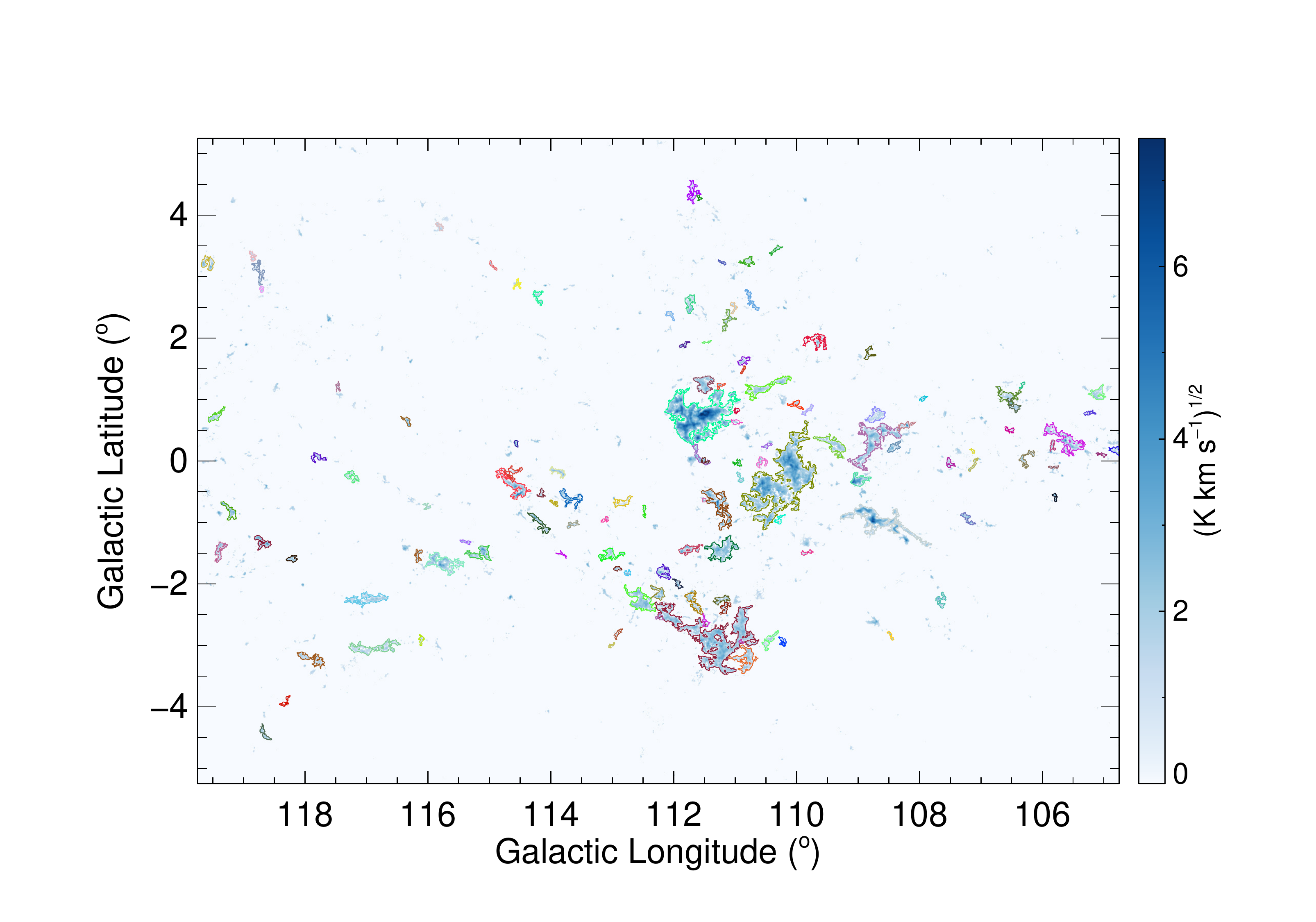}}
	\subfigure[]{
		\label{fig22b}
		\includegraphics[trim=5cm 1cm 4cm 2cm, width = 0.8\linewidth , clip]{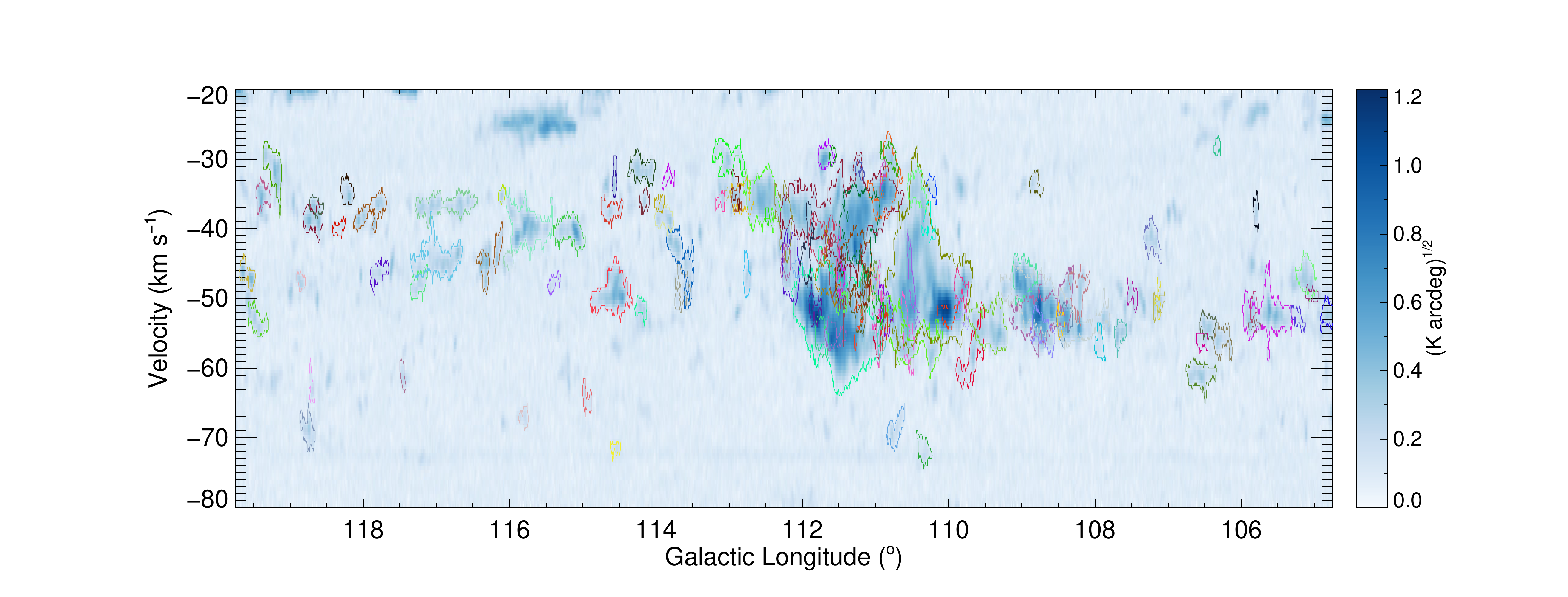}}
	\caption{Same as Figure \ref{fig20}, but for \COl clouds in the Perseus arm.}
	\label{fig22}
\end{figure*}

\begin{figure*}[tb!]
	\centering
	\subfigure[]{
	\label{fig23a}
	\includegraphics[trim=0cm 4cm 4cm 4cm, width = 0.6\linewidth, clip]{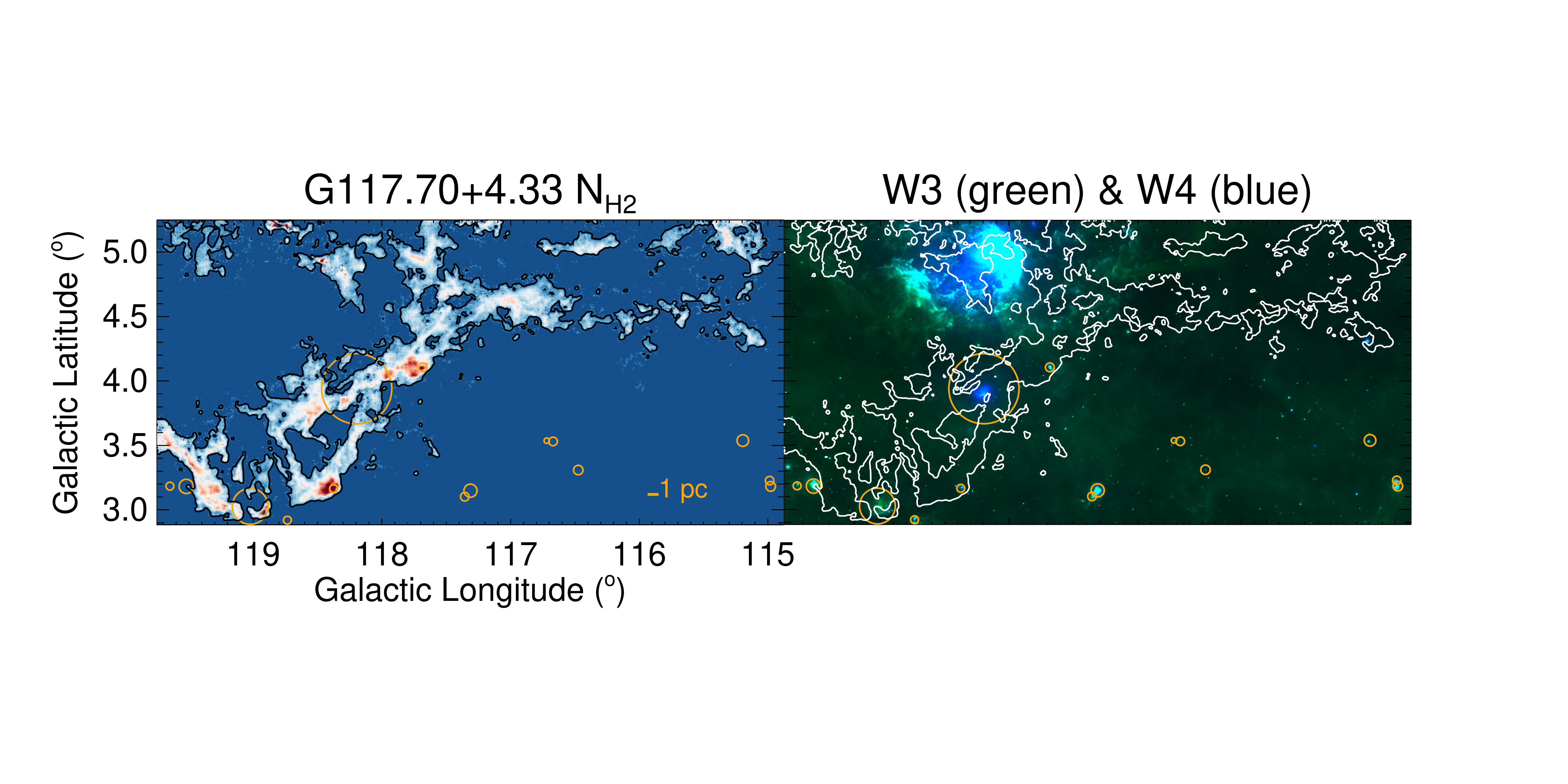}}
	\subfigure[]{
	\label{fig23b}
	\includegraphics[trim=0cm 2.5cm 4cm 2.8cm, width = 0.6\linewidth , clip]{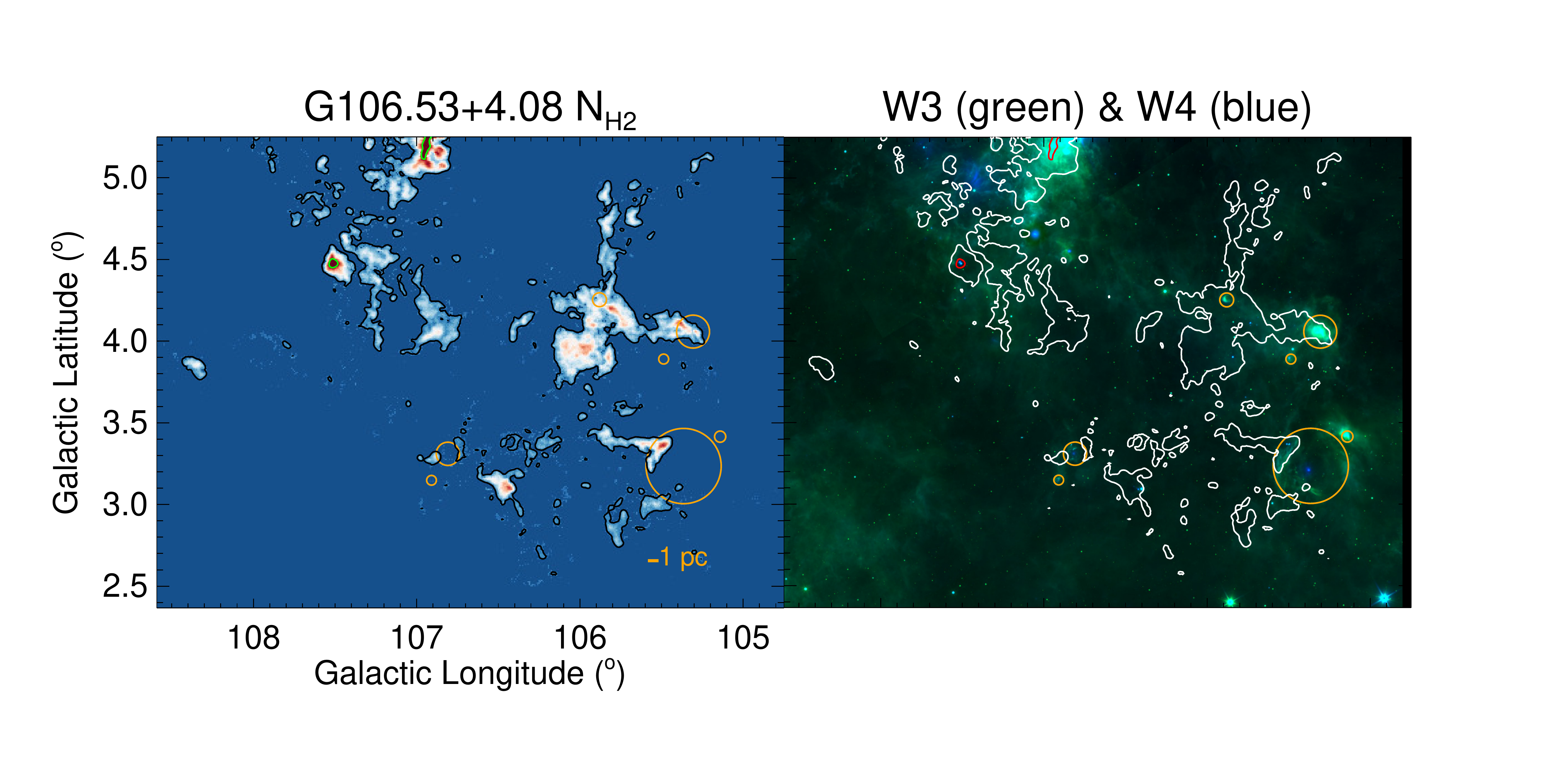}}
	\subfigure[]{
	\label{fig23c}
	\includegraphics[trim=0cm 0cm 2cm 0cm, width = 0.6\linewidth, clip]{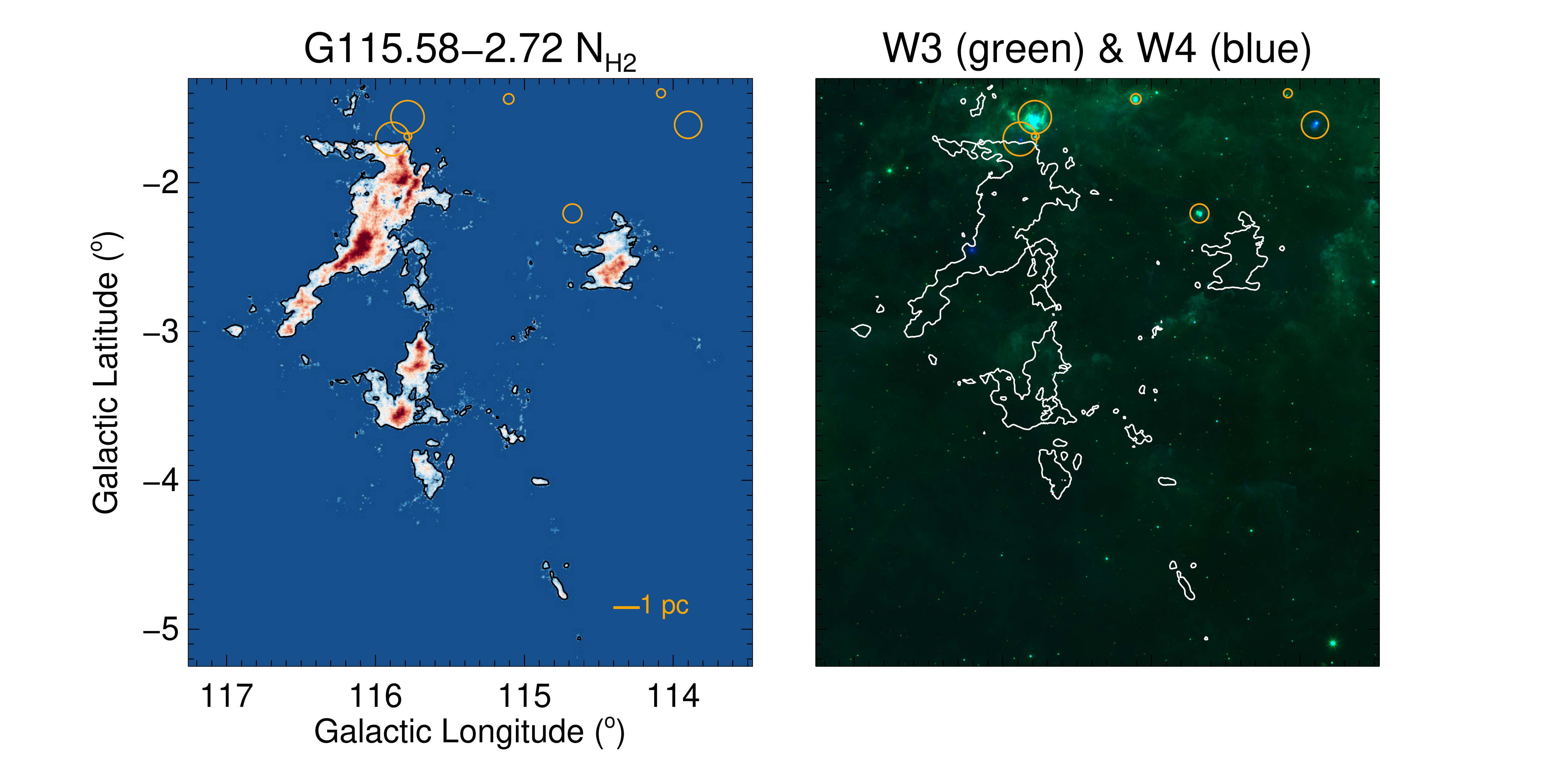}}
	\subfigure[]{
	\label{fig23d}
	\includegraphics[trim=0cm 2cm 3cm 3cm, width = 0.6\linewidth, clip]{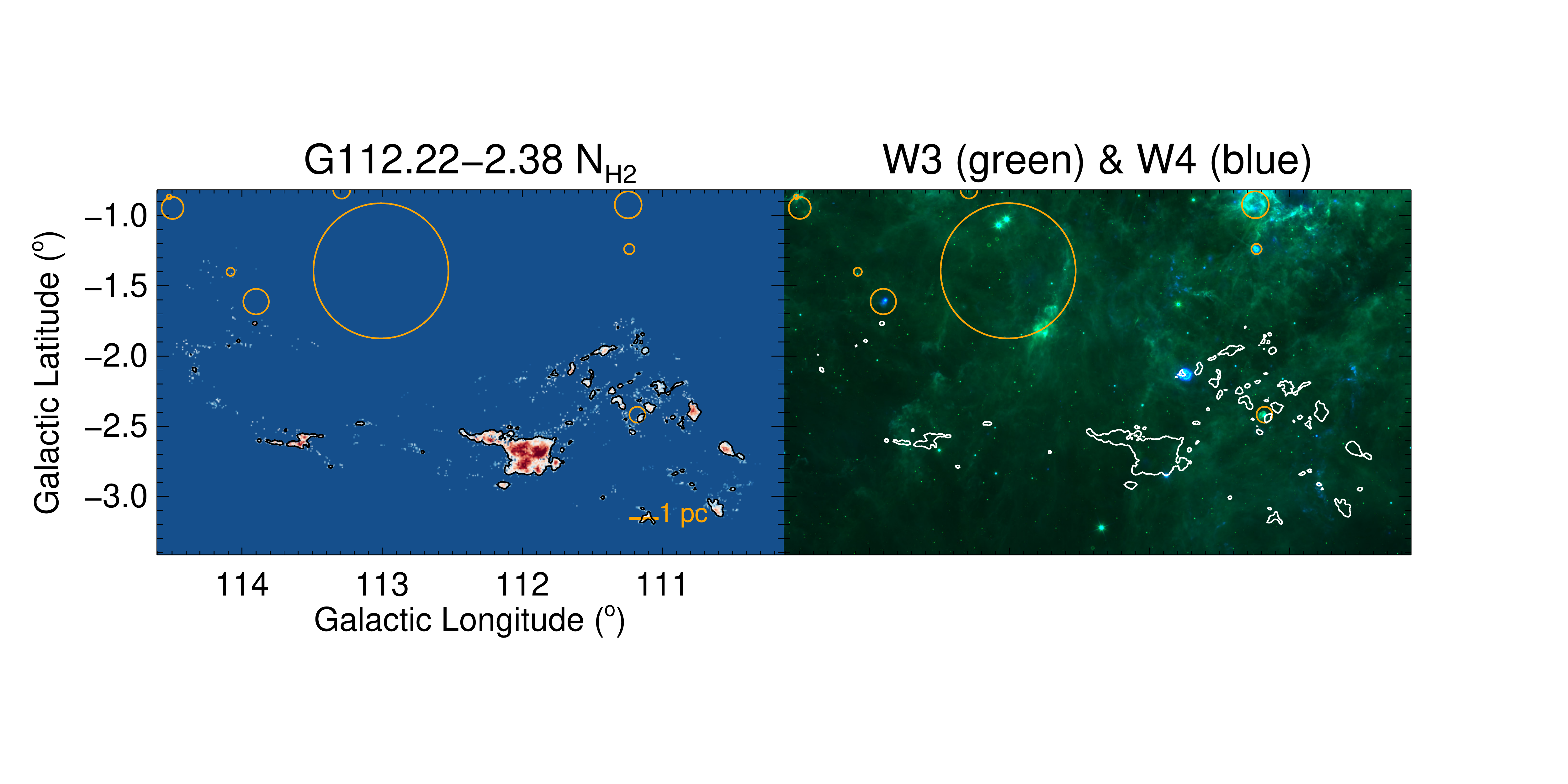}}
	\caption{Same as Figure \ref{fig18}, but for clouds G117.70+4.33 etc.}\label{fig23}
\end{figure*}

\begin{figure*}[tb!]
	\centering
	\subfigure[]{
	\label{fig24a}
	\includegraphics[trim=0cm 4cm 4cm 4cm, width = 0.6\linewidth , clip]{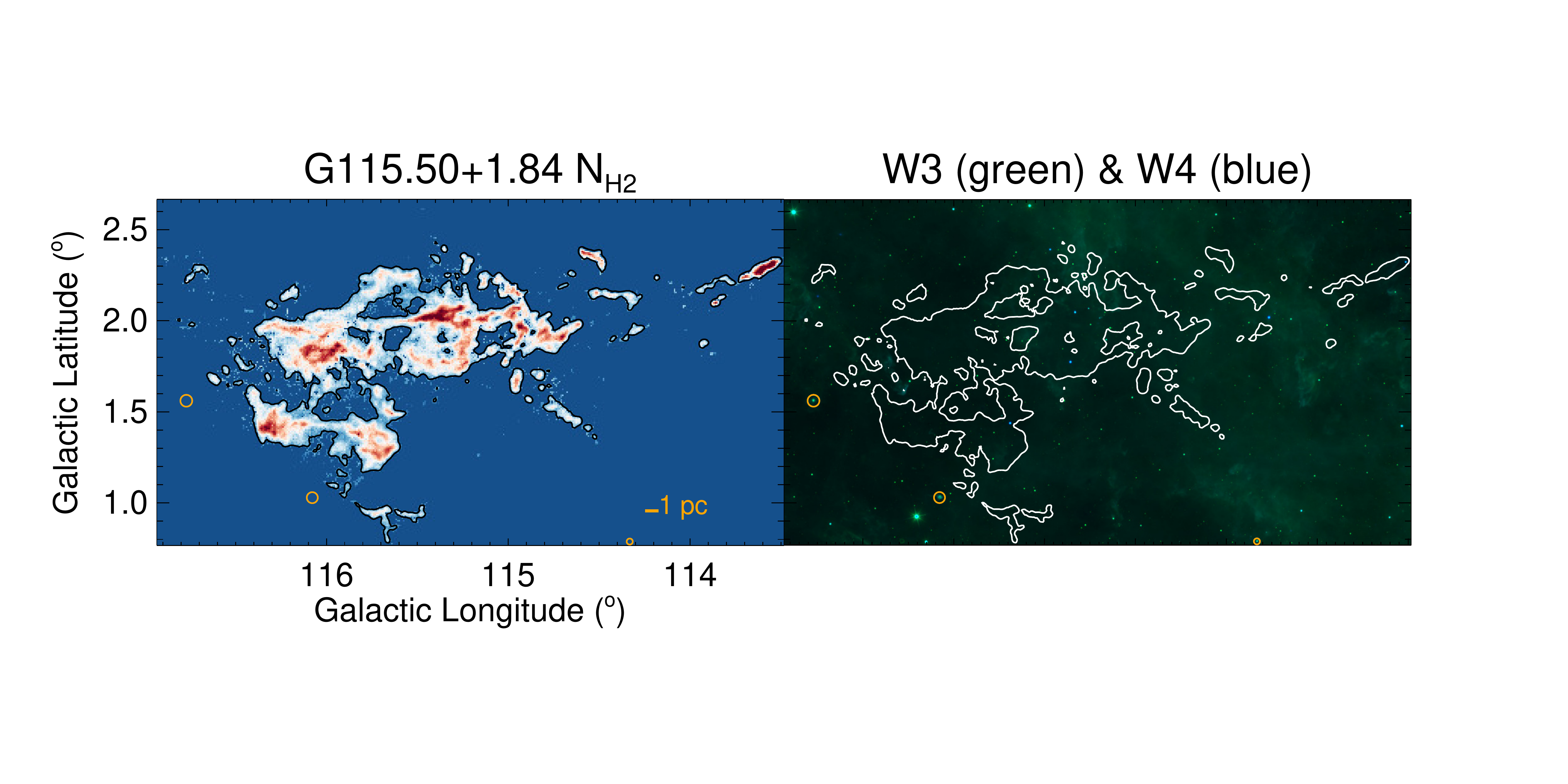}}
	\subfigure[]{
	\label{fig24b}
	\includegraphics[trim=0cm 4cm 4cm 4cm, width = 0.6\linewidth , clip]{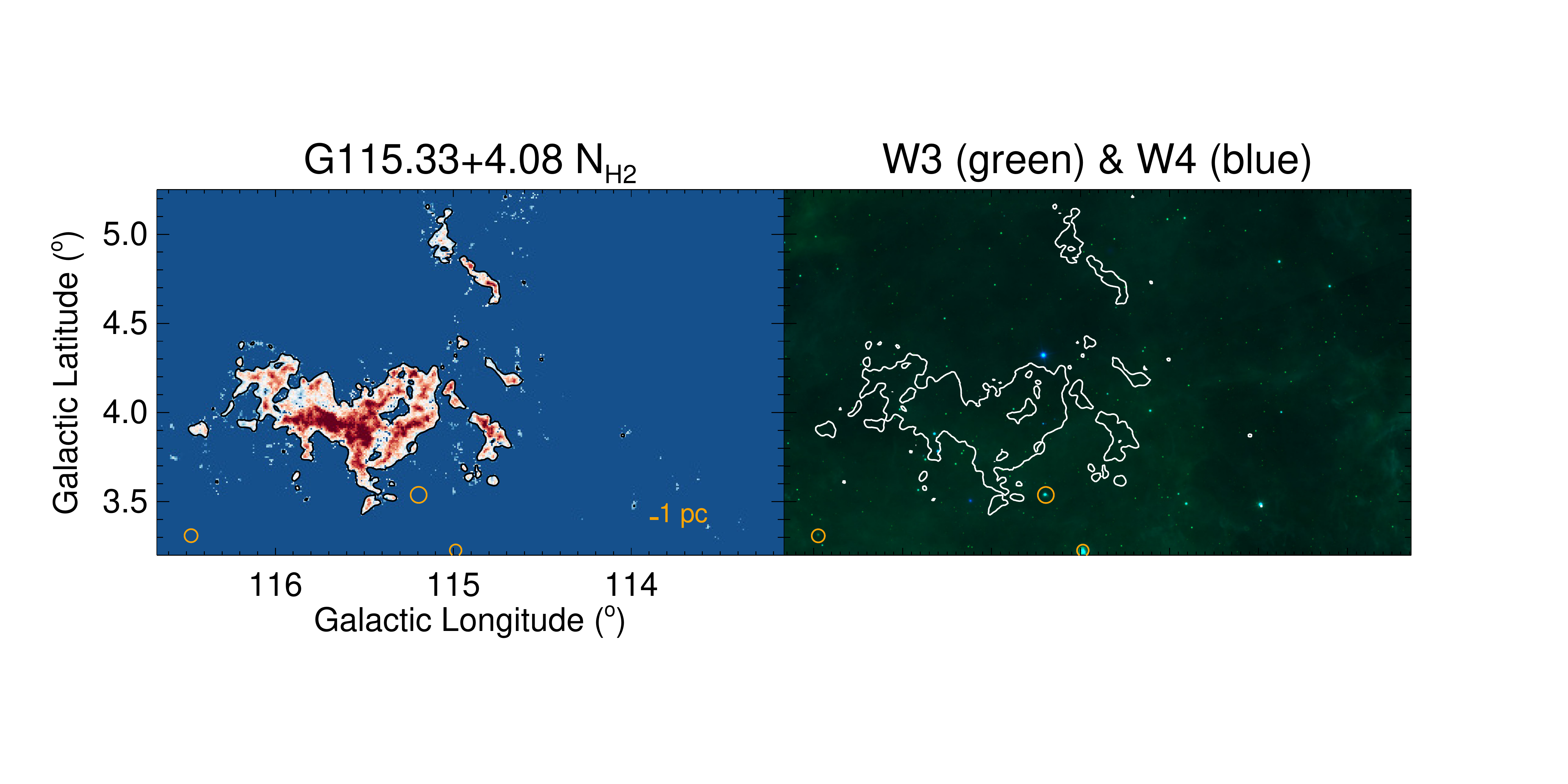}}
	\subfigure[]{
	\label{fig24c}
	\includegraphics[trim=0cm 2cm 2cm 2cm, width = 0.6\linewidth , clip]{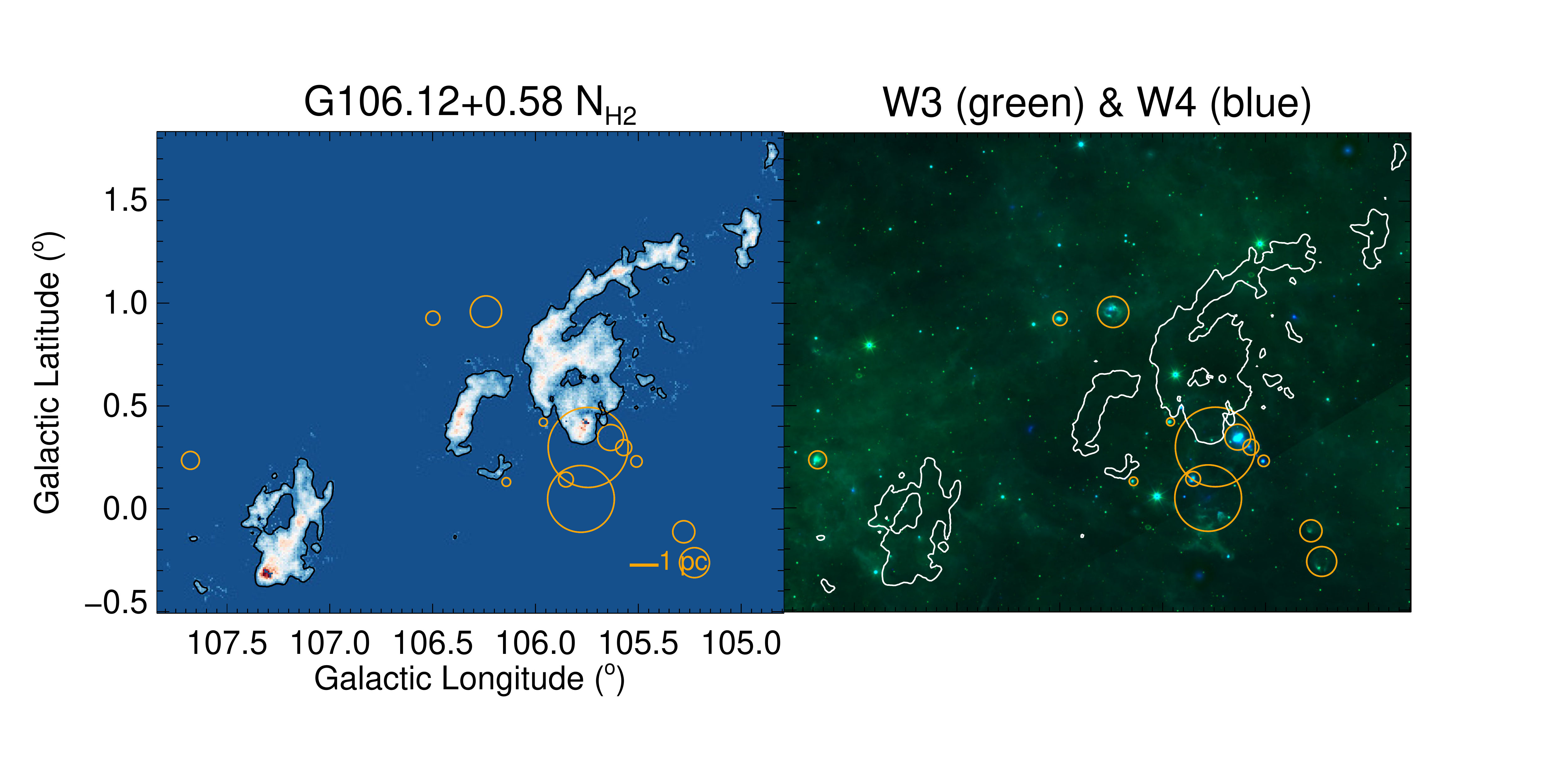}}
	\subfigure[]{
	\label{fig24d}
	\includegraphics[trim=0cm 0cm 3cm 0cm, width = 0.6\linewidth , clip]{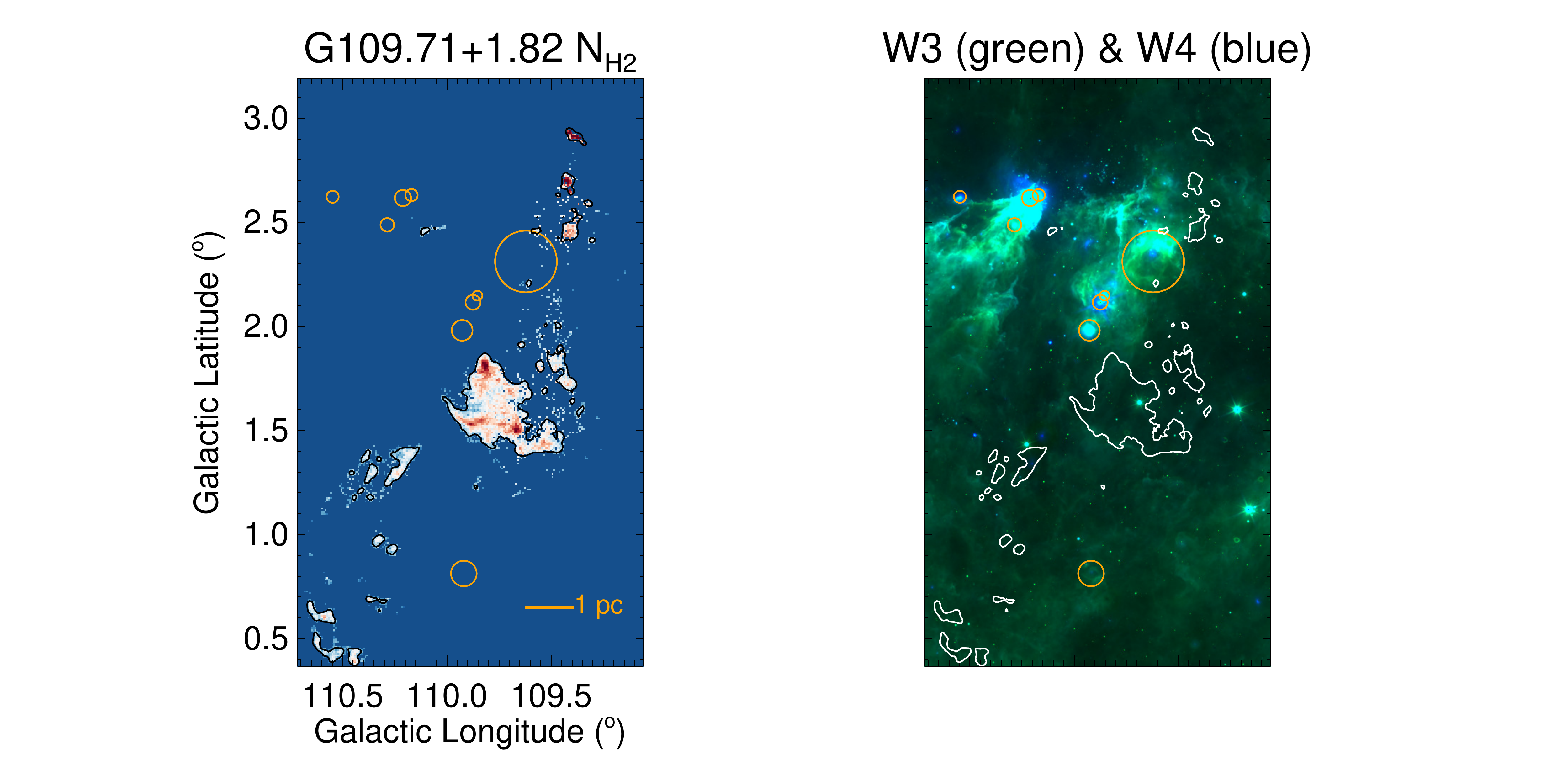}}
	\caption{Same as Figure \ref{fig18}, but for clouds G115.50+1.84 etc.}\label{fig24}
\end{figure*}

\begin{figure*}[tb!]
	\centering
	\subfigure[]{
	\label{fig25a}
		\includegraphics[trim=0cm 2cm 4cm 3cm, width = 0.6\linewidth , clip]{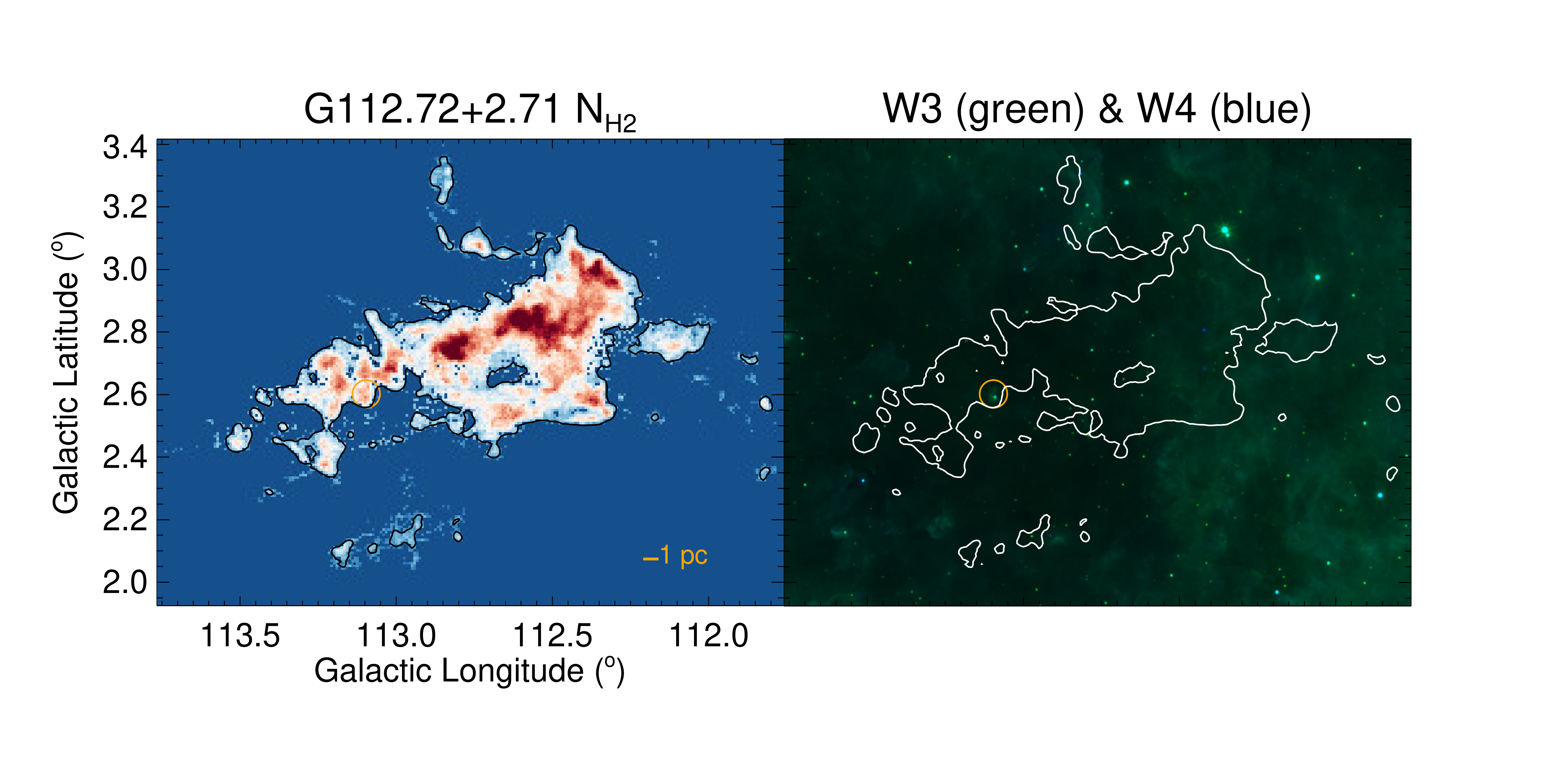}}
	\subfigure[]{
	\label{fig25b}
		\includegraphics[trim=0cm 3cm 4cm 3cm, width = 0.6\linewidth , clip]{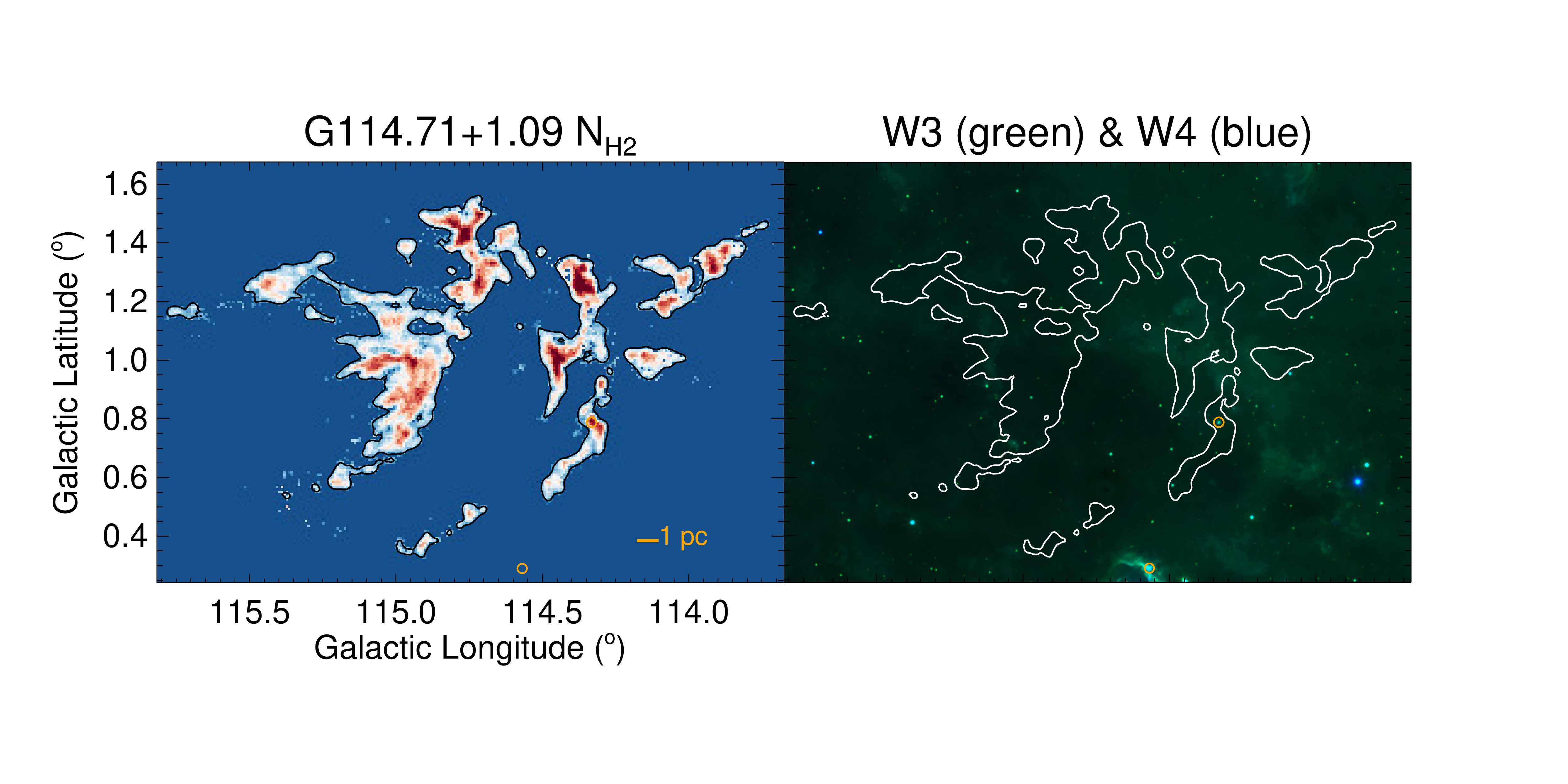}}
	\subfigure[]{
	\label{fig25c}
		\includegraphics[trim=0cm 0cm 2cm 0cm, width = 0.6\linewidth , clip]{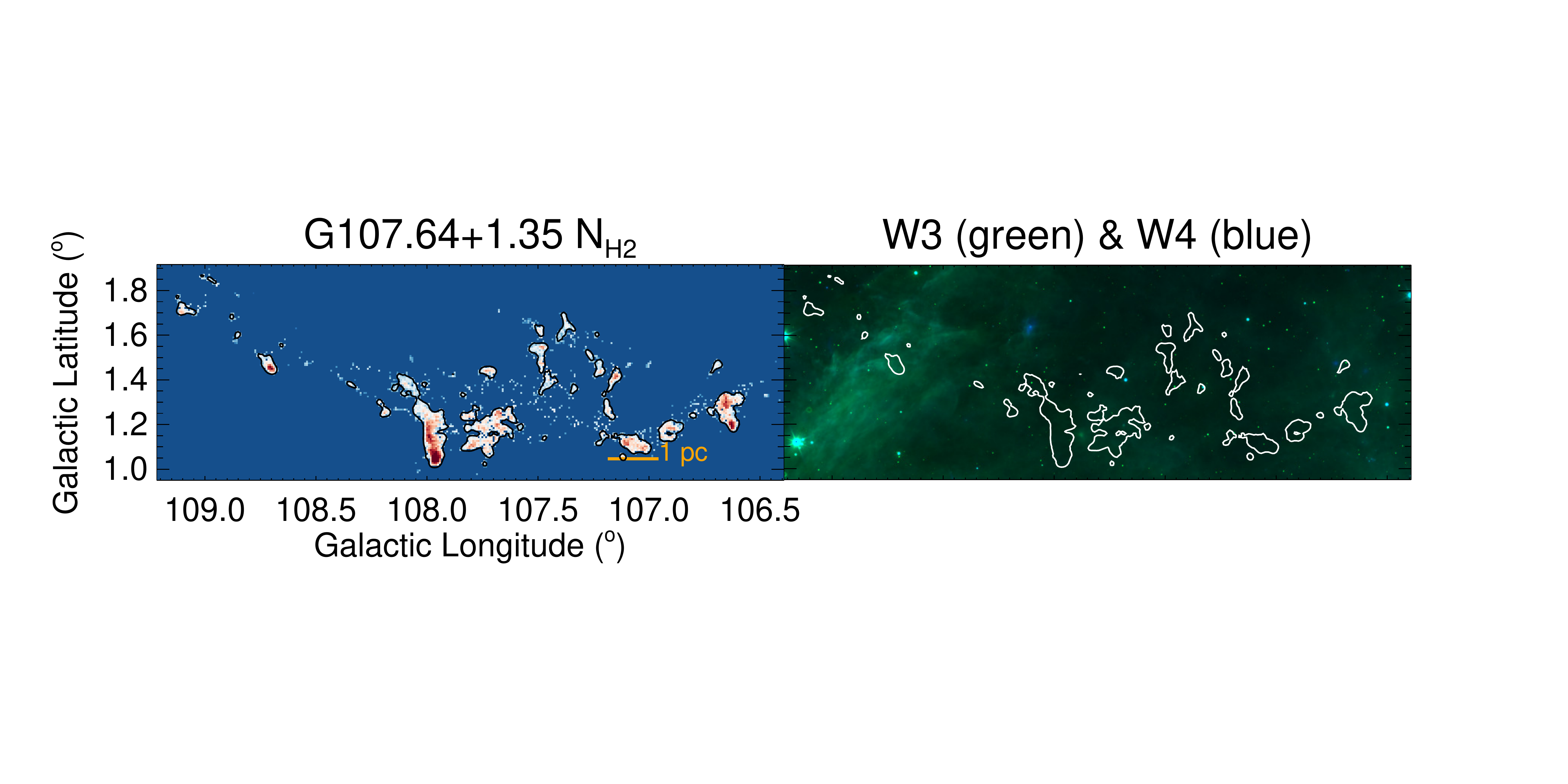}}
	\subfigure[]{
	\label{fig25d}
		\includegraphics[trim=0cm 0cm 3cm 0cm, width = 0.6\linewidth , clip]{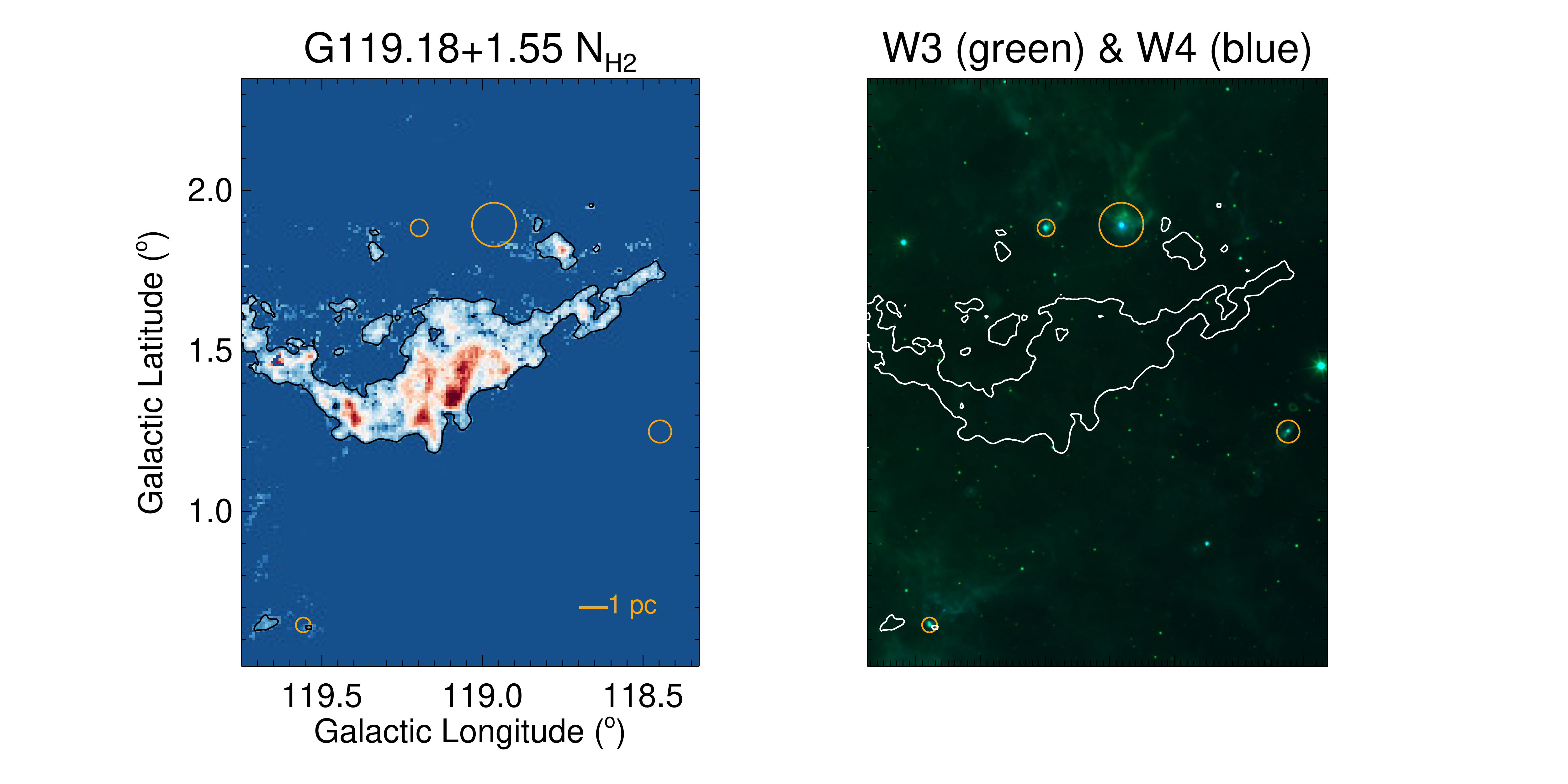}}
	\caption{Same as Figure \ref{fig18}, but for clouds G112.72+2.71 etc.}\label{fig25}
\end{figure*}

\begin{figure*}[tb!]
	\centering
	\subfigure[]{
	\label{fig26a}
		\includegraphics[trim=0cm 0cm 4cm 0cm, width = 0.6\linewidth , clip]{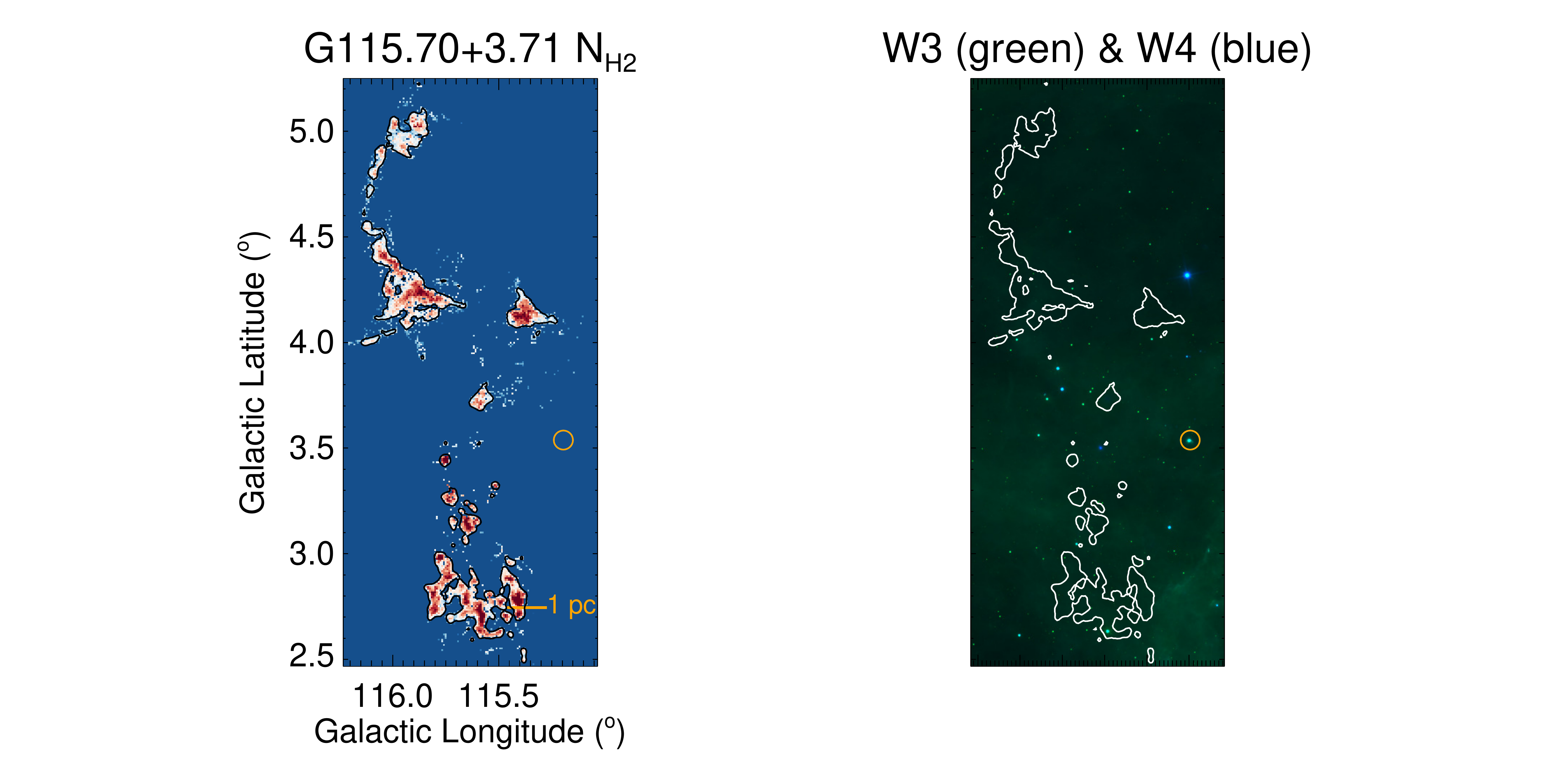}}
	\subfigure[]{
	\label{fig26b}
		\includegraphics[trim=0cm 4cm 4cm 4cm, width = 0.6\linewidth , clip]{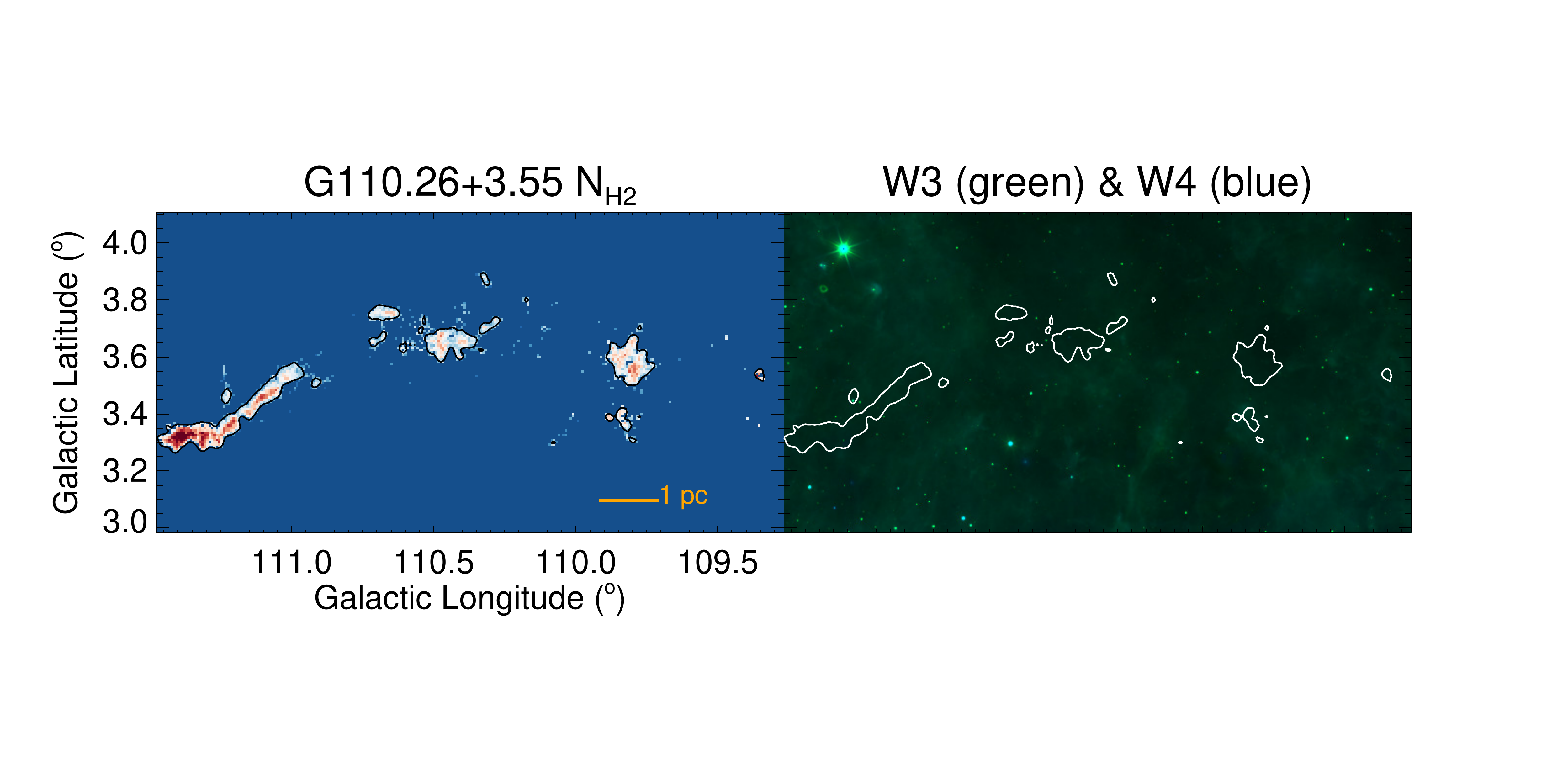}}
	\subfigure[]{
	\label{fig26c}
		\includegraphics[trim=0cm 0cm 2cm 1cm, width = 0.6\linewidth , clip]{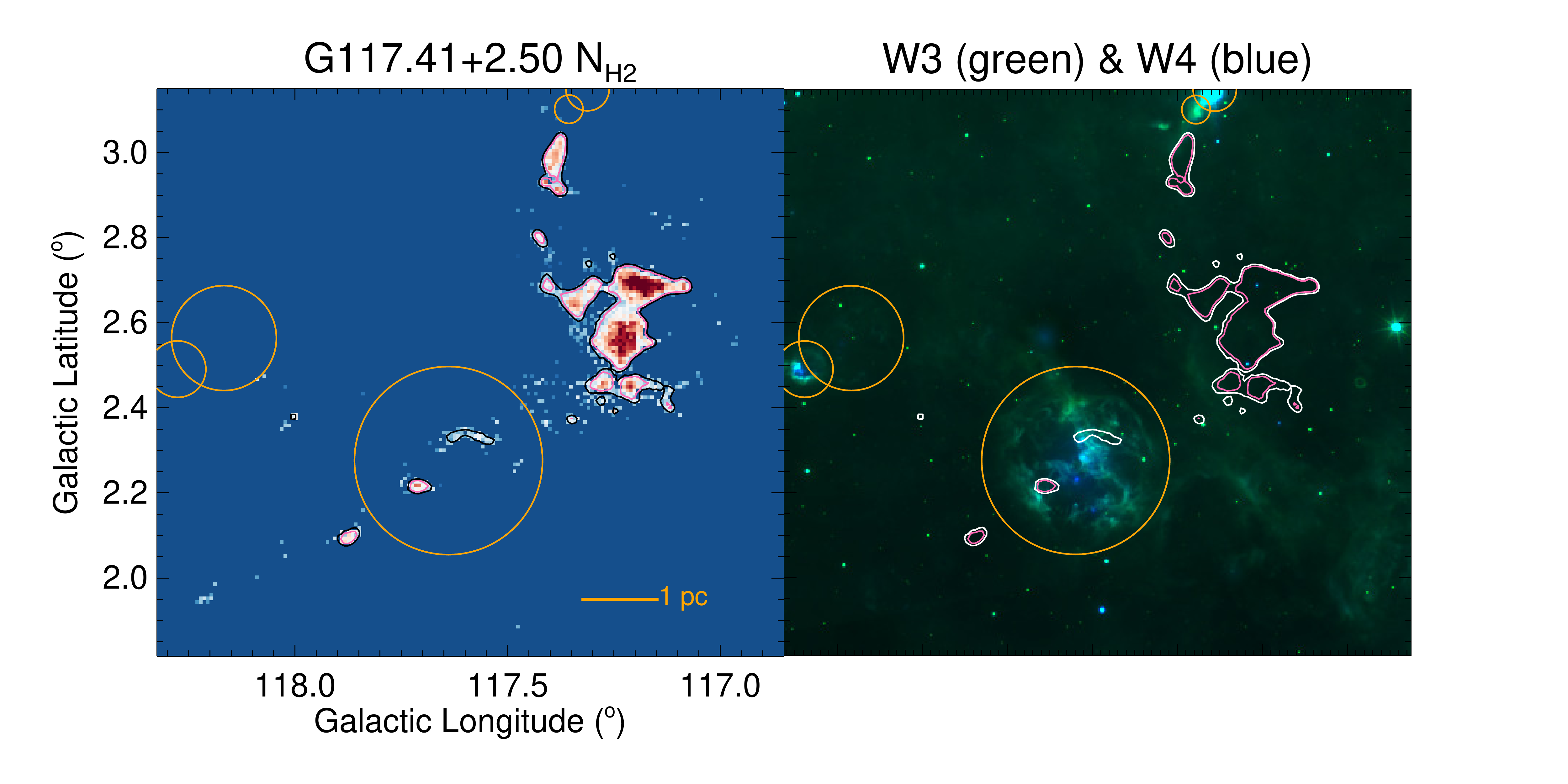}}
	\subfigure[]{
	\label{fig26d}
		\includegraphics[trim=0cm 2cm 3cm 3cm, width = 0.6\linewidth , clip]{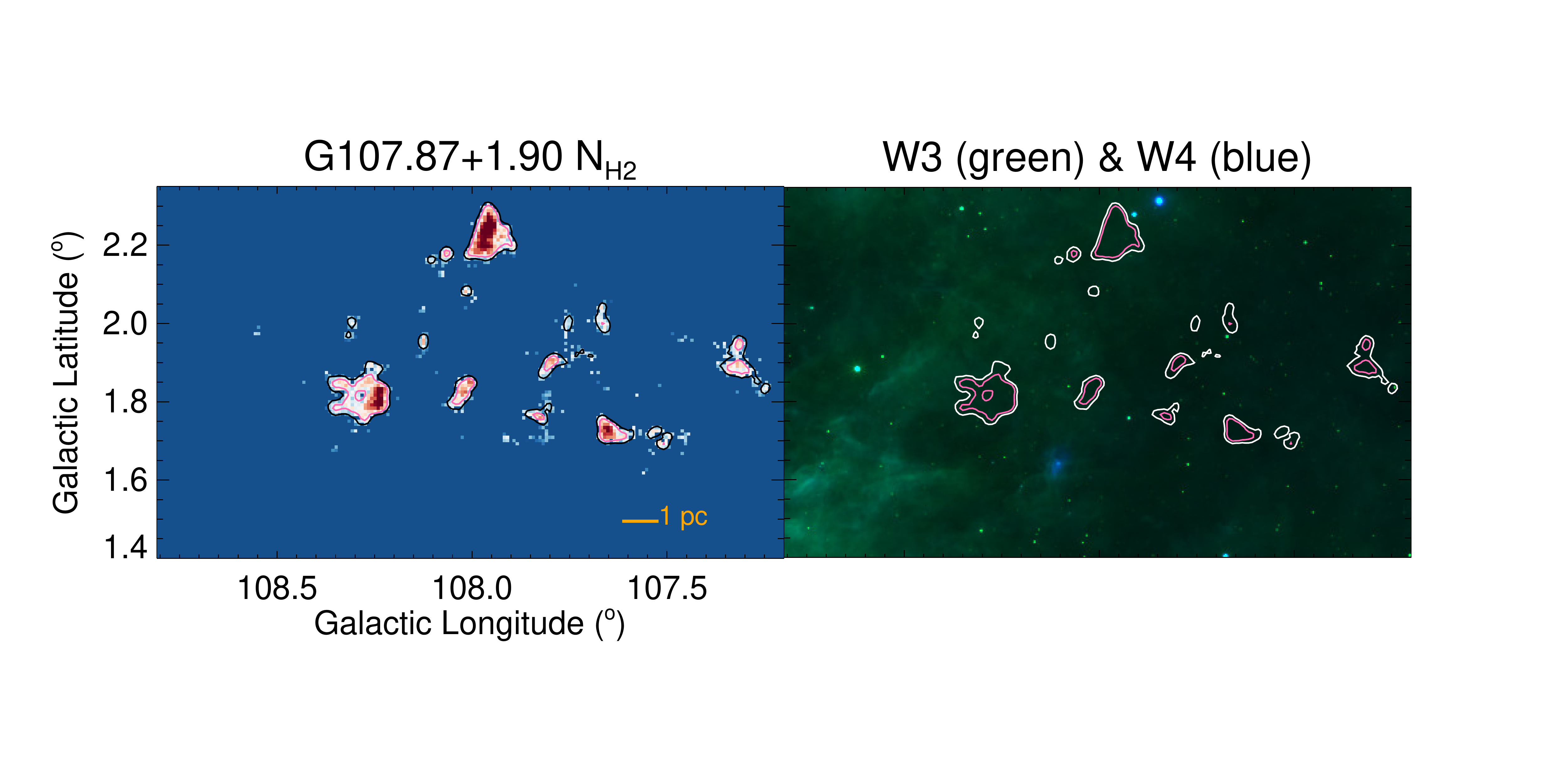}}
	\caption{Same as Figure \ref{fig18}, but for clouds G115.70+3.71 etc.}\label{fig26}
\end{figure*}

\begin{figure*}[tb!]
	\centering
	\subfigure[]{
	\label{fig27a}
		\includegraphics[trim=0cm 3cm 4cm 3cm, width = 0.6\linewidth , clip]{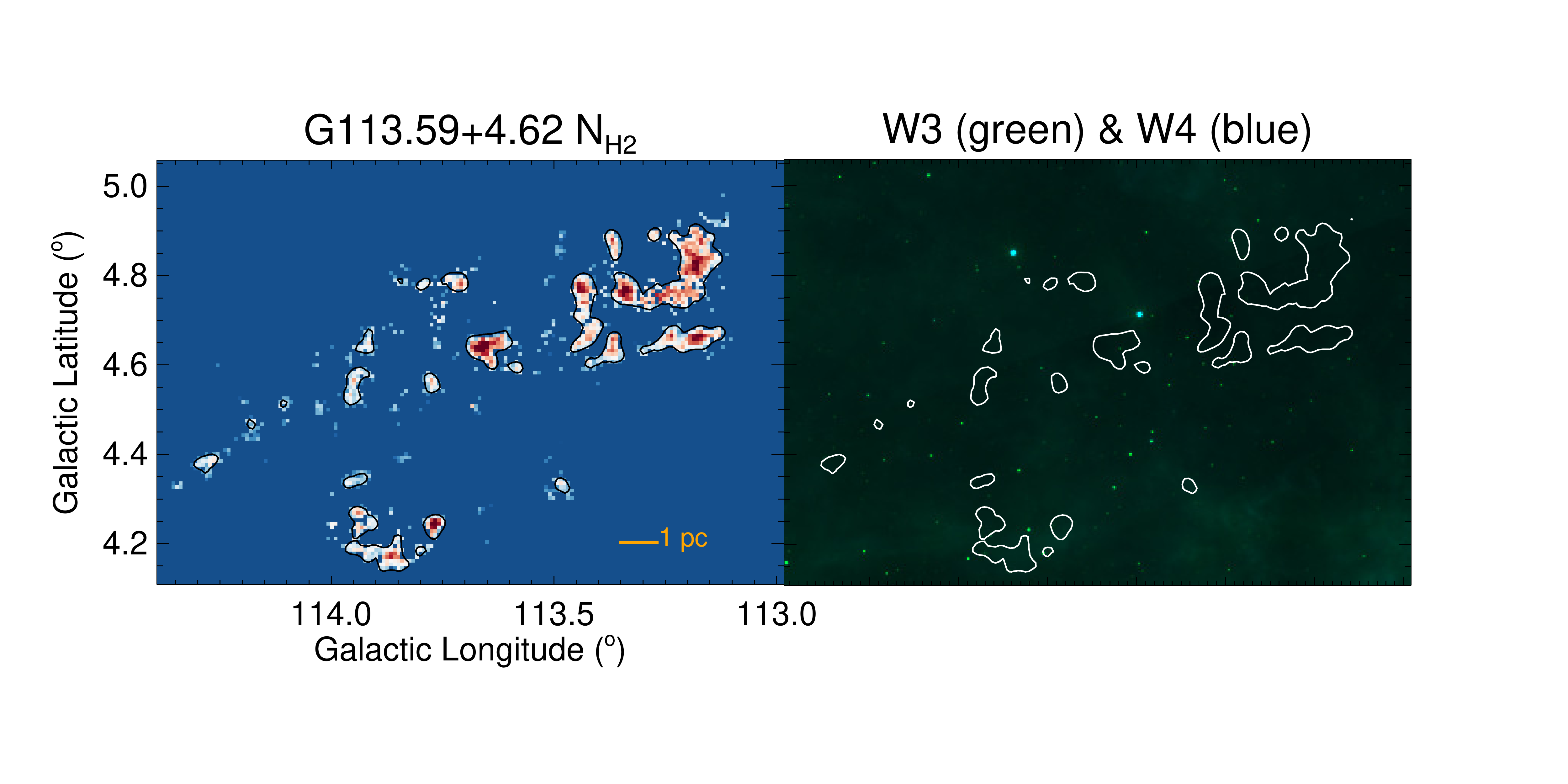}}
		\subfigure[]{
		\label{fig27b}
		\includegraphics[trim=0cm 3cm 4cm 3cm, width = 0.6\linewidth , clip]{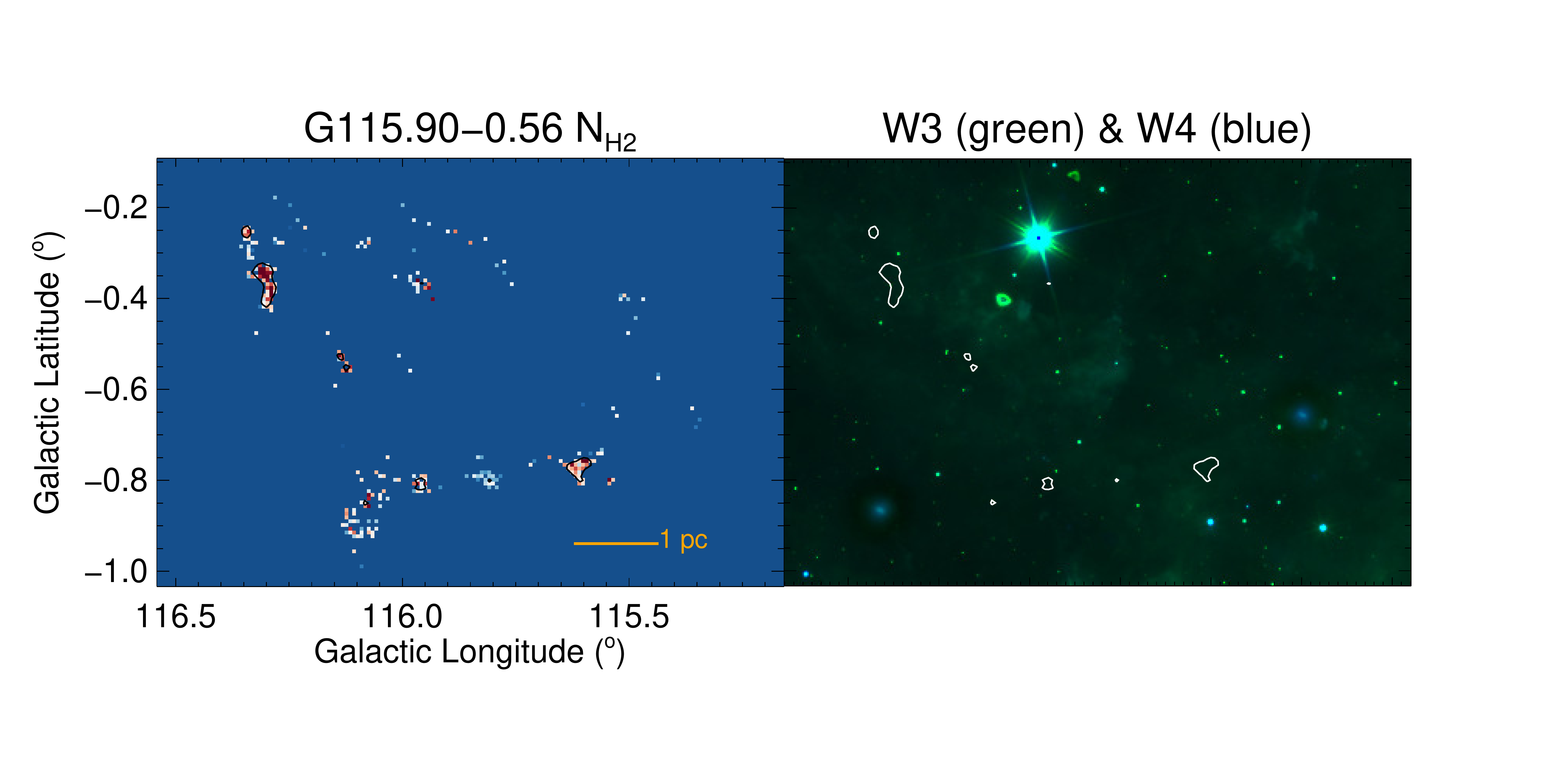}}
	\subfigure[]{
	\label{fig27c}
		\includegraphics[trim=0cm 2cm 2cm 2cm, width = 0.6\linewidth , clip]{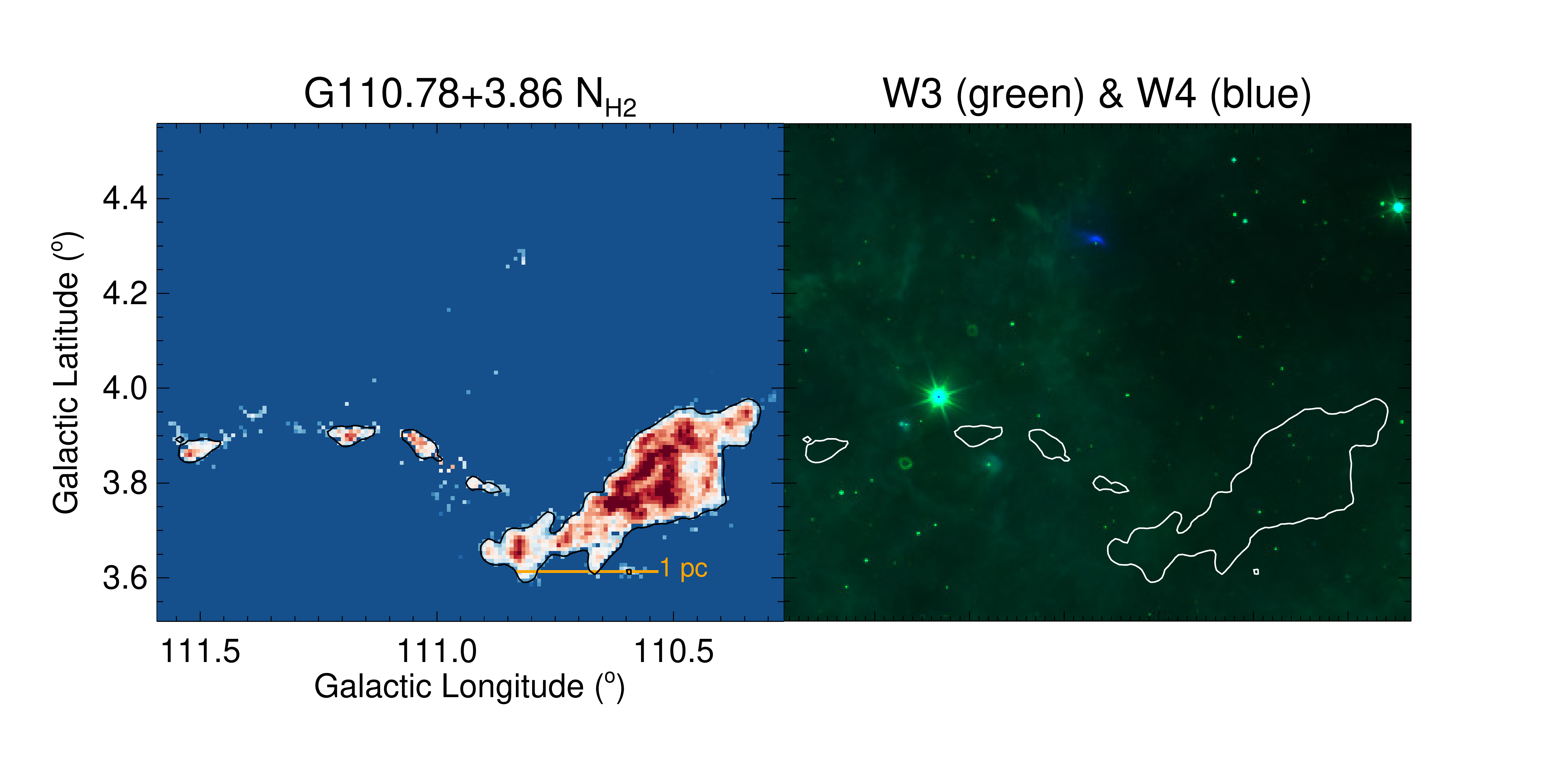}}
	\subfigure[]{
	\label{fig27d}
		\includegraphics[trim=0cm 0cm 3cm 0cm, width = 0.6\linewidth , clip]{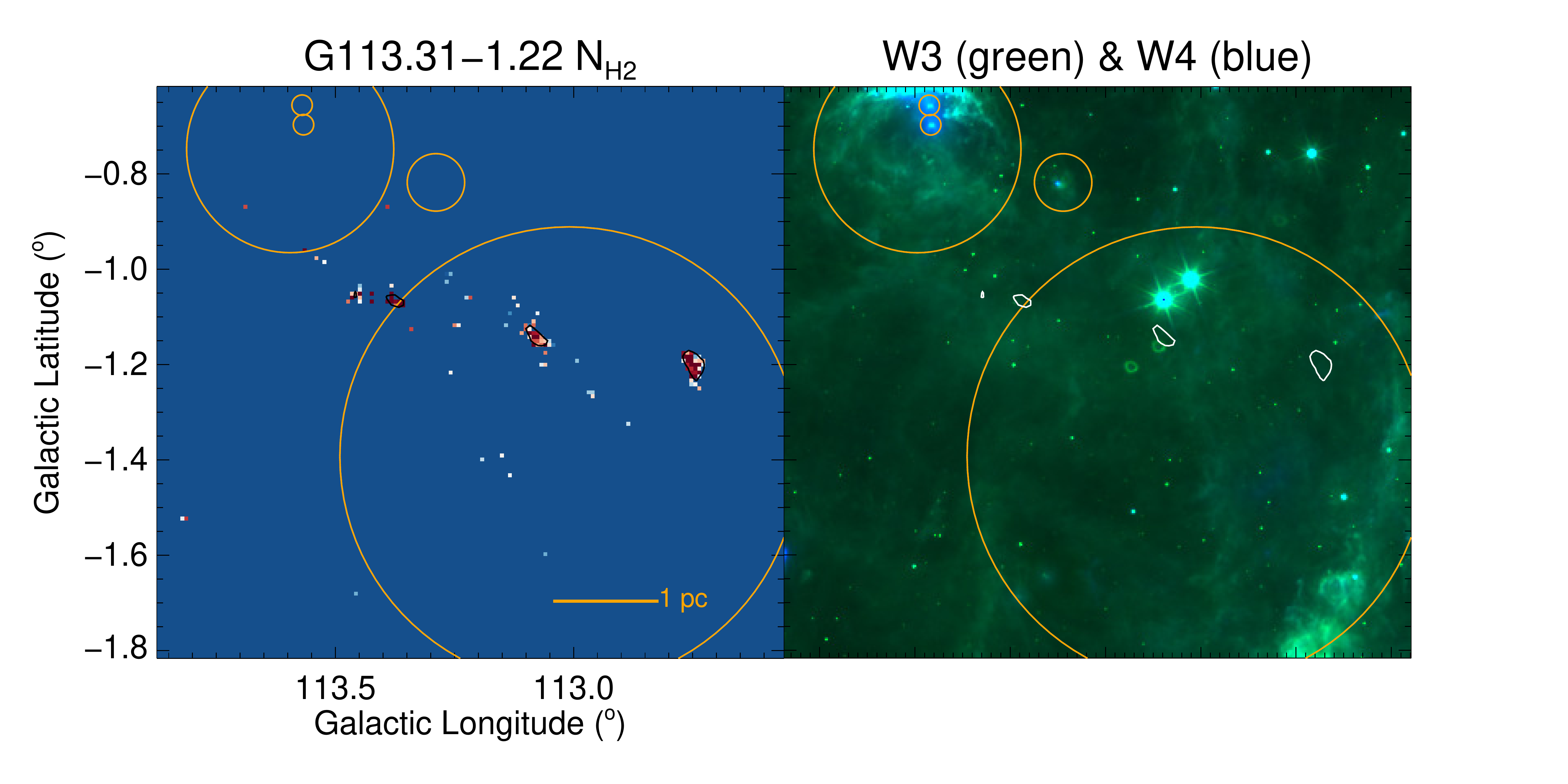}}
	\caption{Same as Figure \ref{fig18}, but for clouds G113.59+4.62 etc.}\label{fig27}
\end{figure*}

\begin{figure*}[tb!]
	\centering
	\subfigure[]{
	\label{fig28a}
		\includegraphics[trim=0cm 4cm 4cm 4cm, width = 0.6\linewidth , clip]{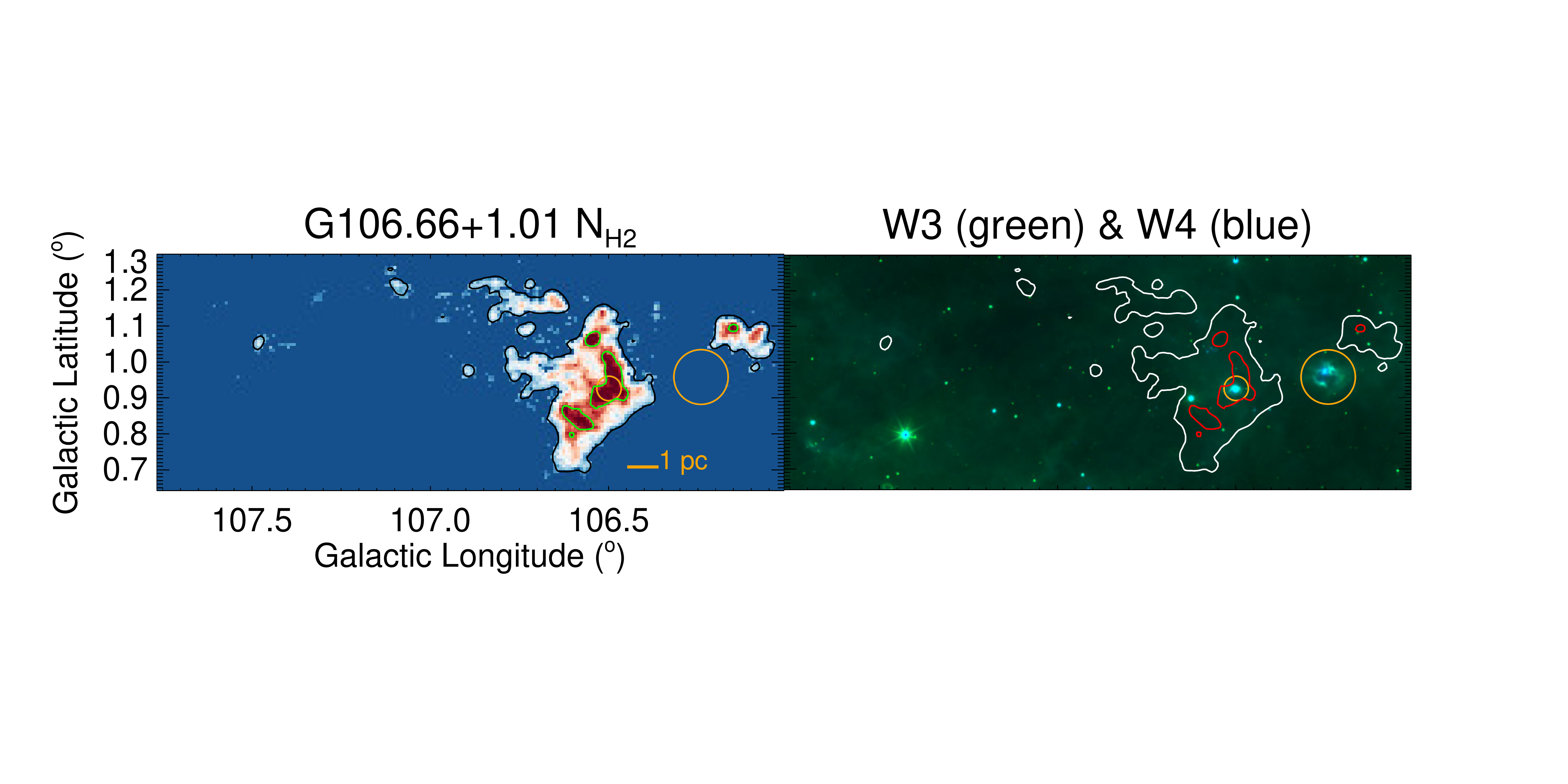}}
		\subfigure[]{
		\label{fig28c}
		\includegraphics[trim=0cm 0cm 2cm 0cm, width = 0.6\linewidth , clip]{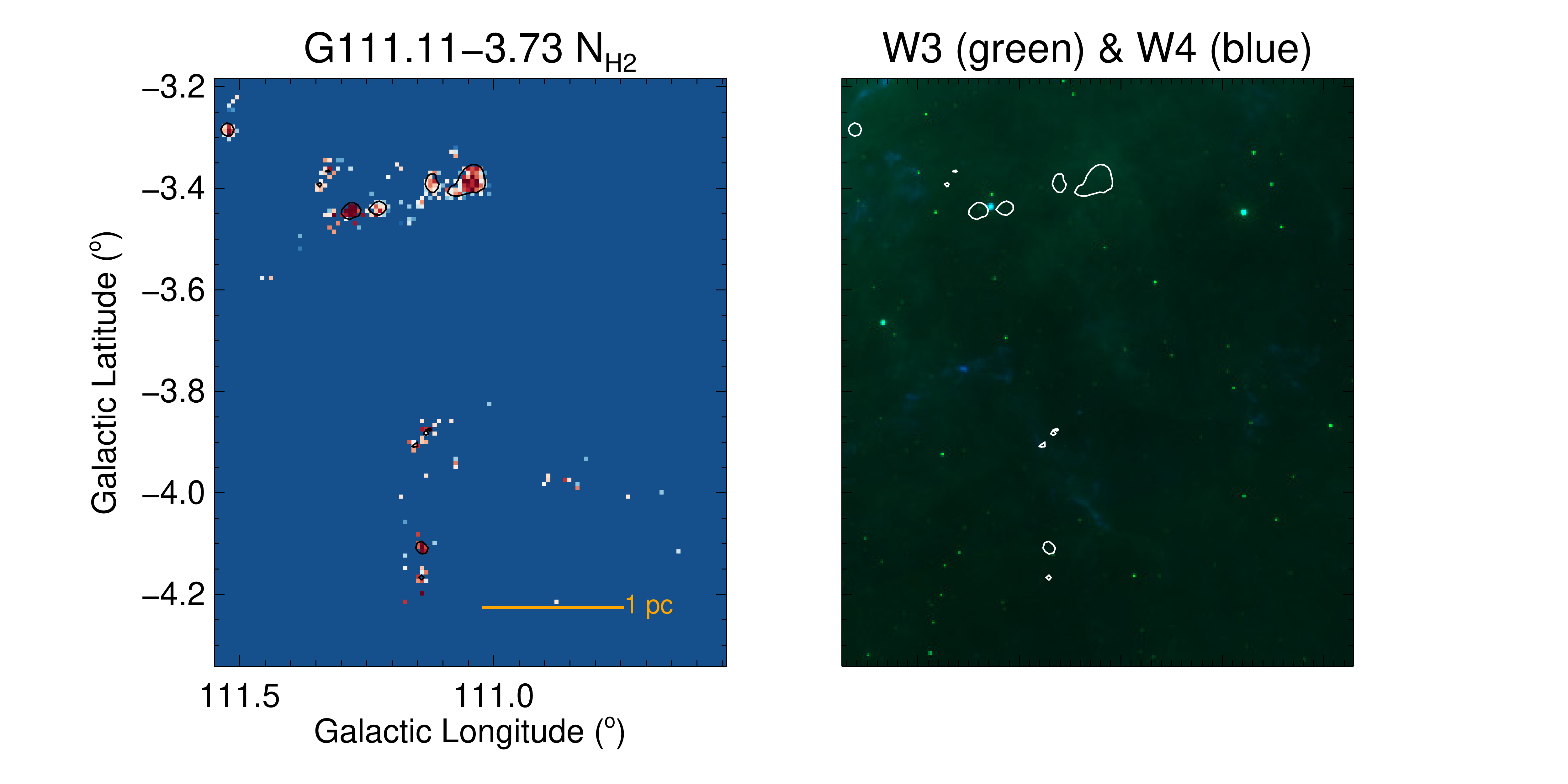}}
	\subfigure[]{
	\label{fig28d}
		\includegraphics[trim=0cm 0cm 3cm 0cm, width = 0.6\linewidth , clip]{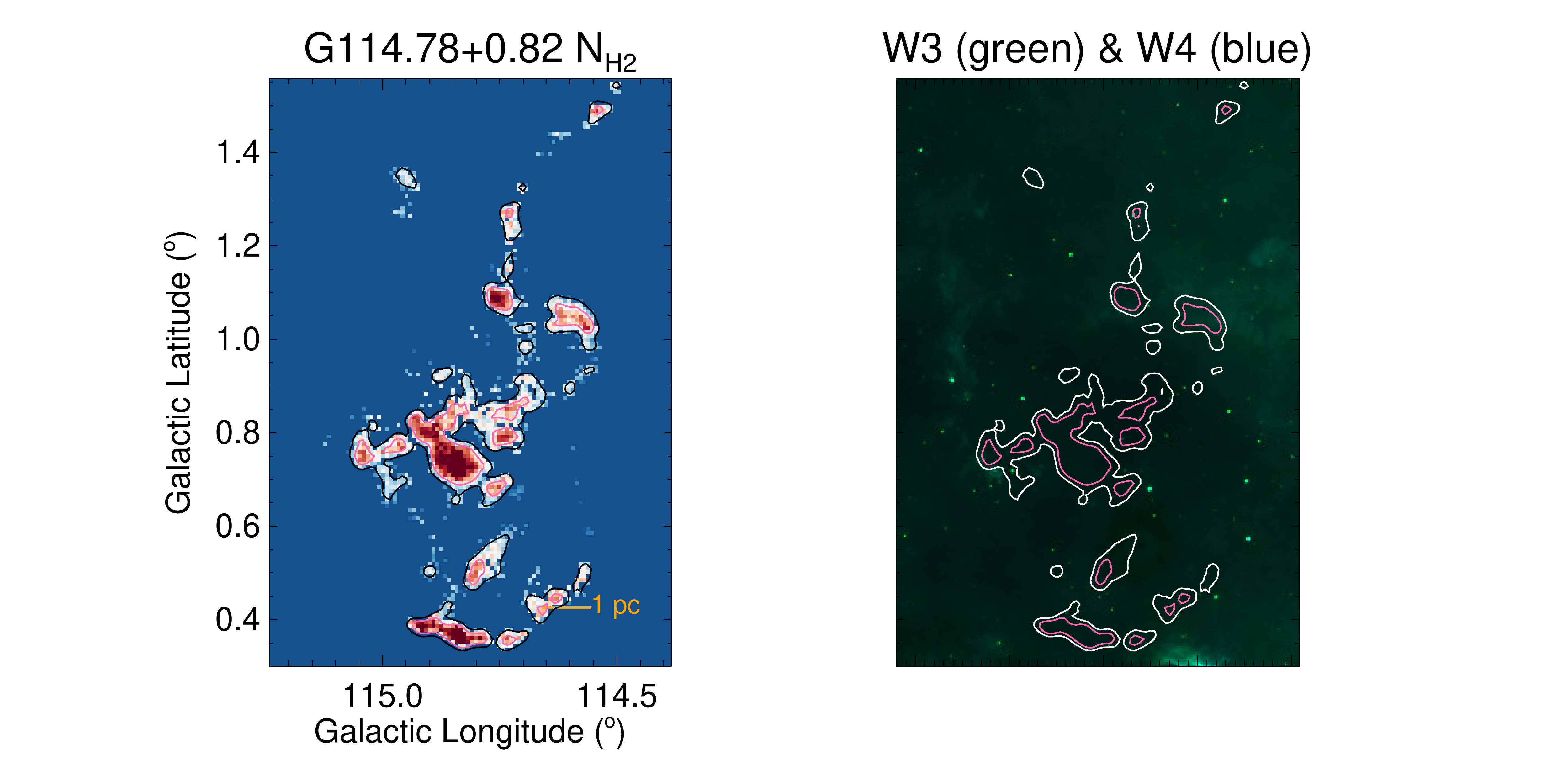}}
	\caption{Same as Figure \ref{fig18}, but for clouds G106.66+1.01 etc.}\label{fig28}
\end{figure*}

\begin{figure*}[htb!]
	\centering
	\subfigure[]{
	\label{fig29a}
		\includegraphics[trim=0cm 4cm 4cm 4cm, width = 0.6\linewidth , clip]{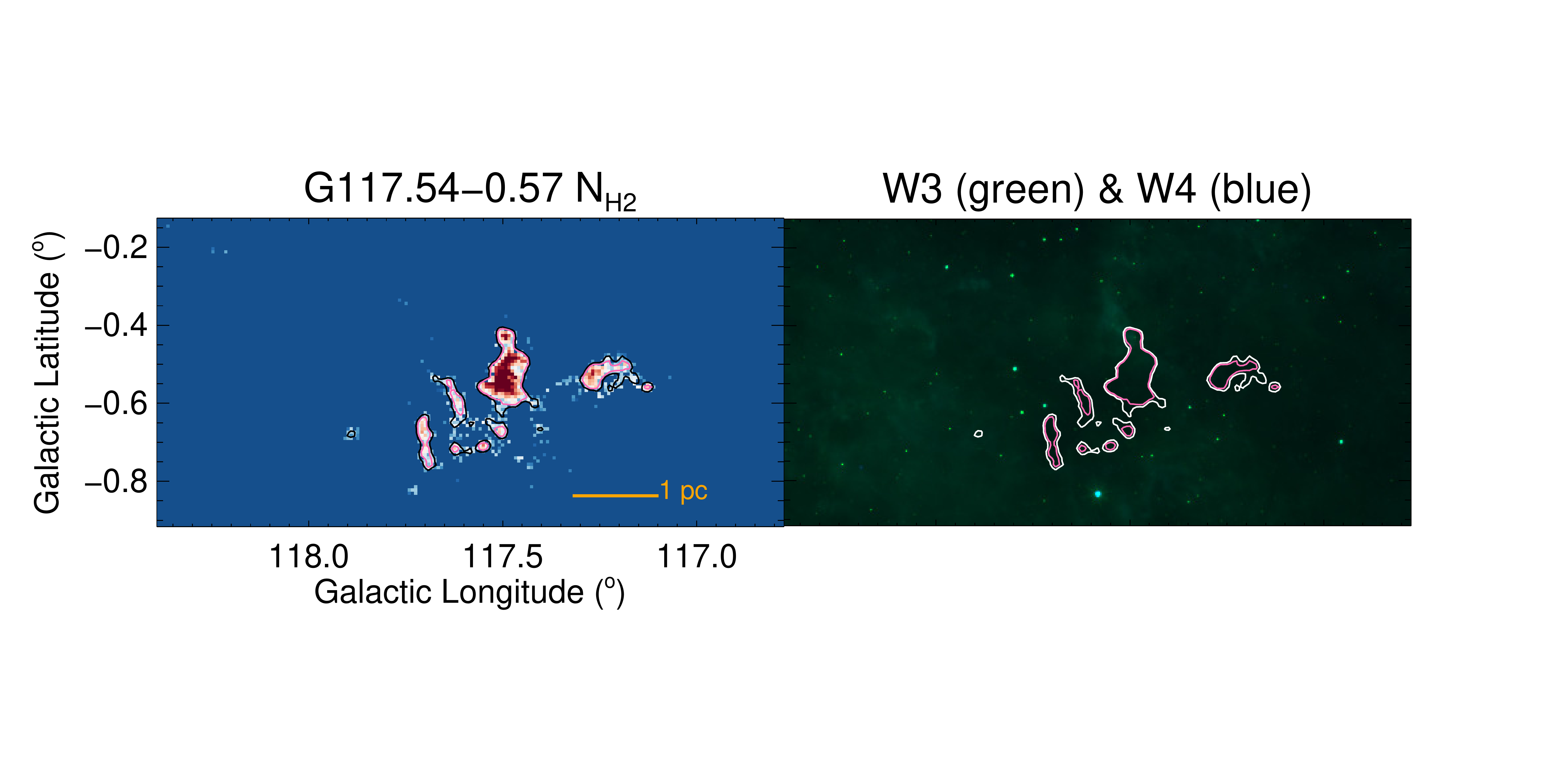}}
		\subfigure[]{
		\label{fig29b}
		\includegraphics[trim=0cm 1cm 4cm 1cm, width = 0.6\linewidth , clip]{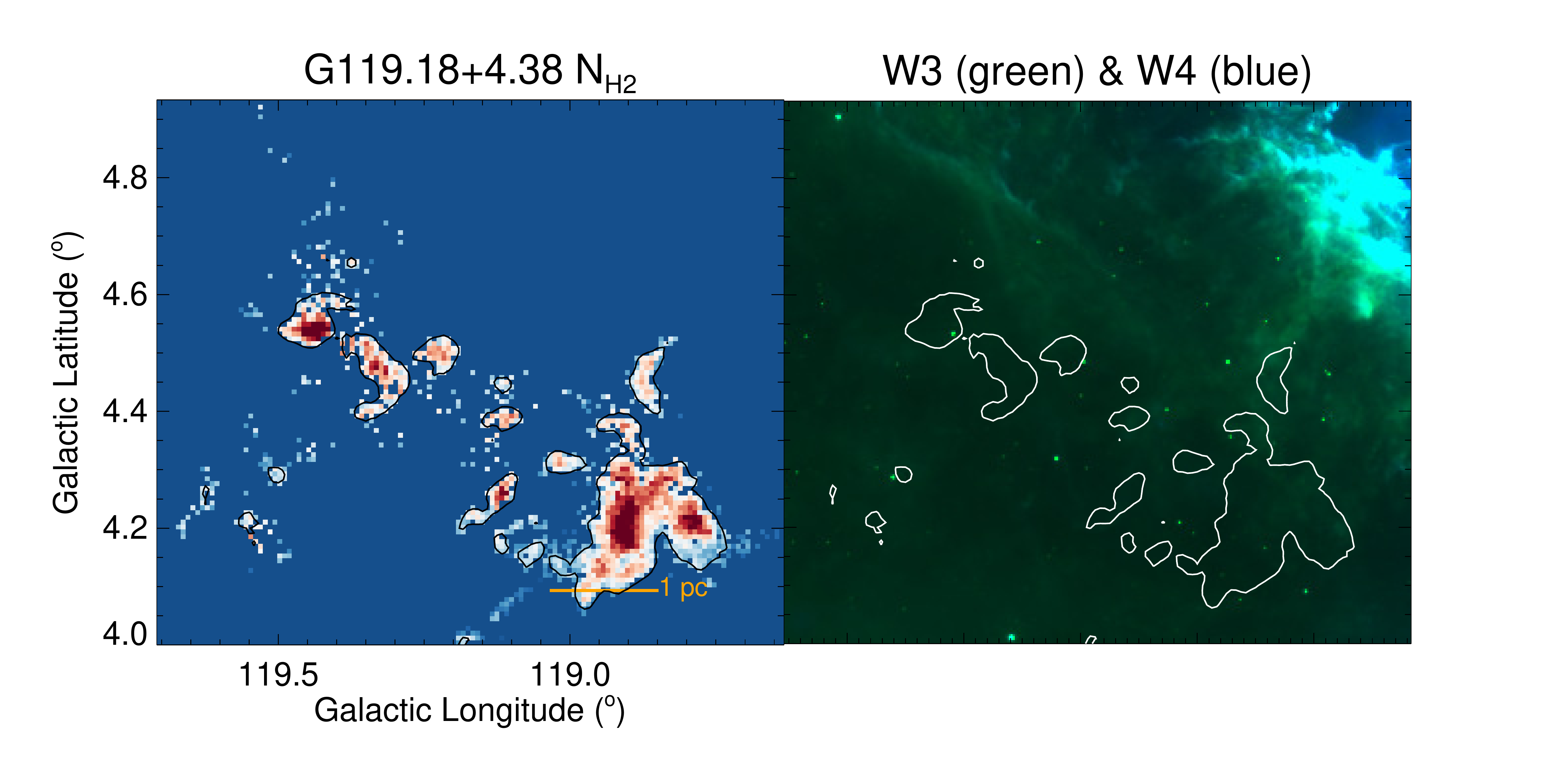}}
	\caption{Same as Figure \ref{fig18}, but for clouds G117.54-0.57 etc.}\label{fig29}
\end{figure*}

\begin{figure*}[tb!]
	\centering
	\subfigure[]{
	\label{fig30a}
		\includegraphics[trim=0cm 2cm 4cm 2.5cm, width = 0.6\linewidth , clip]{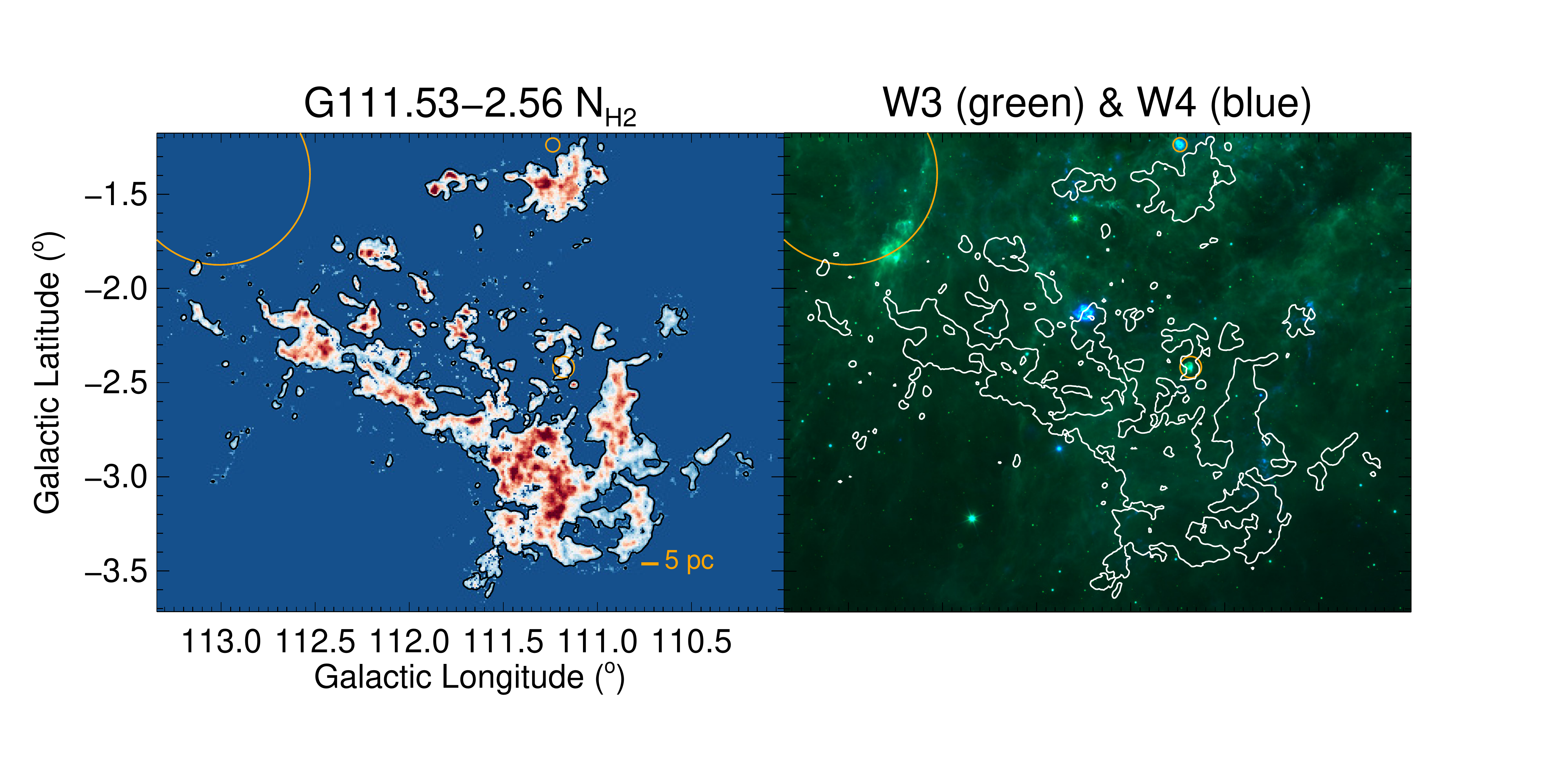}}
	\subfigure[]{
		\label{fig30b}
		\includegraphics[trim=0cm 1cm 4cm 2cm, width = 0.6\linewidth , clip]{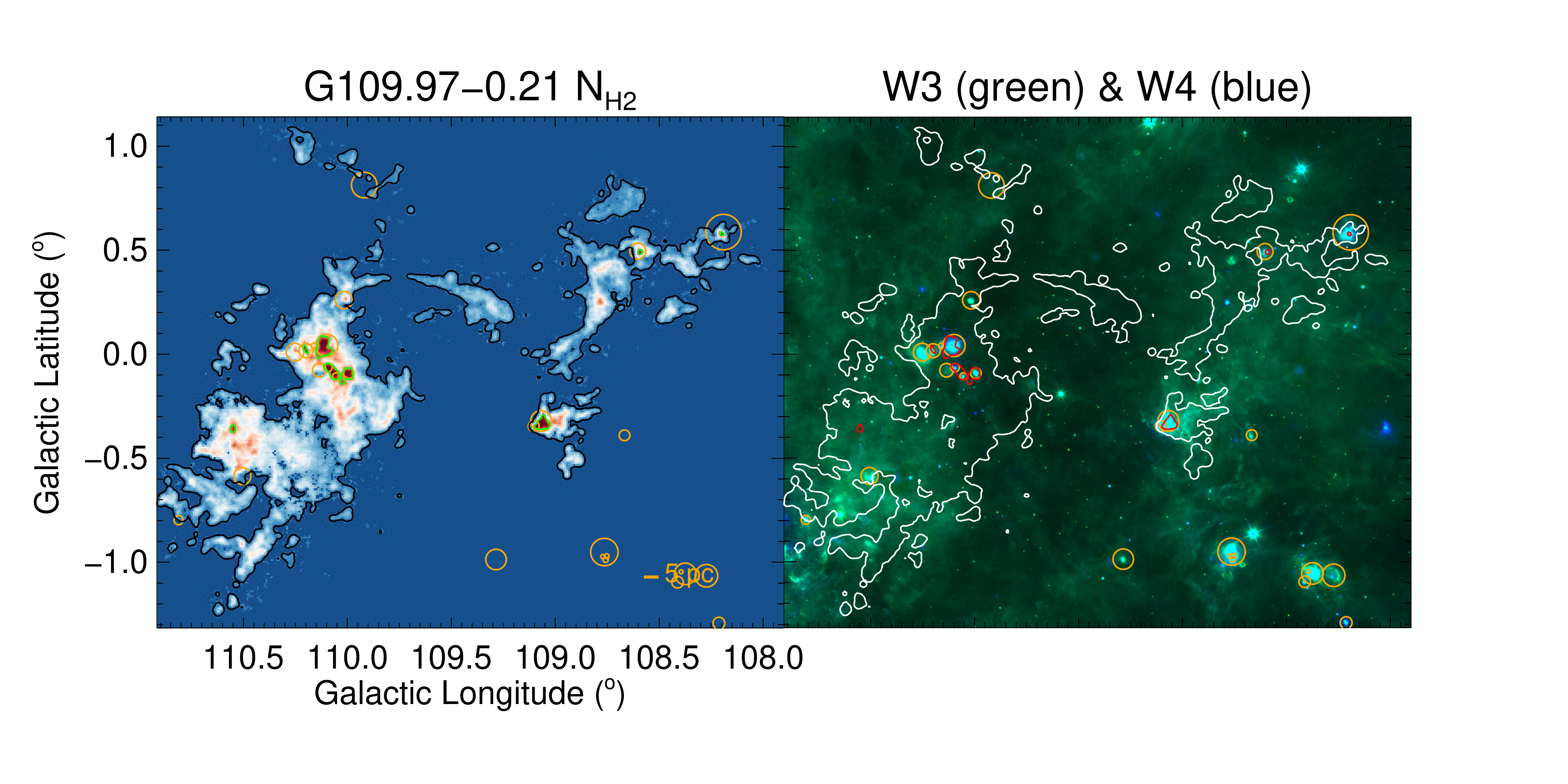}}
	\subfigure[]{
		\label{fig30c}
		\includegraphics[trim=0cm 2cm 2cm 2cm, width = 0.6\linewidth , clip]{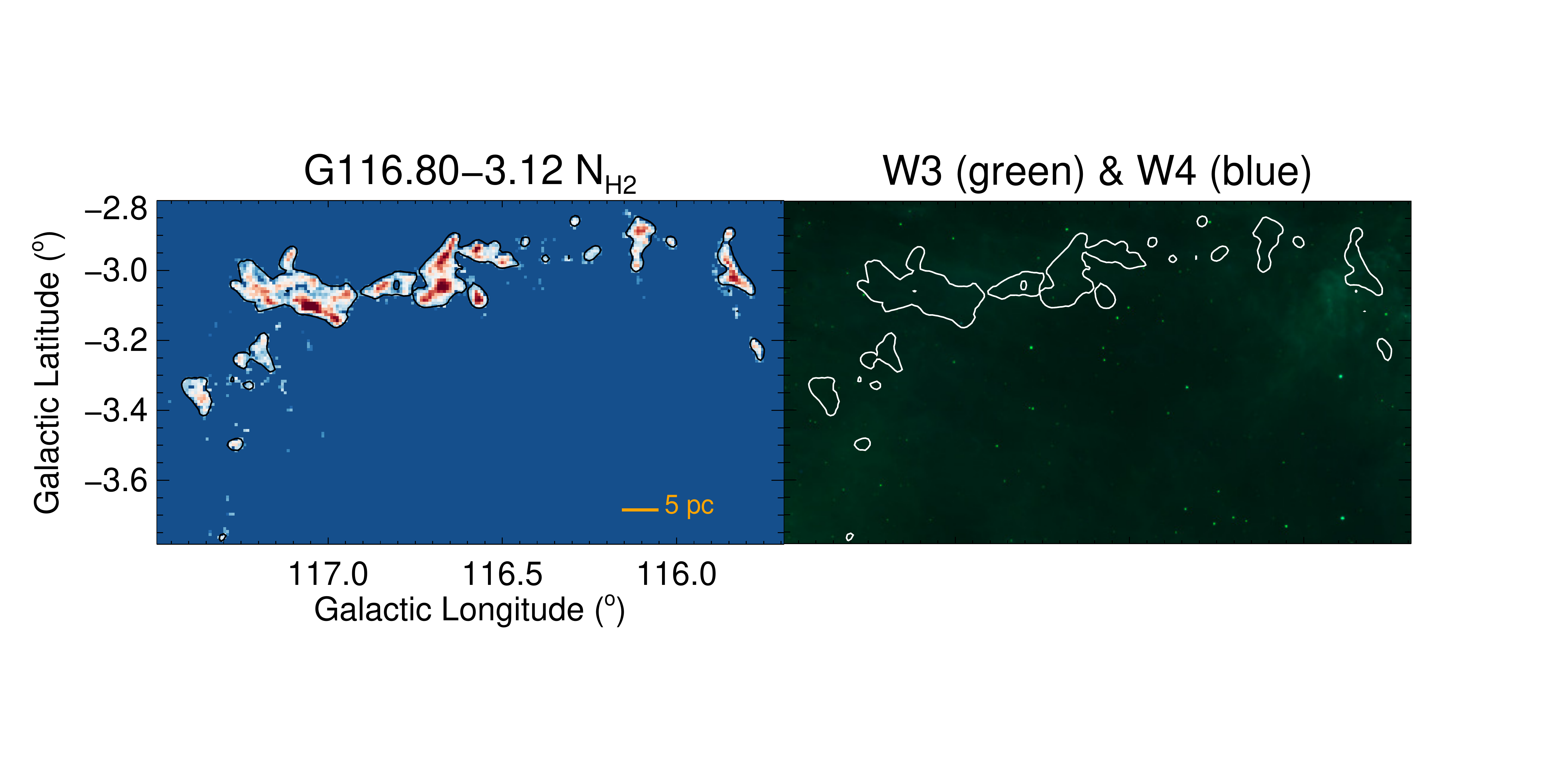}}
	\subfigure[]{
		\label{fig30d}
		\includegraphics[trim=0cm 4cm 4cm 4cm, width = 0.6\linewidth , clip]{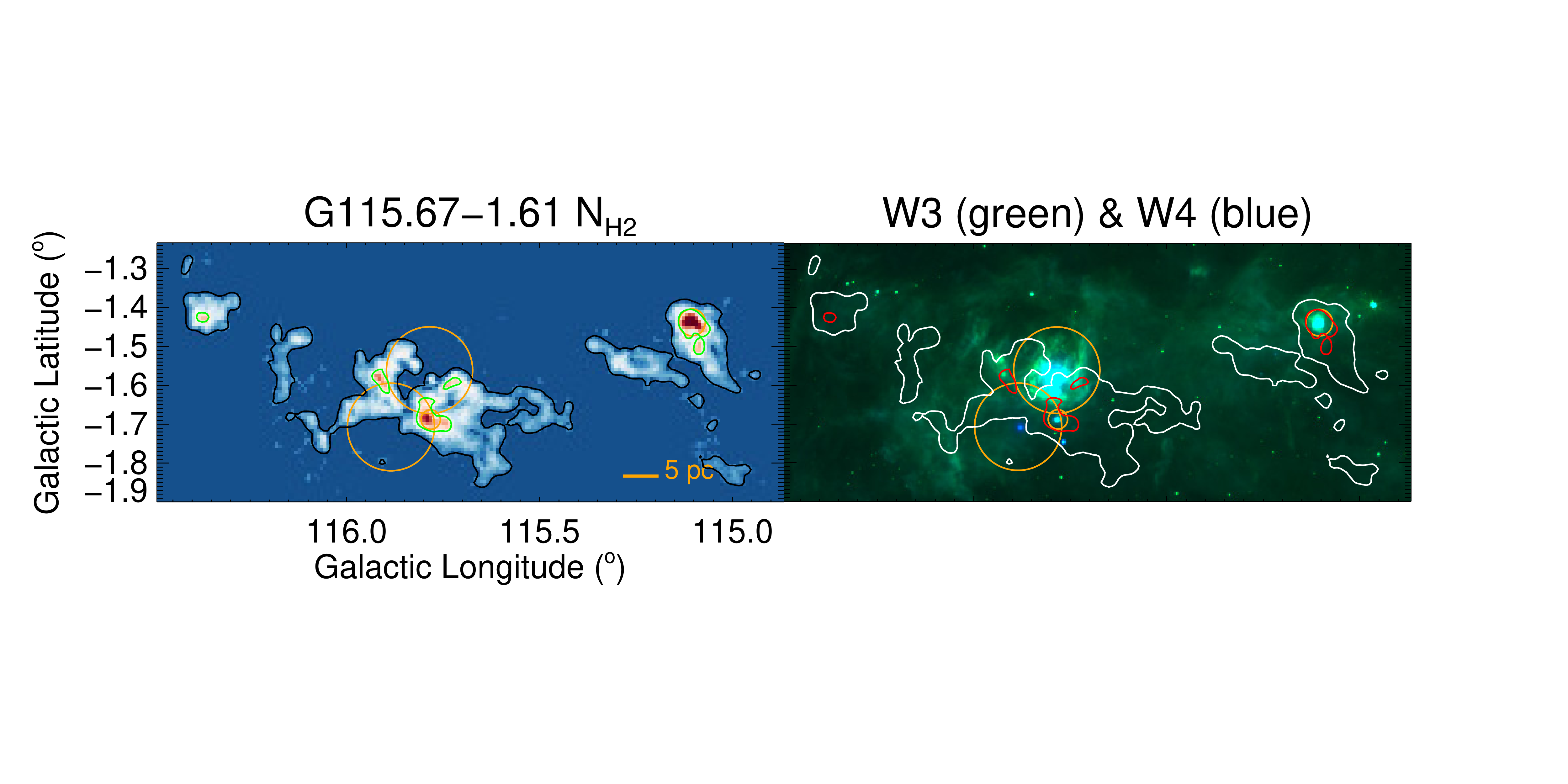}}
	\caption{Same as Figure \ref{fig18}, but for clouds G111.53-2.56 etc.}\label{fig30}
\end{figure*}

\begin{figure*}[tb!]
	\centering
	\subfigure[]{
		\label{fig31b}
		\includegraphics[trim=0cm 0cm 4cm 0cm, width = 0.6\linewidth , clip]{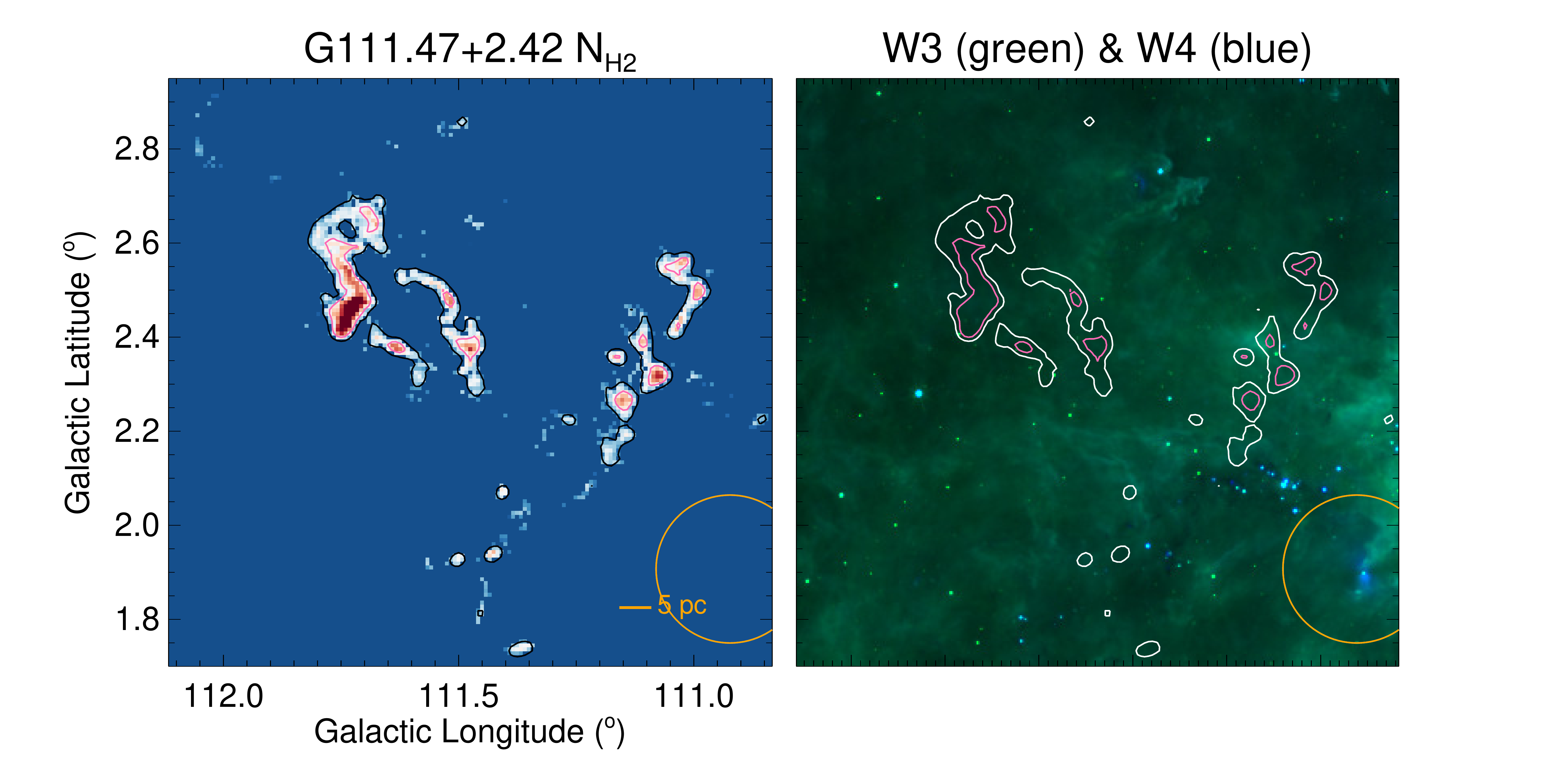}}
		\subfigure[]{
			\label{fig31c}
		\includegraphics[trim=0cm 0cm 2cm 0cm, width = 0.6\linewidth , clip]{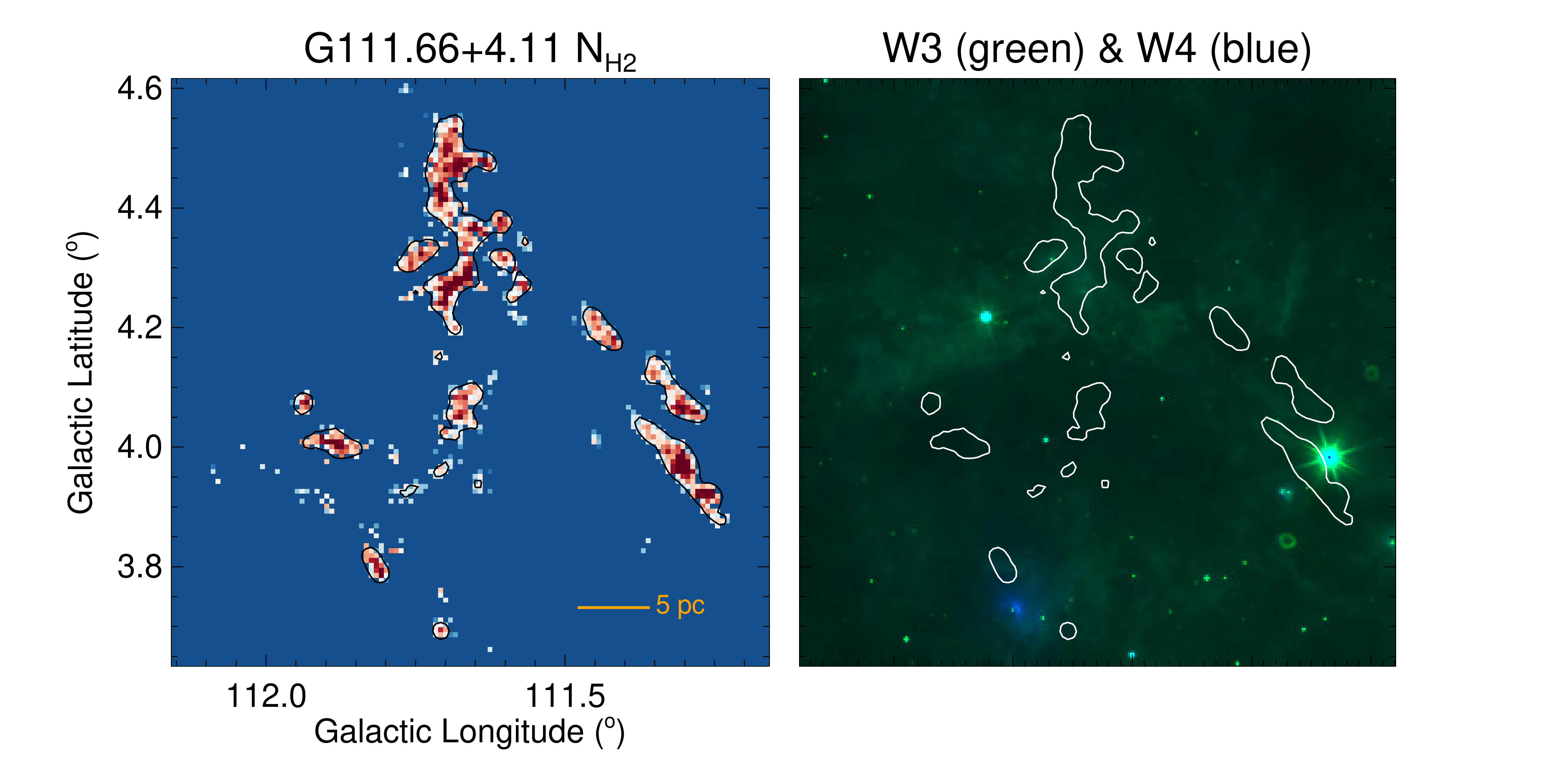}}
		\subfigure[]{
			\label{fig31d}
		\includegraphics[trim=0cm 0cm 3cm 0cm, width = 0.6\linewidth , clip]{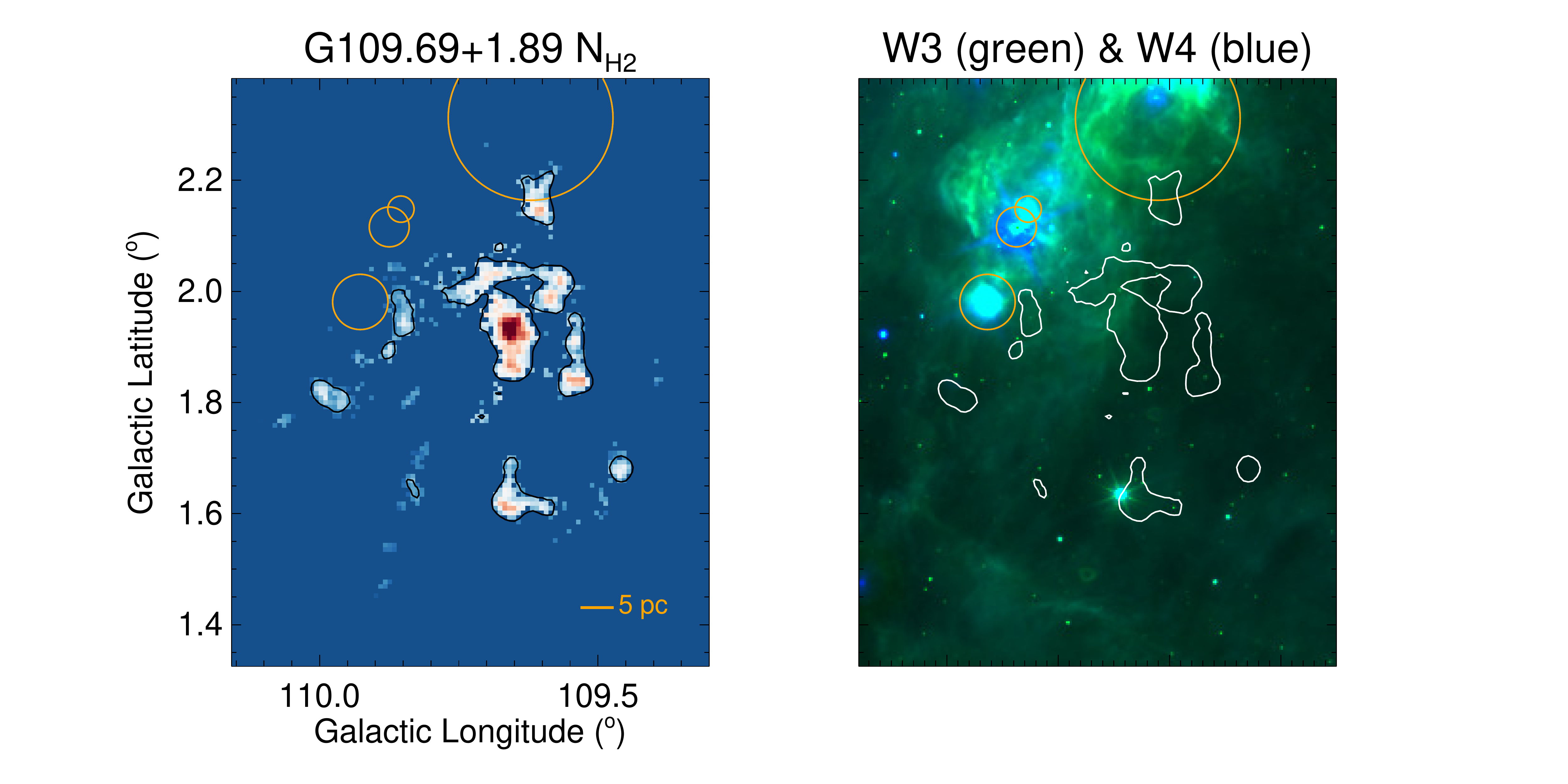}}
	\caption{Same as Figure \ref{fig18}, but for clouds G108.61-1.01 etc.}\label{fig31}
\end{figure*}

\appendix
Figures \ref{fig20}-\ref{fig22} give the demonstration of the cloud identification using the DENDROGRAM+SCIMES algorithms for different tracers and different spiral arms. Figures \ref{fig23}-\ref{fig31} present the column density maps and the WISE 3/4 band images of the selected molecular clouds.

\clearpage
\bibliographystyle{aasjournal}
\bibliography{references}

\end{document}